\documentclass[runningheads,envcountsame,envcountsect]{llncs}

\usepackage{silence}
\ExplSyntaxOn
\msg_redirect_name:nnn { unicode-math } { mathtools-overbracket } { none }
\msg_redirect_name:nnn { unicode-math } { mathtools-colon } { none }
\ExplSyntaxOff

\PassOptionsToPackage{dvipsnames}{xcolor}
\PassOptionsToPackage{hyphens}{url}
\PassOptionsToPackage{pdfversion=1.7}{hyperref}

\usepackage{amsmath}
\usepackage{mathtools}
\usepackage{amssymb}  \usepackage[operators,sets]{cryptocode}  \makeatletter

\renewcommand*{\theH@pclinenumber}{\arabic{H@pclinenumber}}

\def\we@pc@label#1{\zcsetup{currentcounter=pclinenumber}\ltx@label{#1}}
\AddToHook{begindocument/end}[we/cryptocode-label]{\let\@pc@original@label\we@pc@label}
\DeclareHookRule{begindocument/end}{we/cryptocode-label}{after}{amsmath}
\makeatother

    \usepackage{newcomputermodern}
  
\makeatletter
  \IfFontExistsTF{NewCMMath-Bold.otf}{ \def\we@mathfont{NewCMMath-Book.otf}
    \setmathfont{\we@mathfont}[BoldFont=NewCMMath-Bold.otf]
    \setmathfont{NewCMMath-Bold.otf}[version=bold]
    \setmathfont{\we@mathfont}[range={\mathscr,\mathbfscr},StylisticSet=1,BoldFont=NewCMMath-Bold.otf]
  }{}
  \makeatother

  \usepackage[protrusion=true,expansion=false]{microtype}

\usepackage{booktabs}
\usepackage{graphicx}
\usepackage[section]{placeins}  \usepackage{flafter}  \makeatletter
\AtBeginDocument{\expandafter\renewcommand\expandafter\subsection\expandafter{\expandafter\FloatBarrier\subsection}}
\let\we@FloatBarrier\FloatBarrier
\renewcommand*{\FloatBarrier}{\par\if@nobreak\else\addpenalty\@secpenalty\fi\we@FloatBarrier}
\newcount\we@barrierskips
\newcommand*{\weSkipFloatBarriers}[1]{\ifx\FloatBarrier\we@SkipFloatBarrier\else\global\let\we@HeldFloatBarrier\FloatBarrier\fi
  \global\we@barrierskips=#1\relax
  \ifnum\we@barrierskips>\z@ \global\let\FloatBarrier\we@SkipFloatBarrier
  \else \global\let\FloatBarrier\we@HeldFloatBarrier \fi}
\newcommand*{\we@SkipFloatBarrier}{\global\advance\we@barrierskips\m@ne
  \ifnum\we@barrierskips>\z@\else\global\let\FloatBarrier\we@HeldFloatBarrier\fi}
\makeatother

\usepackage{enumitem}
\usepackage{siunitx}
  \usepackage{xspace}

\usepackage{csquotes}

  \PassOptionsToPackage{style=alphabetic,giveninits,maxalphanames=6,maxnames=6}{biblatex}

  \PassOptionsToPackage{backref}{biblatex}

\usepackage[backend=biber,mincrossrefs=999,arxiv=pdf]{biblatex}  \DefineBibliographyStrings{english}{backrefpage  = {cited on p.},backrefpages = {cited on pp.},}

  \appto{\bibsetup}{\raggedright}

\DeclareFieldFormat{eprint:fullversion}{Full version\addcolon\space\href{https://#1}{\nolinkurl{#1}}}

\ExplSyntaxOn
\tl_new:N \l__we_venue_tl
\cs_new_protected:Npn \we_short_venue_year:
  {
    \iffieldundef { booktitle } { }
      {
        \savefield { booktitle } \l__we_venue_tl
        \regex_replace_all:nnN { ,\s*Part(?:\~|\s)+[IVXL]+\b } { } \l__we_venue_tl
        \regex_replace_all:nnNT { (?:\~|\ |')(?:19|20)?(\d\d)\b }
          { \c{nobreakspace}'\1 } \l__we_venue_tl
          { \clearfield { year } }
        \restorefield { booktitle } \l__we_venue_tl
      }
  }
\NewDocumentCommand \WeShortVenueYear { } { \we_short_venue_year: }
\ExplSyntaxOff
\renewbibmacro*{in:}{}
\AtEveryBibitem{\clearname{editor}\clearname{editora}\clearname{editorb}\clearname{editorc}\clearlist{publisher}\clearlist{location}\clearfield{series}\clearfield{volume}\clearfield{number}\clearfield{pages}\clearfield{month}\clearfield{day}\clearfield{isbn}\clearfield{issn}\iffieldundef{eprint}{\iffieldundef{url}{}{\clearfield{doi}}}{\clearfield{doi}\clearfield{url}}\WeShortVenueYear}

  \makeatletter
  \renewcommand{\contentsname}{Contents}
  \renewcommand{\tableofcontents}{\section*{\contentsname}\@starttoc{toc}}
  \def\authcount#1{}  \renewcommand*{\l@section}[2]{\ifnum\c@tocdepth>\z@
      \addpenalty\@secpenalty
      \addvspace{0.6em \@plus\p@}\begingroup
        \parindent\z@ \rightskip\@pnumwidth \parfillskip -\@pnumwidth
        \leavevmode\bfseries
        \@tempdima 1.5em  \advance\leftskip\@tempdima \hskip -\leftskip
        #1\nobreak\hfil\nobreak\hb@xt@\@pnumwidth{\hss #2}\par
      \endgroup
    \fi}
  \renewcommand*{\l@subsection}{\@dottedtocline{2}{1.5em}{2.3em}}
  \makeatother

\usepackage{etoolbox,xpatch}
\AtBeginDocument{\thispagestyle{empty}}
\makeatletter
\let\paper@endproof\endproof
\renewcommand{\endproof}{\qed\paper@endproof}
\AddToHook{begindocument/end}{\let\paper@addcontentsline\addcontentsline
\renewcommand{\addcontentsline}[3]{\ifstrequal{#2}{title}{}{\ifstrequal{#2}{author}{}{\paper@addcontentsline{#1}{#2}{#3}}}}}
\def\paper@TestPunct#1{\ifx\paper@TestPunct#1\paper@TestPunct\else\paper@Puncttrue\fi}
\def\paper@TestDot#1.\paper@TestEnd#2\paper@TestStop{\paper@TestPunct{#2}}
\def\paper@TestQuestM#1?\paper@TestEnd#2\paper@TestStop{\paper@TestPunct{#2}}
\def\paper@TestExclaM#1!\paper@TestEnd#2\paper@TestStop{\paper@TestPunct{#2}}
\def\paper@TestColonM#1:\paper@TestEnd#2\paper@TestStop{\paper@TestPunct{#2}}
\newif\ifpaper@Punct
\def\paper@MaybeAddPunct#1{\paper@Punctfalse\paper@TestDot#1\paper@TestEnd.\paper@TestEnd\paper@TestStop\paper@TestQuestM#1\paper@TestEnd?\paper@TestEnd\paper@TestStop\paper@TestExclaM#1\paper@TestEnd!\paper@TestEnd\paper@TestStop\paper@TestColonM#1\paper@TestEnd:\paper@TestEnd\paper@TestStop#1\ifpaper@Punct \else .\fi}

\def\paper@ignorespacesandimplicitepars{\begingroup
  \catcode13=10
  \@ifnextchar\relax
    {\endgroup}{\endgroup}}

\newcommand{\parheadnoperiod}[1]{\smallskip \noindent {\bfseries\boldmath\ignorespaces {#1}}\hskip 0.9em plus 0.3em minus 0.3em \paper@ignorespacesandimplicitepars}
\newcommand{\parhead}[1]{\parheadnoperiod{\paper@MaybeAddPunct{#1}}\paper@ignorespacesandimplicitepars}

\makeatother
\usepackage{hyperref}
\usepackage{hyperxmp}
\makeatletter
\renewcommand*{\hyxmp@xmp@basic@schema}{\hyxmp@add@simple{xmp:CreateDate}{\hyxmp@today@xmp}}
\makeatother

\hypersetup{colorlinks=true, linkcolor=CadetBlue, citecolor=Periwinkle,
              urlcolor=Periwinkle}  
\hypersetup{
  breaklinks = true,
}

  \hypersetup{
    pdfauthor={David Tse, Nikhil Vanjani, Psi Vesely},
    pdfsubject={Public-key cryptography},
    pdfkeywords={witness encryption, generic group model, Module-SIS, Groth16, pairings, formal verification},
    pdflang={en-US}, keeppdfinfo,
    pdfcopyright={This work is licensed under CC BY 4.0.},
    pdflicenseurl={https://creativecommons.org/licenses/by/4.0/},
    pdfmetalang={en-US}, pdfcreator={}
  }

  \hypersetup{pdfpagemode=UseOutlines,bookmarksnumbered=true,bookmarksopen=true,
              bookmarksopenlevel=2,pdfstartview=FitH}
  \AddToHook{cmd/maketitle/before}{\pdfbookmark[1]{Abstract}{abstract}}

\usepackage{bookmark}
\AddToHook{cmd/appendix/after}{}
\usepackage{zref-clever}
\zcsetup{cap, nameinlink}
\newcommand{\cref}[1]{\zcref{#1}}
\newcommand{\Cref}[1]{\zcref[S]{#1}}
\zcsetup{countertype = {pclinenumber = line, pcrlinenumber = line}}

\ExplSyntaxOn
\clist_map_inline:nn
  { lemma, corollary, proposition, definition, remark, example, claim,
    conjecture, case, note, problem, property, question, exercise,
    solution, assumption, construction, fact, observation, notation }
  { \AddToHook { env/#1/begin } { \zcsetup { countertype = { theorem = #1 } } } }
\ExplSyntaxOff

    \usepackage[letterpaper,margin=1in]{geometry}

\newcommand{\newc}[2]{\newcommand{#1}{\ensuremath{#2}\xspace}}

\newcommand{\schemename}{Praxis}
\newcommand{\scheme}{\textsf{\schemename}\xspace}
\newc{\defeq}{:=}

\newc{\poly}{\mathsf{poly}}  \newc{\bits}{\{0,1\}}

\newc{\Ogg}{\mathcal{O}_{\mathsf{GG}}}
\newc{\gadd}{\mathsf{add}}
\newc{\gsmul}{\mathsf{smul}}
\newc{\gpair}{\mathsf{pair}}

\newc{\Hkey}{H_{\mathsf{key}}}
\newc{\Hseed}{H_{\mathsf{seed}}}

\providecommand{\hdr}{\tilde{m}}  \newc{\seed}{\mathsf{seed}}
\newc{\res}{\mathrm{res}}          \newc{\wt}{\operatorname{wt}}      

\newc{\Ncmp}{N_{\mathrm{cmp}}}
\newc{\sect}{\mathsf{s}}           
\newcommand{\usablestack}{usable stack}
\newcommand{\honeststack}{honest stack}
\newcommand{\testmask}{test mask}
\newcommand{\tiemask}{tie mask}
\newcommand{\symmetryterm}{symmetry term}
\newcommand{\keycopy}{key copy}    

\newcommand{\keyscalar}{key scalar}
\newcommand{\keyelement}{key element}
\newcommand{\maskedmessage}{masked message}
\newcommand{\signclass}{sign class}
\newcommand{\searchvector}{search vector}
\newcommand{\searchbound}{search bound}
\newcommand{\centeredsampler}{centered sampler}
\newcommand{\plainsector}{plain sector}
\newcommand{\alphasector}{alpha sector}
\newcommand{\qapcriterion}{QAP criterion}
\newcommand{\shortmultiple}{short multiple}
\newcommand{\liftingstep}{lifting step}
\newcommand{\privatetestpoint}{private test point}
\newcommand{\recoverypoint}{recovery point}
\newcommand{\probevector}{probe vector}
\newcommand{\stoppedcoupling}{stopped coupling}
\newcommand{\privatemonitor}{private monitor}
\newcommand{\markstack}{mark stack}
\newcommand{\marksector}{mark sector}
\newcommand{\cancellationbases}{cancellation bases}
\newcommand{\tiebases}{tie bases}
\newcommand{\crossbases}{cross bases}

\providecommand{\gp}{\operatorname{gp}}

\providecommand{\Setup}{\mathsf{Setup}}
\providecommand{\Enc}{\mathsf{Enc}}
\providecommand{\Dec}{\mathsf{Dec}}
\providecommand{\Comb}{\mathsf{Comb}}
\providecommand{\Forms}{\mathsf{Forms}}
\providecommand{\pp}{\mathsf{pp}}
\providecommand{\name}{\mathsf{name}}
\providecommand{\lseed}{\ell_{\mathsf{seed}}}
\providecommand{\nc}{n_{\mathsf{c}}}
\providecommand{\leg}{\mathrm{leg}}
\providecommand{\kleg}{k_{\mathrm{leg}}}
\providecommand{\ckey}{c_{\mathrm{key}}}
\providecommand{\xikey}{\xi_{\mathrm{key}}}
\providecommand{\hkey}{h_{\mathrm{key}}}
\providecommand{\hoth}{h_{\mathrm{oth}}}
\providecommand{\sF}{\mathsf{F}}   \providecommand{\sH}{\mathsf{H}}   \providecommand{\sS}{\mathsf{S}}   \providecommand{\sY}{\mathsf{Y}}   \providecommand{\sB}{\mathsf{B}}   \providecommand{\sP}{\mathsf{P}}   \providecommand{\sK}{\mathsf{K}}   \providecommand{\Pquo}{\mathsf{P}^{\mathrm{quo}}}
\providecommand{\Prem}{\mathsf{P}^{\mathrm{rem}}}
\providecommand{\Bquo}{\mathsf{B}^{\mathrm{quo}}}
\providecommand{\Brem}{\mathsf{B}^{\mathrm{rem}}}
\providecommand{\cI}{\mathcal{I}}  
\providecommand{\rank}{\operatorname{rank}}


\providecommand{\val}{\operatorname{val}}
\providecommand{\Nfree}{N_{\mathsf{free}}}
\providecommand{\epsf}{\varepsilon_f} 

\title{\scheme: Practical Witness Encryption for NP}
\titlerunning{\scheme: Practical Witness Encryption for NP}
\hypersetup{pdftitle={\schemename: Practical Witness Encryption for NP}}  
  \author{David Tse\inst{1,2} \and Nikhil Vanjani\inst{1} \and Psi Vesely\inst{1}}
  \authorrunning{D.~Tse, N.~Vanjani, and P.~Vesely}
  \institute{Byzantine Research \and Stanford University}

\usepackage{newunicodechar}
\newunicodechar{^^^^00a0}{~}
\newunicodechar{^^^^202f}{\,}
\makeatletter
\patchcmd{\blx@bblfile}{\blx@ifsigned{\jobname}{bbl}}{\IfFileExists{\jobname.bbl}}{}{\errmessage{Bibliography loader mismatch}}
\makeatother
\begin{document}

\maketitle
  
\begin{abstract}
Witness encryption encrypts a message to an NP statement so that whoever holds a witness can decrypt. Over a decade after its introduction, known constructions for general NP statements are far from practical or rest on heuristic or broken assumptions. We propose \textsf{\schemename}, a practical scheme for NP, and prove it extractable in the bilinear generic group and random oracle models under Module-SIS. Our instantiation encrypts to a Groth16 verifier on BN254, set up honestly, which serves every NP relation at the same cost: at $100$-bit security, its ciphertext is $7.88$ GiB, encryption takes $2^{34.2}$ base-field multiplications, and decryption with a Groth16 proof $2^{33.0}$ on average. We give Lean 4 proofs of correctness and extraction at this instantiation’s parameters.

\textsf{\schemename} starts from the rank-gap template of the recent feasibility result [Hair and Sahai, arXiv 2026]. Where they raise the gap to keep decryption polynomial, we fix it, so decryption ends with a search that is exponential but concretely fast: $O(2^{\lambda/4} N^{1/4})$ expected operations for a statement in $N$ Boolean unknowns at $\lambda$-bit security. A quadratic arithmetic program makes our hidden matrix low rank, so where their ciphertext encodes a basis of a matrix space, ours holds factors of width two, which a pairing multiplies. Weights and identities from an extension field tie copies of this matrix, so an adversary telling encryptions apart yields a witness or a Module-SIS solution. For the same Groth16 statement and $100$-bit security level, \textsf{\schemename} reduces the ciphertext size and the cost of encryption of either of their constructions by over ten orders of magnitude, and that of decryption by over six.
\end{abstract}

  \begingroup
    \linespread{0.8}\footnotesize
    \tableofcontents
  \endgroup

\section{Introduction}\label{sec:introduction}

Witness encryption~\cite{STOC:GGSW13} encrypts a message to a statement instead of a key: anyone holding a witness of the statement can decrypt, and if the statement is false the message stays hidden. The statement can be any NP statement, for instance that a program accepts some input, and any input it accepts is a witness. A ciphertext carries its own release condition, with no key to distribute and no party trusted to release the message. Its applications include public-key, identity-based, and attribute-based encryption~\cite{STOC:GGSW13}, time-lock encryption that opens once a blockchain reaches a given length~\cite{DCC:LJKW18}, one-time programs~\cite{TCC:GoyGoy17}, and, on Bitcoin, bridges without a custodian~\cite{Hio22}, proof verification~\cite{GKSST26}, and covenants~\cite{ACEKKRSTTT26}. Garg, Gentry, Sahai, and Waters introduced witness encryption in 2013, with a puzzle as the statement, so that whoever solves it can read the message even after the encryptor is gone. Later that year, Goldwasser et al.~\cite{C:GKPVZ13} strengthened it to \emph{extractable} witness encryption, in which an adversary that tells encryptions of two messages apart yields a witness.

Since then, every construction has had at most two of three properties: generality, for all of NP; proved security in well-studied models and under well-studied assumptions; and practical costs. The general constructions with such proofs are feasibility results, far from practical. Witness encryption follows from indistinguishability obfuscation~\cite{FOCS:GGHRSW13}, which exists under well-founded assumptions~\cite{STOC:JaiLinSah21}, and, in the generic group model (GGM), from a hypothesis on linear codes~\cite{C:BIOW20} and, by Hair and Sahai, from a \emph{rank gap} that they prove~\cite{HS26}. In their construction, a witness passes a set of linear tests with a matrix of rank one, while on a false statement every nonzero matrix that passes has rank at least $d$, a gap of one versus $d$. The practical schemes with proofs handle restricted relations, such as signatures or openings of commitments~\cite{FC:DHMW23,AC:FleHalSim24}. The other constructions for NP lack the second property, as they rest on heuristic arguments~\cite{ITCS:BIJMSZ20} or on purpose-built assumptions~\cite{C:GenLewWat14,C:Tsabary22}, many of them since attacked~\cite{EC:CHLRS15,C:DJMMV25}, or have been broken~\cite{ICISC:CWLHTM25} (\cref{sec:related-work}).

\scheme has all three properties, and is the first witness encryption scheme that does. It encrypts to any NP statement, and we prove it extractable in the bilinear generic group and random oracle (RO) models under Module-SIS (MSIS). It is also practical. At the $100$-bit security level, with one fixed Groth16~\cite{EC:Groth16} key whose verifier serves every NP relation, its ciphertext is $7.88$\,GiB, and encryption takes seconds on a laptop, as does decryption given a Groth16 proof. A witness leaves the decryptor one secret integer to find, by a search exponential in the security parameter yet concretely short; anyone without a witness must guess four such integers at once, an exponentially longer search, unless they solve MSIS. \scheme starts from Hair and Sahai's construction. For the same statement and security level, their basic construction~\cite[Theorem~5.1]{HS26}, a variant of a construction proved extractable~\cite{Cam26}, needs over $1.1 \times 10^{11}$\,GiB to encrypt a $100$-bit key (\cref{tab:sizes}).

\begin{table}[!t]
\caption{\scheme and Hair and Sahai's basic construction~\cite[Theorem~5.1]{HS26}, a variant of a construction proved extractable~\cite{Cam26}, both on BN254, for the same statement and a $100$-bit key at $100$-bit security, with growth at $\lambda$-bit security. No extraction proof is known for their main construction~\cite[Theorem~6.1]{HS26}, and our lower bounds on its costs exceed the basic construction's, at over ten, ten, and seven orders of magnitude above ours (\cref{sec:results}). Their theorems give polynomial costs at quasi-polynomial security; at $\lambda$-bit security, their parameters grow with $\lambda$ (\cref{sec:results}). We set their scheme up beyond what their theorems allow, and prove in \cref{app:hair-sahai} that it still meets the level: at the least threshold $R$ that their bound certifies ($R = 4$ at $\lambda = 100$, where their theorem takes $18$), and in a group whose order has $\Theta(\lambda)$ bits (BN254's, whose elements take $32$ bytes where their theorems need over $9.7$\,TiB). We also give them our algorithms for encryption and decryption, faster than those of their paper, at our costs per operation: our search takes over eighteen orders of magnitude fewer group operations than their list of multiples, but makes their decryption's time expected, as ours is. Their values are lower bounds. Ciphertext growth is in bits, and encryption and decryption growth in field operations; concrete times count field multiplications, with additions $1.0$ to $1.8$ times as many. Work besides field arithmetic is small for us and, for them, would only add to the gap. For Hair and Sahai, $N$ is polynomial in $\lambda$ and at least $\lambda$; our growth is by \cref{cor:asymptotics}, and our values are from \cref{sec:costs}.}\label{tab:sizes}
\def\tsSep{5pt}\def\tsA{91.5pt}\def\tsB{66.6pt}\def\tsC{57.3pt}\def\tsD{56.5pt}\def\tsE{93.1pt}\def\tsF{53.5pt}
\centering\small\setlength{\tabcolsep}{\tsSep}
\begin{tabular}{@{}>{\raggedright\arraybackslash}p{\tsA}>{\raggedright\arraybackslash}p{\tsB}>{\raggedright\arraybackslash}p{\tsC}>{\raggedright\arraybackslash}p{\tsD}>{\raggedright\arraybackslash}p{\tsE}>{\raggedright\arraybackslash}p{\tsF}@{}}
\toprule
Construction & Security rests on & Ciphertext & Encryption & Decryption & Security level\\
\midrule
Hair and Sahai~\cite{HS26},\newline Theorem 5.1 & GGM & $N^3\, 2^{\Theta(\sqrt{\lambda})}$\newline $1.1 \times 10^{11}$\,GiB & $N^3\, 2^{\Theta(\sqrt{\lambda})}$\newline $2.6 \times 10^{20}$ & $N^3\, 2^{\Theta(\sqrt{\lambda})}$\newline $1.2 \times 10^{16}$ expected & $\lambda$ bits\\
\addlinespace
\scheme (this paper) & \mbox{bilinear GGM,}\newline ROs, MSIS & $O(\lambda^2 N)$\newline $7.88$\,GiB & $O(\lambda^2 N / \log \lambda)$\newline $2.0 \times 10^{10}$ & \mbox{$O(\lambda N \log \lambda + 2^{\lambda/4} N^{1/4})$}\newline $8.6 \times 10^{9}$ expected & $\lambda$ bits\\
\bottomrule
\end{tabular}
\end{table}

\let\savedFloatBarrier\FloatBarrier \let\FloatBarrier\relax
\subsection{Our Results}\label{sec:results}

\scheme encrypts to \emph{Boolean quadratic systems} (BQS): systems of quadratic equations over a prime field $\FF_q$ ($q$ is the order of \scheme's groups) in $N$ Boolean unknowns, the bits of an NP statement's witness and of the intermediate values of its verification (\cref{sec:relations}). Such systems express every NP statement. We instantiate \scheme on the curve BN254~\cite{USENIX:BCTV14} with the Groth16 verifier of a universal argument of knowledge (\cref{sec:groth16}), following Goldwasser et al.~\cite{C:GKPVZ13} and Liu et al.~\cite{DCC:LJKW18} in encrypting to a verifier. This verifier is fully succinct, its size independent of the program, and its one fixed key serves every relation. Written as a BQS, with $N \approx 2^{18.6}$, it gives every relation the same $N$, ciphertext size, and costs. To encrypt, one fixes the instance, a program with its public inputs and outputs, as constants of this system; a proof for that instance then decrypts, and knowledge soundness turns an extracted proof into a witness of the relation. That step assumes an honest Groth16 setup, whose secret randomness would let whoever kept it decrypt every ciphertext made for that key (\cref{sec:groth16}). With this verifier, \scheme builds on the modern infrastructure of zero-knowledge proofs: statements written as R1CS~\cite{C:BCGTV13}, CCS~\cite{EPRINT:SetThaWah23a}, Plonkish~\cite{EPRINT:GabWilCio19}, or AIR~\cite{EPRINT:BBHR18} are proved by arguments whose proofs the universal argument verifies, and zero-knowledge virtual machines whose proofs end in Groth16~\cite{SP1,RISC0} prove the execution of programs written in ordinary languages. So one can even encrypt to a program written in Rust: whoever knows an input that the program accepts can decrypt.

\begin{trivlist}
\item[\hskip\labelsep\bfseries Main result (informal).]\itshape We represent a Groth16 verifier on BN254 as a BQS. For this system, at security level $\lambda = 100$, \scheme has these properties.
\begin{enumerate}[label=\textup{(\alph*)}, leftmargin=*, nosep]
\item \textup{Efficiency.} The ciphertext is $7.88$\,GiB. Encryption takes $2^{34.2}$ field multiplications, and decryption with a proof $2^{33.0}$ in expectation.
\item \textup{Correctness (\cref{thm:correct}).} Decryption with a witness returns the message with probability above $1 - 2^{-252}$.
\item \textup{Extraction (\cref{thm:extraction}).} In the bilinear GGM and RO model, one extractor, fixed before any adversary, reads the adversary's group operations and outputs a witness or nothing. Let an adversary make $Q < 2^{100}$ group operations and at most $2^{100}$ queries to each RO. If it guesses which of two messages was encrypted with probability $\frac{1}{2} + \varepsilon$, the extractor outputs a witness with probability at least $2\varepsilon - (Q+1)/2^{100}$ minus the average success probability of two reductions that use the adversary to solve MSIS at \scheme's parameters.
\end{enumerate}
\end{trivlist}
The main result states these properties at our parameters; \cref{prop:efficiency,thm:correct,thm:extraction} give them in general, for systems of every size and all admissible parameters.

Decryption ends with a search for a secret integer in a short known range. While the rest of decryption takes time linear in $N$, the search is exponential in $\lambda$, but concretely fast (\cref{cor:asymptotics}). At $\lambda = 100$, the search of \cref{sec:optimizations} compares about $2^{29.4}$ candidates in expectation, at $6$ field multiplications and $11$ additions each by a linear recurrence~\cite{C:LenVer00}. Its $2^{32.0}$ multiplications are about half of decryption's expected cost. Decryption with a proof takes at most $2^{34.4}$ except with probability below $2^{-40}$ (\cref{fig:tail}). At these costs, encryption needs about $12$\,GB of memory and decryption about $5$\,GB, with the ciphertext streamed rather than held (\cref{sec:costs}).

\parhead{Formal verification}
We give Lean~4 proofs, at our instantiation's parameters, of correctness (b) and of the extraction theorem behind (c) (\cref{sec:formal-verification}; \href{https://github.com/psivesely/praxis-lean}{\nolinkurl{github.com/psivesely/praxis-lean}}).

\parhead{Witness key encapsulation}
Decryption recovers a group element from which an RO derives a key of any length, so \scheme is a witness key encapsulation mechanism (WKEM)~\cite{ICISC:ChoVau21}. A symmetric cipher under that key then encrypts messages of any length.

\parhead{Comparison with Hair and Sahai}
Hair and Sahai give two constructions, a basic one~\cite[Theorem~5.1]{HS26} and their main construction~\cite[Theorem~6.1]{HS26}. \Cref{tab:sizes} compares \scheme with the basic construction, a variant of a construction proved extractable (below), on encrypting a $100$-bit key, which theirs encrypts one bit at a time. Their theorems give polynomial-time algorithms and security against adversaries of quasi-polynomial size, those making at most $n^{(\log n)^c}$ group operations, for a constant $c$ and $n$ the larger of the security parameter and the statement's size~\cite[Theorem~1.1]{HS26}. Practitioners instead compare schemes at $\lambda$-bit security, where breaking them takes work about $2^\lambda$: $Q < 2^\lambda$ group operations give an advantage of at most $(Q+1)/2^\lambda$ without a witness, by their bound for them and by (c) for us, where any larger advantage requires solving MSIS, whose parameters are set by the cost of its best known attacks (\cref{sec:parameters}). At that level, for $N$ polynomial in $\lambda$, all three costs of each of their constructions grow faster than any polynomial in $\lambda$, while \scheme's ciphertext and encryption are polynomial in $\lambda$, and only the search that ends its decryption is exponential. At $\lambda = 100$, for the same statement, our ciphertext and encryption are over ten orders of magnitude below the basic construction's, and our decryption over six, even though we give their scheme our algorithms and set it up beyond what their theorems allow, which we prove still meets the security level (\cref{app:hair-sahai}). With the key in the corner that we use (\cref{sec:rank-template}), their scheme could likely carry a key in one or two of these ciphertexts, which no one has proved; it would still be over eight orders of magnitude above ours in size and encryption and over four in decryption (\cref{app:hair-sahai}).

Hair and Sahai prove hiding on false statements, which extraction implies. Encrypting to a relation's own system, whose witness decrypts, keeps hiding meaningful, but encrypting to a proof, as our instantiation does, gives every relation the same system and the same costs, and there hiding says nothing: the Groth16 verifier's system has accepting proofs for every public input. For the construction of their first arXiv version, of which the basic construction is a variant with other parameters, Campanelli later proved extraction in a GGM with no further assumption~\cite{Cam26}, and Chen and Vaikuntanathan for statements sampled with public coins, under the knowledge assumption KOALA~\cite{CV26}. Neither changes the scheme or its costs. \Cref{tab:sizes} prices the basic construction at parameters outside these proofs, where \cref{app:hair-sahai} bounds only its hiding. Their main construction~\cite[Theorem~6.1]{HS26} is not in \cref{tab:sizes}: no extraction proof is known for it, and at $\lambda = 100$ our lower bounds on its costs exceed the basic construction's. At its least parameters for $\lambda = 100$, a gap of one versus five on matrices of $2^{53.6}$ rows, its ciphertext takes over $7.0 \times 10^{11}$\,GiB, its encryption over $1.5 \times 10^{21}$ field multiplications, and its decryption over $4.5 \times 10^{17}$, lower bounds over ten, ten, and seven orders of magnitude above ours. With its degree parameter set to minimize growth, to $\Theta(\log\lambda)$, all three of its costs grow as $N^{\Theta(\log\lambda)}$, again for $N$ polynomial in $\lambda$ (\cref{app:hair-sahai}).

\parhead{Jin's construction}
Jin~\cite{Jin26} builds witness encryption for NP in a generic group without a pairing, hiding on false statements. He combines the code-based scheme of Barta et al.~\cite{C:BIOW20}, which he proves extractable for circuits of polylogarithmic size, with a succinct argument whose soundness holds against subexponential-time provers in the standard model, a guarantee that no succinct argument used in practice is known to have. Extraction for NP would follow if the argument were also knowledge-sound at that strength, which he does not claim. For a verifier of our statement's size, his theorems need $\lambda$ above $2^{44}$, and his ciphertext for a $100$-bit key exceeds $10^{17}$\,GiB (\cref{app:hair-sahai}, last paragraph). 
\let\FloatBarrier\savedFloatBarrier
\subsection{From the Rank Template to \scheme}\label{sec:key-ideas}

Both of Hair and Sahai's constructions~\cite{HS26} rest on what we call the \emph{rank template}: the conditions on a witness are linear tests on its \emph{moment matrix}, the matrix of the products of pairs of its entries, and a witness passes them with a matrix of rank one. On a false statement every nonzero matrix that passes, a \emph{cheat}, has higher rank. The template's \emph{table} is the hidden matrix $y r^\top$, with an entry for each pair of unknowns, for a hidden uniform vector $y$ and a vector $r$ of small secret integers, and decryption ends with a bounded search for one integer, $r(w) = \sum_i w_i r_i$ for the witness $w$. The search is why the unknowns are Boolean: confining each $w_i$ to $0$ and $1$, with one quadratic test, is the cheapest way to keep this integer in a short known range. For a cheat of rank $d$, $d$ combinations of the secret integers must be guessed together, which succeeds with probability at most $g^{-d}$ when each secret integer has a range of $g$ values (\cref{lem:gp-rank}), and for some cheats a generic search finds them in about $g^{d/2}$ group operations. So the least rank $d$ of a cheat sets the range, $g \approx 2^{2\lambda/d}$, and the range sets the search with a witness, about $\sqrt{g}\,N^{1/4} \approx 2^{\lambda/d} N^{1/4}$ group operations.

Hair and Sahai keep the range $g$ polynomial in the size of their ciphertexts, so that decryption takes time polynomial in that size, and make the rank of every cheat high. With one table, a cheat can have rank two, so they stack many copies of the table, each weighted by a product of linear forms in the witness, until every cheat has rank above a threshold $R$. A ciphertext carries one bit, and in both of their constructions three features set its size; we give the counts for their basic construction~\cite[Theorem~5.1]{HS26}, the construction of \cref{tab:sizes}. First, with a polynomial range, $R$ must grow with the security level, at more copies for each list of forms (\cref{app:hair-sahai}). Second, the ciphertext holds one element for each matrix in a basis of the passing matrices, on the order of $N^2$ for one table and a system of linear size~\cite[Sections~3.1 and~4]{HS26}. Third, to rule out a matrix of rank at most $R$ as a cheat, their proof needs a list of $R$ forms that are not all zero at any nonzero vector of its row space. The forms are fixed before a cheat is known, so the copies come in $2NR + 1$ lists, and for fixed $R$ the ciphertext grows as $N^3$~\cite[Sections~5.2 and~5.3]{HS26} (\cref{prop:hs-dimension}).

\begin{table}[!t]
\caption{The steps from one table to \scheme, for the statement, the key, the security level, and the conventions of \cref{tab:sizes}. Only the last row is a scheme with a security theorem (\cref{thm:extraction}). Growth is for systems of linear size with $N \ge \lambda^2$ (\cref{cor:asymptotics}). Ciphertext growth is in bits, and encryption and decryption growth in field operations; concrete times count field multiplications. Decryption is the part outside the search plus the search. Bold marks the values that a step improves over the row above, each by a factor of over twenty.}\label{tab:ladder}
\def\tlSep{7pt}\def\tlA{88.9pt}\def\tlB{66.6pt}\def\tlC{54pt}\def\tlD{56.5pt}\def\tlE{111.5pt}
\centering\small\setlength{\tabcolsep}{\tlSep}
\begin{tabular}{@{}>{\raggedright\arraybackslash}p{\tlA}>{\raggedright\arraybackslash}p{\tlB}>{\raggedright\arraybackslash}p{\tlC}>{\raggedright\arraybackslash}p{\tlD}>{\raggedright\arraybackslash}p{\tlE}@{}}
\toprule
Stage & Security rests on & Ciphertext & Encryption & Decryption, expected\\
\midrule
Step 1: one table & GGM,\newline RO & $O(\lambda N^2)$\newline $4.6 \times 10^{3}$\,GiB & $O(\lambda N^2 / \log \lambda)$\newline $2^{43.3}$ & $O(N^2) + O(2^{\lambda/2} N^{1/4})$\newline $2^{37.8} + 2^{57.4}$\\
\addlinespace
Step 2: two factors & \mbox{bilinear GGM,}\newline RO & $O(\lambda N)$\newline \textbf{0.14\,GiB} & $O(\lambda N / \log \lambda)$\newline $\symbf{2^{30.4}}$ & \mbox{$O(\lambda N) + O(2^{\lambda/2} N^{1/4})$}\newline $\symbf{2^{31.0}} + 2^{57.3}$\\
\addlinespace
Step 3: tied copies & \mbox{bilinear GGM,}\newline ROs, MSIS & $O(\lambda^3 N)$\newline $8.1 \times 10^{2}$\,GiB & $O(\lambda^3 N / \log \lambda)$\newline $2^{40.9}$ & $O(\lambda^2 N) + O(2^{\lambda/4} N^{1/4})$\newline $2^{36.6} + \symbf{2^{32.0}}$\\
\addlinespace
Step 4: shared factors,\newline \scheme & \mbox{bilinear GGM,}\newline ROs, MSIS & $O(\lambda^2 N)$\newline \textbf{7.88\,GiB} & $O(\lambda^2 N / \log \lambda)$\newline $\symbf{2^{34.2}}$ & \mbox{$O(\lambda N \log \lambda) + O(2^{\lambda/4} N^{1/4})$}\newline $\symbf{2^{32.0}} + 2^{32.0}$\\
\bottomrule
\end{tabular}
\end{table}

\scheme replaces these three features in four steps, and \cref{tab:ladder} gives the costs after each.
\begin{enumerate}[leftmargin=*, topsep=3pt]
\item \emph{One table: a fixed rank gap, a wide range, and a key in the corner} (\cref{sec:rank-template}). Where Hair and Sahai raise the rank, we fix the rank gap and take a range $g$ exponential in $\lambda$. Size and encryption then stay polynomial in $\lambda$, at the price of a search exponential in $\lambda$, by baby steps and giant steps~\cite{Sha71}. We publish their table entry by entry, the form that step~2 factors. Anyone can combine its entries, with any matrix as coefficients, and a \emph{\testmask{}}, a random combination of the tests, makes the result useless unless that matrix passes. We put a key in the table's corner, where a witness's moment matrix has the entry $1$, so a ciphertext carries a key and not one bit. The key itself is not published, so a second copy of the ciphertext, scaled by a hidden factor, comes with its scaled key, which the decryptor subtracts before the search. With this one table, cheats of rank two remain, so $g \approx 2^{\lambda}$ and the search takes about $2^{\lambda/2} N^{1/4}$ operations, while the ciphertext is quadratic in $N$.
\item \emph{Two factors, from a quadratic arithmetic program (QAP) that takes sums of products} (\cref{sec:thin-table,sec:qap}). The \testmask{} can have high rank, so the table takes one element for each entry. We replace the \testmask{} and the key by one term of rank one: the values, at a hidden point, of the polynomials of a QAP~\cite{EC:GGPR13} that encodes the tests. The table then has rank at most two, and we publish two factors of it, of width two, in the two source groups of a pairing. A pairing, which Hair and Sahai's model lacks, multiplies them, so its two factors take $4(N+1)$ elements in place of $(N+1)^2$ entries. The table is never formed. The encryptor computes only the factors, and a decryptor with a witness combines each factor with it before pairing, so the table's share of encryption and decryption is linear in $N$ too. This puts no limit on the rank of a cheat, which is a matrix of coefficients for the table's entries, not the table. The factored table is not symmetric, so a matrix that is not symmetric could pass with rank one; a new \emph{\symmetryterm{}} masks every combination along such a matrix, without widening the factors. A hidden change of basis between the factors, as in dual pairing vector spaces~\cite{PAIRING:OkaTak08,AC:OkaTak09} and canceling pairings~\cite{EC:Freeman10}, cancels only in whole entries, so an adversary cannot weight an entry's two products differently. A BQS constraint sums many products, as in customizable constraint systems (CCS) of degree two, while the QAP of Gennaro et al.\ has one product in each constraint, as a rank-one constraint system (R1CS) does. Rewriting our constraints in that form would need an auxiliary unknown, a field element of $254$ bits, for each product but one, so the unknowns, and the costs with them, can grow by a factor of order $\lambda$, to $2^{29.1}$ Boolean unknowns in place of $2^{18.6}$ for our statement. Our QAP takes sums of products directly: it places a constraint's products on a coset of a multiplicative subgroup, over which their terms must sum to zero, and checks these sums by allowing its division a remainder with no exponent divisible by the coset's size. A test that an unknown is Boolean takes one point, not a coset, keeping the number of points $O(N)$ for systems of linear size. Our QAP builds on Aurora's sum over a coset~\cite{EC:BCRSVW19}, read for all cosets at once~\cite{Moh25}, and on a vanishing sum enforced through the reference string~\cite{AC:LipSiiZaj22,PKC:ZhaSunGu24}. One factored table takes $0.14$\,GiB, but its search is about as long as step~1's.
\item \emph{Tied copies: weights and ties under MSIS, with the weights from an extension field} (\cref{sec:weights-ties,sec:ext-weights}). To shorten the search, we raise the least rank of an easily guessed cheat from two to four. Like Hair and Sahai~\cite[Section~4]{HS26}, we weight copies of the table by polynomials in the witness, built from linear forms. One table lets a cheat mix the moment matrices of two assignments so that their violations of each constraint cancel, and under the weights they would have to cancel in every copy. Three things are ours. First, an adversary can combine each of our copies along a matrix of its own, even one cheat repeated in every copy, so we \emph{tie} the copies by linear identities between these matrices, enforced by hidden masks. Second, in place of their many lists, we take one list of $n = \num{4608}$ forms, derived from a random MSIS matrix, its length set by the hardness of MSIS, of order $\lambda^2$ whatever $N$. Third, the weights have degree two, since degree one leaves a cheat of rank three, but all products of two forms would be $n(n+1)/2$ copies, of order $\lambda^4$ and over $2^{23}$ here. We take them from an extension field of $\FF_q$ of degree $n$ instead: the $n$ forms become the coefficients of one form with values in that field, one product of this form with a conjugate gives $n$ weights at once, and two such products, the second with only $n/2$ nonzero coefficients, give all $3n/2 = \num{6912}$ weights, of order $\lambda^2$, with identities of that field as the ties. Cheats of rank two and three still exist, but with these weights and ties we prove that each has in its row space a nonzero vector at which every form vanishes. For an easily guessed cheat, an inverse theorem that we prove modulo $q$, after Tao and Vu~\cite{TV09} and Nguyen and Vu~\cite{NV11}, gives a multiple of that vector with small integer entries, which solves MSIS. The idea is that a linear form easy to guess on a box of small integers is, up to a scalar, a form with small integer coefficients. A cheat of rank four or more is guessed with probability at most $g^{-4}$, a rank gap of one versus four. So $g \approx 2^{\lambda/2}$ suffices, and the search takes about $2^{\lambda/4} N^{1/4}$ operations. The copies are the price, and we pay it by choice: they multiply step~2's ciphertext and encryption by a factor of order $\lambda^2$, whatever $N$, raise the rest of decryption from $2^{31.0}$ multiplications to $2^{36.6}$, and cut the search, until now nearly all of decryption, from $2^{57.3}$ to $2^{32.0}$.
\item \emph{Shared factors} (\cref{sec:shared-factors,sec:copies}). With its own factors, a copy costs six elements for each unknown, its scaled copies included, $\num{41472}$ over the copies. But apart from the masks that enforce the ties, the copies' tables differ only in a scalar, the product of a left and a right scalar. So the copies are the cells of a grid, those in one row or one column share a factor, and the factors take $528$ elements for each unknown. The price, none in growth, is in the masks: a column's mask serves every row, and one more hidden coin marks the mixed products that this leaves, so that published elements remove them but not the terms that enforce the ties. The grid returns a factor of order $\lambda$ in size and in encryption, and of order $\lambda/\log\lambda$ in decryption outside the search. In all, the copies cost \scheme a factor $\lambda$ in size and in encryption over step~2's one factored table, for a search about $2^{\lambda/4}$ times shorter.
\end{enumerate} 
\subsection{Related Work}\label{sec:related-work}

\begingroup\tolerance=1000\pretolerance=100\emergencystretch=3em\rightskip=0pt plus .5pt
\parhead{Feasibility results}
Besides obfuscation~\cite{FOCS:GGHRSW13}, which exists under well-founded assumptions~\cite{STOC:JaiLinSah21,TCC:RagVafVai24}, witness encryption follows from subexponentially hard learning with errors (LWE), with encryption time $2^{m/2}$ for witnesses of $m$ bits~\cite{SCN:BJKPW18}. The approach of building witness encryption in a generic group from NP-hardness of approximation is due to Barta et al.~\cite{C:BIOW20}. In their scheme a witness is a nonzero codeword of low Hamming weight, which passes a hidden linear test by missing the ciphertext's sparse noise; security needs a gap of $\omega(\log n)$ for the minimum distance of linear codes, which is not known to be NP-hard. Hair and Sahai, who name it their closest generic-group precursor, replace Hamming weight by matrix rank and sparse noise by small secret integers and a bounded search, prove the gap they need, and so need no further assumption in prime-order generic groups~\cite{HS26}; Jin builds on Barta et al.~\cite{Jin26} (\cref{sec:results}).
\par\endgroup

\begingroup\emergencystretch=1em
\parhead{Restricted relations}
Other schemes, most of them pairing-based, handle restricted relations or interfaces: algebraic languages~\cite{DCC:DerSla18}, openings of KZG commitments~\cite{AC:FleHalSim24}, signatures~\cite{FC:DHMW23,AC:ADMSW24}, functional commitments~\cite{PKC:CamFioKho24}, linearly verifiable succinct arguments~\cite{C:GHKKP25}, invalid proofs of pairing-based SNARKs~\cite{EL26}, proof verification on Bitcoin~\cite{GKSST26}, and the SAT instances that resist mixed-input attacks, with a variant for all instances in which the encryptor or a third party takes part in decryption~\cite{WWZ24}. Groth shows that, in his model, two-move linear interactive proofs with exact correctness cannot have a linear decision procedure for relations with hard decisional problems~\cite[Theorem~3]{EC:Groth16}; our decryption, ending with a bounded search for one integer, is not such a procedure.
\par\endgroup

\begingroup\emergencystretch=1em
\parhead{Candidates on heuristic or purpose-built assumptions}
The first constructions of witness encryption~\cite{STOC:GGSW13,C:GenLewWat14} use multilinear maps, whose candidates met zeroizing and annihilation attacks~\cite{EC:CHLRS15,EC:HuJia16,C:MilSahZha16}; one such scheme was implemented, for Sudoku puzzles~\cite{ULCG21}. Lattice candidates~\cite{C:CheVaiWee18,EPRINT:Kluczniak20,C:Tsabary22,AC:VaiWeeWic22} rest on purpose-built assumptions, some of them on evasive LWE, for which counterexamples are known~\cite{AC:BrzUnaWoo24,C:DJMMV25,TCC:AMYY25}. The scheme built from affine determinant programs~\cite{ITCS:BIJMSZ20} rests on heuristic arguments, and so does its arithmetic variant~\cite{SREKKAT26}, whose authors estimate a ciphertext of about $338$\,TiB at $100$-bit security. Attacks recover the message from the ciphertext for the original candidate~\cite{Jo26} and, for sparse circuits, in both~\cite{EGRSS26}; the latter attack is by authors of the variant. The variant's first author has since proposed a patch meant to resist it~\cite{Sou26}, which also rests on heuristic arguments and claims no ciphertext size.
\par\endgroup

\parhead{Broken proposals}
Pairing-based proposals for general NP statements come with security proofs. Yet in the WKEM of Chang et al.~\cite{ICISC:CWLHTM25}, built from a QAP and the verification equation of Groth16, the key can be computed from the statement and the ciphertext without a witness~\cite{Vang25}. The same write-up states that this also holds for the similar scheme of Wang and Zhang~\cite{WZ25}. A scheme from smooth projective hashing~\cite{ISC:WanZha22} encrypts to a Groth16 proof that the decryptor has already computed, and decryption needs that proof's randomness, so a ciphertext is addressed to the one prover that made it, not to every holder of a witness. Its hash function is also not smooth, since the projection key contains the hash key in the exponent.

\parhead{Relaxed notions}
Offline witness encryption~\cite{ACNS:AbuFucPie16} makes encryption fast but not its obfuscation-based setup or decryption, and in Cassiopeia~\cite{FCW:SaeZin23} a committee releases the message when shown a witness.  
\section{Technical Overview}\label{sec:overview}

A statement is a system of quadratic equations over the field $\FF$ of prime order $q$ in Boolean unknowns, and \scheme works in a bilinear group of order $q$. We first turn the system into linear tests on a matrix, which a witness passes with a matrix of rank one (\cref{sec:tests-rank}). We restate the rank template of Hair and Sahai~\cite{HS26} with a \testmask{}, a random combination of the tests, and give it a fixed rank gap, a wide range, and a key in the table's corner (\cref{sec:rank-template}). We then replace the \testmask{} with a quadratic arithmetic program (QAP), so that the table is virtual, defined by two factors of width two, which a pairing multiplies, and let the QAP take sums of products (\cref{sec:thin-table}). We weight and tie many copies of the table so that, under Module-SIS (MSIS), no adversary forms an easily guessed cheat of rank below four (\cref{sec:weights-ties}), and let the copies share their factors (\cref{sec:shared-factors}). \Cref{sec:rank-template,sec:thin-table,sec:weights-ties,sec:shared-factors} follow the four steps of \cref{tab:ladder}, and \cref{tab:sizes} compares \scheme with Hair and Sahai's construction. For ease of exposition, three constructions that \cref{tab:ladder} prices are given in full only in later sections: the QAP for constraints that sum products (\cref{sec:qap}), the weights from an extension field (\cref{sec:ext-weights}), and the masks that enforce the ties on shared factors (\cref{sec:copies}).

\parhead{Notation}
Vectors with an entry for each index $0, \dots, N$ are lowercase, as in $w = (w_0, \dots, w_N)$, and matrices are uppercase, as in $W$ with entries $W_{ij}$; the rows $F_i$ and columns $H_j$ of the two factors of the table and the vectors $A(\tau)$ and $B(\tau)$ of values of the QAP's polynomials are the exceptions. For an integer $k \ge 1$, $[k] = \{1, \dots, k\}$, and $x \sample \mathcal{S}$ samples $x$ independently and uniformly from a set $\mathcal{S}$. A family of matrices, one for each copy of the table, is bold, as in $\mathbf{W} = (W_c)_c$. The polynomials of the QAP are written with their formal variable, as in $Z_0(X)$, which identities may omit, and $Z_0(\tau)$ is the value at a point $\tau \in \FF$; sets of such points are calligraphic, as in $\mathcal{P}$. The polynomials of degree below $d$ form $\FF^{<d}[X]$, and the \emph{vanishing polynomial} $Z_{\mathcal{S}}(X) \defeq \prod_{p \in \mathcal{S}} (X - p)$ of a point set $\mathcal{S}$ divides exactly the polynomials vanishing on $\mathcal{S}$. The encoding of $v \in \FF$ is $[v]$ while the group has no pairing, as in the template, and $[v]_s$ in the group $\GG_s$ of a bilinear group, for $s \in \{1, 2, T\}$, once the table is factored.

\subsection{Tests and Rank}\label{sec:tests-rank}

This subsection fixes the statements that \scheme encrypts to, systems of quadratic constraints with a Boolean witness, and turns every constraint into a linear test on a matrix, which a witness passes with a matrix of rank one. It ends with the passing matrices that come from no witness, which the rest of the section must deal with.

A \emph{rank-one constraint system} (R1CS) $S$ over $\FF$ with a Boolean witness, a \emph{system} for short, has unknowns $w_1,\dots,w_N$ and a list of quadratic constraints. Vectors in $\FF^{N+1}$ are indexed from $0$, and an \emph{assignment} is a vector $w = (w_0, w_1, \dots, w_N) \in \FF^{N+1}$ with $w_0 = 1$, so that a linear form $v(w) \defeq \langle v,w \rangle = v_0 + \sum_{i \ge 1} v_i w_i$, for $v \in \FF^{N+1}$, can have a constant term; $e_i$ is the unit vector of index $i$. Constraint $t$ is an equation $a_t(w)\,b_t(w) = c_t(w)$ with $a_t, b_t, c_t \in \FF^{N+1}$. For each unknown $w_j$ the list includes its \emph{Booleanity}, the constraint $w_j \cdot w_j = w_j$, so a \emph{witness}, an assignment that satisfies every constraint, is in $\bits^{N+1}$.

The statements we target have constraints $\sum_f a_{t,f}(w)\,b_{t,f}(w) = c_t(w)$ with sums of products: customizable constraint systems (CCS) of degree two~\cite{EPRINT:SetThaWah23a} with a Boolean witness, whose one-product case is R1CS. Rewriting such a constraint as R1CS needs an auxiliary unknown for the value of every product but one, a field element, which a Boolean witness spells out in $\lceil \log_2 q \rceil = \Theta(\lambda)$ bits, $254$ at our parameters (\cref{tab:params}). So the number of unknowns, and the costs with it, can grow by a factor of $\Theta(\lambda)$. In general form, our QAP takes sums of products directly (\cref{def:qap}); the overview focuses on R1CS for ease of exposition.

Since $w_0 = 1$, constraint $t$ can be written $a_t(w)\,b_t(w) = w_0\,c_t(w)$, and Booleanity of $w_j$ as $w_j^2 = w_0 w_j$. In this form every term is a product of two entries of $w$, so every requirement on a witness is linear in the products $w_i w_j$. Let $W \in \FF^{(N+1) \times (N+1)}$, a symmetric matrix, stand for the \emph{moment matrix} $w w^\top$ of an assignment. Constraint $t$ becomes the test $a_t^\top W b_t = e_0^\top W c_t$, which for the Booleanity of $w_j$ reads $e_j^\top W e_j = e_0^\top W e_j$; $W$ \emph{passes} when every test holds (\cref{fig:tests}). For a witness $w$ the matrix $w w^\top$ passes, with corner $W_{00} = 1$. Conversely, a nonzero passing symmetric matrix of rank one equals $W_{00}\, w w^\top$ for a witness $w$, with $W_{00} \ne 0$ (\cref{lem:rank-one}).

\begin{figure}[htb]
\centering
\begin{tikzpicture}[font=\footnotesize, line width=0.5pt, x=0.4cm, y=0.4cm,
    every node/.append style={execute at begin node={\hyphenpenalty=10000 \exhyphenpenalty=10000 }}]
  \definecolor{figblue}{HTML}{3871C1}
  \definecolor{figrust}{HTML}{7A3D00}
  \definecolor{figbluefill}{HTML}{DCE7F5}
  \definecolor{figrustlight}{HTML}{F3DDCB}
  \definecolor{figrustfill}{HTML}{D9A57E}
\node[anchor=north west, inner sep=0pt, align=left] (tab) at (0,4.95) {\renewcommand{\arraystretch}{1.25}\begin{tabular}{@{}l@{\hspace{0.9em}}l@{}}
      (a)\enspace constraint & test on $W$\\[-1pt]
      \noalign{\vskip 1pt\hrule height 0.4pt\vskip 1pt}
      $w_1 w_3 = 1$ & $W_{13} = W_{00}$\\
      $w_2(w_1 + w_3) = 0$ & $W_{21} + W_{23} = 0$\\
      $w_j^2 = w_j$ & $W_{jj} = W_{0j}$
    \end{tabular}};
  \node[anchor=north west, inner sep=0pt, text width=4.05cm, align=left, text=black!70] (note) at ([yshift=-5pt]tab.south west)
    {$W$ stands for the matrix of products $w_i w_j$, with $w_0 = 1$.};
  \node[anchor=north west, inner sep=0pt, text width=4.05cm, align=left, text=figrust] at ([yshift=-5pt]note.south west)
    {False system, used in (c): $w_2(w_1 + w_3) = 1$, with test $W_{21} + W_{23} = W_{00}$.};
\begin{scope}[xshift=5.35cm]
    \node[anchor=south west, inner sep=0pt, align=left] at (-1.0,4.75) {(b)\enspace $W = w w^\top$,\\ \phantom{(b)}\enspace $w = (1,1,0,1)$};
    \foreach \r/\row in {0/{1,1,0,1}, 1/{1,1,0,1}, 2/{0,0,0,0}, 3/{1,1,0,1}} {
      \foreach \v [count=\c from 0] in \row {
        \ifnum\v=1 \fill[figbluefill] (\c,3-\r) rectangle ++(1,1); \fi
        \node at (\c+0.5,3.5-\r) {$\v$};
      }
      \node[text=black!60, font=\scriptsize] at (-0.45,3.5-\r) {\r};
      \node[text=black!60, font=\scriptsize] at (\r+0.5,4.3) {\r};
    }
    \draw[black!55, line width=0.4pt] (0,0) grid[step=1] (4,4);
    \node[anchor=north west, inner sep=0pt, text width=2.85cm, align=left] at (-0.2,-0.3)
      {\textcolor{figblue}{\textbf{rank one:}} every row is $w$ or zero. Passes every test.};
  \end{scope}
\begin{scope}[xshift=8.95cm]
    \node[anchor=south west, inner sep=0pt] at (-1.0,4.75) {\vphantom{$w^\top_j$}(c)\enspace $W = \tfrac12(u u^\top + u' u'^\top)$};
    \foreach \r in {0,...,3} {
      \foreach \c in {0,...,3} {
        \pgfmathtruncatemacro{\half}{(\r==2) || (\c==2)}
        \ifnum\half=1
          \fill[figrustfill] (\c,3-\r) rectangle ++(1,1); \node at (\c+0.5,3.5-\r) {\scriptsize$\frac12$};
        \else
          \fill[figrustlight] (\c,3-\r) rectangle ++(1,1); \node at (\c+0.5,3.5-\r) {$1$};
        \fi
      }
      \node[text=black!60, font=\scriptsize] at (-0.45,3.5-\r) {\r};
      \node[text=black!60, font=\scriptsize] at (\r+0.5,4.3) {\r};
    }
    \draw[black!55, line width=0.4pt] (0,0) grid[step=1] (4,4);
    \node[anchor=north west, inner sep=0pt, text width=2.85cm, align=left] at (-0.2,-0.3)
      {\textcolor{figrust}{\textbf{rank two:}} no witness behind it, yet it passes every test of the false system.};
  \end{scope}
\end{tikzpicture} 
\caption{Equations as linear tests on a matrix, for the systems of \cref{ex:toy-system}. (a) The constraints and Booleanity as linear conditions on a symmetric matrix $W$ that stands for the products $w_i w_j$. (b) The witness's matrix $w w^\top$, of rank one. (c) On the false system, the average of $u u^\top$ and $u' u'^\top$: a cheat of rank two.}\label{fig:tests}
\end{figure}

Every test is a linear equation in the entries of $W$ with no constant term, so the passing matrices form a linear space. It contains the linear combinations of the moment matrices $w w^\top$ of witnesses, and can contain matrices that come from no witness. A \emph{cheat} is a passing matrix that is not such a combination; it has rank at least two, and on a system without witnesses every nonzero passing matrix is a cheat. Two cheats of rank two, on a system without witnesses, show what \scheme has to deal with. The first is a \emph{mixture} $(1-\chi)\, u u^\top + \chi\, u' u'^\top$ of two Boolean assignments that are not witnesses: it passes when, in every constraint, the residuals of $u$ and $u'$ cancel with the weights $1-\chi$ and $\chi$, where the \emph{residual} of $u$ in constraint $t$ is $\res_t(u) \defeq a_t(u)\,b_t(u) - c_t(u)$. The second needs no assignment, only two unit vectors: when no constraint has a term in $w_i w_j$, for two distinct unknowns $w_i$ and $w_j$, the matrix $e_i e_j^\top + e_j e_i^\top$ passes every test. Sums of cheats pass; on a system without witnesses every nonzero sum is again a cheat, and its rank can exceed two.

\begin{example}[A toy system]\label{ex:toy-system}
Over $\FF_{101}$, the system $w_1 w_3 = 1$, $w_2(w_1 + w_3) = 0$, with the Booleanity of its three unknowns, has the single witness $w = (1,1,0,1)$, with $w_0$ written first. Its tests on a symmetric $W$ are $W_{jj} = W_{0j}$ for $j = 1,2,3$, $W_{13} = W_{00}$, and $W_{21} + W_{23} = 0$. If the second constraint becomes $w_2(w_1 + w_3) = 1$, no witness exists. The assignments $u = (1,1,0,1)$ and $u' = (1,1,1,1)$ then have residuals $-1$ and $1$ in that constraint and $0$ in the others, so $\frac12 (u u^\top + u' u'^\top)$ passes every test.
\end{example}

The next subsections show how \scheme deals with these cheats. 
\subsection{The Rank Template}\label{sec:rank-template}

This subsection recalls how Hair and Sahai turn the tests into witness encryption, and restates their scheme in the form that \scheme builds on: a table $T$ under a \testmask{} $M$, with a key $\kappa$ in its corner and a bounded search in decryption. We then show that the least rank $d$ of a cheat sets the range $g$ of the secret integers $r_i$, and with it the cost of decryption, and that one table has a ciphertext quadratic in $N$; the next two subsections remove these two costs.

Hair and Sahai build witness encryption from a \emph{rank gap}~\cite[Section~3]{HS26}: a witness $w$ gives the passing matrix $w w^\top$, of rank one, and on a system without witnesses every nonzero passing matrix $W$ has higher rank. We call their construction the \emph{rank template}. It works in a group $\GG$ of prime order $q$, without a pairing, in which the encoding $[v]$ of $v \in \FF$ is $v$ times the generator. When a ciphertext contains the encodings $[T_{ij}]$ of the entries of a hidden matrix $T$, anyone can \emph{combine} them \emph{along} a matrix $W$ of its choice, that is, add them with the entries of $W$ as coefficients:
\[
  \sum_{i,j} W_{ij}\, [T_{ij}] = [\langle T,W \rangle], \qquad \langle T,W \rangle \defeq \sum_{i,j} T_{ij} W_{ij} .
\]
The template arranges that a combination along a passing matrix is of use only with a guess of secret integers $r_i$, as many as the matrix has rank: one for the matrix of a witness, which a search finds, and at least two for a cheat.

\parhead{The idea of Hair and Sahai}
Their scheme encrypts to a linear space of matrices, given by a basis $W^{(1)}, W^{(2)}, \dots$, under a promise they call \emph{bounded-factor MinRank}: a witness yields coefficients $a_j$ and vectors $u$ and $v$, with $v$ Boolean, such that $\sum_j a_j W^{(j)} = u v^\top \ne 0$, and on a false statement every nonzero matrix of the space has rank at least a threshold~\cite[Theorem~3.1]{HS26}. A witness-preserving reduction computes the space from a Boolean circuit, through quadratic equations in Boolean unknowns, as in \cref{sec:tests-rank}~\cite[Section~4]{HS26}. The simplest such space is that of the passing matrices of the previous subsection, where a witness $w$ gives $u = v = w$ and the gap is one versus two; their reductions build larger spaces with a wider gap, in which $u$ stacks weighted copies of $v$ (as will ours, \cref{sec:weights-ties}). We describe the scheme on the simplest space. The passing matrices are the solutions of public linear equations, so the encryptor computes the basis from the system alone: it needs no witness, and need not know whether one exists. It samples a uniform $y \in \FF^{N+1}$ and a vector $r$ of small integers, and publishes $[y_i]$ for every $i$ and $[y^\top W^{(j)} r]$ for every $j$. Since $y^\top W r = \sum_{i,j} W_{ij}\, y_i r_j = \langle y r^\top, W \rangle$, these are combinations of a \emph{table} $y r^\top$, the matrix of the products $y_i r_j$, whose entries stay hidden: the ciphertext gives the combination of the table along every passing matrix, and along no other. A decryptor with a witness computes $a$, $u$, and $v$, and then $\sum_j a_j [y^\top W^{(j)} r] = [y^\top u v^\top r] = [y(u)\, r(v)]$ and $\sum_i u_i [y_i] = [y(u)]$. The first element is $r(v)$ times the second, for the integer $r(v) = \sum_i v_i r_i$, and since $v$ is Boolean and the entries of $r$ are small, the decryptor can try every value of $r(v)$: it tests whether the first element is one of the multiples of the second up to the largest possible value of $r(v)$. This encrypts one bit; for the other bit, uniform elements replace the $[y^\top W^{(j)} r]$, and the test fails except with small probability~\cite[Section~3.1]{HS26}.

\parhead{Our table}
Unlike Hair and Sahai, we do not reduce the statement to bounded-factor MinRank: \scheme{} takes an R1CS, or in general a CCS of degree two, and works with its tests directly (\cref{sec:tests-rank}). We publish their table $y r^\top$ in another form, from which the next subsection starts: every entry of it, made symmetric, under a \emph{\testmask{}} $M$ that cancels on the passing matrices, so that again only the combinations along passing matrices are of use. We also make two changes, each explained where it is used: a key $\kappa$ in the corner $T_{00}$ of the table, so that a ciphertext carries a message and not one bit, and a second copy of the ciphertext, scaled by a hidden $\alpha$, for the search.

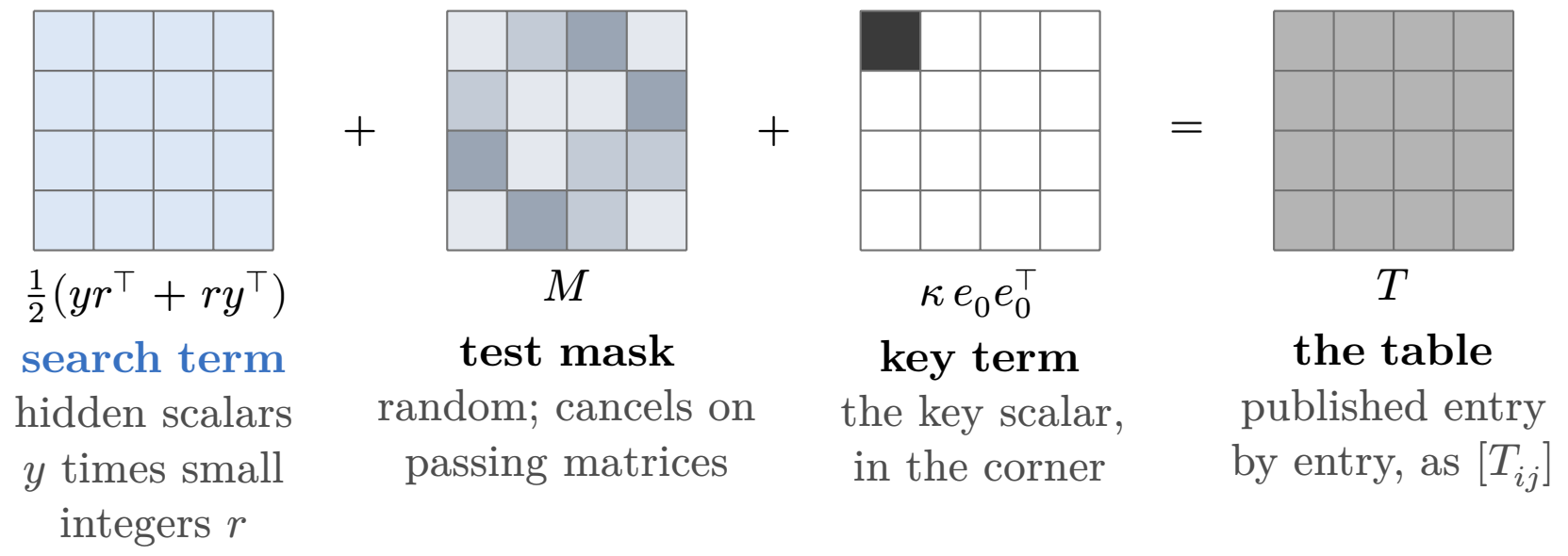
\begin{figure}[htb]
\centering
\begin{tikzpicture}[font=\footnotesize, line width=0.5pt, x=0.42cm, y=0.42cm,
    lab/.style={anchor=north, align=center, text width=2.75cm, inner sep=0pt,
      execute at begin node={\hyphenpenalty=10000 \exhyphenpenalty=10000 }}]
  \definecolor{figblue}{HTML}{3871C1}
  \definecolor{figbluefill}{HTML}{DCE7F5}
  \definecolor{figkey}{HTML}{3A3A3A}
  \definecolor{figmaskA}{HTML}{E4E8EE}
  \definecolor{figmaskB}{HTML}{C3CBD6}
  \definecolor{figmaskC}{HTML}{9AA5B4}
  \definecolor{figtable}{HTML}{B4B4B4}
\fill[figbluefill] (0,0) rectangle (4,4);
  \draw[black!55, line width=0.4pt] (0,0) grid[step=1] (4,4);
\begin{scope}[xshift=2.9cm]
    \foreach \r/\row in {0/{A,B,C,A}, 1/{B,A,A,C}, 2/{C,A,B,B}, 3/{A,C,B,A}} {
      \foreach \v [count=\c from 0] in \row { \fill[figmask\v] (\c,3-\r) rectangle ++(1,1); }
    }
    \draw[black!55, line width=0.4pt] (0,0) grid[step=1] (4,4);
  \end{scope}
\begin{scope}[xshift=5.8cm]
    \fill[figkey] (0,3) rectangle (1,4);
    \draw[black!55, line width=0.4pt] (0,0) grid[step=1] (4,4);
  \end{scope}
\begin{scope}[xshift=8.7cm]
    \fill[figtable] (0,0) rectangle (4,4);
    \draw[black!55, line width=0.4pt] (0,0) grid[step=1] (4,4);
  \end{scope}
  \foreach \x/\s in {5.45/{$+$}, 12.36/{$+$}, 19.26/{$=$}} { \node at (\x,2) {\s}; }
  \foreach \x/\f/\n/\d in {
      2/{$\tfrac12(y r^\top + r y^\top)$}/{\textcolor{figblue}{search term}}/{hidden scalars $y$ times small integers $r$},
      8.9/{$M$}/{\testmask{}}/{random; cancels on passing matrices},
      15.8/{$\kappa\, e_0 e_0^\top$}/{key term}/{the \keyscalar{}, in the corner},
      22.7/{$T$}/{the table}/{published entry by entry, as $[T_{ij}]$}} {
    \node[lab] at (\x,-0.3) {\f\\[2pt] \textbf{\n}\\ \textcolor{black!70}{\d}};
  }
\end{tikzpicture} 
\caption{The template's table, a sum of three symmetric matrices. In a combination along a symmetric $W$, the search term gives $y^\top W r$, the \testmask{} gives zero if $W$ passes the tests and a uniform scalar if not, and the key term gives $\kappa\, W_{00}$, which is $\kappa$ for $W = w w^\top$.}\label{fig:template}
\end{figure}

Each test is one linear equation in the entries of $W$. With every term on the left, it is a sum of terms $a^\top W b$, one for each \emph{pair} $(a,b)$ of vectors of the test, which stands for the product $a(w)\, b(w)$ of two linear forms. The test of constraint $t$ has two pairs, $(a_t, b_t)$ for its product and $(e_0, -c_t)$ for its right-hand side:
\begin{equation}\label{eq:test-products}
  a_t^\top W\, b_t + e_0^\top W\, (-c_t) = 0 .
\end{equation}
For a symmetric $W$, $a^\top W b = \langle \frac12 ( a b^\top + b a^\top ), W \rangle$. So the test of constraint $t$ reads $\langle C_t,W \rangle = 0$ for the public symmetric matrix
\[
  C_t \defeq \tfrac12 \bigl( a_t b_t^\top + b_t a_t^\top \bigr) - \tfrac12 \bigl( e_0 c_t^\top + c_t e_0^\top \bigr) .
\]
The encryptor samples a uniform $y \in \FF^{N+1}$; an integer vector $r = (r_0, r_1, \dots, r_N)$, the \emph{\searchvector{}}, of \emph{secret integers}, independent and uniform on a range of $g$ consecutive integers; a uniform \emph{\keyscalar{}} $\kappa \in \FF$; and the \testmask{} $M \defeq \sum_t \mu_t C_t$, with independent uniform coefficients $\mu_t \in \FF$, so that $\langle M,W \rangle = \sum_t \mu_t \langle C_t,W \rangle$ is zero when $W$ passes, and uniform in $\FF$ when a symmetric $W$ fails a test. Like the basis above, the \testmask{} is made from the tests alone, without a witness. The table is
\begin{align*}
  T &\defeq \tfrac12 \bigl( y\, r^\top + r\, y^\top \bigr) + M + \kappa\, e_0 e_0^\top , \quad\text{so that} \\
  \langle T,W \rangle &= y^\top W r + \langle M,W \rangle + \kappa\, W_{00}
\end{align*}
for every symmetric $W$ (\cref{fig:template}). The first term of $T$ is the table of Hair and Sahai, made symmetric; we call it the \emph{search term}, after the search that removes it in decryption. The last term puts the \keyscalar{} in the corner, where $w w^\top$ has the entry $w_0^2 = 1$.

The encryptor also samples a uniform hidden scalar $\alpha \ne 0$. The ciphertext is
\[
  ct \defeq \Bigl( \bigl\{ [T_{ij}] \bigr\}_{0 \le i \le j \le N},\ \bigl\{ [y_i] \bigr\}_{0 \le i \le N},\ \bigl\{ [\alpha T_{ij}] \bigr\}_{0 \le i \le j \le N},\ \bigl\{ [\alpha y_i] \bigr\}_{0 \le i \le N},\ [\alpha\kappa],\ \hdr \Bigr) .
\]
The entries $[T_{ij}]$ with $i \le j$ give every entry of the symmetric $T$, and the $[y_i]$ are the encodings of the entries of $y$. The \emph{\maskedmessage{}} $\hdr \defeq m \oplus \Hkey([\kappa])$ is the exclusive or of the message $m$ with a key that a hash function $\Hkey$, modeled as a random oracle (RO), the \emph{key oracle}, derives from the \emph{\keyelement{}} $[\kappa]$, which is not published. The \emph{$\alpha$-copies} $[\alpha T_{ij}]$ and $[\alpha y_i]$ and the element $[\alpha\kappa]$ serve the search.

\begin{remark}[Why the table is symmetric]\label{rem:symmetric-table}
Hair and Sahai's table $y r^\top$ is not symmetric, so the table $y r^\top + M + \kappa\, e_0 e_0^\top$ would be published with all its $(N+1)^2$ entries, and anyone could combine it along a matrix that is not symmetric. Take $e_i e_j^\top$ for two unknowns $w_i$ and $w_j$, $i \ne j$, that no constraint multiplies together. Along it the \testmask{} gives $M_{ij} = 0$ and the key term gives zero, so the combination is $[y_i r_j]$: a cheat of rank one, whose search term hides the single secret integer $r_j$. Its corner is zero, but a search of about $\sqrt{g}$ group operations reveals $r_j$. The symmetric table rules this out. For every $W$, $\langle T,W \rangle = \langle T, \frac12 (W + W^\top) \rangle$, and the published $[T_{ij}]$ with $i \le j$ give exactly the combinations along symmetric matrices. On a system without witnesses, every nonzero symmetric passing matrix has rank at least two (\cref{lem:rank-one}), so no cheat of rank one is left. Cheats of rank two remain, such as $e_i e_j^\top + e_j e_i^\top$, with the search term $y_i r_j + y_j r_i$; \cref{sec:weights-ties} deals with them. Looking ahead, in \cref{sec:thin-table}, a symmetric table would double the width of its factors; there, a combination along a matrix that is not symmetric is masked instead, by a term that adds no width.
\end{remark}

\parhead{Decryption}
A decryptor with a witness $w$ combines the $[T_{ij}]$ along $w w^\top$. Since $w$ is Boolean, this is the sum of the $[T_{ij}]$ over all $i$ and $j$ with $w_i = w_j = 1$, each published entry off the diagonal counted twice, for $(i,j)$ and $(j,i)$. It encodes $\langle T, w w^\top \rangle = w^\top T\, w$, which we compute term by term, writing $y(w) = \sum_i w_i y_i$ and $r(w) = \sum_i w_i r_i$ as before:
\begin{align*}
  w^\top T\, w
  &= \tfrac12 (w^\top y)(r^\top w) + \tfrac12 (w^\top r)(y^\top w) + w^\top M w + \kappa\, w_0^2 \\
  &= y(w)\, r(w) + \langle M, w w^\top \rangle + \kappa\, w_0^2 \\
  &= \kappa + \psi\, r(w) , \qquad\text{where}\qquad \psi \defeq y(w) .
\end{align*}
The \testmask{} vanishes because $w$ is a witness: it satisfies every constraint, so for every $t$, $\langle C_t, w w^\top \rangle = a_t(w)\, b_t(w) - c_t(w) = 0$, and $\langle M, w w^\top \rangle = \sum_t \mu_t \langle C_t, w w^\top \rangle$, a random linear combination of these zeros, is zero too. And $\kappa\, w_0^2 = \kappa$ because $w_0 = 1$. The decryptor now holds $[\kappa + \psi\, r(w)]$, the \keyscalar{} shifted by a product of two numbers that it does not know: the \emph{multiplier} $\psi$, a hidden scalar, and $r(w)$, a sum of entries of $r$. It can compute $[\psi] = \sum_i w_i [y_i]$, so if it knew the integer $r(w)$, it would subtract $r(w) \cdot [\psi]$ and be left with $[\kappa]$. It does not, and nothing so far tells it for which integer $\hat r$ the candidate $[\kappa + \psi\, r(w)] - \hat r \cdot [\psi]$ is $[\kappa]$.

The $\alpha$-copies supply this check. The same combination of the $[\alpha T_{ij}]$ gives $[\alpha\kappa + \alpha\psi\, r(w)]$, and here the \keyscalar{} can be removed, because $[\alpha\kappa]$ is published. What is left is
\[
  [\alpha\psi\, r(w)] = r(w) \cdot [\alpha\psi] , \qquad\text{with the base}\qquad [\alpha\psi] = \sum_i w_i\, [\alpha y_i] .
\]
The decryptor thus holds a base and $r(w)$ times that base, and searches for the integer between them, a discrete logarithm known to lie in a short range. The plain entries allow no such step, since removing $\kappa$ from them would take $[\kappa]$ itself.

The search is short because the decryptor knows roughly where $r(w)$ lies. Up to rounding, the integer $r(w)$ differs from its mean, the \emph{center} $c_w$, which the decryptor computes from $g$ and the number $\wt(w)$ of ones among $w_1, \dots, w_N$, by a sum of independent terms, and so typically by a distance of the order of $g \sqrt{N}$ at most. A baby-step giant-step search outward from the center finds an integer $\hat r$ with $\hat r \cdot [\alpha\psi] = r(w) \cdot [\alpha\psi]$ in $O(\sqrt{\lvert r(w) - c_w \rvert} + 1)$ group operations, so in $O(\sqrt{g}\, N^{1/4})$ group operations in expectation, against $O(\sqrt{gN})$ for a search of the whole range. As $\alpha \ne 0$, also $\hat r \cdot [\psi] = r(w) \cdot [\psi]$, so $[\kappa + \psi\, r(w)] - \hat r \cdot [\psi]$ is the \keyelement{} $[\kappa]$, and the key derived from it unmasks the message.

\parhead{Why the unknowns are Boolean}
The search needs $r(w)$ in a short known range. For $w \in \bits^{N+1}$, $r(w)$ is a sum of entries of $r$, an integer near its center (\cref{lem:range}):
\[
  r(w) = \sum_{i \,:\, w_i = 1} r_i \in \ZZ , \qquad\text{at most } (N+1)(g+1) \text{ from the center,}
\]
and much closer for most $r$. For an unrestricted assignment $w \in \FF^{N+1}$, $r(w) = \sum_i w_i r_i$ is an element of $\FF$ for which no short range is guaranteed, so a test must confine each unknown to a few small integers. One test on a single unknown is a quadratic equation in it, with at most two roots, so $w_j^2 = w_0 w_j$, with the roots $0$ and $1$, is the cheapest test that keeps $r(w)$ in a short known range; a larger set of small values would take more tests and auxiliary unknowns. No generality is lost, since every NP relation can be written as a system with Boolean unknowns (\cref{sec:quadratic-systems}), and Booleanity costs one test for each unknown.

\begin{figure}[t]
\centering
\begin{tikzpicture}[
    font=\footnotesize, line width=0.5pt, >=stealth,
    box/.style={draw=black!55, rounded corners=2pt, align=left, inner sep=3.5pt,
      minimum height=1.15cm, anchor=west,
      execute at begin node={\hyphenpenalty=10000 \exhyphenpenalty=10000 }},
  ]
  \definecolor{figblue}{HTML}{3871C1}
  \definecolor{figrust}{HTML}{7A3D00}
  \foreach \y/\col/\a/\b/\c in {
      0/black!55/{$W$ fails a test}/{the \testmask{} does not cancel: a uniform group element}/{of no use},
      -1.4/figblue/{\textcolor{figblue}{\textbf{a witness:}} \mbox{$W = w w^\top$}, rank one}/{$[\kappa + \psi\, r(w)]$: the \keyscalar{}, shifted by a multiple of \textbf{one} unknown integer $r(w)$}/{a search: about $\sqrt{g}\,N^{1/4}$ group operations},
      -2.9/figrust/{\textcolor{figrust}{\textbf{a cheat:}} a passing $W$ of rank $d \ge 2$, on a false system}/{$[\kappa\, W_{00} + \sum_{\ell \le d} (y^\top u_\ell)(v_\ell^\top r)]$: $W_{00}$ times the \keyscalar{}, shifted by $d$ products with $d$ unknown numbers $v_\ell^\top r$}/{a guess of $d$ numbers together: probability at most $g^{-d}$}} {
    \node[box, draw=\col, text width=2.3cm] (a) at (0,\y) {\a};
    \node[box, draw=\col, text width=5.2cm] (b) at (3.0,\y) {\b};
    \node[box, draw=\col, text width=3.0cm] (c) at (8.9,\y) {\c};
    \draw[->] (a) -- (b);
    \draw[->] (b) -- (c);
  }
\end{tikzpicture} 
\caption{What a combination of the ciphertext along a symmetric matrix $W$ gives, and what it costs to remove the search term. The rank $d$ of a passing $W$ is the number of unknown numbers $v_\ell^\top r$ that must be guessed together.}\label{fig:combination}
\end{figure}
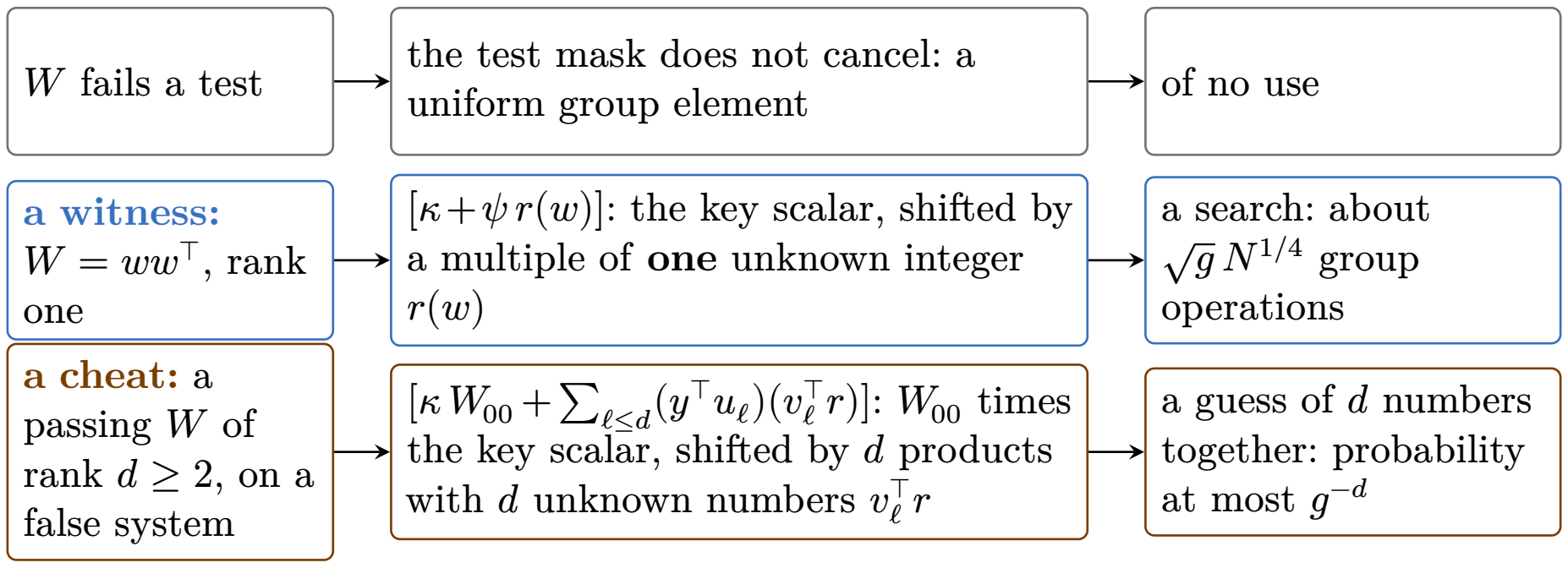

\parhead{Rank decides what must be guessed}
Anyone can combine the ciphertext along any symmetric $W$, not only along the matrix of a witness, and gets $[\kappa\, W_{00} + y^\top W r + \langle M,W \rangle]$. If $W$ fails a test, the \testmask{} makes this a uniform element, of no use. If $W$ passes, it is $[\kappa\, W_{00} + y^\top W r]$, and the search term stands between its holder and the key. Let $d$ be the rank of $W$, and write $W$ as a sum of $d$ matrices of rank one, $W = \sum_{\ell=1}^{d} u_\ell v_\ell^\top$, with $u_1, \dots, u_d$ linearly independent and $v_1, \dots, v_d$ a basis of the row space of $W$. Then
\[
  y^\top W r = \sum_{\ell=1}^{d} \bigl( y^\top u_\ell \bigr) \bigl( v_\ell^\top r \bigr) .
\]
Each of the $d$ products has the form of the honest $\psi\, r(w)$: a hidden multiplier $y^\top u_\ell$, whose encoding $\sum_i u_{\ell,i}\, [y_i]$ anyone can compute, times a number $v_\ell^\top r$, a combination of the entries of $r$ whose coefficients are known and whose value is not. As $y$ is uniform and the $u_\ell$ are linearly independent, the $d$ multipliers are independent uniform scalars, so, apart from chance coincidences in $\FF$, no product cancels another: the search term is removed only by subtracting $\sum_\ell (v_\ell^\top r) \cdot [y^\top u_\ell]$ with the right value of every $v_\ell^\top r$. A passing matrix of rank $d$ thus hides the key behind $d$ unknown numbers, and using it costs a guess of all of them together, since a comparison of group elements succeeds only when the whole term is gone (\cref{fig:combination}). For $d = 1$ the matrix is a multiple of $w w^\top$ for a witness $w$, and the one number is $r(w)$. A cheat has $d \ge 2$: of the two cheats of the previous subsection, the mixture $(1-\chi)\, u u^\top + \chi\, u' u'^\top$ of two Boolean assignments $u$ and $u'$ has the search term $(1-\chi)\, y(u)\, r(u) + \chi\, y(u')\, r(u')$, with the unknown integers $r(u)$ and $r(u')$, and $e_i e_j^\top + e_j e_i^\top$ has $y_i r_j + y_j r_i$, with two of the secret integers themselves.

The \emph{guessing probability} $\gp(W) \defeq \max_x \Pr_r[Wr = x]$ is the probability that the best guess is right: the vector $W r = \sum_\ell (v_\ell^\top r)\, u_\ell$ carries the same information as the $d$ numbers. When the $v_\ell$ are unit vectors $e_i$ with $i \ge 1$, as in the second cheat above, the $d$ numbers are $d$ of the secret integers, which together are uniform on $g^d$ values: $\gp(W) = g^{-d}$, and a generic search finds them in about $g^{d/2}$ group operations. No matrix of rank $d$ is easier to guess: $\gp(W) \le g^{-\rank W}$ for every $W$ (\cref{lem:gp-rank}). Even when $Wr$ concentrates near a center, as $r(w)$ does in the honest search, each single guess is right with probability at most $g^{-\rank W}$.

\parhead{The least rank of a cheat sets the range}
The argument of Hair and Sahai~\cite[Lemma~3.4 and Theorem~3.5]{HS26} in the generic group model (GGM), adapted to our table, says that on a system without witnesses, apart from identities, coincidences in $\FF$, and encodings guessed at random, a comparison of two combinations of the published elements succeeds only if it amounts to a correct guess of $W r$ for a nonzero passing matrix $W$, that is, for a cheat. So an adversary gains only by choosing a cheat and guessing its numbers. One guess is right with probability at most $\gp(W)$, and $Q$ group operations, once $Q$ is at least the number of published elements, give a number of comparisons quadratic in $Q$. If every cheat has $\gp(W) \le \theta$, for a \emph{threshold} $\theta$ of about $2^{-2\lambda}$, all comparisons together succeed with probability at most about $Q^2 \theta$, which is at most about $Q/2^\lambda$ for $Q \le 2^\lambda$.

A cheat of rank $d$ has $\gp(W) \le g^{-d}$. So if every cheat has rank at least $d$, a range with $g^d \ge 1/\theta$ suffices, and the least such range is $g \approx 2^{2\lambda/d}$; and a smaller range would not do if some cheat of rank $d$ rests on unit vectors, since a generic search would find its $d$ secret integers in about $g^{d/2} < 2^\lambda$ group operations. The least rank of a cheat thus sets the range $g$, and the range sets the cost of honest decryption, a search of about $\sqrt{g}\, N^{1/4} \approx 2^{\lambda/d} N^{1/4}$ group operations (\cref{fig:costs}). Hair and Sahai keep the range polynomial in the size of their ciphertext and raise the least rank with many weighted copies, so at $\lambda$-bit security, for $N$ polynomial in $\lambda$ and at least $\lambda$, their costs grow faster than any polynomial in $\lambda$ (\cref{app:hair-sahai}). We fix the least rank and take a range exponential in $\lambda$: size and encryption stay polynomial in $\lambda$ and $N$, and only the search is exponential in $\lambda$.

\begin{table}[tb]
\caption{If only cheats of rank at least $d$ are left to guessing, the range $g$ of each secret integer needs $g^d \ge 1/\theta$, with $\theta \approx 2^{-2\lambda}$, and the range sets the honest search, counted in group operations. With one table, cheats of rank two exist, so $d = 2$. \scheme takes $d = 4$; cheats of rank two and three still exist, and the next subsections show how \scheme deals with them.}\label{fig:costs}
\centering\small
\begin{tabular}{@{}c@{\hspace{2em}}c@{\hspace{2em}}c@{\hspace{2em}}l@{}}
\toprule
rank $d$ & range $g$ & honest search & \\
\midrule
$2$ & $2^{\lambda}$ & $2^{\lambda/2}\, N^{1/4}$ & one table (this subsection)\\
$3$ & $2^{2\lambda/3}$ & $2^{\lambda/3}\, N^{1/4}$ & \\
$4$ & $2^{\lambda/2}$ & $2^{\lambda/4}\, N^{1/4}$ & \scheme\\
\midrule
$d$ & $2^{2\lambda/d}$ & $2^{\lambda/d}\, N^{1/4}$ & \\
\bottomrule
\end{tabular} 
\end{table}

\parhead{Ciphertext size and search time with one table}
With a single table and a system of linear size, the template's ciphertext, encryption, and decryption outside the search are quadratic in the number of unknowns. One \testmask{} serves all the tests, so the size of the ciphertext does not depend on the number $\nc$ of constraints other than the $N$ Booleanity constraints. But the table is published entry by entry: about $N^2/2$ entries, each twice, plainly and as an $\alpha$-copy. And a cheat can have rank two: when no constraint multiplies $w_i$ and $w_j$ together, $e_i e_j^\top + e_j e_i^\top$ passes, with guessing probability $\gp = g^{-2}$. Its corner is zero, so guessing it does not give the key, but it reveals $r_i$ and $r_j$, which later comparisons can use; the analysis therefore follows every cheat, whatever its corner, and requires $g^2 \ge 1/\theta$. So $g \approx 2^\lambda$, and the search takes about $2^{\lambda/2} N^{1/4}$ group operations. \Cref{sec:thin-table} shrinks the ciphertext, and \cref{sec:weights-ties} shortens the search.

\begin{example}[The template on the toy system]\label{ex:toy-template}
On the true system of \cref{ex:toy-system}, take $g = 8$ and the secret integers $r = (8,9,3,4)$; and take $y = (22,2,16,46)$, $\kappa = 13$, and $\alpha = 7$. Whatever \testmask{} is drawn, the entries of $T$ that $w w^\top$ selects sum to $\kappa + \psi\, r(w) = 69$, with $\psi = y(w) = 70$ and $r(w) = 21$. The $\alpha$-copies give $[\alpha\psi\, r(w)] = [89]$ with the base $[\alpha\psi] = [86]$; the search, around the center $16$ (\cref{fig:dec}), returns $21$, and $69 - 21 \cdot 70 = 13$ is the \keyscalar{}. On the false system, the cheat $\frac12 (u u^\top + u' u'^\top)$ gives $\kappa + \frac12 \bigl( y(u)\, r(u) + y(u')\, r(u') \bigr) = 63$, which hides the \keyscalar{} behind the two integers $r(u) = 21$ and $r(u') = 24$, with the multipliers $\frac12 y(u) = 35$ and $\frac12 y(u') = 43$.
\end{example} 
\subsection{Reducing Ciphertext Size from Quadratic to Linear via a QAP and Pairings}\label{sec:thin-table}

This subsection makes the template's ciphertext linear in $N$. We replace the \testmask{} $M$ and the key term $\kappa\, e_0 e_0^\top$ by a single term $A(\tau)\, B(\tau)^\top$ of rank one built from a QAP, so that our table $T$ has two factors $F$ and $H$ of width two, which a pairing multiplies; we then build the QAP's polynomials $A(X)$ and $B(X)$ so that this term tests a matrix $W$ as $M$ does, and give the ciphertext and decryption.

The template's ciphertext is quadratic in $N$ because of one term of its table $T$. In
\[
  T = \underbrace{\tfrac12 \bigl( y\, r^\top + r\, y^\top \bigr)}_{\text{search term}} + \underbrace{M}_{\text{\testmask{}}} + \underbrace{\kappa\, e_0 e_0^\top}_{\text{key term}} , \qquad M = \sum_t \mu_t C_t ,
\]
the search term has rank at most two and the key term rank at most one, but the \testmask{} $M$ can have high rank, so the template publishes the table $T$ entry by entry. Yet each $C_t$ sums a matrix $\frac12 ( a b^\top + b a^\top )$ for each pair $(a,b)$ of vectors of the test of constraint $t$ (\cref{eq:test-products}), and each $a b^\top$ has rank at most one.

We replace $M$ and the key term by a single matrix of rank at most one, $A(\tau)\, B(\tau)^\top \in \FF^{(N+1) \times (N+1)}$, made of the values at a hidden point $\tau \in \FF$ of two public vectors of polynomials $A(X), B(X) \in \FF[X]^{N+1}$, which interpolate the pairs' left and right vectors. With the template's $y$ and $r$ and a hidden scalar $\gamma \in \FF^*$, our table $T \in \FF^{(N+1) \times (N+1)}$ is
\begin{align*}
  T &\defeq \gamma\, \bigl( \underbrace{y\, r^\top}_{\text{search}} + \underbrace{y\, A(\tau)^\top - A(\tau)\, y^\top}_{\text{symmetry}} + \underbrace{A(\tau)\, B(\tau)^\top}_{\text{QAP}} \bigr) \\
    &= \gamma\, \bigl( y\, ( r + A(\tau) )^\top + A(\tau)\, ( B(\tau) - y )^\top \bigr) .
\end{align*}
The second line is a sum of two outer products, whose columns and rows form two \emph{factors} of width two, $F \in \FF^{(N+1) \times 2}$ and $H \in \FF^{2 \times (N+1)}$: with $\gamma = \pi s$ for hidden scalars $\pi, s \in \FF^*$,
\[
  T = \underbrace{\pi \begin{pmatrix} y & A(\tau) \end{pmatrix}}_{F} \; \underbrace{s \begin{pmatrix} ( r + A(\tau) )^\top \\ ( B(\tau) - y )^\top \end{pmatrix}}_{H} .
\]
The ciphertext holds the left factor $F$ in $\GG_1$ and the right factor $H$ in $\GG_2$: $4(N+1)$ elements, not $(N+1)^2$. Our table factors because it has rank at most two; that puts no limit on the rank of a matrix along which an adversary combines its entries, and the least rank of a cheat still sets the range (\cref{sec:rank-template}). Row $i$ of $F$ and column $j$ of $H$ give $[T_{ij}]_T$ with two pairings, one for each of the two products summing to $T_{ij}$. Pairing a factor with a generator keeps only its scale, $\pi$ or $s$, not $\gamma$, and so cancels no term of the table.

Our table $T$ is \emph{virtual}: its factors $F$ and $H$ define it, and no one forms it. Factors alone shrink only the ciphertext, while the template's table also makes encryption and decryption quadratic in $N$. Here our table's share of each is linear, each for its own reason: the encryptor computes only $F$ and $H$, and a decryptor with a witness $w$ combines along $w w^\top$, of rank one, computing $\langle T, w w^\top \rangle = (w^\top F)(H w)$ with two sums and two pairings.

The search term $y\, r^\top$ is the template's~\cite{HS26}, no longer made symmetric: along a symmetric $W$, both give $y^\top W r$. Along one that fails a test, the \emph{QAP term} $A(\tau)\, B(\tau)^\top$ acts as the mask, leaving a nonzero polynomial in $\tau$ that no published element removes. Along one that passes, it acts as the key term, giving $\kappa\, W_{00}$ plus parts the holder of $W$ can remove. Its polynomials form a \emph{QAP}, a variant of those of Gennaro et al.~\cite{EC:GGPR13}. The \emph{\symmetryterm{}} is new. The template's table is symmetric, so a combination along any matrix is one along its symmetric part (\cref{rem:symmetric-table}). Ours could be made symmetric too, but $\frac12 \bigl( y\, r^\top + r\, y^\top \bigr) + \frac12 \bigl( A(\tau)\, B(\tau)^\top + B(\tau)\, A(\tau)^\top \bigr)$ has rank four in general, so its factors would have width four, doubling the ciphertext. So our search and QAP terms are not symmetric, and the factors give an element for $T_{ij}$ and another for $T_{ji}$, which an adversary can weight differently: a non-symmetric matrix such as $e_i e_j^\top$ can then pass every test with rank one. The \symmetryterm{} tests symmetry instead, without widening the factors: it is zero along a symmetric $W$ and masks a combination along any other. It remains to build $A(X)$ and $B(X)$ so that the QAP term tests $W$ as the \testmask{} does.

\parhead{What our table must test}
As in the template, each constraint gives a test, a linear equation in the entries of $W$. Symmetry adds its own tests. Each test sums the terms of two pairs (\cref{eq:test-products}). For the Booleanity of $w_j$, these are $(e_j, e_j)$ and $(e_0, -e_j)$.
\begin{align*}
  &\text{test of constraint } t \text{:} & a_t^\top W\, b_t + e_0^\top W\, (-c_t) &= 0 , \\
  &\text{symmetry test at } i < j \text{:} & e_i^\top W\, e_j - e_j^\top W\, e_i &= 0 .
\end{align*}
With $\nc$ constraints besides Booleanity, there are $\nc + N$ tests of constraints and $N(N+1)/2$ of symmetry. As before, $W$ passes when the tests of constraints hold, now whether symmetric or not. The QAP term tests the constraints, whose pairs $A(X)$ and $B(X)$ interpolate. Its cost grows with the number of tests, so the symmetry tests would make the ciphertext quadratic in $N$ again: symmetry is the barrier, and the \symmetryterm{} handles it.

\parhead{From R1CS to a QAP}
The \testmask{} gives each test a hidden coefficient $\mu_t$, but can have high rank. A QAP~\cite{EC:GGPR13} checks all the constraints at once with polynomials, and, along a matrix, gives such hidden coefficients with a single matrix of rank one. Write the vectors of the constraints as the rows of three matrices, with $a_t^\top$, $b_t^\top$, and $c_t^\top$ as row $t$, so that column $i$ of each holds the coefficients of $w_i$ in all the constraints. Give constraint $t$ its own \emph{evaluation point} $p_t \in \FF$; the \emph{evaluation domain} $\mathcal{P}$ is the set of these $D$ points, and the \emph{Lagrange basis polynomial} $L_p(X) \in \FF^{<D}[X]$ of $p \in \mathcal{P}$ is $1$ at $p$ and $0$ elsewhere on $\mathcal{P}$. Interpolating column $i$ of each matrix over $\mathcal{P}$ gives, for $i = 0, \dots, N$,
\[
  A_i(X) \defeq \sum_t a_{t,i}\, L_{p_t}(X) , \qquad B_i(X) \defeq \sum_t b_{t,i}\, L_{p_t}(X) ,
\]
and likewise $C_i(X)$ from the entries $c_{t,i}$, so that the vectors $A(X)$, $B(X)$, and $C(X)$, with the entries $A_i(X)$, $B_i(X)$, and $C_i(X)$, have the values $a_t$, $b_t$, and $c_t$ at $p_t$. For an assignment $w$, the polynomial $A(X)^\top w = \sum_i w_i\, A_i(X)$ thus has the value $a_t(w)$ at $p_t$, and likewise for $B$ and $C$. So $w$ satisfies every constraint exactly when $(A(X)^\top w)\, (B(X)^\top w) - C(X)^\top w$ vanishes on $\mathcal{P}$, that is, when the vanishing polynomial $Z_{\mathcal{P}}(X)$ divides it: when, for some \emph{quotient} $h(X) \in \FF^{<D-1}[X]$,
\[
  \bigl( A(X)^\top w \bigr) \bigl( B(X)^\top w \bigr) - C(X)^\top w = h(X)\, Z_{\mathcal{P}}(X) .
\]

\parhead{The QAP along a matrix}
As in \cref{sec:tests-rank}, every term is linear in $W = w w^\top$: $(A(X)^\top w)\, (B(X)^\top w) = A(X)^\top W\, B(X)$ and, as $w_0 = 1$, $C(X)^\top w = e_0^\top W\, C(X)$. For any matrix $W$, the products give the polynomial $P_W(X) \in \FF[X]$,
\[
  P_W(X) \defeq A(X)^\top W\, B(X) = \sum_{i=0}^{N} \sum_{j=0}^{N} W_{ij}\, A_i(X)\, B_j(X) ,
\]
and at a hidden point $\tau \in \FF$, $P_W(\tau) = \langle A(\tau)\, B(\tau)^\top, W \rangle$ is the combination along $W$ of the one matrix $A(\tau)\, B(\tau)^\top$, of rank at most one: the QAP term of our table. Since each test sums the terms of pairs, we let each point carry one pair: the \emph{pair at $p$} is $(a_p, b_p)$, with $A(p) = a_p$ and $B(p) = b_p$, so that $A_i(X) = \sum_{p \in \mathcal{P}} a_{p,i}\, L_p(X)$ for the entry $a_{p,i}$ of $a_p$, and likewise for $B_i(X)$; above, the pair at $p_t$ is $(a_t, b_t)$. Then $P_W(p) = a_p^\top W\, b_p$ at each $p$. As $P_W(X)$, of degree at most $2D - 2$, agrees on $\mathcal{P}$ with $\sum_p ( a_p^\top W\, b_p )\, L_p(X)$, there is $h_W(X) \in \FF^{<D-1}[X]$ with
\begin{equation}\label{eq:qap-lagrange}
  P_W(X) = \sum_{p \in \mathcal{P}} \bigl( a_p^\top W\, b_p \bigr)\, L_p(X) + h_W(X)\, Z_{\mathcal{P}}(X) .
\end{equation}
At $X = \tau$, along every $W$, the QAP term gives the terms of all pairs with the hidden coefficients $L_p(\tau)$, plus $h_W(\tau)\, Z_{\mathcal{P}}(\tau)$, which whoever knows $W$ can remove, given encodings of powers of $\tau$.

\parhead{Pairs of points for the constraints}
The right sides need a change. In the QAP above they form a third vector $C(X)$, which would put the term $e_0\, C(\tau)^\top$ beside $A(\tau)\, B(\tau)^\top$ and widen our table's factors to three. Instead, each right side becomes a pair of its own, $(e_0, -c_t)$, as in \cref{eq:test-products}, at a second point. The test of constraint $t$ gets its own pair of points, $\delta_t$ and $-\delta_t$ for a $\delta_t \in \FF^*$: the product's pair $(a_t, b_t)$ sits at $\delta_t$, and the right side's $(e_0, -c_t)$ at $-\delta_t$. The domain $\mathcal{P}$ is now the union of these pairs of points, the $D = 2\, (\nc + N)$ \emph{tests' points} $\pm\delta_t$, and the test of constraint $t$ sets the sum of $P_W(X)$ over its two points to zero, $P_W(\delta_t) + P_W(-\delta_t) = 0$. A passing $W$ zeroes only the sum of each test's two terms, not each term, so the division of $P_W(X)$ by $Z_{\mathcal{P}}(X)$ is no longer exact: the sum in \cref{eq:qap-lagrange} remains. To remove it at $\tau$, the holder of $W$ needs encoded powers of $\tau$. The encryptor chooses which powers it publishes, so we choose points whose test sums depend on coefficients it can require to vanish.

One sum over all the tests' points would check only the total of their terms. The sum of a polynomial over $\delta$ and $-\delta$ is twice the value at $\delta$ of its even part, its terms of even exponent, which has the same value at $-\delta$. So a polynomial of degree below $2\, (\nc + N)$ sums to zero over each of these pairs of points exactly when its even part vanishes at all the tests' points, that is, when it has odd exponents only. We call such a polynomial \emph{allowed}.

\Cref{eq:qap-lagrange} divides $P_W(X)$ by the vanishing polynomial $Z_{\mathcal{P}}(X)$: the quotient is $h_W(X)$, and the \emph{remainder}
\[
  R_W(X) \defeq \sum_{p \in \mathcal{P}} \bigl( a_p^\top W\, b_p \bigr)\, L_p(X) \in \FF^{<D}[X]
\]
has the terms as its values, $R_W(p) = a_p^\top W\, b_p$. Whoever knows $W$ removes $R_W(\tau)$ exactly when $R_W(X)$ is allowed: when $W$ passes.

\parhead{A point for the key}
The QAP term must carry the key. The template's key term gives $\kappa\, W_{00}$, where the corner $W_{00}$ is $1$ for a witness's matrix. So the corner gets its own point, the \emph{key point} $\rho$, not among the tests' points $\pm\delta_t$, with the pair $(e_0, e_0)$ and the term $P_W(\rho) = W_{00}$; $\mathcal{P}$ is now the tests' points and $\rho$, and the sum in \cref{eq:qap-lagrange} gains $W_{00}\, L_\rho(X)$:
\[
  P_W(X) = W_{00}\, L_\rho(X) + \sum_{p \in \mathcal{P} \setminus \{\rho\}} \bigl( a_p^\top W\, b_p \bigr)\, L_p(X) + h_W(X)\, Z_{\mathcal{P}}(X) ,
\]
with $L_p(X)$ and $Z_{\mathcal{P}}(X)$ now those of the larger domain. This pair is in no test, so nothing asks its term to vanish: for a witness's matrix, the QAP term keeps $L_\rho(\tau)$, which only the encryptor knows. The \keyscalar{} is $\gamma\, L_\rho(\tau)$.

\parhead{The \symmetryterm{}}
The $N(N+1)/2$ symmetry tests are too many for evaluation points: any placement takes at least one point each, and each point costs the ciphertext one to two powers of $\tau$. Instead they get a term of our table, the \symmetryterm{} $y\, A(\tau)^\top - A(\tau)\, y^\top$, made of $y$ and $A(\tau)$, which the search and QAP terms already use, so the factors need no new vectors. Along $W$ it gives
\[
  y^\top ( W - W^\top )\, A(\tau) = \sum_{i < j} \bigl( y_i\, A_j(\tau) - y_j\, A_i(\tau) \bigr)\, ( W_{ij} - W_{ji} ) ,
\]
a combination of the symmetry tests with hidden coefficients. It is zero for symmetric $W$, and otherwise a nonzero polynomial in $y$ and $\tau$ that no published element removes: the values $e_0$ of $A(X)$ at the key point and nonzero multiples of $e_j$ at the point $\delta_t$ of the Booleanity of each $w_j$ span $\FF^{N+1}$, so $( W - W^\top )\, A(X) \ne 0$.

\parhead{Putting the pieces together}
The QAP consists of $A(X)$ and $B(X)$ on the domain $\mathcal{P}$ of $D = 2 + 2\, (\nc + N)$ points: the tests' points $\pm\delta_t$ and, outside them, the \emph{single points} $\mathcal{P}_0 \defeq \{d_0, \rho\}$. The point $d_0$ is in no test: its pair $(0, e_0)$ gives $A(d_0) = 0$, for the security argument alone. With $Z_0(X) \defeq Z_{\mathcal{P}_0}(X)$, let $R_W(X) \in \FF^{<2 (\nc + N)}[X]$ now have the value $P_W(p) / Z_0(p)$ at each of the tests' points $p$, where $Z_0(p) \ne 0$. The sum in \cref{eq:qap-lagrange} splits by kind of point: for every $W$, omitting $X$,
\begin{equation}\label{eq:qap-expansion}
  P_W = \underbrace{W_{00}\, L_\rho}_{\text{corner}} + \underbrace{Z_0\, R_W}_{\text{tests}} + \underbrace{h_W\, Z_{\mathcal{P}}}_{\text{quotient}} .
\end{equation}
The first part is the term of $\rho$ with its Lagrange polynomial; the second vanishes on $\mathcal{P}_0$ and equals $P_W(X)$ on the tests' points. At $\rho$ and $d_0$, the pairs give $W_{00}$ and $0$; at one of the tests' points $p$, the pair's left vector is scaled by $Z_0(p)$, so that $R_W(p) = P_W(p)/Z_0(p)$ is the term of the test's pair there. So $W$ passes exactly when $R_W(X)$ is allowed.

\parhead{The \qapcriterion{}}
The \emph{\qapcriterion{}} (\cref{lem:qap}) combines the tests into one identity: a symmetric $W$ passes exactly when there are a polynomial $h(X)$, the quotient, and an allowed $R(X)$, the remainder, with
\[
  P_W(X) - W_{00}\, L_\rho(X) = h(X)\, Z_{\mathcal{P}}(X) + Z_0(X)\, R(X) .
\]
For a passing $W$, this is \cref{eq:qap-expansion}, with $h = h_W$ and $R = R_W$. Passing does not normalize the corner $W_{00}$ to $1$ (\cref{sec:tests-rank}). The identity puts $h(X)$ in $\FF^{<D-1}[X]$, bounding the powers of $\tau$ needed for the quotient (below). Being allowed bounds the degree of $R(X)$, which the powers for the remainder (below) enforce by stopping below $2\, (\nc + N)$; security rests on their skipping the even exponents.

\parhead{The table}
The QAP is public: a deterministic setup with no secret computes the evaluation points, $A(X)$, and $B(X)$ from the system alone. The encryptor samples $y$ and $r$ as in the template, a hidden point $\tau \sample \FF \setminus \mathcal{P}$, and hidden scalars $\pi, s \sample \FF^*$. With $\gamma \defeq \pi s$, these define the opening's virtual table $T$, and for every $W$,
\[
  \langle T,W \rangle = \gamma\, \bigl( y^\top W r + y^\top ( W - W^\top )\, A(\tau) + P_W(\tau) \bigr) .
\]
The \keyscalar{} is no longer sampled but computed: $\kappa = \gamma\, L_\rho(\tau)$. For a symmetric $W$ that passes, the \symmetryterm{} is zero, and the criterion at $X = \tau$ gives $P_W(\tau) = W_{00}\, L_\rho(\tau) + h(\tau)\, Z_{\mathcal{P}}(\tau) + Z_0(\tau)\, R(\tau)$, so that
\[
  \langle T,W \rangle = \underbrace{\gamma\, y^\top W r}_{\text{search term}} + \underbrace{\kappa\, W_{00}}_{\text{corner term}} + \underbrace{\gamma\, h(\tau)\, Z_{\mathcal{P}}(\tau)}_{\text{quotient part}} + \underbrace{\gamma\, Z_0(\tau)\, R(\tau)}_{\text{remainder part}} .
\]

\parhead{The two factors}
Our ciphertext holds our table through its two factors $F$ and $H$ above, with $T_{ij} = F_i H_j$. Published as they stand, $[F_i]_1$ and $[H_j]_2$ would let an adversary pair any coordinate of a row with any of a column, and so weight an entry's two products differently, breaking the security argument's reading of every combination as $\langle T,W \rangle$ for one $W$. So the encryptor publishes $[F_i \Omega]_1 \in \GG_1^2$ and $[\Omega^{-1} H_j]_2 \in \GG_2^2$ for a hidden random invertible matrix $\Omega \in \FF^{2 \times 2}$ with first column $(1, 0)^\top$. The sum $(F_i \Omega)(\Omega^{-1} H_j) = T_{ij}$ of an entry's two products is free of $\Omega$. Dual pairing vector spaces~\cite{PAIRING:OkaTak08,AC:OkaTak09} and canceling pairings~\cite{EC:Freeman10} hide bases so that they cancel only in the pairings a scheme intends; ours, of dimension two, sits between the factors of a virtual table, so that $\Omega$ cancels only in whole entries. By the first column of $\Omega$, the first coordinate of $[F_i \Omega]_1$ is $[\pi y_i]_1$.

\parhead{Removing the quotient and remainder parts}
These parts depend on $W$, so the encryptor cannot cancel them. For each, it publishes encoded powers of $\tau$ in $\GG_1$ and a \emph{base} in $\GG_2$: combining the powers with the coefficients of $h(X)$ or $R(X)$ and pairing with the base gives $[\gamma\, h(\tau)\, Z_{\mathcal{P}}(\tau)]_T$ or $[\gamma\, Z_0(\tau)\, R(\tau)]_T$. Each family of powers carries a hidden scalar of its own, which only its base divides out, so neither family can remove the other's part. Here the pairs of points pay off: the remainder's powers stop below $2\, (\nc + N)$ and skip the even exponents, so every removable remainder is allowed.

\parhead{Decryption}
Besides the factors and the powers with their bases, the ciphertext holds the \emph{search base} $[s]_2$ and, as in the template, the \maskedmessage{} $\hdr$, $[\alpha\kappa]_T$ for a hidden $\alpha \sample \FF^*$, and $\alpha$-copies for the search, here only of the published elements of $\GG_2$, fewer than those of $\GG_1$. From the setup's public polynomials $A_i(X)$ and $B_j(X)$, a decryptor with a witness $w$ computes $A_w(X) \defeq \sum_i w_i A_i(X)$ and $B_w(X) \defeq \sum_j w_j B_j(X)$, whose product is $P_{w w^\top}(X)$. As $w w^\top$ is symmetric and passes with the corner $1$, the criterion gives an allowed remainder $R(X)$, interpolated from its values $A_w(p)\, B_w(p) / Z_0(p)$ on the tests' points as in \cref{eq:qap-expansion}, and a quotient $h(X) \in \FF^{<D-1}[X]$, from dividing $A_w B_w - L_\rho - Z_0 R$ by $Z_{\mathcal{P}}$. Pairing $\sum_i w_i [F_i \Omega]_1 = [w^\top F\, \Omega]_1$ with $\sum_j w_j [\Omega^{-1} H_j]_2 = [\Omega^{-1} H w]_2$, first coordinate with first and second with second, gives $[(w^\top F)(H w)]_T = [\langle T, w w^\top \rangle]_T$. By the display of the four parts of $\langle T,W \rangle$, at $W = w w^\top$, where $y^\top W r = y(w)\, r(w)$, and with the multiplier $\psi \defeq \gamma\, y(w)$,
\[
  \langle T, w w^\top \rangle = \kappa + \psi\, r(w) + \gamma\, h(\tau)\, Z_{\mathcal{P}}(\tau) + \gamma\, Z_0(\tau)\, R(\tau) .
\]
Removing the last two parts leaves $[\kappa + \psi\, r(w)]_T$, and the same on the $\alpha$-copies, less $[\alpha\kappa]_T$, gives $[\alpha\psi\, r(w)]_T$. Pairing $\sum_i w_i [\pi y_i]_1$ with $[s]_2$ and $[\alpha s]_2$ gives $[\psi]_T$ and $[\alpha\psi]_T$. The search and key are as in the template. For systems of linear size, the ciphertext, encryption, and decryption outside the search are now linear in $N$, but the search is no shorter: one table still has cheats of rank two, so $g \approx 2^\lambda$.

\parhead{Sums of products}
In general form (\cref{def:qap}), the pair of points $\pm\delta_t$, a coset of $\{1, -1\}$, becomes a coset of the subgroup of order $\nu$ of $\FF^*$, for a \emph{coset size} $\nu$ above the bound $F_{\max}$ on the number of products in a constraint. All pairs of a constraint's test sit on its coset, and an allowed remainder, now one with no exponent divisible by $\nu$, checks that their terms sum to zero, so a constraint sums many products with no auxiliary unknown (\cref{sec:tests-rank}). A Booleanity test's two pairs merge into one, at a single point, and with $\nu \le 2F_{\max}$ the points stay $O(N)$ for systems of linear size (\cref{sec:qap}). 
\subsection{Weights and Ties}\label{sec:weights-ties}

This subsection shortens the search. Honest decryption has two parts: the search, of about $2^{\lambda/d} N^{1/4}$ group operations for the least rank $d$ of an easily guessed cheat (\cref{fig:costs}), exponential in $\lambda$; and everything else, the sums, multi-scalar multiplications (MSMs), and pairings, polynomial in $\lambda$, about $2^{32.0}$ field multiplications at our parameters (\cref{sec:costs}). The exponential part cannot be removed, but it can be made concretely about as costly as the rest. With one table, $d = 2$, and the search would take about $2^{57}$ multiplications, $2^{25}$ times the rest. We raise $d$ in two steps, by publishing weighted copies of the table and tying them together: linear weights give $d = 3$, a search of about $2^{40}$ multiplications, still a few hundred times the rest; quadratic weights, which \cref{sec:ext-weights} takes from an extension field to keep the ciphertext and the rest of decryption small, give, under MSIS, $d = 4$, and $g \approx 2^{\lambda/2}$, a search of about $2^{32}$ multiplications, about as costly as the rest. For intuition, we first build the copies on the template of \cref{sec:rank-template}, where a tie is enforced as a test is, and only at the end publish them on the factors of \cref{sec:thin-table}.

One table $T$ tests one linear condition $\langle C_t, W \rangle = 0$ for each constraint $t$, which many matrices $W$ of rank two meet. Hair and Sahai stack copies of the table: copy $c$ is multiplied by a \emph{weight} $\xi_c(w)$, a public polynomial in the unknowns $w$. Their weights are built from \emph{linear forms}, the linear polynomials $\varphi(w) = \langle \varphi, w \rangle$ of \cref{sec:tests-rank}, each written with its coefficient vector $\varphi \in \FF_q^{N+1}$. For a pair of forms $(\varphi_1, \varphi_2)$, the six monomials of degree at most two in $\varphi_1(w)$ and $\varphi_2(w)$, from $1$ to $\varphi_2(w)^2$, serve as weights, and with them no cheat of rank two is left~\cite[Section~4]{HS26}; higher degrees exclude higher ranks~\cite[Section~5]{HS26}. Their ciphertext can be combined only along linear combinations of the stacks $(\xi_c(v)\, v v^\top)_c$ of Boolean assignments $v$, and they take $J$ pairs $(\varphi_{j,1}, \varphi_{j,2})$, with $J$ growing with $N$, so that on every subspace $\mathcal{V} \subseteq \FF_q^{N+1}$ of low dimension some pair gives an injective map $v \mapsto (\varphi_{j,1}(v), \varphi_{j,2}(v))$. Three things are new here. An adversary can combine our copies along any family $(W_c)_c$ of matrices, one for each copy, so linear relations between the copies, the \emph{ties}, are enforced in the exponent. We take $n$ forms $\varphi_0, \dots, \varphi_{n-1}$, with $n$ set by the hardness of MSIS, and turn an easily guessed $(W_c)_c$ whose row space contains a nonzero $v$ with $\varphi_e(v) = 0$ for every $e$ into a solution of MSIS. And \scheme's weights of degree two are not all the products $\varphi_e \varphi_{e'}$ but $3n/2$ quadratic forms from an extension field (\cref{sec:ext-weights}).

\parhead{Copies on the template}
Copy $c$ has a public \emph{weight} $\xi_c$, a polynomial in the unknowns, and a hidden scalar $\gamma_c \in \FF_q^*$. Its table is the template's, with the search term scaled by $\gamma_c$:
\[
  T_c \defeq \gamma_c\, \tfrac12 \bigl( y\, r^\top + r\, y^\top \bigr) + M_c ,
\]
plus the key term $\kappa\, e_0 e_0^\top$ in one copy alone, the \emph{\keycopy{}}, of weight $\xikey$. All the copies share $y$ and $r$, and each has a \testmask{} $M_c$ with coefficients of its own. The ciphertext holds every copy as the template holds its table, the entries $[T_{c,ij}]$ with $i \le j$, with $[\gamma_c\, y_i]$ for every $c$ and $i$, the $\alpha$-copies of all of these, and $[\alpha\kappa]$. The \emph{stack of a vector} $v \in \FF_q^{N+1}$ is $\mathbf{H}(v) \defeq (\xi_c(v)\, v v^\top)_c$, and the stack of a witness $w$ is its \emph{\honeststack{}}: the honest decryptor combines copy $c$ along $\xi_c(w)\, w w^\top$. The \testmask{}s vanish along it, so the combinations add up to $\xikey(w)\, \kappa + \psi\, r(w)$, with the single multiplier $\psi = y(w) \sum_c \gamma_c\, \xi_c(w)$, whose encoding is $\sum_c \xi_c(w) \sum_i w_i\, [\gamma_c\, y_i]$: one search still serves, and the decryptor divides by $\xikey(w)$.

A \emph{stack} $\mathbf{W} = (W_c)_c$ is a family of matrices with $N+1$ columns, one for each copy. Its \emph{row space} is the span of the rows of all the $W_c$, and its rank is the dimension of its row space. As the tables are symmetric, an adversary's object is now a stack of symmetric $(N+1) \times (N+1)$ matrices (\cref{rem:symmetric-table}). Along it, the search terms give $\sum_c \gamma_c\, y^\top W_c\, r$, and the hidden products $\gamma_c\, y_i$ are distinct monomials, so using it takes every vector $W_c r$: $\gp(\mathbf{W}) \defeq \max_v \Pr_r[(W_c r)_c = v]$, which is at most $g^{-\rank \mathbf{W}}$ (\cref{lem:gp-rank}); the stack is \emph{light} if $\gp(\mathbf{W}) \le \theta$, for the threshold $\theta \approx 2^{-2\lambda}$ of \cref{sec:rank-template}, and \emph{heavy} otherwise. A stack \emph{passes} when every $W_c$ is symmetric and passes the tests (\cref{def:tests}). Passing alone gains nothing over one table, since a cheat $W$ of one table can be repeated in every copy, $W_c = W$. So the copies are tied together: the \emph{ties} are linear relations between the $W_c$ that the stack $\mathbf{H}(v)$ of every vector satisfies, and the encryptor enforces them as the template enforces the tests, by adding to each table $T_c$ a hidden matrix $\Theta_c$, its share of a random combination of the ties. What the ties are depends on the weights; the warm-up below gives them, and $\Theta_c$, for linear weights. A cheat is now a passing stack that satisfies the ties and is not a combination of \honeststack{}s; on a system without witnesses, every nonzero passing stack that satisfies the ties is one.

\parhead{Warm-up: linear weights and their ties}
In the warm-up the weights are linear. There are $n$ copies, $c = 0, \dots, n-1$, and the weight of copy $c$ is a public linear form, $\xi_c(w) = \varphi_c(w) = \langle \varphi_c, w \rangle$. Copy $0$ is the \keycopy{}, with $\varphi_0 = e_0$, so that $\varphi_0(w) = w_0 = 1$ for every assignment; the coefficient vectors $\varphi_1, \dots, \varphi_{n-1} \in \FF_q^{N+1}$ of the other forms are random.

To find the ties, look for linear relations between the matrices of the stack of a vector $v$, whose copy $c$ holds $\varphi_c(v)\, v v^\top$. Multiply copy $c$ by the coefficient vector $\varphi_{c'}$ of another copy's form; as $v v^\top \varphi_{c'} = \varphi_{c'}(v)\, v$,
\[
  \bigl( \varphi_c(v)\, v v^\top \bigr)\, \varphi_{c'} = \varphi_c(v)\, \varphi_{c'}(v)\, v = \bigl( \varphi_{c'}(v)\, v v^\top \bigr)\, \varphi_c ,
\]
which is symmetric in $c$ and $c'$. The ties require the same of every stack: for all copies $c$ and $c'$,
\[
  W_c\, \varphi_{c'} = W_{c'}\, \varphi_c ,
\]
an equation between vectors of $\FF_q^{N+1}$. A cheat repeated in every copy, $W_c = W$, breaks them unless $W (\varphi_{c'} - \varphi_c) = 0$ for all $c$ and $c'$.

\parhead{Antisymmetric coins enforce the ties}
For each pair of copies $c < c'$, the encryptor samples a uniform hidden coin vector $\eta_{cc'} \in \FF_q^{N+1}$, and sets $\eta_{c'c} \defeq -\eta_{cc'}$: the coins are \emph{antisymmetric} in the pair. The table $T_c$ of copy $c$ gains the symmetric matrix
\[
  \Theta_c \defeq \tfrac12 \sum_{c' \ne c} \bigl( \eta_{cc'}\, \varphi_{c'}^\top + \varphi_{c'}\, \eta_{cc'}^\top \bigr) .
\]
For a symmetric $W_c$, $\langle \Theta_c, W_c \rangle = \sum_{c' \ne c} \eta_{cc'}^\top W_c\, \varphi_{c'}$. Summed over the copies, each coin vector $\eta_{cc'}$ appears twice, in copy $c$ against $\varphi_{c'}$ and, with the opposite sign, in copy $c'$ against $\varphi_c$:
\[
  \sum_c \langle \Theta_c, W_c \rangle = \sum_{c < c'} \eta_{cc'}^\top \bigl( W_c\, \varphi_{c'} - W_{c'}\, \varphi_c \bigr) .
\]
This is a random combination of the ties. It is zero along every stack that satisfies them, the \honeststack{}s among them, so decryption is unchanged; along a stack that breaks a tie, it is a nonzero linear form in uniform coins, hence uniform, and masks the combination as the \testmask{} does. Antisymmetry is what makes the sum a combination of the ties: with independent coins in each copy, the sum would ask every $W_c\, \varphi_{c'}$ to vanish, which the \honeststack{} does not do.

\parhead{Linear weights exclude rank two}
Let a passing stack satisfy these ties, and let its row space $\mathcal{V}$ have dimension two and contain no nonzero $v$ with $\varphi_c(v) = 0$ for every $c$. In three steps, the stack is a combination of the \honeststack{}s of at most two witnesses.

First, the ties make the stack one object. Entry $(i,j)$ of the stack of $v$ in copy $c$ is $\varphi_c(v)\, v_i\, v_j$, the value of the tensor $v \otimes v \otimes v$ on the forms $\varphi_c$, $e_i$, and $e_j$. A symmetric matrix allows the exchange of $e_i$ and $e_j$, and the ties that of $\varphi_c$ and $e_j$; and on $\mathcal{V}$ the forms $\varphi_c$ span all linear forms, because no nonzero $v \in \mathcal{V}$ has $\varphi_c(v) = 0$ for every $c$. So the matrices of the stack are slices of one symmetric tensor of order three over $\mathcal{V}$.

Second, Booleanity, which says that the tensor vanishes on $\varphi$, $e_j$, and $e_j - e_0$ for every form $\varphi$ and every $j$, leaves only the cubes $u \otimes u \otimes u$ of the Boolean assignments $u$ in $\mathcal{V}$, of which a plane contains at most two: $W_c = \chi\, \varphi_c(u)\, u u^\top + \chi'\, \varphi_c(u')\, u' u'^\top$.

Third, the weights let each constraint see each assignment by itself. The test of constraint $t$ in copy $c$ reads $\chi\, \varphi_c(u)\, \res_t(u) + \chi'\, \varphi_c(u')\, \res_t(u') = 0$. In the copy with the weight $w_0$ this is the one condition of a single table, which a mixture meets (\cref{sec:tests-rank}). Over all copies, the pairs $(\varphi_c(u), \varphi_c(u'))$ span $\FF_q^2$, again because no nonzero combination $v$ of $u$ and $u'$ has $\varphi_c(v) = 0$ for every $c$. So $\chi\, \res_t(u) = \chi'\, \res_t(u') = 0$ in every constraint: an assignment with a nonzero coefficient is a witness (\cref{fig:weights}).

\begin{figure}[tb]
\centering
\begin{tikzpicture}[font=\footnotesize, line width=0.5pt]
  \definecolor{figrust}{HTML}{7A3D00}
  \node[anchor=north west, inner sep=0pt] (tab) at (0,0) {\renewcommand{\arraystretch}{1.45}\begin{tabular}{@{}l@{\hspace{1.2em}}l@{\hspace{1.2em}}l@{}}
      \textbf{copy and its form} & \textbf{values at $u$, $u'$} & \textbf{test of constraint 2}\\
      \noalign{\vskip 1pt\hrule height 0.5pt\vskip 1pt}
      $w_0$, the form $(1, 0, 0, 0)$ & $1,\ 1$ & $-\chi + \chi' = 0$\\
      $(0, 55, 33, 6)$ & $61,\ 94$ & $-61\chi + 94\chi' = 0$
    \end{tabular}};
  \node[draw=figrust, rounded corners=2pt, anchor=north west, text width=2.4cm, align=left, inner sep=4pt,
      execute at begin node={\hyphenpenalty=10000 \exhyphenpenalty=10000 }]
    at ([xshift=0.3cm]tab.north east)
    {The copy with weight $w_0$ alone allows $\chi = \chi'$, the unweighted mixture. A second copy with an independent pair of values forces \textcolor{figrust}{$\chi = \chi' = 0$}.};
\end{tikzpicture} 
\caption{Linear weights against a mixture, over $\FF_{101}$. On the false system of \cref{ex:toy-system}, $u$ and $u'$ have residuals $-1$ and $1$ in the second constraint, so the stack $\chi\, \varphi_c(u)\, u u^\top + \chi'\, \varphi_c(u')\, u' u'^\top$ passes that constraint's test in copy $c$ only if $-\chi\, \varphi_c(u) + \chi'\, \varphi_c(u') = 0$.}\label{fig:weights}
\end{figure}

The hypothesis on $\mathcal{V}$ is where MSIS enters: when the row space of a heavy stack contains a nonzero $v$ with $\varphi_c(v) = 0$ for every $c$, a \emph{\shortmultiple{}} of $v$, an integer vector with small entries whose reduction modulo $q$ is a nonzero multiple of $v$, solves MSIS, as the paragraph \emph{From a heavy stack to a short vector} below explains for our weights. So under MSIS no adversary forms a heavy cheat of rank two.

\parhead{Linear weights leave rank three}
Linear weights thus give a gap of one versus three: under MSIS, no adversary forms an easily guessed cheat of rank below three, and $g^3 \ge 1/\theta$ suffices, $g \approx 2^{2\lambda/3}$. This is progress: at our parameters the search falls from about $2^{57}$ to about $2^{40}$ multiplications. But it is still a few hundred times the rest of decryption, so we go one step further, to rank four, and for that a cheat of rank three must be excluded too. Linear weights cannot exclude it. Let no constraint multiply two of the unknowns $w_1$, $w_2$, and $w_3$. The symmetric tensor with the value one on $e_1$, $e_2$, and $e_3$ in any order, and zero on every triple with a repeated index, has the slices
\[
  W_c = \varphi_c(e_3)\, ( e_1 e_2^\top + e_2 e_1^\top ) + \varphi_c(e_2)\, ( e_1 e_3^\top + e_3 e_1^\top ) + \varphi_c(e_1)\, ( e_2 e_3^\top + e_3 e_2^\top ) .
\]
This stack satisfies the ties, and it passes: Booleanity involves only its diagonal and its row $0$, which are zero, and no test of a constraint has a coefficient at its other entries. When no nonzero combination $v$ of $e_1$, $e_2$, and $e_3$ has $\varphi_c(v) = 0$ for every $c$, it has rank three and guessing probability $g^{-3}$ (\cref{lem:gp-rank}), and the reduction to MSIS does not apply to it, so linear weights need $g^3 \ge 1/\theta$. In general, weights of degree $d - 2$ leave such a stack on $d$ unknowns that no constraint multiplies in pairs: the alternating sum $\sum_J (-1)^{d - |J|}\, \mathbf{H}(u_J)$ over the subsets $J$ of these unknowns, where $u_J$ is the Boolean assignment whose ones are the unknowns in $J$; the signs leave only entries $(i,j)$ with $i \ne j$ among these unknowns. So the next step takes weights of degree two, for which it has rank four when no nonzero combination $v$ of the unit vectors of its four unknowns has $\varphi_e(v) = 0$ for every form $\varphi_e$ below.

\parhead{Quadratic weights}
The direct weights of degree two are the products $\varphi_e(w)\, \varphi_{e'}(w)$, one copy for each pair $e \le e'$: the stack of $v$ is then the tensor $v \otimes v \otimes v \otimes v$ on two forms. As in the warm-up, multiplying copy $(e,e')$ of the stack of $v$ by a form's coefficient vector $\varphi_f$ gives $\varphi_e(v)\, \varphi_{e'}(v)\, \varphi_f(v)\, v$, symmetric in $e'$ and $f$, and the ties ask $W_{(e,e')}\, \varphi_f = W_{(e,f)}\, \varphi_{e'}$ of every stack, enforced by antisymmetric coins as there. These weights remove the stack that linear weights leave, the alternating sum $\sum_J (-1)^{3 - |J|}\, \mathbf{H}(u_J)$, whose entry $(i,j)$ in a copy of weight $\xi$ is the coefficient of $w_1 w_2 w_3$ in $\xi(w)\, w_i\, w_j$, reduced with $w_l^2 = w_l$. For a linear $\xi$, only the entries with $i \ne j$ among the three unknowns survive. For a quadratic $\xi$, row $0$ and the diagonal survive too, and the sum remains a combination of stacks of Boolean assignments, which the constraints reject on a system without witnesses, as in the third step of the warm-up. A passing alternating sum now needs four unknowns, and has rank four; that no other cheat of rank at most three survives is what the ties give, which we prove for the weights of \scheme (\cref{sec:ext-weights}).

As $n = O(\lambda^2)$ (\cref{sec:efficiency}), these weights are $n(n+1)/2 = O(\lambda^4)$ copies, more than $2^{23}$ for our $n = \num{4608}$. Even on a shared grid (\cref{sec:shared-factors}), their factors cost at least about $4n$ elements for each unknown, against our $528$, about $35$ times as many. \scheme takes $3n/2 = O(\lambda^2)$ weights of degree two from an extension field instead, for which \cref{app:proofs} proves the gap of one versus four (\cref{sec:ext-weights}; \cref{lem:heavy,lem:classification}).

\parhead{What the ties give}
Call a nonzero vector $v$ a \emph{kernel vector} if $\varphi_e(v) = 0$ for every $e$. For the quadratic weights of \scheme, \cref{lem:classification} shows, along the outline of \cref{sec:ext-weights}, that the three steps of the warm-up carry over, one order higher, to a row space $\mathcal{V}$ of dimension at most three without kernel vectors: a passing stack of rank at most three that satisfies the ties and whose row space has no kernel vector is a combination $\sum_w \chi_w\, \mathbf{H}(w)$ of the \honeststack{}s of witnesses $w$ in its row space, so every cheat of rank at most three has a kernel vector in its row space.

\parhead{From a heavy stack to a short vector}
The first form is $\varphi_0(w) = w_0$, so a kernel vector $v$ has $v_0 = 0$. The other nonzero forms come from a uniformly random matrix $\Lambda$ over the ring $R_q$ of \cref{sec:module-sis} (the matrix $M$ of \cref{def:forms}), which the ciphertext carries or a seed in it determines (\cref{sec:optimizations}): with the entries $w_1, \dots, w_N$ packed as the coefficients of a vector $\hat w$ of ring elements, the forms are the coefficients of the entries of $\Lambda\, \hat w$, plus random multiples of $w_0$. A vector $v$ with $v_0 = 0$ is therefore a kernel vector exactly when $\Lambda\, \hat v = 0$. Kernel vectors exist, since the forms are fewer than the unknowns, and the weights do not see them: two Boolean assignments $u \ne u'$ with $\varphi_e(u) = \varphi_e(u')$ for every $e$ have the same weight in every copy, so a mixture of their stacks passes whenever the mixture of their matrices passes in one table. But $u - u'$ is then a kernel vector with entries in $\{-1, 0, 1\}$, and short kernel vectors are solutions of MSIS.

In general, if a heavy stack $\mathbf{W}$ has a kernel vector $v$ in its row space, then $\gp(v^\top) \ge \gp(\mathbf{W}) > \theta$, because $v^\top r$ is a function of the vectors $W_c r$. An inverse theorem (\cref{lem:inverse}) turns this into a \shortmultiple{} $\bar v$ of $v$ with $\lVert \bar v \rVert_\infty \le \beta$, the norm bound of the MSIS instance of \cref{sec:attack-costs}, hence a solution of that instance. The idea is that a linear form easy to guess on a box of small integers is, up to a scalar, a form with small integer coefficients. If two distinct points $(r_1, r_2)$ and $(r_1', r_2')$ give the same $v_1 r_1 + v_2 r_2$ modulo $q$, their difference $(\Delta_1, \Delta_2)$ is nonzero, has entries below $g$ in absolute value, and satisfies $v_1 \Delta_1 + v_2 \Delta_2 = 0$, so in two coordinates $v$ is a multiple of the short integer vector $(\Delta_2, -\Delta_1)$. In $N$ coordinates one collision does not determine $v$, but probability above $\theta$ means very many collisions. They confine $v$, modulo $q$ and up to scaling, to an integer lattice of rank at most three and bounded height, because four independent integer conditions on $r$ hold with probability at most $g^{-4} \le \theta$, and that lattice contains the \shortmultiple{} $\bar v$.

\parhead{The range}
Under MSIS, then, no adversary forms a heavy cheat of rank at most three, and a stack of rank at least four has $\gp(\mathbf{W}) \le g^{-4}$. \scheme takes $g$ with $g^4 \ge 1/\theta$, so that every stack of rank at least four is light: $g \approx 2^{\lambda/2}$, and the honest search takes about $2^{\lambda/4} N^{1/4}$ operations. At our parameters this is about $2^{29.4}$ steps and $2^{32.0}$ multiplications in expectation, as many as the rest of decryption (\cref{sec:costs}): the goal of this subsection.

\parhead{Publishing on the factors}
On the template, every copy is a table published entry by entry: the ciphertext is quadratic in $N$ again, times the number of copies. So \scheme publishes every copy as \cref{sec:thin-table} publishes one table, with the hidden scalar $\gamma_c$ in place of $\gamma$ and the same $y$, $r$, $\tau$, and $\Omega$. The factors publish no table to which the $\Theta_c$ could be added, so the tie coins move into the factors as \emph{\tiemask{}s} in the right factors' first coordinates, and the ties are enforced as symmetry is (\cref{sec:thin-table}): along a stack that breaks a tie, the masks leave a nonzero polynomial in fresh hidden coins, and along one that satisfies them, published bases remove the masks' terms. \Cref{sec:scheme} gives the masks and their bases for the weights of \scheme. With these elements, the argument of \cref{sec:rank-template} carries over to the copies: apart from identities, coincidences in $\FF_q$, and guessed encodings, a comparison succeeds only on a correct guess of every $W_c r$ for a nonzero passing stack that satisfies the ties (\cref{lem:view}). A light stack is guessed with probability at most $\theta$, and a heavy one yields a witness or an MSIS solution (\cref{thm:extraction}). Published on factors of their own, the $3n/2$ copies multiply the ciphertext, encryption, and decryption's sums of one factored table by a factor $O(\lambda^2)$, the price of the shorter search; \cref{sec:shared-factors} saves a factor $\lambda$. 
\subsection{Shared Factors}\label{sec:shared-factors}
This subsection cuts the copies' factors from $O(\lambda^2 N)$ elements to $O(\lambda N)$ by sharing them, leaving the search and the rank gap as they are. After \cref{sec:weights-ties}, each of the $3n/2 = \num{6912}$ copies, named by their weights $U_k$ for $k < n$ and $V_j$ for $j < n/2$ (\cref{sec:ext-weights}), has factors of its own (\cref{sec:thin-table}). For each unknown, a copy then costs six elements, two of $\GG_1$ and, with their $\alpha$-copies, four of $\GG_2$, so the copies cost $9n = \num{41472}$.

\begin{figure}[!htb]
\centering
\begin{tikzpicture}[font=\footnotesize, line width=0.5pt,
    rule/.style={black!55, line width=0.4pt},
    virtual/.style={black!40, dashed, line width=0.4pt},
    every node/.append style={execute at begin node={\hyphenpenalty=10000 \exhyphenpenalty=10000 }}]
  \definecolor{figgone}{HTML}{DCE7F5}   \definecolor{figgtwo}{HTML}{E6E6E6}   \definecolor{figkey}{HTML}{3A3A3A}    \node[anchor=south west, inner sep=0pt] at (-0.05,0.62) {(a)\enspace a factor pair for each copy};
  \foreach \r/\row in {0/{U_0,U_1,V_0}, 1/{U_2,U_3,V_1}} {
    \foreach \name [count=\c from 0] in \row {
      \pgfmathsetmacro{\x}{0.34 + 1.62*\c}
      \pgfmathsetmacro{\y}{-1.22*\r}
      \pgfmathtruncatemacro{\iskey}{(\r==0) && (\c==0)}
      \fill[figgtwo] (\x,\y) rectangle ++(1.0,0.34);
      \draw[rule] (\x,\y) rectangle ++(1.0,0.34);
      \fill[figgone] (\x-0.34,\y) rectangle ++(0.34,-0.62);
      \draw[rule] (\x-0.34,\y) rectangle ++(0.34,-0.62);
      \ifnum\iskey=1
        \fill[figkey] (\x,\y) rectangle ++(1.0,-0.62);
        \node[text=white] at (\x+0.5,\y-0.31) {$\name$};
      \else
        \draw[virtual] (\x,\y-0.62) -- ++(1.0,0) -- ++(0,0.62);
        \node at (\x+0.5,\y-0.31) {$\name$};
      \fi
    }
  }
  \node[anchor=north west, inner sep=0pt, text width=5.0cm, align=left] at (-0.05,-1.98)
    {\textbf{36} elements for each unknown: six copies with $2 + 4$ each};
\begin{scope}[xshift=6.45cm]
    \node[anchor=south west, inner sep=0pt] at (-1.05,0.62) {(b)\enspace a grid of shared factors};
    \foreach \k [count=\c from 0] in {E,O,V} {
      \fill[figgtwo] (0.34+1.0*\c,0) rectangle ++(1.0,0.34);
      \draw[rule] (0.34+1.0*\c,0) rectangle ++(1.0,0.34);
      \node at (0.84+1.0*\c,0.17) {$\mathrm{\k}$};
    }
    \foreach \r in {0,1} {
      \fill[figgone] (0,-0.62*\r) rectangle ++(0.34,-0.62);
      \draw[rule] (0,-0.62*\r) rectangle ++(0.34,-0.62);
      \node[anchor=east] at (-0.08,-0.31-0.62*\r) {$k_1 = \r$};
    }
    \foreach \r/\row in {0/{U_0,U_1,V_0}, 1/{U_2,U_3,V_1}} {
      \foreach \name [count=\c from 0] in \row {
        \pgfmathsetmacro{\x}{0.34 + 1.0*\c}
        \pgfmathsetmacro{\y}{-0.62*\r}
        \pgfmathtruncatemacro{\iskey}{(\r==0) && (\c==0)}
        \ifnum\iskey=1
          \fill[figkey] (\x,\y) rectangle ++(1.0,-0.62);
          \node[text=white] at (\x+0.5,\y-0.31) {$\name$};
        \else
          \node at (\x+0.5,\y-0.31) {$\name$};
        \fi
      }
    }
    \draw[virtual] (1.34,0) -- (1.34,-1.24) (2.34,0) -- (2.34,-1.24) (0.34,-0.62) -- (3.34,-0.62);
    \draw[virtual] (0.34,-1.24) -- (3.34,-1.24) -- (3.34,0);
    \node[anchor=north west, inner sep=0pt, text width=5.0cm, align=left] at (-1.05,-1.98)
      {\textbf{16} elements for each unknown: $2 \cdot 2$ in the left legs, $3 \cdot 4$ in the right legs};
  \end{scope}
\end{tikzpicture} 
\caption{Shared factors for four forms ($n = 4$): six copies, named by their weights, with $K_1 = 2$ and $K_2 = 1$. For each unknown, a left factor (blue) holds two elements of $\GG_1$, and a right factor (gray) two of $\GG_2$ and their $\alpha$-copies; the copies' tables (dashed) are virtual, and the \keycopy{} $U_0$ is dark. (a) A left and a right factor for each copy: $6 \cdot 6 = 36$ elements for each unknown. (b) The grid: left legs $k_1 = 0, 1$ and right legs of kinds E, O, and V, with $U_{2k_1}$, $U_{2k_1+1}$, and $V_{k_1}$ in row $k_1$: $2K_1 + 12K_2 = 16$ elements for each unknown.}\label{fig:grid}
\end{figure}

Apart from their \tiemask{}s, the copies' tables differ only in their hidden scalars $\gamma_c$, each the product of a left and a right scalar (\cref{sec:thin-table}), so any left factor with any right factor gives a table. The copies become the \emph{cells} of a $K_1 \times 3K_2$ \emph{grid}, where $K_1 K_2 = n/2$, with one left factor for each row $k_1$, a \emph{left leg}, and one right factor for each column $\leg$, a \emph{right leg}; cell $c = (k_1, \leg)$ has $\gamma_c = \pi_{k_1} s_{\leg}$. The \keycopy{}, the one cell whose quotient base keeps the corner term and so the key, is $U_0$, and \cref{fig:grid} shows the layout, which \cref{sec:copies} derives from the weights' exponents.

The $K_1$ left and $3K_2$ right factors cost $2K_1 + 12K_2$ elements for each unknown, $528$ at $(K_1, K_2) = (192, 12)$ (\cref{tab:params}). With $K_1$ and $K_2$ in $O(\lambda)$, for systems of linear size with $N \ge \lambda^2$, the ciphertext thus has $O(\lambda N)$ elements, not $O(\lambda^2 N)$. Encryption computes these elements and decryption sums them, so encryption takes $O(\lambda^2 N / \log \lambda)$ field operations, not $O(\lambda^3 N / \log \lambda)$, and decryption outside the search $O(\lambda N \log \lambda)$, not $O(\lambda^2 N)$, now dominated by evaluating the forms at $w$. \Cref{cor:asymptotics} proves the new growths. The search, set by the rank gap of \cref{sec:weights-ties}, is unchanged.

The price, none in growth, is in the \tiemask{}s: a right leg's mask serves its whole column, but the terms that enforce the ties change from row to row by two different factors that only the left scalar can supply. So each $\pi_{k_1}$ is a sum of two parts, and its product with a mask has mixed products, the \emph{cross terms}; one more hidden coin, $\varepsilon$, marks them, so that published \crossbases{} remove them along every stack but no term that enforces the ties (\cref{sec:copies}, \emph{Shared factors} and \emph{Setup and encryption}).

In all, \scheme's costs are at most linear in $N$ and, apart from the search, polynomial in $\lambda$, while all three costs of Hair and Sahai's construction in \cref{tab:sizes} grow as $O(N^3\, 2^{\Theta(\sqrt{\lambda})})$.  
\section{Preliminaries}\label{sec:preliminaries}

\subsection{Notation}\label{sec:notation}

Throughout, $\lambda$ is the security parameter. For a positive integer $n$, $[n] \defeq \{1,\dots,n\}$, and $[a,b] \defeq \{a,\dots,b\}$. For a finite set $S$, $x \sample S$ samples $x$ independently and uniformly from $S$, $y \gets \mathcal{A}(x)$ runs a probabilistic algorithm $\mathcal{A}$ on $x$ with fresh coins, and $\mathcal{A}^{\mathcal{O}}$ is $\mathcal{A}$ with oracle access to $\mathcal{O}$. Logarithms are to base~$2$. For an integer $d \ge 0$, $\FF_q^{<d}[X]$ denotes the polynomials over $\FF_q$ of degree below~$d$. 
\subsection{Generic Bilinear Group Model}\label{sec:generic}

\parhead{Bilinear groups}
A \emph{bilinear group} $(\GG_1,\GG_2,\GG_T,q,G_1,G_2,e)$ consists of cyclic groups $\GG_1$, $\GG_2$, and $\GG_T$ of prime order $q$, generators $G_1$ of $\GG_1$ and $G_2$ of $\GG_2$, and a non-degenerate bilinear map $e\colon \GG_1 \times \GG_2 \to \GG_T$. We write the groups additively and let $G_T \defeq e(G_1,G_2)$. For $s \in \{1,2,T\}$, $[x]_s \defeq x\,G_s$ is the encoding of $x \in \FF_q$ in $\GG_s$, taken entrywise for vectors and matrices, so that $e\bigl([x]_1,[y]_2\bigr) = [x\,y]_T$. Our groups are of Type~III~\cite{GPS08}: no efficiently computable homomorphism between $\GG_1$ and $\GG_2$ is known, in either direction.

\parhead{Generic bilinear group model}
We prove security in Shoup's generic group model (GGM)~\cite{EC:Shoup97}, as Boneh and Boyen extended it to bilinear groups~\cite{EC:BonBoy04a} but without their isomorphism oracle from $\GG_2$ to $\GG_1$, matching Type~III groups. Our group oracle also returns the \emph{name} of an element $V$ of $\GG_T$, the set of the labels of $V$ and $-V$, which \scheme's search uses (\cref{sec:optimizations}).

\begin{definition}[Generic bilinear group model]\label{def:ggm}
Fix the order $q$ and a label length $\ell_{\mathsf{lab}}$ with $2^{\ell_{\mathsf{lab}}} \ge q$. At the start of each run of an experiment, the model samples independent, uniformly random injections $\mathsf{lab}_1,\mathsf{lab}_2,\mathsf{lab}_T\colon \FF_q \to \bits^{\ell_{\mathsf{lab}}}$; the \emph{label} of $[v]_s$ is $\mathsf{lab}_s(v)$. Every group element that an algorithm receives or outputs is a label, and every algorithm receives the labels of $G_1$, $G_2$, and $G_T$. For $s \in \{1,2,T\}$, $a, b \in \bits^{\ell_{\mathsf{lab}}}$, and $c \in \FF_q$, the group oracle $\Ogg$ answers
\begin{align*}
  \gadd(s,a,b)  &\defeq \mathsf{lab}_s\bigl(\mathsf{lab}_s^{-1}(a) + \mathsf{lab}_s^{-1}(b)\bigr),\\
  \gsmul(s,c,a) &\defeq \mathsf{lab}_s\bigl(c\,\mathsf{lab}_s^{-1}(a)\bigr),\\
  \gpair(a,b)   &\defeq \mathsf{lab}_T\bigl(\mathsf{lab}_1^{-1}(a)\,\mathsf{lab}_2^{-1}(b)\bigr),\\
  \mathsf{name}(a) &\defeq \bigl\{\mathsf{lab}_T(v), \mathsf{lab}_T(-v)\bigr\} \quad \text{for } v \defeq \mathsf{lab}_T^{-1}(a),
\end{align*}
and returns $\bot$ if an argument is not a label of its group. It also returns $\bot$ for $\mathsf{name}(a)$ unless the algorithm has received $a$ as a label of $\GG_T$ in its input or as the answer to an $\gadd$ or $\gsmul$ call in $\GG_T$ or to a $\gpair$ call; a label that it has seen only in a name does not count (its name is that name). A call to $\Ogg$ counts also when it returns $\bot$.
\end{definition}

An algorithm may call $\Ogg$, or query a random oracle, on a string that it has not received: a string fixed before the injections are sampled is a label with probability $q/2^{\ell_{\mathsf{lab}}}$. A guessed label is that of an element whose exponent the algorithm does not know, as in oblivious sampling, and \cref{app:monitor} treats it as a new variable. 
\subsection{Random Oracle Model}\label{sec:random-oracle}

\scheme uses two random oracles (ROs)~\cite{CCS:BelRog93}: a \emph{key oracle} $\Hkey$, which derives the key from an element of $\GG_T$, and a \emph{seed oracle} $\Hseed$, which expands a short seed into the matrix of \scheme's public forms, one entry per query. Neither outputs group elements. The reductions and the extractor of \cref{thm:extraction} see none of the adversary's RO queries (\cref{app:reduction,app:extractor}), and only the reductions program an oracle: $\Hseed$, at one fresh seed, where they place their Module-SIS challenge. So they are non-observing~\cite{PROVSEC:AnaBha13}, and for $\Hkey$ also non-programming, a restriction Nielsen introduced for simulators~\cite{C:Nielsen02a}.

\begin{definition}[Random oracle model]\label{def:rom}
At the start of each run of an experiment, the model samples independent, uniformly random functions $H_1, \dots, H_{n_H}$ from bit strings to finite sets. Every algorithm may query them, with inputs encoded as bit strings.
\end{definition} 
\subsection{Lattices and MSIS}\label{sec:module-sis}

Let $d_R$ be a power of two, $R \defeq \ZZ[Y]/(Y^{d_R}+1)$, and $R_q \defeq R/qR = \FF_q[Y]/(Y^{d_R}+1)$ for the group order $q$, with $q \equiv 1 \bmod 2d_R$. For $a \in R$, $\|a\|_\infty$ is the largest absolute value of a coefficient of $a$, and for a vector over $R$ the largest over its entries. Module-SIS (MSIS), the module version of Ajtai's short integer solution problem~\cite{STOC:Ajtai96}, is due to Langlois and Stehl\'e, who for suitable parameters reduce to it the approximate shortest independent vectors problem on module lattices~\cite[Theorem~3.6]{DCC:LanSte15}.

\begin{definition}[MSIS]\label{def:msis}
Let $k, \ell \in \NN$ and $0 < \beta < q/2$. A \emph{solution} of MSIS for a matrix $A \in R_q^{k \times \ell}$ is a vector $\mathbf{z} \in R^\ell$ with $0 < \|\mathbf{z}\|_\infty \le \beta$ and $A \mathbf{z} = 0 \bmod q$. The advantage of an adversary $\mathcal{A}$ against $\mathsf{MSIS}_{k,\ell,\beta}$ over $R_q$ is $\Pr\bigl[\, \mathbf{z} \text{ is a solution for } A : A \sample R_q^{k \times \ell};\ \mathbf{z} \gets \mathcal{A}(A) \,\bigr]$.
\end{definition}

\subsection{Witness Encryption}\label{sec:witness-encryption}

Let $\mathcal{R}$ be an NP relation: a set of pairs $(S,w)$ of an instance $S$ and a witness $w$, decidable in time $\poly(|S|)$, with $|w| \le \poly(|S|)$, where $|S|$ is the length of an encoding of $S$. An indexed relation, of triples $(i, x, w)$ with an index $i$ such as a circuit or a program and a public input $x$ that may include claimed outputs, is used through its pairs $((i, x), w)$, so that its instances are $S = (i, x)$. For \scheme, $\mathcal{R}$ pairs Boolean quadratic systems (BQS) of a fixed shape (\cref{sec:relations}) with their witnesses, the Boolean assignments that satisfy every constraint (\cref{def:bqs}). An index and a public input enter such a system only as its constants.

\begin{definition}[Witness encryption]\label{def:we}
A \emph{witness encryption scheme} for an NP relation $\mathcal{R}$ and messages in $\bits^\mu$ is a pair $\mathsf{WE} = (\mathsf{Enc}, \mathsf{Dec})$ of algorithms with access to the oracles $\mathcal{O} \defeq (\Ogg, H_1, \dots, H_{n_H})$ of \cref{def:ggm,def:rom}. The probabilistic algorithm $\mathsf{Enc}^{\mathcal{O}}(S, m)$ outputs, in time $\poly(|S|, \lambda)$, a ciphertext $ct$ of a message $m \in \bits^\mu$ for an instance $S$, and $\mathsf{Dec}^{\mathcal{O}}(S, ct, w)$, whose running time is not bounded, outputs a message or $\bot$.
\end{definition}

The original definition~\cite{STOC:GGSW13} asks that $\mathsf{Dec}$, too, run in polynomial time; in \scheme it ends with a search that is exponential in $\lambda$ but concretely fast (\cref{cor:asymptotics,sec:costs}).

\begin{definition}[Correctness]\label{def:we-correct}
$\mathsf{WE}$ is \emph{correct with error $\varepsilon_{\mathrm{cor}}$} if for every $(S,w) \in \mathcal{R}$ and every $m \in \bits^\mu$,
\[
  \Pr\bigl[\, \mathsf{Dec}^{\mathcal{O}}(S, ct, w) \ne m \;:\; ct \gets \mathsf{Enc}^{\mathcal{O}}(S, m) \,\bigr] \le \varepsilon_{\mathrm{cor}},
\]
where $S$, $w$, and $m$ are fixed before $\mathcal{O}$ is sampled, and the probability is over $\mathcal{O}$ and the coins of $\mathsf{Enc}$ and $\mathsf{Dec}$.
\end{definition}

\begin{definition}[Extractability]\label{def:we-ext}
$\mathsf{WE}$ is \emph{extractable with error $\delta$ and extraction time $t_{\mathcal{E}}$} if there is an extractor $\mathcal{E}$, which makes no oracle call, such that for every instance $S$, all $m_0, m_1 \in \bits^\mu$, and every adversary $\mathcal{A}$, with any advice, that runs in time $t$ and makes at most $Q$ calls to $\Ogg$ and at most $Q_H$ queries to each RO, $\mathcal{E}$ runs in time at most $t_{\mathcal{E}}(|S|, Q)$ and
\[
  \Pr\left[
    \begin{array}{c}
      b' = b \\
      \wedge\ (S, w) \notin \mathcal{R}
    \end{array}
    :
    \begin{array}{r}
      b \sample \bits \\
      ct \gets \mathsf{Enc}^{\mathcal{O}}(S, m_b) \\
      b' \gets \mathcal{A}^{\mathcal{O}}(ct) \\
      w \gets \mathcal{E}(S, ct, \mathsf{tr})
    \end{array}
  \right] \le \frac{1}{2} + \delta(t, Q, Q_H),
\]
where $\mathsf{tr}$ is the transcript of $\mathcal{A}$'s answered calls to $\Ogg$, in order, with their answers. The experiment fixes $S$, $m_0$, $m_1$, and $\mathcal{A}$ before it samples $\mathcal{O}$, and the probability is over $\mathcal{O}$, $b$, and the coins of $\mathsf{Enc}$, $\mathcal{A}$, and $\mathcal{E}$.
\end{definition}

So if $\mathcal{A}$ guesses $b$ with probability $\frac12 + \varepsilon$, then $\mathcal{E}$ outputs a witness with probability at least $\varepsilon - \delta(t, Q, Q_H)$. The extractor is universal and straight-line: it is fixed before the adversary and runs once, on the transcript of the finished run.

Extractable witness encryption with arbitrary auxiliary input~\cite{C:GKPVZ13} is implausible in the standard model: under an obfuscation assumption, the auxiliary input can be an obfuscated program that decrypts with a witness that no extractor recovers~\cite[full version, Theorem~6.1]{C:GGHW14}. This counterexample does not apply in our model, where the extractor sees every group-oracle call of $\mathcal{A}$, including such a program's; like any generic-group result, ours does not cover such programs run on a concrete group.

\begin{lemma}[Extractability implies hiding]\label{lem:we-hiding}
If $\mathsf{WE}$ is extractable with error $\delta$ and any extraction time, then for every instance $S$ that has no witness, all $m_0, m_1 \in \bits^\mu$, and every adversary $\mathcal{A}$ as in \cref{def:we-ext}, the probabilities $p_b \defeq \Pr\bigl[ \mathcal{A}^{\mathcal{O}}(ct) = 1 : ct \gets \mathsf{Enc}^{\mathcal{O}}(S, m_b) \bigr]$, for $b \in \bits$, satisfy $|p_1 - p_0| \le 2\,\delta(t, Q, Q_H)$. That is, $\mathsf{WE}$ is hiding in the sense of Garg, Gentry, Sahai, and Waters~\cite{STOC:GGSW13}, with error $2\delta$.
\end{lemma}

\begin{proof}
Replacing every output of $\mathcal{A}$ other than $1$ by $0$ changes neither the probabilities $p_b$ nor the resources, so let $\mathcal{A}$ output a bit $b'$. As $S$ has no witness, the event of \cref{def:we-ext} is $b' = b$, so $\frac12 + \frac12 (p_1 - p_0) = \Pr[b' = b] \le \frac12 + \delta(t, Q, Q_H)$; likewise with $p_0 - p_1$ for the adversary that outputs $1 - b'$.
\end{proof}  
\section{Relations as Boolean Quadratic Systems}\label{sec:relations}

\subsection{Boolean Quadratic Systems}\label{sec:quadratic-systems}

Vectors in $\FF_q^{N+1}$ have indices $0, \dots, N$; $e_i$ is the unit vector of index $i$, and each vector $v$ defines the linear form $v(w) \defeq \langle v, w \rangle = \sum_{i=0}^{N} v_i w_i$.

\begin{definition}[Boolean quadratic system]\label{def:bqs}
A \emph{Boolean quadratic system} (BQS) $S$ over $\FF_q$ in $N$ unknowns is a finite list of \emph{constraints}. Constraint $t$ has $F_t \ge 1$ \emph{products} and vectors $a_{t,f}, b_{t,f}, c_t \in \FF_q^{N+1}$ for $f \in [F_t]$, and reads
\[
  \sum_{f \in [F_t]} a_{t,f}(w)\, b_{t,f}(w) = w_0\, c_t(w).
\]
An \emph{assignment} is a $w \in \FF_q^{N+1}$ with $w_0 = 1$, and a \emph{witness} of $S$ is an assignment in $\bits^{N+1}$ that satisfies every constraint. $S$ is \emph{true} if it has a witness and \emph{false} otherwise.
\end{definition}

As $w_0 = 1$, entry $0$ of a vector is its form's constant term, and the factor $w_0$ makes every constraint homogeneous quadratic in $w$.

A BQS is a customizable constraint system (CCS) of degree two~\cite{EPRINT:SetThaWah23a} with a Boolean witness. With one product per constraint, $a_{t,1}(w)\, b_{t,1}(w) = c_t(w)$, it is a rank-one constraint system (R1CS)~\cite{C:BCGTV13} with a Boolean witness, to which the same work reduces Boolean circuits. So BQS express every NP relation: a witness-checking circuit gives one constraint per gate, such as $e_i(w)\, e_j(w) = w_0\, e_k(w)$ for an AND gate $w_k = w_i w_j$, plus $e_o(w)\, e_0(w) = w_0\, e_0(w)$, setting the output $w_o$ to $1$; a witness is an accepting input with the values of all gates.

\parhead{Shape}
A system has \emph{shape} $(N, n_{\mathsf{c}}, F_{\max})$ if it has $N$ unknowns, at most $n_{\mathsf{c}}$ constraints, and at most $F_{\max}$ products per constraint. Padding with constraints $0 = 0$ (one product, zero vectors) keeps the witnesses, so we assume exactly $n_{\mathsf{c}}$ constraints. The systems of one shape and their witnesses form the NP relation of \cref{sec:witness-encryption}. A system has \emph{linear size} if $F_{\max}\,\nc = O(N)$ and it has $O(N)$ nonzero coefficients; the systems of Boolean circuits above have linear size, with $F_{\max} = 1$ and $\nc \le N + 1$. 
\subsection{Moment Matrices and Linear Tests}\label{sec:linear-tests}

Rows and columns of matrices in $\FF_q^{(N+1) \times (N+1)}$ have indices $0, \dots, N$. For an assignment $w$, every condition on a witness is linear in its \emph{moment matrix} $w w^\top$: $w_j \in \bits$ exactly when $w_j^2 = w_0 w_j$, and constraint $t$ holds exactly when $\sum_f a_{t,f}^\top w w^\top b_{t,f} = e_0^\top w w^\top c_t$. The tests ask the same of any symmetric matrix $W$.

\begin{definition}[Linear tests]\label{def:tests}
Let $S$ be a system in $N$ unknowns. A symmetric matrix $W \in \FF_q^{(N+1) \times (N+1)}$ passes the \emph{Booleanity test} of $j \in [N]$ if $W_{jj} = W_{0j}$, and the \emph{test of constraint $t$} if
\[
  \sum_{f \in [F_t]} a_{t,f}^\top W\, b_{t,f} = e_0^\top W c_t .
\]
$W$ \emph{passes} if it passes every test of $S$.
\end{definition}

The passing matrices form a subspace, and $w w^\top$ passes exactly when the assignment $w$ is a witness. A \emph{cheat} is a passing matrix that is not a linear combination of moment matrices of witnesses; a false system can have cheats too (\cref{ex:toy-system,sec:guessing-probability}). None has rank one, which is the rank gap of Hair and Sahai~\cite[Section~4]{HS26}, the Boolean version of the reduction of Bl\"aser et al.\ from quadratic equations to rank-one matrices~\cite[full version, Theorem~8.7]{SODA:BILPS21}:

\begin{lemma}[Rank gap]\label{lem:rank-one}
Let $W$ be a nonzero passing matrix of rank one. Then $W_{00} \ne 0$ and $W = W_{00}\, w w^\top$ for a witness $w$ of $S$. In particular, if $S$ is false, every nonzero passing matrix has rank at least two.
\end{lemma}

\begin{proof}
The row space of $W$ is a line $\FF_q u$, so $W = y u^\top$ for some $y \in \FF_q^{N+1}$, and as $W$ is symmetric and nonzero, $W = \chi\, u u^\top$ for some $\chi \ne 0$. Booleanity gives $\chi u_j^2 = \chi u_0 u_j$, so $u_j \in \{0, u_0\}$ for every $j$; as $u \ne 0$, $u_0 \ne 0$. Then $w \defeq u / u_0$ is an assignment in $\bits^{N+1}$, and $W = \chi u_0^2\, w w^\top = W_{00}\, w w^\top$ with $W_{00} = \chi u_0^2 \ne 0$. The test of constraint $t$ reads $W_{00} \bigl( \sum_f a_{t,f}(w)\, b_{t,f}(w) - c_t(w) \bigr) = 0$, so $w$ satisfies every constraint.
\end{proof} 
\subsection{Rank and Guessing Probability}\label{sec:guessing-probability}

\scheme hides a vector $r = (r_0, r_1, \dots, r_N)$ of small integers, and decryption with a witness $w$ ends by searching for the integer $r(w) = \sum_i w_i r_i$ (\cref{sec:construction}). Fix an integer $g \ge 1$ with $g + 1 < q$, the \emph{range}. The entries $r_1, \dots, r_N$ are independent and uniform on $\{2, \dots, g+1\}$, and $r_0$, given them, is uniform on $g$ consecutive integers: \cref{sec:scheme} draws $r_0$ like the others, and \cref{sec:optimizations} centers it (\cref{def:sampler}). The vector $r$ is reduced modulo $q$ wherever it multiplies a matrix over $\FF_q$.

A \emph{stack} is a finite family $\mathbf{W} = (W_c)_{c \in C}$ of matrices with $N+1$ columns, such as one matrix for each copy of \scheme's table (\cref{sec:copy-set}); it \emph{passes} if every $W_c$ is a symmetric $(N+1) \times (N+1)$ matrix that passes (\cref{def:tests}), and its \emph{rank} $\operatorname{rank} \mathbf{W}$ is the dimension of its \emph{row space}, the sum of the row spaces of the $W_c$.

\begin{definition}[Guessing probability]\label{def:gp}
The \emph{guessing probability} of a stack $\mathbf{W} = (W_c)_{c \in C}$ is
\[
  \gp(\mathbf{W}) \defeq \max_{u} \Pr_{r}\bigl[ (W_c r)_{c \in C} = u \bigr],
\]
for $r$ drawn by the sampler of the scheme at hand: independent entries (\cref{fig:enc}) or $\mathcal{D}_g$ (\cref{def:sampler}), the sampler of \cref{app:spec} and of the theorems. That of a matrix is that of the stack of one. For a threshold $\theta \in (0,1)$, $\mathbf{W}$ is \emph{light} if $\gp(\mathbf{W}) \le \theta$ and \emph{heavy} otherwise.
\end{definition}

For each comparison of expressions in $\GG_T$ that the adversary makes, the security proof reads a stack $\mathbf{W}$ and vectors $u_c$, both fixed by the comparison, and the comparison holds identically in the coins of $\Enc$ other than $r$ only if $\mathbf{W}$ passes and $W_c r = u_c$ for every $c$ (\cref{lem:view}), which has probability at most $\gp(\mathbf{W})$. The next lemma holds for both samplers.

\begin{lemma}[Rank bound]\label{lem:gp-rank}
For every stack $\mathbf{W} = (W_c)_{c \in C}$:
\begin{enumerate}[label=(\alph*)]
  \item $\gp(\mathbf{W}) \le g^{-\operatorname{rank} \mathbf{W}}$, with equality if the row space of $\mathbf{W}$ is spanned by unit vectors $e_i$ with $i \in [N]$;
  \item $\gp(v^\top) \ge \gp(\mathbf{W})$ for every $v$ in the row space of $\mathbf{W}$.
\end{enumerate}
\end{lemma}

\begin{proof}
(a) Let $M$ be the matrix of all rows of all $W_c$, of rank $d \defeq \operatorname{rank} \mathbf{W}$, and fix $u$. Choose $d$ linearly independent columns of $M$, column $0$ among them unless it is zero. Given the entries of $r$ at the other indices in $[N]$, at most one value modulo $q$ of its entries at the chosen indices solves $M r = u$. Each of the independent entries $r_1, \dots, r_N$ takes a given value modulo $q$ with probability at most $1/g$, as $g + 1 < q$; so does $r_0$ given all of them. Hence $\Pr[M r = u] \le g^{-d}$, as in Odlyzko's lemma~\cite[Lemma~3.1]{NW22}. If the row space is spanned by $d$ unit vectors $e_i$ with $i \in [N]$, then $(W_c r)_c$ is an injective linear function of the $d$ entries $r_i$, which take each of their $g^d$ values with probability $g^{-d}$.
(b) Write $v = \sum_c W_c^\top z_c$. Then $v^\top r = \sum_c z_c^\top W_c r$ is a function of $(W_c r)_c$, so it takes one value whenever $(W_c r)_c$ takes its likeliest value, and $\gp(v^\top) \ge \gp(\mathbf{W})$.
\end{proof}

\parhead{Heavy and light stacks}
By \cref{lem:gp-rank}, every heavy stack has rank at most three when $g^{-4} \le \theta$. Passing stacks of rank two and three can be heavy, even when $S$ is false: if no vector $a_{t,f}$ or $b_{t,f}$ has a nonzero entry $i$, $j$, or $k$, for distinct $i, j, k \in [N]$, then $e_i e_j^\top + e_j e_i^\top$ passes, and so does its stack with $e_i e_k^\top + e_k e_i^\top$; their guessing probabilities, $g^{-2}$ and $g^{-3}$, exceed $\theta$ at our parameters (\cref{sec:parameters}). \scheme adds ties among a stack's matrices (\cref{sec:construction}), and the security proof (\cref{sec:security}, planned in \cref{app:bound}) turns a nonzero heavy passing stack satisfying the ties into a witness or, through a vector of its row space, heavy by part (b), an MSIS solution. 
\subsection{The QAP of a System}\label{sec:qap}

The quadratic arithmetic program (QAP) of Gennaro et al.~\cite{EC:GGPR13} is an R1CS in polynomial form: each constraint, with its one product, sits at its own point, and exact division by the points' vanishing polynomial checks them all. Ours places at each point a pair $(a, b)$ of vectors, which stands for the product $a(w)\, b(w)$; the pair's \emph{term} on a matrix $W$ is $a^\top W\, b$, and each test sets a sum of terms to zero. The pairs of each constraint's test sit on its own coset of one subgroup, and an allowed remainder replaces exact division, so a constraint can sum many products without the auxiliary unknowns of an R1CS rewrite, which can multiply the number of unknowns, and so the costs, by $\Theta(\lambda)$ (\cref{sec:tests-rank}). \Cref{sec:thin-table} worked out only the case of an R1CS, with the subgroup $\{1, -1\}$: a pair of points for each constraint and for each Booleanity test. The pieces are known: the coset sum behind Aurora's univariate sumcheck~\cite{EC:BCRSVW19}, read for all cosets at once~\cite{Moh25}, and a vanishing sum enforced through the reference string~\cite{AC:LipSiiZaj22,PKC:ZhaSunGu24}; our QAP combines them. Fix a \emph{coset size} $\nu$ that divides $q - 1$ and exceeds $F_{\max}$; we take the least power of two above $F_{\max}$, so that $\nu \le 2F_{\max}$. Write $D$ for the number $N + 2 + \nu\,\nc$ of evaluation points.

\begin{definition}[QAP of a system]\label{def:qap}
Let $S$ be a system of shape $(N, \nc, F_{\max})$. A choice of \emph{evaluation points} for $S$ consists of points $d_0, \dots, d_N$ and $\rho$ of $\FF_q$, of cosets $\mathcal{K}_1, \dots, \mathcal{K}_{\nc}$ of the subgroup of order $\nu$ of $\FF_q^*$, and, in each $\mathcal{K}_t$, of points $p_{t,1}, \dots, p_{t,F_t}$ and $p'_t$. It is \emph{admissible} if the cosets are pairwise distinct, the points $d_0, \dots, d_N, \rho$ are pairwise distinct and in no $\mathcal{K}_t$, and the $F_t + 1$ points chosen in $\mathcal{K}_t$ are pairwise distinct, for every $t$. For an admissible choice, let $\mathcal{P}_0 \defeq \{d_0, \dots, d_N, \rho\}$ and $\mathcal{P} \defeq \mathcal{P}_0 \cup \bigcup_t \mathcal{K}_t$, a set of $D$ points; let $Z_0(X) \defeq \prod_{p \in \mathcal{P}_0} (X - p)$ and $Z(X) \defeq \prod_{p \in \mathcal{P} \setminus \{\rho\}} (X - p)$; and let $L_\rho(X) \in \FF_q^{<D}[X]$ be the polynomial with value $1$ at $\rho$ and $0$ on $\mathcal{P} \setminus \{\rho\}$. The \emph{QAP} of $S$ for this choice consists of the polynomials $A_i(X)$ and $B_i(X)$ in $\FF_q^{<D}[X]$, for $i \in [0,N]$, with the values on $\mathcal{P}$ in the table below, where $A(X) \defeq (A_i(X))_{i \in [0,N]}$ and $B(X) \defeq (B_i(X))_{i \in [0,N]}$. To a matrix $W \in \FF_q^{(N+1) \times (N+1)}$ the QAP assigns $P_W(X) \defeq \sum_{i,j} W_{ij}\, A_i(X)\, B_j(X) = A(X)^\top W\, B(X)$, whose values on $\mathcal{P}$ the last column gives.
\begin{center}\upshape
\begin{tabular}{@{}l@{\hspace{1.4em}}l@{\hspace{1.4em}}l@{\hspace{1.4em}}l@{}}
\toprule
point $p$ & $A(p)$ & $B(p)$ & $P_W(p) = A(p)^\top W\, B(p)$ \\
\midrule
$d_j$ with $j \in [N]$ & $e_j - e_0$ & $e_j$ & $W_{jj} - W_{0j}$ \\
$d_0$ & $0$ & $e_0$ & $0$ \\
$\rho$ & $e_0$ & $e_0$ & $W_{00}$ \\
$p_{t,f}$ with $f \in [F_t]$ & $Z_0(p_{t,f})\, a_{t,f}$ & $b_{t,f}$ & $Z_0(p_{t,f}) \cdot a_{t,f}^\top W\, b_{t,f}$ \\
$p'_t$ & $Z_0(p'_t)\, e_0$ & $-c_t$ & $-Z_0(p'_t) \cdot e_0^\top W c_t$ \\
other points of $\mathcal{K}_t$ & $0$ & $0$ & $0$ \\
\bottomrule
\end{tabular}
\end{center}
A polynomial is \emph{allowed} if it lies in $\FF_q^{<\nu\nc}[X]$ and none of its exponents is divisible by $\nu$.
\end{definition}

By the table, $P_W(d_j)$ is the term of the Booleanity test of $j$, the \emph{key point} $\rho$ carries the corner $W_{00}$, and $d_0$, where every $A_i$ is zero, serves the security proof (\cref{lem:view}). On $\mathcal{K}_t$, where $Z_0(p) \ne 0$, the values $P_W(p)/Z_0(p)$ are the terms of the test of constraint $t$.

Over a coset $\delta \mathcal{G}$ of the subgroup $\mathcal{G}$ of order $\nu$, the monomial $X^k$ sums to $\nu\, \delta^k$ if $\nu$ divides $k$ and to $0$ otherwise~\cite{EC:BCRSVW19}. So for a polynomial of degree below $\nu\,\nc$, the sums over the cosets $\mathcal{K}_t = \delta_t \mathcal{G}$ are, up to the nonzero factor $\nu$, the $\nc \times \nc$ Vandermonde matrix of the values $\delta_t^\nu$, distinct as the cosets are, applied to the polynomial's coefficients at the exponents divisible by $\nu$; they all vanish exactly when the polynomial is allowed. Hence a symmetric matrix $W$ passes exactly when $P_W(d_j) = 0$ for every $j$ and the polynomial $R_W$ of degree below $\nu\,\nc$ with the values $P_W(p)/Z_0(p)$ on the cosets is allowed.

\begin{lemma}[\qapcriterion{}]\label{lem:qap}
For an admissible choice of evaluation points and a symmetric matrix $W \in \FF_q^{(N+1) \times (N+1)}$, the following are equivalent:
\begin{enumerate}[label=(\alph*)]
  \item $W$ passes the tests of $S$ (\cref{def:tests});
  \item $P_W(X) - W_{00}\, L_\rho(X) = h(X)\, Z(X)\, (X - \rho) + Z_0(X)\, R(X)$ for a polynomial $h(X)$ and an allowed polynomial $R(X)$;
  \item $P_W(X) = h'(X)\, Z(X) + Z_0(X)\, R(X)$ for a polynomial $h'(X)$ and an allowed polynomial $R(X)$.
\end{enumerate}
In (b) and (c), $R(X)$ is the polynomial $R_W(X) \in \FF_q^{<\nu\nc}[X]$ with $R_W(p) = P_W(p)/Z_0(p)$ for every $p$ in the cosets, $h(X) \in \FF_q^{<D-1}[X]$, and $h'(X) \in \FF_q^{<D}[X]$.
\end{lemma}

\Cref{app:calc} has the proof. For a witness $w$, $w w^\top$ passes, with corner $1$, and $P_{w w^\top} = A_w B_w$ for $A_w \defeq \sum_i w_i A_i$ and $B_w \defeq \sum_i w_i B_i$. By \cref{lem:qap} there are unique polynomials $\hkey$ and $\hoth$ and a unique allowed $R_w \defeq R_{w w^\top}$, the \emph{quotients} of $w$ for the \keycopy{} and the other copies of \scheme (\cref{sec:copy-set}) and the \emph{remainder} of $w$, with
\begin{equation}\label{eq:qap-w}
  A_w B_w - L_\rho = \hkey \cdot Z \cdot (X - \rho) + Z_0\, R_w
  \qquad\text{and}\qquad
  A_w B_w = \hoth \cdot Z + Z_0\, R_w .
\end{equation}
Conversely, an assignment satisfying one of these identities with any polynomial as quotient and any allowed polynomial as remainder is a witness.

A Booleanity test takes a single point because it has one term. The $N$ tests thus take $N$ points, not the $\nu N$ of a coset for each: about $2^{22}$ points in all at our parameters (\cref{tab:params}), against about $2^{30}$. With $\nu \le 2F_{\max}$ and $F_{\max}\,\nc = O(N)$, as for systems of linear size (\cref{sec:quadratic-systems}), the single points keep $D = O(N)$, while cosets would give $\Theta(F_{\max} N)$. A constraint whose test has one term, a single product equal to zero, such as a linear constraint $\ell(w) = 0$ written as $w_0\, \ell(w) = 0$, could take a single point in the same way. Every constraint of the system of \cref{sec:groth16} has at least $254$ products, so we give each constraint a coset.

The theorems of \cref{sec:construction} hold for every admissible choice of evaluation points. We place all of them in the subgroup $\mathcal{H}$ of $\FF_q^*$ of order $2^{\lceil \log_2 D \rceil}$, which must divide $q - 1$, each $\mathcal{K}_t$ a coset of its subgroup of order $\nu$: the cosets first, then $d_0, \dots, d_N, \rho$, in order from the powers of a generator of $\mathcal{H}$ in bit-reversed order. Fast Fourier transforms over $\mathcal{H}$ then serve the QAP's polynomials~\cite[Section~1]{Pol71}, in $O(D \log D)$ operations in $\FF_q$ (\cref{prop:efficiency}), against $O(D \log^2 D)$ with arbitrary points.  
\section{Witness Encryption for Boolean Quadratic Systems}\label{sec:construction}

We give \scheme for BQS of a fixed shape, over an abstract bilinear group, ring, and parameters, which \cref{sec:parameters} instantiates.

\subsection{The Construction}\label{sec:scheme}

\Cref{fig:enc,fig:dec} state \scheme, $\mathsf{WE} = (\Enc, \Dec)$; \cref{sec:optimizations} adds what saves size and time, and a sampler that gives the search, up to rounding, one law for every witness; \cref{app:spec} states the result in full. Three parts give its size and its search (\cref{cor:asymptotics}; \cref{tab:sizes} counts bits): the factored table, $O(N)$ elements for each copy (\cref{sec:thin-table}); $3n/2 = O(\lambda^2)$ weights, a copy each, from one list of $n$ forms, which with their ties give a rank gap of one versus four, allowing $g \approx 2^{\lambda/2}$ and a search of $O(2^{\lambda/4} N^{1/4})$ operations (\cref{sec:weights-ties,sec:ext-weights}); and a grid in which the copies share factors, $O(\lambda N)$ elements in all, not $O(\lambda^2 N)$ (\cref{sec:shared-factors}).

\parhead{Parameters}
We fix a bilinear group $(\GG_1,\GG_2,\GG_T,q,G_1,G_2,e)$ (\cref{sec:generic}), a shape $(N, \nc, F_{\max})$ of systems (\cref{sec:quadratic-systems}), and a message length $\mu$. The QAP of a system has coset size $\nu$ and $D = N + 2 + \nu\,\nc < q$ evaluation points (\cref{def:qap}). The \emph{key oracle} $\Hkey$ maps bit strings to $\bits^\mu$ (\cref{sec:random-oracle}). The range $g$ (\cref{sec:guessing-probability}) and the \emph{\searchbound{}} $B \defeq (N+1)(g+1)$ satisfy $2B + 1 < q$.

\parhead{Forms and weights}
The forms (\cref{sec:weights-ties}) come from a uniform matrix over the ring $R_q$ of degree $d_R$ (\cref{sec:module-sis}), with $k$ rows and columns $0, \dots, \ell$, for $\ell \defeq \lceil N/d_R \rceil$; its columns $1$ to $\ell$ form an MSIS instance. The \emph{number of forms} $n \defeq (k+1)\,d_R$ is even and divides $q - 1$.

\begin{definition}[Forms of a matrix]\label{def:forms}
Let $M = (M_{i,j})$ be a matrix over $R_q$ with rows $i \in [k]$ and columns $j \in [0,\ell]$. For $w \in \FF_q^{N+1}$ and $j \in [\ell]$, let $\hat w_j \in R_q$ be the polynomial $\sum_{s < d_R} w_{(j-1) d_R + s + 1}\, Y^s$, where $w_c \defeq 0$ for $c > N$. The \emph{forms} of $M$ are the linear forms $\varphi_0, \dots, \varphi_{n-1}$ on $\FF_q^{N+1}$ with $\varphi_0(w) \defeq w_0$, with $\varphi_e \defeq 0$ for $0 < e < d_R$, and with
\[
  \varphi_{i d_R + s}(w) \defeq \text{the coefficient of } Y^s \text{ in }\; M_{i,0}\, w_0 + \sum_{j \in [\ell]} M_{i,j}\, \hat w_j
\]
for $i \in [k]$ and $0 \le s < d_R$.
$\Forms(M)$ denotes the list of their coefficient vectors.
\end{definition}

A vector $v$ with $v_0 = 0$ is in the kernel of every form exactly when the columns $1, \dots, \ell$ of $M$, a matrix in $R_q^{k \times \ell}$, map $(\hat v_1, \dots, \hat v_\ell)$ to zero.

Let $c_0 \in \FF_q$ be such that $P(T) \defeq T^n - c_0$ is irreducible, and write the elements of $\FF_{q^n} \defeq \FF_q[T]/(P)$ as polynomials in $\FF_q^{<n}[T]$. Its Frobenius automorphism $\sigma$ maps $T$ to $\omega T$, where $\omega \defeq c_0^{(q-1)/n}$ has order $n$; so $\omega^{n/2} = -1$, and $\sigma^j$ maps $a(T)$ to $a(\omega^j T)$. Together the forms are one form $z(w) \defeq \sum_{e<n} \varphi_e(w)\, T^e$ with values in $\FF_{q^n}$ and coefficient vector $z = \sum_{e<n} \varphi_e T^e \in \FF_{q^n}^{N+1}$. For $x \in \FF_q$ or a formal polynomial $x$, $z[x] \defeq \sum_{e<n} x^e \varphi_e$ is $z$ with $x$ substituted for $T$. The weights (\cref{sec:ext-weights}) are the coefficients of two products in $\FF_{q^n}$:
\begin{align*}
  U(w) &\defeq z(w) \cdot \sigma^{n/2-1}(z(w)) = \sum_{k<n} U_k(w)\, T^k, \\
  V(w) &\defeq z(w) \cdot \sigma^{n/2}(z(w)) = \sum_{j<n/2} V_j(w)\, T^{2j} .
\end{align*}

\newsavebox{\wemaskbox}
\newsavebox{\weomegabox}
\begin{figure}[!tb]
\centering\small
\sbox{\weomegabox}{$\bigl(\begin{smallmatrix} 1 & \Omega_{12} \\ 0 & \Omega_{22} \end{smallmatrix}\bigr)$}
\sbox{\wemaskbox}{$\begin{cases}
  \upsilon^{\kleg} \bigl( \eta_1 z_{\circ} + \eta_2 z_{+} \bigr) - \varepsilon^{-1} (-\omega\upsilon)^{\kleg} \bigl( \eta_2 z_{-} + \eta_3 z_{\circ} \bigr) & \text{if } \mathrm{X} \in \{\mathrm{E}, \mathrm{O}\} \\
  \upsilon^{\kleg} \bigl( \eta_3 z_{+} - \eta_1 z_{-} \bigr) & \text{if } \mathrm{X} = \mathrm{V}
\end{cases}$}
\fbox{\begin{minipage}{\dimexpr\linewidth-2\fboxsep-2\fboxrule\relax}
\procedure[linenumbering, width=\linewidth]{$\Setup(S)$, for a system $S$ of shape $(N, \nc, F_{\max})$}{
  \text{choose admissible evaluation points for } S \text{ (\cref{def:qap})} \\
  \text{compute } A_i(X) \text{ and } B_i(X), \text{ for } i \in [0,N], \text{ and } Z_0(X),\ Z(X), \text{ and } L_\rho(X) \\
  \pcreturn \pp \defeq \bigl( S, \text{ the evaluation points}, (A_i, B_i)_{i}, Z_0, Z, L_\rho \bigr)
}

\medskip
\procedure[linenumbering, width=\linewidth]{$\Enc(\pp, m)$, for a message $m \in \bits^\mu$}{
  M \sample R_q^{k \times (\ell + 1)};\quad (\varphi_e)_{e<n} \gets \Forms(M) \\
  r_0, \dots, r_N \sample \{2, \dots, g+1\};\quad y \sample \FF_q^{N+1};\quad \tau \sample \FF_q \setminus \mathcal{P};\quad \Omega_{12} \sample \FF_q \\
  \Omega_{22},\, \alpha,\, \pi,\, s_{\mathrm{E}},\, s_{\mathrm{O}},\, s_{\mathrm{V}},\, \upsilon,\, \zeta,\, \zeta',\, \zeta'',\, \eta_1,\, \eta_2,\, \eta_3 \sample \FF_q^* \\
  \varepsilon \sample \FF_q^* \setminus \{ -\omega^{-2k_1} : k_1 \in [0, K_1 - 1] \};\quad \Omega \gets \usebox{\weomegabox} \\
  \pi_{k_1} \gets \pi\, \upsilon^{2k_1} (1 + \varepsilon\, \omega^{2k_1});\quad s_{\leg} \gets s_{\mathrm{X}}\, \upsilon^{\kleg};\quad \gamma_c \gets \pi_{k_1} s_{\leg};\quad \kappa \gets \gamma_{\ckey}\, L_\rho(\tau) \label{line:scalars} \\
  z_{-} \gets z[-\upsilon/\omega];\quad z_{\circ} \gets z[-\upsilon];\quad z_{+} \gets z[-\omega\upsilon] \label{line:conj} \\
  \vartheta_{\leg} \gets \usebox{\wemaskbox} \label{line:mask} \\
  F^{k_1}_i \gets \pi_{k_1} \bigl( y_i,\ A_i(\tau) \bigr);\quad
  H^{\leg}_j \gets \bigl( s_{\leg} (r_j + A_j(\tau)) + \vartheta_{\leg,j},\ \ s_{\leg} (B_j(\tau) - y_j) \bigr)^{\top} \label{line:factors} \\
  ct_1,\, ct_2 \gets \text{the elements of } \GG_1 \text{ and of } \GG_2 \text{ below};\quad ct_2^\alpha \gets \alpha \cdot ct_2 \label{line:publish} \\
  \sK^\alpha \gets [\alpha\kappa]_T;\quad \hdr \gets m \oplus \Hkey\bigl( [\kappa]_T \bigr) \\
  \pcreturn ct \defeq \bigl( M,\, ct_1,\, ct_2,\, ct_2^\alpha,\, \sK^\alpha,\, \hdr \bigr)
}

\medskip
\renewcommand{\arraystretch}{1.15}
\begin{tabular*}{\linewidth}{@{\extracolsep{\fill}}l@{\hspace{0.6em}}l@{\hspace{0.6em}}l@{\hspace{0.6em}}l@{}}
 & element & indices & role \\
\midrule
$ct_1$ & $\sF^{k_1}_i \defeq [F^{k_1}_i\, \Omega]_1 \in \GG_1^2$ & $k_1$, $i$ & left factors \\
 & $\Pquo_k \defeq [\zeta \tau^k]_1$ & $k \in [0, D-1]$ & quotient powers \\
 & $\Prem_k \defeq [\zeta' \tau^k]_1$ & $k \in [0, \nu\nc - 1]$, $\nu \nmid k$ & remainder powers \\
 & $\sY_i \defeq [\zeta'' y_i]_1$ & $i$ & cancellation column \\
\midrule
$ct_2$ & $\sH^{\leg}_j \defeq [\Omega^{-1} H^{\leg}_j]_2 \in \GG_2^2$ & $\leg$, $j$ & right factors \\
 & $\sS^{\leg} \defeq [s_{\leg}]_2$ & $\leg$ & search bases \\
 & $\Bquo_c \defeq [\gamma_c\, Z_c(\tau)/\zeta]_2$ & $c$ & quotient bases \\
 & $\Brem_c \defeq [\gamma_c\, Z_0(\tau)/\zeta']_2$ & $c$ & remainder bases \\
 & $\sB^{0}_{a,e} \defeq [\pi\, \eta_a\, P(\upsilon)\, \upsilon^e/\zeta'']_2$ & $a \in [3]$, $e \in [0, n-2]$ & \tiebases{} \\
 & $\sB^{\iota}_{a,e} \defeq [\pi\, \eta_a\, \varepsilon^{\iota} \upsilon^e/\zeta'']_2$ & $(a, \iota) \in \cI$, $\iota \ne 0$, $e \in [0, 2n-2]$ & \crossbases{} \\
\end{tabular*}
\end{minipage}}
\caption{Setup and encryption. A line with free indices holds for all left legs $k_1$, right legs $\leg$, copies $c = (k_1, \leg) \in C$, and $i, j \in [0, N]$, with $\mathrm{X}$ the kind of $\leg$. $Z_c(X)$ is $Z(X)(X - \rho)$ for the \keycopy{} and $Z(X)$ for every other copy. The product $\alpha \cdot ct_2$ is entry by entry, and a superscript $\alpha$ on an element of $ct_2$ denotes its $\alpha$-copy in $ct_2^\alpha$.}\label{fig:enc}
\end{figure}

\begin{figure}[!tb]
\centering\small
\fbox{\begin{minipage}{\dimexpr\linewidth-2\fboxsep-2\fboxrule\relax}
\procedure[linenumbering, width=\linewidth]{$\Dec(\pp, ct, w)$, for a witness $w$ of $S$ and $ct = (M, ct_1, ct_2, ct_2^\alpha, \sK^\alpha, \hdr)$}{
  (\varphi_e)_{e<n} \gets \Forms(M);\quad \text{evaluate } \xi_c(w) \text{ for } c \in C \label{line:weights} \\
  \pcif \xikey(w) = 0 \pcthen \pcreturn \bot \\
  \hkey,\, \hoth,\, R_w \gets \text{the quotients and the remainder of } w \text{ (\cref{eq:qap-w})} \\
  x^\iota_a \gets \text{the cancellation polynomials of } w \text{ (\cref{eq:cancel})},\ \text{for } (a, \iota) \in \cI \\
  \sF^{k_1}_w \gets \textstyle\sum_i w_i\, \sF^{k_1}_i;\quad \sF^{\leg}_w \gets \textstyle\sum_{k_1} \xi_{(k_1, \leg)}(w)\, \sF^{k_1}_w;\quad \sY_w \gets \textstyle\sum_i w_i\, \sY_i \label{line:sums} \\
  \sP^{\mathrm{key}}_w \gets \textstyle\sum_k h_{\mathrm{key},k}\, \Pquo_k;\quad
  \sP^{\mathrm{oth}}_w \gets \textstyle\sum_k h_{\mathrm{oth},k}\, \Pquo_k;\quad
  \sP^{\mathrm{rem}}_w \gets \textstyle\sum_k R_{w,k}\, \Prem_k \label{line:msm} \\
  (\Lambda, \Psi) \gets \Comb(ct_2);\quad (\Lambda^\alpha, \Psi^\alpha) \gets \Comb(ct_2^\alpha) \label{line:comb} \\
  \Lambda' \gets \Lambda^\alpha - \xikey(w) \cdot \sK^\alpha - c_w \cdot \Psi^\alpha \label{line:target} \\
  t \gets \text{the integer of least absolute value with } |t| \le B \text{ and } t \cdot \Psi^\alpha = \Lambda' \label{line:search} \\
  \pcif \text{there is no such integer} \pcthen \pcreturn \bot \\
  K \gets \xikey(w)^{-1} \cdot \bigl( \Lambda - (c_w + t) \cdot \Psi \bigr) \label{line:key} \\
  \pcreturn \hdr \oplus \Hkey(K)
}

\medskip
\procedure[linenumbering, width=\linewidth]{$\Comb(ct_2)$; on $ct_2^\alpha$, every element of $ct_2$ is replaced by its $\alpha$-copy}{
  \sH^{\leg}_w \gets \textstyle\sum_i w_i\, \sH^{\leg}_i;\quad
  \sB_w \gets \textstyle\sum_{(a,\iota) \in \cI} \sum_e x^\iota_{a,e}\, \sB^\iota_{a,e} \\
  \Lambda \gets \textstyle\sum_{\leg} e\bigl( \sF^{\leg}_w, \sH^{\leg}_w \bigr)
    - e\bigl( \sP^{\mathrm{key}}_w,\ \xikey(w)\, \Bquo_{\ckey} \bigr) \pcskipln \\
  \t\t {} - e\bigl( \sP^{\mathrm{oth}}_w,\ \textstyle\sum_{c \ne \ckey} \xi_c(w)\, \Bquo_c \bigr)
    - e\bigl( \sP^{\mathrm{rem}}_w,\ \textstyle\sum_{c} \xi_c(w)\, \Brem_c \bigr)
    - e\bigl( \sY_w, \sB_w \bigr) \\
  \Psi \gets \textstyle\sum_{\leg} e\bigl( \sF^{\leg}_{w,1}, \sS^{\leg} \bigr) \\
  \pcreturn (\Lambda, \Psi)
}
\end{minipage}}
\caption{Decryption. Indices are as in \cref{fig:enc}, and $\Comb$ uses the sums of \cref{line:sums,line:msm} of $\Dec$. The pairing of a pair in $\GG_1^2$ with a pair in $\GG_2^2$ is the sum of the pairings of their coordinates, $\sF^{\leg}_{w,1}$ is the first coordinate of $\sF^{\leg}_w$, and $x^\iota_{a,e}$, $h_{\mathrm{key},k}$, $h_{\mathrm{oth},k}$, and $R_{w,k}$ are the coefficients of $T^e$ in $x^\iota_a$ and of $X^k$ in $\hkey$, $\hoth$, and $R_w$. The \emph{center} of the search is $c_w \defeq \bigl\lfloor (1 + \wt(w))(g+3)/2 \bigr\rfloor$, where $\wt(w)$ is the number of ones among $w_1, \dots, w_N$.}\label{fig:dec}
\end{figure}

\parhead{Copies}\phantomsection\label{sec:copy-set}Each weight has a copy of the table, whose \emph{exponent} is that of the weight's power of $T$: copy $(\mathrm{U}, k)$, for $k \in [0, n-1]$, has the weight $U_k$ and the exponent $k$, and copy $(\mathrm{V}, j)$, for $j \in [0, n/2-1]$, the weight $V_j$ and the exponent $2j$. The copies of $U$ form $C_U$, those of $V$ form $C_V$, and each of the $3n/2$ copies $c \in C \defeq C_U \cup C_V$ has weight $\xi_c$ and exponent $k_c$. Only the \emph{\keycopy{}} $\ckey \defeq (\mathrm{U}, 0)$, with weight $\xikey \defeq U_0$, carries the key.

For a stack $\mathbf{W} = (W_c)_{c \in C}$, let $W_U(T) \defeq \sum_{c \in C_U} W_c\, T^{k_c}$ and $W_V(T) \defeq \sum_{c \in C_V} W_c\, T^{k_c}$. The stack \emph{satisfies the ties} (\cref{sec:weights-ties,sec:ext-weights}) if $P(T)$ divides its \emph{tie polynomials} $t_1, t_2, t_3 \in \FF_q[T]^{N+1}$:
\begin{equation}\label{eq:tie-poly}
\begin{aligned}
  t_1(T) &\defeq W_U(T)\, z[-T] - W_V(T)\, z[-T/\omega], \\
  t_2(T) &\defeq W_U(T)\, z[-\omega T] - W_U(-\omega T)\, z[-T/\omega], \\
  t_3(T) &\defeq W_V(T)\, z[-\omega T] - W_U(-\omega T)\, z[-T] .
\end{aligned}
\end{equation}
The stack $\mathbf{H}(v) \defeq (\xi_c(v)\, v v^\top)_{c \in C}$ of every vector $v$ satisfies the ties; that of a witness is its \emph{\honeststack{}}.

\parhead{Weights from an extension field}\phantomsection\label{sec:ext-weights}The quadratic weights of \cref{sec:weights-ties} are all the products $\varphi_e \varphi_{e'}$ of two forms, $n(n+1)/2 = O(\lambda^4)$ copies; $U$ and $V$ give weights of degree two from $3n/2 = O(\lambda^2)$ copies, and \cref{app:proofs} proves the gap of one versus four for them (\cref{lem:heavy,lem:classification}). The idea has three parts. First, the $n$ forms are the coefficients of the one form $z$, and $z(v) = 0$ exactly when $\varphi_e(v) = 0$ for every $e$. Second, a conjugate $\sigma^j(z) = z[\omega^j T]$ rescales coefficient $e$ by $\omega^{je}$, so it is again a public form, and, on a space of dimension at most three where $z$ vanishes only at $0$, three consecutive conjugates of $z$ do what three independent forms do in the warm-up of \cref{sec:weights-ties}. Third, one product of $z$ with a conjugate has $n$ coefficients over $\FF_q$, each a quadratic form in $w$. The three consecutive conjugates $\sigma^{n/2-1}(z)$, $\sigma^{n/2}(z)$, and $\sigma^{n/2+1}(z)$, that is $z[-T/\omega]$, $z[-T]$, and $z[-\omega T]$, take the place of the warm-up's linear weights, each multiplied by $z$, which makes the degree two. The third product, $z \cdot \sigma^{n/2+1}(z)$, is the conjugate under $\sigma^{n/2+1}$ of $U$, as $\sigma^n$ is the identity, so it needs no copies of its own; and $V$ is fixed by $\sigma^{n/2}$, which maps $T$ to $-T$, so it has only even powers of $T$. Each weight is a fixed combination of products of two forms, not one product: $3n/2 = \num{6912}$ weights at $n = \num{4608}$, against over $2^{23}$ products.

\begin{example}[Six weights from four forms]\label{ex:toy-weights}
For $n = 4$, $\omega^2 = -1$ and $\sigma^2(z) = \varphi_0 - \varphi_1 T + \varphi_2 T^2 - \varphi_3 T^3$, so
\[
  V = z \cdot \sigma^2(z) = \bigl( \varphi_0^2 + c_0 \varphi_2^2 - 2 c_0 \varphi_1 \varphi_3 \bigr) + \bigl( 2 \varphi_0 \varphi_2 - \varphi_1^2 - c_0 \varphi_3^2 \bigr)\, T^2
\]
gives two weights and $U = z \cdot \sigma(z)$ gives four: six quadratic forms, where four forms have ten products.
\end{example}

The ties are those of the warm-up for the three conjugates, one for each pair. Read $W_U$ and $W_V$ as matrices over $\FF_{q^n}$, and let $\sigma$ act on them entry by entry; for the stack of $v$, $W_U$, $W_V$, and $\sigma^{n/2+1}(W_U) = W_U(-\omega T)$ are $v v^\top$ times the three products at $v$. Multiplied by a third conjugate, two of them give the same product of three conjugates, grouped in two ways, and the ties ask every stack to allow both groupings:
\begin{align*}
  W_U\, \sigma^{n/2}(z)   &= W_V\, \sigma^{n/2-1}(z), \\
  W_U\, \sigma^{n/2+1}(z) &= \sigma^{n/2+1}(W_U)\, \sigma^{n/2-1}(z), \\
  W_V\, \sigma^{n/2+1}(z) &= \sigma^{n/2+1}(W_U)\, \sigma^{n/2}(z) ,
\end{align*}
identities in $\FF_{q^n}^{N+1}$ whose differences, before reduction modulo $P(T)$, are the tie polynomials of \cref{eq:tie-poly}. On the stack of $v$ both sides of each are the same product of three conjugates of $z(v)$ times $v$; for the first, $z(v)\, \sigma^{n/2-1}(z(v))\, \sigma^{n/2}(z(v))\, v$.

The three steps of the warm-up carry over to a row space $\mathcal{V}$ of dimension at most three without kernel vectors $v$, those with $z(v) = 0$. On such a space $z$ is injective, and then as many consecutive conjugates of $z$ as the dimension are linearly independent over $\FF_{q^n}$, by a Moore determinant (\cref{lem:moore}). Three consecutive conjugates of the one form $z$ thus do on $\mathcal{V}$ what independent random forms do in the warm-up, and the ties make $W_U$, $W_V$, and $\sigma^{n/2+1}(W_U)$ slices of one symmetric tensor over $\FF_{q^n}$. Booleanity reduces it to the cubes of the Boolean assignments in $\mathcal{V}$, at most four, with coefficients in $\FF_{q^n}$. The first and third slices are conjugates of each other and the second is fixed by $\sigma^{n/2}$, which forces the coefficient of the cube of $u$ to be $z(u)$ times a scalar of $\FF_q$ and excludes the tensor without repeated index that linear weights leave. The constraints, read copy by copy, then remove every $u$ that is not a witness. So a passing stack of rank at most three that satisfies the ties and whose row space has no kernel vector is a combination $\sum_w \chi_w\, \mathbf{H}(w)$ of the \honeststack{}s of witnesses $w$ in its row space (\cref{lem:classification}).

\parhead{Shared factors}\phantomsection\label{sec:copies}As in \cref{sec:shared-factors}, the copies' tables, unlike their masks, differ only in their scalars $\gamma_c$, products of a left and a right scalar, so any left factor with any right factor gives a table. For positive integers $K_1$ and $K_2$ with $K_1 K_2 = n/2$, $K_1$ left and $3K_2$ right factors serve all $3n/2$ copies with $2K_1 + 12K_2$ elements for each unknown, $528$ at $(K_1, K_2) = (192, 12)$ (\cref{tab:params}), against six for each copy with its own factors (\cref{fig:grid}), $9n = \num{41472}$ in all. With $K_1, K_2 = O(\lambda)$, for systems of linear size with $N \ge \lambda^2$, the grid saves a factor $\lambda$ in $\Enc$ and $\lambda / \log \lambda$ in $\Dec$ outside the search (\cref{sec:shared-factors,cor:asymptotics}).

The copies are the \emph{cells} of a $K_1 \times 3K_2$ grid (\cref{fig:grid}), with one left factor for each row, a \emph{left leg} $k_1 \in [0, K_1 - 1]$, and one right factor for each column, a \emph{right leg}. As $V$'s exponents are even, the rows step by two, and $U$'s odd exponents need their own columns, of kind O: a right leg $\leg = (\mathrm{X}, k_2) \in \{\mathrm{E}, \mathrm{O}, \mathrm{V}\} \times [0, K_2 - 1]$ has \emph{kind} $\mathrm{X}$ and \emph{exponent} $\kleg \defeq 2K_1 k_2$, or $2K_1 k_2 + 1$ for kind O. The cell $c = (k_1, \leg)$ is the copy of exponent $k_c = 2k_1 + \kleg$: $(\mathrm{U}, k_c)$ on E and O legs and $(\mathrm{V}, k_c/2)$ on V legs; every copy is exactly one cell, and the \keycopy{} is $(0, (\mathrm{E}, 0))$.

Left leg $k_1$ has the left factor $F^{k_1}$, whose scalar $\pi_{k_1}$ the masks fix below, and each right leg $\leg$ has the right factor $H^{\leg}$, with the scalar $s_{\leg} = s_{\mathrm{X}}\, \upsilon^{\kleg}$ for hidden $s_{\mathrm{X}}$ and $\upsilon$. By \cref{line:scalars}, $\gamma_c = \pi_{k_1} s_{\leg}$ has its own monomial $\pi\, s_{\mathrm{X}}\, \upsilon^{k_c}$ in each cell, by which the proof of \cref{lem:view} reads the cell: E and V cells share exponents, so need different scalars, while one scalar would do for E and O, of different parity. By \cref{line:factors},
\[
  F^{k_1}_i H^{\leg}_j = \gamma_c \bigl( y_i\, r_j + y_i\, A_j(\tau) - A_i(\tau)\, y_j + A_i(\tau)\, B_j(\tau) \bigr) + \pi_{k_1}\, y_i\, \vartheta_{\leg,j} ,
\]
entry $(i,j)$ of the table of cell $c$ (\cref{sec:thin-table}) plus the term of the \emph{\tiemask{}} $\vartheta_{\leg} \in \FF_q^{N+1}$ of its right leg.

\parhead{Setup and encryption}
$\Setup$ is deterministic and has no secret, so $\Enc(\pp, m)$ and $\Dec(\pp, ct, w)$ stand for $\Enc(S, m)$ and $\Dec(S, ct, w)$ of \cref{def:we}.

The factors are published under the hidden basis $\Omega$ (\cref{sec:thin-table}). The powers of $\tau$ with the quotient and remainder bases, bound by $\zeta$ and $\zeta'$, remove those parts of the QAP term. If every copy kept the corner term, the QAP term of the \honeststack{} of a witness $w$ would reduce to $L_\rho(\tau) \sum_{c} \gamma_c\, \xi_c(w)$, a mix of the hidden $\gamma_c$ that changes with $w$. So only the quotient base of the \keycopy{} is built on $Z(X)(X - \rho)$ and keeps the corner term; the others are built on $Z(X)$, which divides $L_\rho(X)$, and their quotients absorb it (\cref{lem:qap}). The QAP term then reduces to $\xikey(w)\, \kappa$, a known multiple of the one \keyscalar{} $\kappa = \gamma_{\ckey} L_\rho(\tau)$.

 To enforce the ties (\cref{sec:weights-ties}), copy $c$ needs a mask that combines $z_{-}$, $z_{\circ}$, and $z_{+}$, the values at $T = \upsilon$ of $\sigma^{n/2-1}(z)$, $\sigma^{n/2}(z)$, and $\sigma^{n/2+1}(z)$ (\cref{line:conj}), under a hidden \emph{tie coin} $\eta_a$ for each tie, in a part with $\upsilon^{k_c}$ and, if $c \in C_U$, a second part with $(-\omega\upsilon)^{k_c}$. As \cref{sec:shared-factors} notes, a right leg's mask serves the $K_1$ cells of its column, so its parts carry only $\upsilon^{\kleg}$ and $(-\omega\upsilon)^{\kleg}$ (\cref{line:mask}), and the left scalar must supply $\upsilon^{2k_1}$ to the first and $(-\omega\upsilon)^{2k_1} = (\omega\upsilon)^{2k_1}$ to the second, which differ unless $k_1 = 0$. So $\pi_{k_1} = \pi \bigl( \upsilon^{2k_1} + \varepsilon\, (\omega\upsilon)^{2k_1} \bigr)$ (\cref{line:scalars}) for one more hidden coin $\varepsilon \in \FF_q^* \setminus \{ -\omega^{-2k_1} : k_1 \in [0, K_1 - 1] \}$, and the second part of an E or O mask carries $\varepsilon^{-1}$. The excluded values keep every $\pi_{k_1}$ nonzero. The coin marks the mixed products, the \emph{cross terms}, with $\varepsilon$ or $\varepsilon^{-1}$, so that bases can remove them without removing the terms that enforce the ties.

 The terms of $\pi_{k_1} \vartheta_{\leg}$ without $\varepsilon$ carry $\upsilon^{k_c}$ and $(-\omega\upsilon)^{k_c}$, which, summed with the matrices of a stack $\mathbf{W}$, give $W_U(\upsilon)$, $W_V(\upsilon)$, and the $W_U(-\omega\upsilon)$ of $t_2$ and $t_3$, so along $\mathbf{W}$ these terms contribute $\pi\, y^\top \sum_a \eta_a\, t_a(\upsilon)$ (\cref{eq:mask}). The cancellation column with the \emph{\tiebases{}}, bound by $\zeta''$, removes them, identically in the hidden coins, exactly on the stacks that satisfy the ties. The \emph{\crossbases{}} $\sB^{\iota}_{a,e} = [\pi\, \eta_a\, \varepsilon^{\iota} \upsilon^e/\zeta'']_2$ (\cref{fig:enc}), for $e \in [0, 2n-2]$ and $(a, \iota) \in \cI \defeq ([3] \times \{-1, 0, 1\}) \setminus \{(1,-1)\}$ with $\iota \ne 0$ (no cross term has $\eta_1\, \varepsilon^{-1}$), omit the factor $P(\upsilon)$, so they remove the cross terms of every stack and, carrying $\varepsilon^{\pm 1}$, no term without $\varepsilon$. With the \tiebases{}, they are the \emph{\cancellationbases{}}. Every element of $\GG_2$ has its $\alpha$-copy, for the search. The \keyelement{} $[\kappa]_T$ is not published, only $\sK^\alpha$ and the \maskedmessage{} $\hdr$.

\parhead{Decryption}
$\Dec$ weights the summed left factors for each right leg, $\sF^{\leg}_w = \sum_{k_1} \xi_{(k_1, \leg)}(w)\, \sF^{k_1}_w$ (\cref{line:sums}), and so pairs once for each right leg and coordinate, not for each cell. Along the \honeststack{} of $w$, with $\psi \defeq y(w) \sum_c \gamma_c\, \xi_c(w)$, these pairings encode the search term $r(w)\, \psi$, the QAP term, and the masks' terms. Three pairings with the quotient and remainder bases reduce the QAP term to $\xikey(w)\, \kappa$, by the quotients and the remainder of $w$ (\cref{eq:qap-w}). One with the \cancellationbases{} removes the masks' terms, by the eight \emph{cancellation polynomials} $x^\iota_a$ of \cref{eq:cancel}, where a superscript $\pm$ abbreviates $\pm 1$: $x^{\pm}_a \in \FF_q^{<2n-1}[T]$ for the cross terms and $x^0_a \in \FF_q^{<n-1}[T]$ for the others. On the \honeststack{}, which satisfies the ties, $t_a = P\, x^0_a\, w$, so $x^0_a = (t_a)_0 / P$, as $w_0 = 1$. So $\Comb$ returns $\Lambda = [\xikey(w)\, \kappa + r(w)\, \psi]_T$ and $\Psi = [\psi]_T$, and $\alpha$ times these on the $\alpha$-copies. \Cref{lem:honest,thm:correct} state this identity and correctness for the scheme of \cref{app:spec}.

The search, on the $\alpha$-copies, finds $t = r(w) - c_w$, or $0$ if $\Psi^\alpha = 0$; the center $c_w$ of \cref{fig:dec} is, up to rounding, the mean of $r(w)$, and $|r(w) - c_w| \le B$ always (\cref{lem:range}). \Cref{line:key} then leaves $K = [\kappa]_T$. \Cref{line:search} states the search by its result: any algorithm that finds this integer serves. 
\weSkipFloatBarriers{1}  \subsection{Optimizations}\label{sec:optimizations}

\Cref{sec:efficiency,sec:security} state correctness, costs, and security for \scheme with all the changes of this section (\cref{app:spec}), the scheme we instantiate. These changes bring concrete gains, or gains in lower-order terms: the growths of \cref{cor:asymptotics}, which takes one chunk of powers, rest on none of them.

\parhead{Chunked powers}
The powers of $\tau$ grow in number with the constraints, while each cell has one base per family; chunks trade powers for bases. Let $J, J' \ge 1$ be the numbers of chunks, with $J'$ dividing $\nc$ ($J' = 1$ if $\nc = 0$), and the chunk lengths an integer $\Delta$ with $J \Delta \ge D$ and $\Delta' \defeq \nu\,\nc / J'$. $\Enc$ publishes $\Pquo_k$ only for $k < \Delta$ and $\Prem_k$ only for $k < \Delta'$ with $\nu \nmid k$, and for each cell $c$, in place of its quotient and remainder bases (the case $J = J' = 1$ and $\Delta = D$),
\begin{align*}
  \Bquo_{c,j} &\defeq \bigl[ \gamma_c\, Z_c(\tau)\, \tau^{j \Delta} / \zeta \bigr]_2 && \text{for } j \in [0, J - 1], \\
  \Brem_{c,j} &\defeq \bigl[ \gamma_c\, Z_0(\tau)\, \tau^{j \Delta'} / \zeta' \bigr]_2 && \text{for } j \in [0, J' - 1],
\end{align*}
each with its $\alpha$-copy. $\Dec$ pairs chunk $j$ of each quotient, its coefficients of $X^{j\Delta}$ to $X^{(j+1)\Delta - 1}$, with the bases $j$, and likewise for the remainder (\cref{fig:full-dec}).

For $n \le D$, $n = O(\lambda^2)$ (\cref{sec:efficiency}), and $\nu \le 2F_{\max}$ (\cref{sec:qap}), $J \approx \sqrt{D}/\lambda$ chunks of quotient powers and $J' \approx \sqrt{F_{\max}\,\nc}/\lambda$ of remainder powers cut the powers and bases from $\Theta(D)$ elements to $O(\lambda \sqrt{D} + F_{\max})$. Padding with constraints $0 = 0$ makes $J'$ divide $\nc$, and padding with unknowns makes $\Delta = D/J$ an integer.
At our parameters (\cref{sec:costs}), chunks make the ciphertext $2.25$ percent smaller and, for a witness with $\lfloor N/2 \rfloor$ ones, expected decryption $4.48$ percent slower.

The pairing of $\Pquo_k$ with $\Bquo_{c,j}$ encodes $\gamma_c\, Z_c(\tau)\, \tau^{k + j\Delta}$, and $k + j\Delta$ takes each value below $J\Delta$ once, so the combinations encode $\gamma_c\, Z_c(\tau)\, h(\tau)$ for $h \in \FF_q^{<J\Delta}[X]$, as one chunk of $J\Delta$ powers would; likewise for the allowed remainders, as $\nu$ divides $\Delta'$. For correctness, each sum over the chunks has the exponent of the pairing it replaces. The security theorem holds for all $J$, $J'$, and $\Delta$ that keep the parameters admissible (\cref{sec:security}), including $D \le J\Delta \le D + n$ of \ref{it:p-degree}.

\parhead{A seed for the matrix}
The matrix $M$ has $k(\ell+1)\,d_R$ coefficients in $\FF_q$: $\num{27818368}$ at our parameters, about $0.83$\,GiB, a tenth of the ciphertext. As in Kyber~\cite[Algorithm~4]{Kyber} and Dilithium~\cite[Section~2.3]{Dilithium}, $\Enc$ instead samples a seed $\seed \sample \bits^{\lseed}$, sets $M_{i,j} \defeq \Hseed(\seed, i, j)$ for $i \in [k]$ and $j \in [0,\ell]$ (\cref{sec:random-oracle}), and publishes the seed in place of $M$; $\Dec$ computes $M$ likewise, each with $k(\ell+1)$ queries to $\Hseed$. Correctness is unchanged, as $M$ is again uniform, now over $\Hseed$ (\cref{def:we-correct}). Security still rests on MSIS: a reduction places its challenge in $M$ by programming $\Hseed$ at a fresh seed (\cref{app:reduction}), and Kyber's analysis likewise models its expansion as an RO~\cite[Theorems~1 and~2]{Kyber}.

\parhead{Sign classes}
The search below compares elements of $\GG_T$ up to sign. The \emph{\signclass{}} of $V \in \GG_T$ is $\{V, -V\}$, and its \emph{name} $\name(V)$ identifies it: $\name(V) = \name(V')$ exactly when $V' = \pm V$. On BN254 (\cref{sec:parameters}), names are free: $-V$ is the conjugate of $V$ over $\FF_{p^6}$, and the six coordinates over the base field $\FF_p$ of their shared $\FF_{p^6}$ part determine the class. The generic model therefore gives names to every algorithm (\cref{def:ggm}).

\parhead{A \centeredsampler{}}
So that $r(w) - c_w$ has, up to rounding, one law for every witness, on which the certified bounds of \cref{sec:costs} rest, $\Enc$ draws $r$ from the following sampler, not \cref{fig:enc}'s independent entries.

\begin{definition}[Centered sampler]\label{def:sampler}
For an integer $g \ge 1$ with $g + 1 < q$, the \emph{\centeredsampler{}} $\mathcal{D}_g$ draws $r_1, \dots, r_N, r_\star$ independently and uniformly from $\{2, \dots, g+1\}$ and outputs the integer vector $r = (r_0, r_1, \dots, r_N)$ with
\[
  r_0 \defeq \Bigl\lceil \frac{N(g+1)}{2} \Bigr\rceil + r_\star - \Bigl\lfloor \frac{1}{2} \sum_{i \in [N]} r_i \Bigr\rfloor ,
\]
which is reduced modulo $q$ wherever it multiplies a matrix over $\FF_q$.
\end{definition}

$\Dec$'s center becomes $c_w \defeq \bigl\lfloor \lceil N(g+1)/2 \rceil + \frac{g+3}{2} \bigl( 1 + \wt(w) - N/2 \bigr) \bigr\rfloor$. As $r_0$ subtracts half the sum of the others, each other $r_i$ enters $r(w) - c_w$ as $\pm \frac{1}{2} \bigl( r_i - \frac{g+3}{2} \bigr)$, symmetric about zero, so with the same law for $w_i = 0$ and $1$. Before rounding, the variance of $r(w) - c_w$ is $(1 + N/4)(g^2 - 1)/12$, against $(1 + \wt(w))(g^2 - 1)/12$ for independent entries (the proof of \cref{lem:range}); for $g > 1$, it is smaller above weight $N/4$ and larger below. For such $g$ and weights that grow in proportion to $N$, a normal approximation gives a search about $\bigl( (1 + \wt(w))/(1 + N/4) \bigr)^{1/4}$ times faster, as it costs about the square root of the distance (\cref{prop:efficiency}(d)): $2^{1/4} \approx 1.19$ at weight $N/2$ and $\sqrt{2} \approx 1.41$ at $N$. Through $r_\star$, $r_0$ is uniform on $g$ consecutive integers given the others, so the bounds of \cref{sec:guessing-probability} hold.

\parhead{A faster search}
We search for the integer of \cref{line:full-search} of \cref{fig:full-dec} outward from the center, as \cref{sec:rank-template} does. One fixed table of the names of the baby steps $b \cdot \Psi^\alpha$ for $0 \le b < \bar b$, with $\bar b$ of order $\sqrt{g}\, N^{1/4}$, serves $b$ and $-b$ at once, and the giant steps $\Lambda' - j(2\bar b - 1) \cdot \Psi^\alpha$ for $j = 0, \pm 1, \pm 2, \dots$ cover the integers outward from $0$; it meets the expected bound of \cref{prop:efficiency}(d), whose bound for each result is for growing intervals.

To keep its table small (\cref{sec:costs}), the search compares not names but $64$-bit \emph{fingerprints}, the low bits of a random linear form in the shared coordinates, and checks each match in full. By a linear recurrence along the steps, as for XTR's traces~\cite[Corollary~2.35]{C:LenVer00}, a candidate costs $6$ multiplications and $11$ additions in $\FF_p$, not a multiplication in $\GG_T$.

As the search returns this integer, \cref{thm:correct} applies; capping it at a fixed number of steps adds to the error at most the probability, over the coins of $\Enc$, that $|r(w) - c_w|$ is out of its reach. 
\subsection{Correctness and Efficiency}\label{sec:efficiency}

Correctness rests on two lemmas, proved in \cref{app:calc}.

\begin{lemma}[Honest combination]\label{lem:honest}
Let $w$ be a witness of $S$ and $ct$ an output of $\Enc(\pp, m)$ (\cref{fig:full-enc}), for any coins, with scalars $\gamma_c$ and $\kappa$ as there, with weights $\xi_c(w)$ for the forms of its matrix $M$. Then the divisions of \cref{eq:cancel} are exact, and the combinations defined by $\Comb$ in \cref{fig:full-dec} give
\[
  \Lambda = [\xikey(w)\, \kappa + r(w)\, \psi]_T, \qquad
  \Psi = [\psi]_T, \qquad
  \Lambda^\alpha = \alpha \cdot \Lambda, \qquad
  \Psi^\alpha = \alpha \cdot \Psi,
\]
where $\psi \defeq y(w) \sum_{c \in C} \gamma_c\, \xi_c(w)$.
\end{lemma}

\begin{lemma}[Search range]\label{lem:range}
Let $w \in \bits^{N+1}$ with $w_0 = 1$, and let $r$ and $c_w$ be those of \cref{fig:enc,fig:dec}, or $r \gets \mathcal{D}_g$ and the center of \cref{sec:optimizations}. Then $|r(w) - c_w| \le B$ for every $r$ in the support, and the expectation of $(r(w) - c_w)^2$ is at most $\bigl( g \sqrt{(1 + \wt(w))/12} + 1/2 \bigr)^2$ in the first case and $\bigl( g \sqrt{(N+4)/48} + 3/2 \bigr)^2$ in the second.
\end{lemma}
\begin{theorem}[Correctness]\label{thm:correct}
For every admissible choice of evaluation points, \scheme as specified in \cref{app:spec} is correct with error $2/q$ (\cref{def:we-correct}). For a witness $w$, decryption fails exactly when $\xikey(w) = 0$.
\end{theorem}

\begin{proof}
If $\xikey(w) = 0$, $\Dec$ (\cref{fig:full-dec}) returns $\bot$. Otherwise, by \cref{lem:honest}, $\Lambda' = t_0 \cdot \Psi^\alpha$ for $t_0 \defeq r(w) - c_w$, and $|t_0| \le B$ by \cref{lem:range}. If $\Psi^\alpha \ne 0$, it has order $q > 2B + 1$, so \cref{line:full-search} sets $t = t_0$. If $\Psi^\alpha = 0$, \cref{line:full-search} sets $t = 0$, and $\Psi = 0$, as $\alpha \ne 0$. In both cases $K = \xikey(w)^{-1} \cdot ( \Lambda - r(w) \cdot \Psi ) = [\kappa]_T$, and $\Dec$ returns $\hdr \oplus \Hkey([\kappa]_T) = m$.

Finally, by \cref{def:forms}, as $w_0 = 1$, $z(w) - 1$ is linear, without constant term, in the coefficients of the $M_{i,j}$. So $\xikey(w) = U_0(w)$, a coefficient of $z(w) \cdot \sigma^{n/2-1}(z(w))$, is a polynomial of degree at most two in these coefficients, with value $1$ when all are zero. They are uniform and independent over the choice of $\Hseed$, and independent of $w$, which is fixed before $\Hseed$ (\cref{def:we-correct}), so $\xikey(w) = 0$ with probability at most $2/q$ by the Schwartz--Zippel lemma~\cite{Sch80,Zip79}.
\end{proof}

\begin{proposition}[Efficiency]\label{prop:efficiency}
For \scheme as specified in \cref{app:spec}, with the evaluation points placed in $\mathcal{H}$ as in \cref{sec:qap} and $n \le D$, and with $n_1 \defeq (2K_1 + 1)(N+1) + \Delta + (\nu - 1)\nc/J'$ and $n_2 \defeq 6K_2 (N+1) + 3K_2 + \frac{3n}{2}(J + J') + 13n - 8$:
\begin{enumerate}[label=(\alph*)]
  \item \emph{Size.} A ciphertext consists of $n_1$ elements of $\GG_1$, $2 n_2$ of $\GG_2$, one of $\GG_T$, the seed, and $\mu$ bits.
  \item \emph{Encryption.} $\Enc$ computes one scalar multiplication of a generator for each of these group elements and one more in $\GG_T$.
  \item \emph{Decryption.} Outside \cref{line:full-search}, $\Dec$ computes $18K_2 + 4J + 2J' + 2$ pairings; at most $(2K_1 + 1) N$ additions and MSMs (sums $\sum_{i \le s} a_i P_i$ of length $s$) of total length at most $3n + 2D + (\nu - 1)\nc$ in $\GG_1$; at most $12 K_2 N$ additions and MSMs of total length $3n\,(J + J') + 26n - 16$ in $\GG_2$; and $18K_2 + 4J + 2J' + 1$ additions and four scalar multiplications in $\GG_T$.
  \item \emph{Search.} \Cref{line:full-search} can be computed with $O(\sqrt{|t|} + 1)$ operations in $\GG_T$, where $t$ is its result. On a ciphertext of $\Enc$, this is $O(\sqrt{g}\, N^{1/4})$ operations in expectation over the coins of $\Enc$.
\end{enumerate}
Each of $\Enc$ and $\Dec$ also makes $k(\ell + 1)$ queries to $\Hseed$ and at most one to $\Hkey$, and performs $O(n_1 + n_2 + k(\ell + 1)\, d_R \log d_R + n \log n + D \log D + |S|)$ operations in $\FF_q$. $\Setup$ performs $O(|S| + D \log D + \log q)$.
\end{proposition}

\Cref{app:calc} has the proof, and \cref{sec:costs} the values at our parameters.

\parhead{Asymptotics}
We take $\log q = \Theta(\lambda)$ and groups over fields of $\Theta(\lambda)$ bits or extensions of bounded degree, as for BN254, so a group element is $O(1)$ field elements, a group operation $O(1)$ field operations, a scalar multiplication $O(\lambda)$ group operations, and a pairing $O(\lambda)$ field operations. $\Enc$'s many multiples of one generator take fewer each, with precomputed multiples whose cost is counted (the proof below; \cref{sec:costs} gives the algorithms at our parameters). We count field elements and field operations, and leave out point compression and decompression, hashing, and memory traffic, such as the search's sorting and probing. Let $g = \Theta(2^{\lambda/2})$ (\cref{sec:weights-ties}), $d_R = \Theta(\lambda)$, $K_1, K_2 = O(\lambda)$, $n = O(\lambda^2)$ with $n \le D$, $J = J' = 1$, $\Delta = D$, and $\mu, \lseed = O(\lambda)$, so $k = O(\lambda)$. Lattice-reduction estimates for MSIS (\cref{sec:attack-costs}), not a proof, give this $n$. An encoding of a system of linear size (\cref{sec:quadratic-systems}) that lists its nonzero coefficients has $|S| = O(\lambda N)$. The corollary counts costs: the theorems do not give a secure instance for every $\lambda$ (\cref{def:ledger,lem:inverse}).

\begin{corollary}[Asymptotic costs]\label{cor:asymptotics}
With these parameters and the evaluation points placed in $\mathcal{H}$ as in \cref{sec:qap}, for systems of linear size (\cref{sec:quadratic-systems}) with $N \ge \lambda^2$:
\begin{enumerate}[label=(\alph*)]
  \item \emph{Size.} A ciphertext has $O(\lambda N)$ field elements.
  \item \emph{Encryption.} $\Enc$ takes $O(\lambda^2 N / \log \lambda)$ field operations.
  \item \emph{Decryption.} $\Dec$ takes $O(\lambda N \log \lambda)$ field operations outside \cref{line:full-search}. In it, $\Dec$ takes $O(2^{\lambda/4} N^{1/4})$ field operations in expectation over the coins of $\Enc$.
\end{enumerate}
\end{corollary}

\begin{proof}
As $\nu \le 2F_{\max}$ (\cref{sec:qap}), $D \le N + 2 + 2 F_{\max}\,\nc = O(N)$ and $\log D < \log q = O(\lambda)$. As $N \ge \lambda^2$, $n = O(N)$, and (a) follows. The work in $\FF_q$ is $O(\lambda N \log \lambda)$, by (a) and as $k(\ell + 1)\, d_R = O(\lambda N)$. $\Enc$ also computes $O(\lambda N)$ multiples of each generator, in $O(\lambda N \cdot \lambda / \log(\lambda N))$ group operations, precomputed multiples included~\cite[Section~3]{Pip80}. $\Dec$'s $O(\lambda)$ pairings, $O(\lambda N)$ additions, and MSMs and scalar multiplications, with $O(N)$ scalars at $O(\lambda)$ group operations each, take $O(\lambda N)$ field operations. The search is \cref{prop:efficiency}(d), which compares elements of $\GG_T$, with $\sqrt{g} = \Theta(2^{\lambda/4})$.
\end{proof} 
\subsection{Security}\label{sec:security}
\scheme is extractable (\cref{def:we-ext,cor:ext-error}): when an adversary that makes $Q < 2^\lambda$ calls to the group oracle tells the encryptions of two messages apart, either the extractor reads a witness off these calls or a reduction solves MSIS, up to $(Q+1)/2^{\lambda+1}$.

\begin{theorem}[Extraction]\label{thm:extraction}
Let the parameters be admissible, that is, satisfy \ref{it:p-degree} to \ref{it:p-generic} below. There is an extractor $\mathcal{E}$ (\cref{app:extractor}) such that for every system $S$ of the shape, every choice of evaluation points that meets the conditions of \cref{def:qap}, all $m_0, m_1 \in \bits^\mu$, and every adversary $\mathcal{A}$ that on every run makes at most $Q < 2^\lambda$ calls to $\Ogg$ and at most $2^\lambda$ queries to each RO:
\begin{enumerate}[label=(\alph*),leftmargin=*,nosep]
  \item\label{it:ext-bound} \emph{Extraction.} Let $s_0$ and $s_1$ be the advantages against $\mathsf{MSIS}_{k,\ell,\beta}$ (\cref{def:msis}) of the reductions $\mathcal{B}_0$ and $\mathcal{B}_1$ of \cref{app:reduction}, which run $\mathcal{A}$ once and do not use its output. In the experiment of \cref{def:we-ext},
  \[
    \Pr\bigl[\, \mathcal{A} \text{ queries } \Hkey \text{ at } [\kappa]_T \;\wedge\; \mathcal{E} \text{ outputs no witness of } S \bigm| b \,\bigr] \le \frac{Q+1}{2^\lambda} + s_b .
  \]
  \item\label{it:ext-sound} \emph{Soundness.} On every input, every output of $\mathcal{E}$ is $\bot$ or a witness of $S$.
  \item\label{it:ext-time} \emph{Running time.} $\mathcal{E}$ runs in time $O\bigl( (Q+1) (nN + |S|) + L \bigr)$, where $|S|$ is the length of $S$ (\cref{sec:witness-encryption}) and $L$ counts the labels given to $\mathcal{A}$ (the generators' and the ciphertext's).
\end{enumerate}
\end{theorem}

\parhead{Scope}
The theorem is for \scheme as specified in \cref{app:spec}, for all admissible $J$, $J'$, and $\Delta$, and for any search in $\Dec$, which the experiment never runs. With a published matrix in place of the seed, the reductions place their challenge in it, and the bound is the same.

\parhead{Admissibility}
\emph{Admissible} parameters meet the conditions of \cref{sec:scheme,sec:optimizations} and the five below, as ours do (\cref{sec:parameters}); $\theta$ is the threshold above which a stack is heavy (\cref{def:gp}), and $\beta$ our MSIS norm bound.
\begin{enumerate}[label=(P\arabic*),leftmargin=*,nosep]
  \item\label{it:p-degree} $D \le J\Delta \le D + n$ and $n \le D$, which bounds the degrees of the published exponents in the coins of $\Enc$.
  \item\label{it:p-labels} $3(L + 3 \cdot 2^{\lambda+1} + 1) \le q$, so that a run touches a small part of each group.
  \item\label{it:p-inverse} The conditions of the inverse theorem (\cref{lem:inverse}) on $(q, N, g, \theta, \beta)$, which include $g^{-4} < \theta$ and $2\beta < q$. Under them, a nonzero vector in the row space of a heavy stack on which all forms vanish has a \shortmultiple{}, which solves MSIS.
  \item\label{it:p-lift} $24\beta^4 < q^3$, so that the \liftingstep{} finds a \shortmultiple{} if one exists, in polynomial time (\cref{lem:lift}).
  \item\label{it:p-generic} $E(2) \le 2^{-\lambda}$ and $E(2^{\lambda+1}) - E(2) \le 1 - 2^{-\lambda}$ for the quadratic polynomial $E$ of \cref{def:ledger}, an explicit bound on what is left to chance, so that the proof's generic term $E(2Q+2)$ is at most $(Q+1)/2^\lambda$ (\cref{lem:ledger}); this requires $\theta < 2^{-(2\lambda+1)}$.
\end{enumerate}

\parhead{The bound}
In \ref{it:ext-bound}, the generic term $(Q+1)/2^\lambda$ pays for what the generic model leaves to chance in $Q$ calls: guessed labels, coincidences in $\FF_q$, comparisons that succeed with a light stack, and unlucky coins of the extractor and the reductions (\cref{app:bound}). The term $s_b$ pays for a heavy stack without a witness: $\mathcal{B}_b$ runs $\mathcal{A}$ on an encryption of $m_b$ whose matrix $M$ holds its challenge; flipping the unused output of $\mathcal{A}$ changes neither $s_0$ nor $s_1$ (compare \cref{lem:we-hiding}). As the runs with $m_0$ and $m_1$ have the same law until the query of~\ref{it:ext-bound} (\cref{app:coupling}), an adversary that guesses $b$ with probability $\frac12 + \varepsilon$ makes that query with probability at least $2\varepsilon$, and $\mathcal{E}$ outputs a witness with probability at least $2\varepsilon - (Q+1)/2^\lambda - (s_0 + s_1)/2$. The theorem assumes no hardness; that of MSIS enters only in \cref{sec:parameters}. Nor, by \ref{it:ext-sound}, is it extraction for an augmented relation whose witnesses include MSIS solutions. Lean~4 proves the theorem at our parameters for its formalization of \scheme and of the two reductions (\cref{sec:formal-verification}).

\begin{corollary}[Extractability error]\label{cor:ext-error}
For admissible parameters, \scheme is extractable with error $\delta$ and the extraction time of \cref{thm:extraction}\ref{it:ext-time}, where $\delta(t, Q, Q_H) = (Q+1)/2^{\lambda+1} + \mathrm{Adv}^{\mathsf{MSIS}}(t_R)$ if $Q < 2^\lambda$ and $Q_H \le 2^\lambda$, and $\delta = 1/2$ otherwise. Here $t_R = t + O(Q + d_R\, Q_H + \mu) + \poly(D, \log q)$ bounds the reductions' running time (\cref{prop:reduction-time}), and $\mathrm{Adv}^{\mathsf{MSIS}}(t_R)$ is the largest advantage against $\mathsf{MSIS}_{k,\ell,\beta}$ of an algorithm, with any advice, that runs in time at most $t_R$. Running times are as in \cref{app:reduction}, descriptions and advice included.
\end{corollary}

\begin{proof}
Within the caps, \cref{thm:extraction}\ref{it:ext-bound} and \cref{lem:half} bound the probability of the event of \cref{def:we-ext} by $1/2 + (Q+1)/2^{\lambda+1} + (s_0 + s_1)/4$, and $s_0, s_1 \le \mathrm{Adv}^{\mathsf{MSIS}}(t_R)$, as $\mathcal{B}_0$ and $\mathcal{B}_1$ carry $\mathcal{A}$'s advice, sample the oracles lazily~\ref{it:red-run}, and run in time at most $t_R$. Otherwise $1/2 + \delta = 1$; the proof of \ref{it:ext-time} uses neither cap.
\end{proof}  
\section{Instantiation}\label{sec:instantiation}

\subsection{Witness Encryption for R1CS via Groth16}\label{sec:groth16}
\scheme's ciphertext and costs grow with the system, and so with the relation. As instantiated here, following Goldwasser et al.~\cite{C:GKPVZ13} and Liu et al.~\cite{DCC:LJKW18}, \scheme encrypts to the verifier of a succinct argument of knowledge, with its verification key and public inputs fixed. A decryptor computes a proof from its witness; the proof's bits and those of the verifier's intermediate values solve its system.

Any argument of knowledge with a fully succinct verifier, whose size does not depend on the program, would serve. We use Groth16~\cite{EC:Groth16} on BN254, its verifier over $\FF_q$. Its setup is per circuit, so we fix one key, for a \emph{universal argument of knowledge}, which proves that a program accepts an input and witness within a time bound, the program and input public. All relations share this setup, one system up to constants, and so the ciphertext size and cost bounds given a proof. A universal circuit gives one key for fixed program, input, and time bounds~\cite{USENIX:BCTV14}, so universality needs no assumption beyond Groth16's knowledge soundness. In practice, as in SP1~\cite{SP1}, the circuit instead verifies a faster-proving argument designed in the RO model, and using it would add an assumption, the knowledge soundness of this recursion, for which that model offers no proof~\cite{TCC:BarChiGua25}. Such arguments take statements as CCS~\cite{EPRINT:SetThaWah23a}, Plonkish~\cite{EPRINT:GabWilCio19}, or an algebraic intermediate representation (AIR)~\cite{EPRINT:BBHR18}, and SP1 proves the execution of programs written in languages such as Rust. We assume Groth16's knowledge soundness, with an honest setup (\cref{tab:assumptions}), for the adversary followed by our extractor, as long as the two together cost less than a discrete logarithm on BN254 (\cref{sec:parameters}), which would reveal the setup's trapdoor, with which one can make accepting proofs for every public input. At the extractor's $2^{37}$ operations a call, by Lean's count (\cref{thm:extraction}\ref{it:ext-time}), this covers adversaries that make up to $2^{65}$ calls in time $2^{100}$, precomputation included. Precomputation counts because, with one fixed key, advice could otherwise hold the setup's trapdoor or an accepting proof for a false statement, and every public input has accepting proofs. Whoever ran the setup and kept its trapdoor can also make such proofs, and so decrypt every ciphertext made for this key. \scheme's $\Setup$ is deterministic, with no secret (\cref{sec:scheme}). For concreteness, our counts use SP1's 6.1.0 key as a stand-in for any Groth16 key on BN254: $N = \num{391693}$ Boolean unknowns and \num{1692} constraints in our encoding; other keys give nearly the same. For this key, the encoding is built so that every solution decodes to a proof the verifier accepts, and a computer check confirms the bounds that make its arithmetic in BN254's base field exact. We also tested the encoder on proofs under many other keys against independent implementations of the verifier. 
\subsection{Parameters and Concrete Security}\label{sec:parameters}\label{sec:attack-costs}

\Cref{tab:params} gives the parameters at $\lambda = 100$. The range $g = \num{1736375024658487}$, the least integer with $g^8 \ge 2^{405}$, and the threshold $\theta = (2^{20} - 1)/2^{221}$, just below $2^{-201}$, satisfy $g^{-4} \le 2^{-202} < \theta < g^{-3} < g^{-2}$ (\cref{sec:guessing-probability}).

\begin{table}[!htbp]
\caption{Parameters of the instantiation at $\lambda = 100$.}\label{tab:params}
\centering\small\setlength{\tabcolsep}{4pt}
\begin{tabular}{@{}>{\raggedright\arraybackslash}p{2.0cm}>{\raggedright\arraybackslash}p{3.0cm}>{\raggedright\arraybackslash}p{6.4cm}@{}}
\toprule
Parameter & Value & Role \\
\midrule
$q$ & $\approx 2^{253.6}$ & order of BN254's groups $\GG_1$, $\GG_2$, and $\GG_T$ \\
$\ell_{\mathsf{lab}}$ & $254$ & length of the labels of the generic model \\
$N$, $n_{\mathsf{c}}$, $F_{\max}$ & \num{391693}, \num{1694}, \num{2047} & unknowns, constraints, and products per constraint \\
$\mu$, $\lseed$ & $100$, $100$ & message and seed lengths, at the security level \\
\midrule
$\nu$, $|\mathcal{H}|$ & $2^{11}$, $2^{22}$ & orders of the subgroup whose cosets carry the constraints and of the domain containing all evaluation points \\
$D$ & \num{3861007} & evaluation points, $N + 2 + \nu\, n_{\mathsf{c}}$ \\
$J$, $\Delta$ & $11$, \num{351001} & chunks of the quotient powers, $J \Delta = D + 4$ \\
$J'$, $\Delta'$ & $11$, \num{315392} & chunks of the remainder powers, $\Delta' = \nu\, n_{\mathsf{c}} / J'$ \\
\midrule
$n$, $c_0$ & \num{4608}, $5$ & forms, and $\FF_{q^n} = \FF_q[T]/(T^n - c_0)$ \\
$d_R$, $k$, $\ell$ & $64$, $71$, \num{6121} & degree of $R_q$ and the rows and columns of the MSIS matrix: $n = (k+1)\, d_R$, $\ell = \lceil N / d_R \rceil$ \\
$\beta$ & $\approx 2^{187.31}$ & norm bound of MSIS, from the inverse theorem \\
$K_1$, $K_2$ & $192$, $12$ & left legs, and right legs of each kind: $K_1 K_2 = n/2$, so $3n/2 = \num{6912}$ cells \\
$g$ & $\approx 2^{50.6}$ & range of each secret integer \\
$\theta$ & $\approx 2^{-201}$ & threshold of light stacks \\
\bottomrule
\end{tabular}
\end{table}

These values are admissible (\cref{sec:security}); among the conditions, $g^{-4} < \theta < 2^{-201}$, $2(N+1)(g+1) + 1 < q$, $24\beta^4 < q^3$, $E(2) < 2^{-122}$, and $E(2^{101}) - E(2) \le 1 - 2^{-100}$ (\cref{lem:ledger}). So the extractability error is $(Q+1)/2^{101} + \mathrm{Adv}^{\mathsf{MSIS}}(t_R)$ for $Q < 2^{100}$ and $Q_H \le 2^{100}$ (\cref{cor:ext-error}).

\parhead{Generic search}
The best known attack of the kind that the first term bounds is a generic search for four secret integers at once, in about $2^{101}$ group operations.

\parhead{Pairing groups}
We use BN254~\cite{USENIX:BCTV14}, a Barreto--Naehrig curve~\cite{SAC:BarNae05} with base field $\FF_p$. As $p$ is a polynomial in the curve's seed, the cheapest known attack that uses the curve's representation, which the generic model does not capture, computes a discrete logarithm in $\GG_T$, a subgroup of $\FF_{p^{12}}^*$, with the special variant of the extended tower number field sieve~\cite{C:KimBar16}. With Guillevic and Singh's software~\cite{GS21}, Aranha, El Housni, and Guillevic estimate this for BN254's seed at about $2^{103}$ operations~\cite{DCC:AraElHGui23}.

\parhead{MSIS}
 Our instance is $\mathsf{MSIS}_{k,\ell,\beta}$ over $R_q$ (\cref{def:msis,tab:params}), where $q \equiv 1 \bmod 2d_R$ and $\beta$ is the norm bound that the inverse theorem gives (\cref{lem:inverse}). With the lattice estimator~\cite{APS15,Est} and its default cost model for lattice reduction, MATZOV's~\cite{MATZOV22}, the cheapest lattice reduction that succeeds with constant probability, at block size $242$, costs about $2^{103.5}$ operations. \Cref{thm:extraction} assumes no hardness: its two reductions turn the adversary into solutions of this instance, so its bound has content when the instance is hard at their running time $t_R$, which exceeds the adversary's. Hardness enters only here, in choosing the parameters: as is standard practice~\cite{Kyber,Dilithium}, we set $k$ and $d_R$ by the cost of the best known attack rather than by $t_R$, and state the reductions' loss beside it. Against an adversary that runs in time $2^{100}$, our reductions run in about $2^{108}$, not counting polynomial-time \liftingstep{}s (\cref{prop:reduction-time}). For the proof alone to give $100$ bits, the instance would have to cost about $2^{108}$, $4.5$ bits over this estimate. This limits the proof, not attacks: by the same estimator, a heavy cheat through this lattice costs about $2^{173}$. 
\subsection{Concrete Costs}\label{sec:costs}

For the system of \cref{sec:groth16} at the parameters of \cref{tab:params}, a ciphertext has about $2^{27.2}$ elements of $\GG_1$, $2^{25.8}$ of $\GG_2$, and one of $\GG_T$, $2^{27.6}$ in all (\cref{prop:efficiency}). Koshelev's batch compression~\cite{Kos19,Kos21} packs the factors' elements into about $2^{26.4}$ records of $766$ bits, each with an element of $\GG_2$ and one of $\GG_1$, or three of $\GG_1$; the others take $255$ bits in $\GG_1$ and $512$ in $\GG_2$, and the element of $\GG_T$, the seed, and the \maskedmessage{} $410$ bytes: $7.88$\,GiB in all.

\begin{figure}[!htb]
\centering
\begin{tikzpicture}[x=6cm,y=0.1cm,font=\footnotesize]
\definecolor{figline}{HTML}{3A3A3A}
\begin{scope}[shift={(-32.9,0)}]
  \foreach \y/\lab in {0/1, -10/2^{-10}, -20/2^{-20}, -30/2^{-30}, -40/2^{-40}} {
    \draw[black!15] (32.9,\y) -- (34.45,\y);
    \node[anchor=east] at (32.9,\y) {$\lab$};
  }
  \foreach \x in {33.25, 33.75, 34.25} {
    \draw[black!15] (\x,0) -- (\x,-42);
  }
  \foreach \x/\lab in {33.0/2^{33}, 33.5/2^{33.5}, 34.0/2^{34}} {
    \draw[black!15] (\x,0) -- (\x,-42);
    \node[anchor=north] at (\x,-42) {$\lab$};
  }
  \draw (32.9,-42) rectangle (34.45,0);
  \draw[figline,thick] plot[const plot mark left] file {figure-data.dat};
  \fill[figline] (34.36517,-40.1152) circle (1.4pt);
  \node[anchor=north] at (33.675,-46.5) {field multiplications $x$};
  \node[rotate=90,anchor=south] at (32.77,-21) {probability of needing more than $x$};
\end{scope}
\end{tikzpicture} 
\caption{Certified numerical upper bounds on the probability, over the coins of $\Enc$, that the counted cost of decryption with a witness of $\lfloor N/2 \rfloor$ ones exceeds $x$ field multiplications: $6$ per candidate and about $2^{32.0}$ besides, with the search's setup and the full check of the match, apart from false matches of fingerprints. The dot is the first endpoint below $2^{-40}$.}\label{fig:tail}
\end{figure}
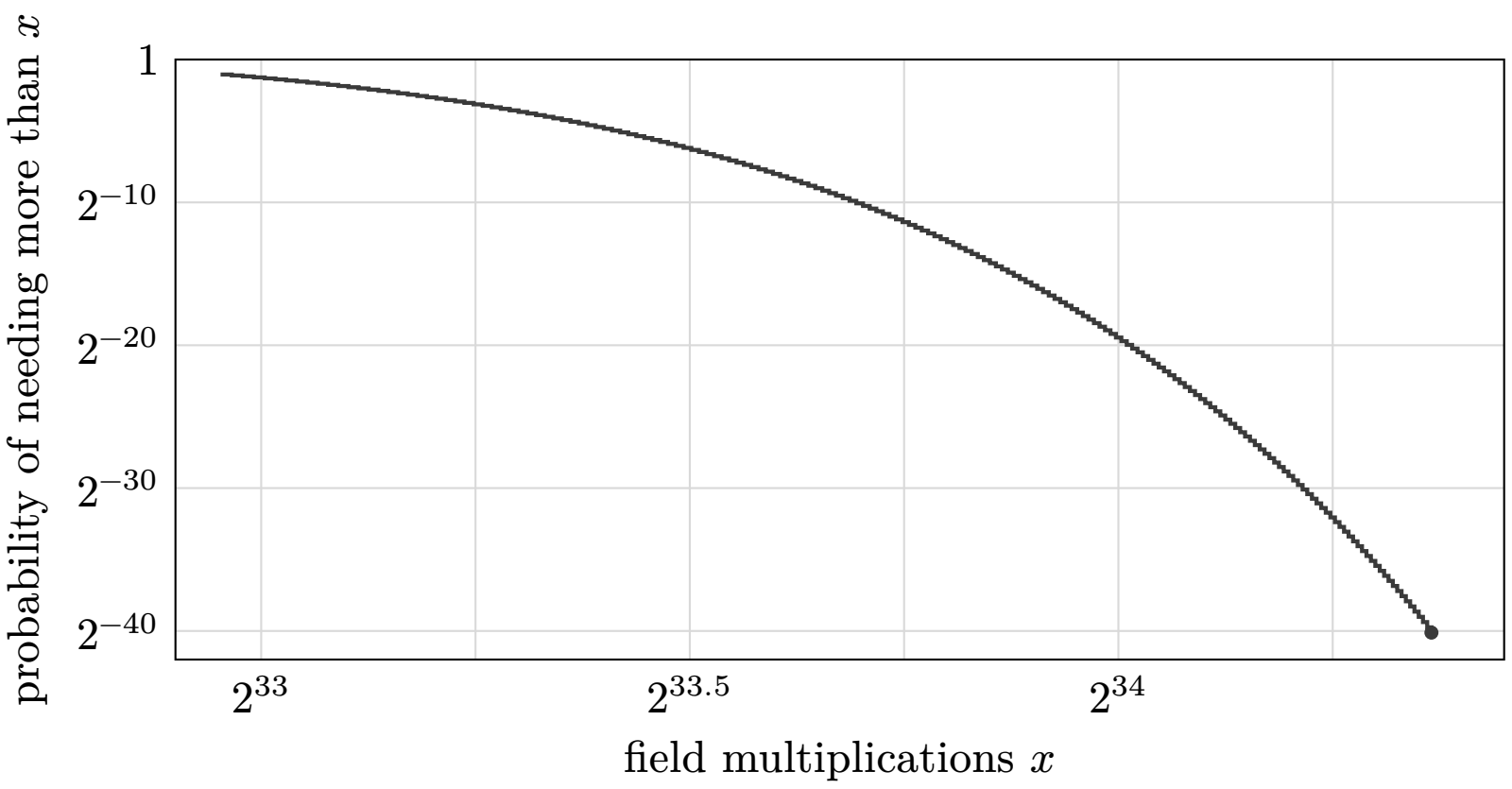

We count multiplications in BN254's base field $\FF_p$, and one in $\FF_q$, also of $254$ bits, as one of them: a squaring as one, as gnark-crypto, the fastest library for BN254's scalar field in a comparison of libraries~\cite{SCN:ECKSJGLO24}, computes it in both fields~\cite{gnark-crypto}, and an inversion as the exponentiation by the field's order minus two~\cite[Algorithm~14.85]{MvOV96}. The counts are exact for the algorithms below, or upper bounds where they depend on data, such as the digits of scalars. We give the additions beside them, subtractions and the like among them: $1.3$ to $1.8$ per multiplication in each total; an addition takes at most a fifth of a multiplication's time~\cite{SCN:ECKSJGLO24,mcl}. As usual, we leave out point compression and decompression. We also leave out hashing, $k(\ell + 1) = \num{434662}$ queries to $\Hseed$ and at most one to $\Hkey$ in each algorithm (\cref{prop:efficiency}), and memory traffic, both small next to the arithmetic. Work done once is counted apart: $\Setup$, about $2^{26.5}$ multiplications and $2^{26.2}$ additions, and $\Enc$'s tables of multiples, $2^{30.4}$ and $2^{31.2}$.

\parhead{Encryption}
$\Enc$ computes one fixed-base scalar multiplication per element of $\GG_1$ and $\GG_2$, with precomputed multiples for signed windows of $22$ bits in $\GG_1$ and $24$ in $\GG_2$~\cite{EC:BGMW92}: at most $11$ and $10$ affine additions~\cite{EFD}, of $6$ and $15$ multiplications besides one inversion shared among $2048$ additions~\cite{Mon87}. With its work in $\FF_q$, mostly a multiplication and an addition for each of the system's about $2^{30.0}$ nonzero coefficients, $\Enc$ takes about $2^{34.2}$ multiplications and $2^{34.6}$ additions.

\parhead{Decryption}
For a witness with $\lfloor N/2 \rfloor$ ones, near the $49.9\%$ of the witness of a published SP1 v6.0.0 proof, $\Dec$ sums $h = (N+1)/2$ of the $N + 1$ entries, counting $w_0$, in each of $2K_1 + 1 + 12K_2 = 529$ arrays, by trees of affine additions batched as in $\Enc$: about $2^{29.7}$ multiplications. Pippenger's MSMs~\cite{Pip80}, with signed digits and buckets in XYZZ coordinates~\cite{EFD}, take about $2^{31.3}$, mostly the $2J + J' = 33$ in $\GG_1$ of length at most $\Delta$, at about $200$ per point. The $284$ pairings, in four sums with one final exponentiation each~\cite{mcl}, take about $2^{20.8}$, and the work in $\FF_q$ about $2^{29.3}$. These total about $2^{32.0}$ multiplications and $2^{32.4}$ additions, fixed per witness, of which $2^{29.2}$ and $2^{31.0}$ do not use the ciphertext. With the search, whose count alone varies, $\Dec$ takes about $2^{33.0}$ multiplications and $2^{33.7}$ additions in expectation, $51\%$ of the multiplications in the search, and, apart from false matches of fingerprints, at most $2^{34.4}$ and $2^{35.2}$ except with probability below $2^{-40}$, $81\%$ in the search (\cref{fig:tail}). Any witness sums at most $N + 1$ entries, which keeps the two counts below $2^{33.2}$ and $2^{34.5}$ multiplications, and $2^{33.9}$ and $2^{35.3}$ additions.

\parhead{Search}
\Cref{sec:optimizations}'s search uses about $2^{28.7}$ baby steps~\cite{Sha71} and compares giant steps in rounds of $12 \cdot 2^{16}$ values of $\lvert j \rvert$, then $2$, $4$, and thereafter $8$ times as many, until a round finds the integer: about $2^{29.4}$ baby and giant steps in expectation and at most $2^{31.5}$ except with probability below $2^{-40}$, by certified bounds on the law of $r(w) - c_w$ for any witness (\cref{lem:range}, \cref{app:search-law}). In expectation, false matches of fingerprints, each a further full check, number at most $0.014$, and $0.13$ in a search of $2^{31.5}$ steps, as two names collide with probability at most $2^{-63} + 1/p$ over the random linear form~\cite{CW79}.

\parhead{Memory}
Neither algorithm needs to hold the whole ciphertext: $\Enc$ writes each element as it computes it, and $\Dec$ reads each record once, in order, into partial sums and Pippenger's buckets. $\Enc$'s peak is its tables of precomputed multiples, the same for every ciphertext, $12.35$\,GB: $2^{21}$ affine points of $64$ bytes for each of its $12$ windows in $\GG_1$, and $2^{23}$ of $128$ bytes for each of its $11$ in $\GG_2$, but $\lfloor (q-1)/2^{240} \rfloor + 1 = \num{12389}$ in the last, enough for scalars below $q$. $\Dec$'s peak is the search's sorted table of $64$-bit fingerprints, each with the $32$-bit index of its baby step, $12$ bytes per baby step: $5.28$\,GB plus a small directory. Holding the whole ciphertext, $8.46$\,GB as sent, works too but is not needed.

\subsection{Formal Verification}\label{sec:formal-verification}

We formalize \scheme, with the key derived from the name of $[\kappa]_T$ rather than from $[\kappa]_T$, at the parameters of \cref{tab:params}, with $256$-bit messages and seeds, in Lean~4~\cite{MU21} (\href{https://github.com/psivesely/praxis-lean}{\nolinkurl{github.com/psivesely/praxis-lean}})\begin{table}[!tbp]
\caption{What the theorems and the $100$-bit level rest on.}\label{tab:assumptions}
\centering\footnotesize\setlength{\tabcolsep}{4pt}
\begin{tabular}{@{}>{\raggedright\arraybackslash}p{3.9cm}>{\raggedright\arraybackslash}p{7.7cm}@{}}
\toprule
Assumption or result & Justification \\
\midrule
bilinear GGM & standard for pairing-based schemes~\cite{EC:Shoup97,EC:BonBoy04a} \\
discrete logarithm in BN254's $\GG_T$ & the best known attack outside that model, about $2^{103}$ by the tower number field sieve~\cite{C:KimBar16,DCC:AraElHGui23} \\
ROs & the standard heuristic~\cite{CCS:BelRog93}; no query is observed, $\Hkey$ is never programmed, and $\Hseed$ only at one fresh seed, to place an MSIS challenge \\
MSIS hardness & standard~\cite{Dilithium}; a worst-case to average-case reduction for suitable parameters~\cite{DCC:LanSte15} \\
parameters by best known attacks & standard practice~\cite{Kyber,Dilithium}; \cref{sec:parameters} states our reductions' loss \\
universal argument of knowledge & Groth16's knowledge soundness on a universal circuit~\cite{USENIX:BCTV14} with an honest setup, a generic-group theorem~\cite{EC:Groth16}; and a correct encoding of the verifier \\
running time of the \liftingstep{} & polynomial by Lenstra's and Kannan and Bachem's algorithms~\cite{Len83,KB79} \\
\bottomrule
\end{tabular}
\end{table}\begin{table}[!tbp]
\caption{What the formalization leaves to the paper, and why.}\label{tab:gaps}
\centering\footnotesize\setlength{\tabcolsep}{4pt}
\begin{tabular}{@{}>{\raggedright\arraybackslash}p{3.9cm}>{\raggedright\arraybackslash}p{7.7cm}@{}}
\toprule
Left to the paper & Reason \\
\midrule
both theorems at other parameters & Lean fixes those of \cref{tab:params}, with $256$-bit messages and seeds, but proves the inverse theorem for all that meet its conditions \\
the reductions' running time & Lean counts only an adversary's oracle calls \\
the paper's bound & Lean proves it for its own reductions \\
faithfulness of the definitions & checked by reading, as for every formalization \\
\bottomrule
\end{tabular}
\end{table}
and prove both theorems there, correctness (\cref{thm:correct}) and extraction (\cref{thm:extraction}), for every system of the shape; the paper proves both for general parameters (\cref{tab:gaps}). Extraction is for every generic adversary with $Q < 2^{100}$ group-oracle calls and at most $2^{100}$ queries to each RO on every run: Lean states, for its own reductions, the guessing bound $1/2 + (Q+1)/2^{101} + (s_0 + s_1)/2$, the form of \cref{cor:ext-error}, and proves it through the bound~\ref{it:ext-bound}; it also proves soundness~\ref{it:ext-sound}, an explicit operation count for the extractor~\ref{it:ext-time}, and Lean's form of the coupling of \cref{app:coupling}, from which, with the bound~\ref{it:ext-bound}, \cref{sec:security} derives the form of extraction that \cref{sec:results} states. Lean's group oracle answers a name request with an opaque reference, not the two labels of \cref{def:ggm}; this excludes no adversary: negating the element takes one call, as the request does, and yields the other label, and Lean's key oracle also accepts the pair. Lean's inverse theorem (\cref{lem:inverse}) has $r_i$ uniform on $\{0, \dots, g-1\}$; as $v_0 = 0$, the shift by $2$ to $[2, g+1]$ keeps $\gp(v^\top)$.

\parhead{Key derivation}
The formalization predates the simpler derivation from $[\kappa]_T$. Correctness is unchanged, as $\Dec$ recovers $[\kappa]_T$ exactly before deriving the key. Extraction depends on the derivation only through queries at the key oracle's target, the name of $[\kappa]_T$ in Lean and $[\kappa]_T$ here; the first of them ends the coupling of the runs on $m_0$ and $m_1$. With a label as the target, the monitor's target rule loses a case, and guessing the target costs at most $Q_H/\Nfree$ (\cref{lem:names}), inside $E$'s fifth term, which pays for Lean's guessed names; so \cref{app:proofs} proves the bound~\ref{it:ext-bound} with the same $E$.

\parhead{Running time}
As usual in the generic model, Lean counts an adversary's oracle calls but not its computation between them, and so cannot state its running time or the reductions'. These enter only in the extractability error (\cref{cor:ext-error}), and the paper proves the reductions' (\cref{prop:reduction-time}). \Cref{thm:extraction} is stated to need neither: its bound~\ref{it:ext-bound} assumes no hardness, and $s_b$ is the probability that the reduction $\mathcal{B}_b$, which runs the adversary once on an encryption of $m_b$ whose matrix holds its challenge, solves the MSIS instance of \cref{tab:params}.

\parhead{The reductions}
 Like those of \cref{app:reduction}, Lean's reductions program $\Hseed$ at one fresh seed; they differ in two ways, neither of which changes the argument. They search exhaustively for the \shortmultiple{}, succeeding exactly when the \liftingstep{} does (\cref{lem:lift}). Their \privatemonitor{} reads the queries to $\Hkey$ from a log, as Pass's model~\cite{C:Pass03} and non-programming reductions~\cite{AC:FLRSST10} allow, where the paper's reads none and gives two expressions; it also gives up at a disagreement in $\GG_1$ or $\GG_2$, stops at the \keyelement{}'s name only if its record of it differs from $\kappa$, and reads an unrecorded operand of $\GG_1$ or $\GG_2$ as zero without recording it. Both proofs use the reductions only at a good first mark (\cref{app:monitor}), where both versions try the mark's expression, up to formal equality, except with a probability counted in the term $(Q+1)/2^\lambda$; at every other first mark that term pays, whatever the reductions do. Lean's success probabilities may thus differ from $s_0$ and $s_1$: Lean checks its own version of the argument, and \cref{app:proofs} gives the paper's.

\parhead{Axioms}
Lean's kernel checks the proofs, built on Mathlib~\cite{mathlib20} and VCVio~\cite{TDWHH26}, with only the three standard axioms that make Lean's type theory classical: propositional extensionality, quotient soundness, and choice.  

\begin{credits}
\subsubsection{\ackname}
This paper is the result of a research process carried out with AI, mostly in the weeks since Hair and Sahai's paper~\cite{HS26} appeared; it began by interrogating their construction and then trying to build on it. We used Claude and GPT to assist with the research and to draft text and proofs. The research ideas are ours, as are some of the directions and techniques and the main ideas for the improvements; for the most part, the models worked out the details of how to carry them out. For example, our earlier work on zero-knowledge proofs led us to suggest the techniques based on QAPs, and we suggested extracting a solution of the system as a way to rule out low ranks; in both cases, the models worked out the concrete approach. We also found a number of simplifications of the scheme, and we distilled the results into the set of techniques and the methodology the paper presents. The models also wrote the Lean proofs and checked the proofs and computations. We take full responsibility for the paper and its contents.
\end{credits} 

\FloatBarrier  \phantomsection\addcontentsline{toc}{section}{\refname}
\printbibliography

\appendix
\section{Full Specification}\label{app:spec}

\Cref{fig:full-enc,fig:full-dec} give \scheme with the optimizations of \cref{sec:optimizations}.

\newsavebox{\wefullmaskbox}
\newsavebox{\wefullomegabox}
\begin{figure}[!tb]
\centering\small
\sbox{\wefullomegabox}{$\bigl(\begin{smallmatrix} 1 & \Omega_{12} \\ 0 & \Omega_{22} \end{smallmatrix}\bigr)$}
\sbox{\wefullmaskbox}{$\begin{cases}
  \upsilon^{\kleg} \bigl( \eta_1 z_{\circ} + \eta_2 z_{+} \bigr) - \varepsilon^{-1} (-\omega\upsilon)^{\kleg} \bigl( \eta_2 z_{-} + \eta_3 z_{\circ} \bigr) & \text{if } \mathrm{X} \in \{\mathrm{E}, \mathrm{O}\} \\
  \upsilon^{\kleg} \bigl( \eta_3 z_{+} - \eta_1 z_{-} \bigr) & \text{if } \mathrm{X} = \mathrm{V}
\end{cases}$}
\fbox{\begin{minipage}{\dimexpr\linewidth-2\fboxsep-2\fboxrule\relax}
\procedure[linenumbering, width=\linewidth]{$\Setup(S)$, for a system $S$ of shape $(N, \nc, F_{\max})$}{
  \text{choose admissible evaluation points for } S \text{ (\cref{def:qap})} \\
  \text{compute } A_i(X) \text{ and } B_i(X), \text{ for } i \in [0,N], \text{ and } Z_0(X),\ Z(X), \text{ and } L_\rho(X) \\
  \pcreturn \pp \defeq \bigl( S, \text{ the evaluation points}, (A_i, B_i)_{i}, Z_0, Z, L_\rho \bigr)
}

\medskip
\procedure[linenumbering, width=\linewidth]{$\Enc(\pp, m)$, for a message $m \in \bits^\mu$}{
  \seed \sample \bits^{\lseed};\quad M \gets \bigl( \Hseed(\seed, i, j) \bigr)_{i \in [k],\, j \in [0, \ell]};\quad (\varphi_e)_{e<n} \gets \Forms(M) \label{line:full-seed} \\
  r \gets \mathcal{D}_g;\quad y \sample \FF_q^{N+1};\quad \tau \sample \FF_q \setminus \mathcal{P};\quad \Omega_{12} \sample \FF_q \\
  \Omega_{22},\, \alpha,\, \pi,\, s_{\mathrm{E}},\, s_{\mathrm{O}},\, s_{\mathrm{V}},\, \upsilon,\, \zeta,\, \zeta',\, \zeta'',\, \eta_1,\, \eta_2,\, \eta_3 \sample \FF_q^* \\
  \varepsilon \sample \FF_q^* \setminus \{ -\omega^{-2k_1} : k_1 \in [0, K_1 - 1] \};\quad \Omega \gets \usebox{\wefullomegabox} \\
  \pi_{k_1} \gets \pi\, \upsilon^{2k_1} (1 + \varepsilon\, \omega^{2k_1});\quad s_{\leg} \gets s_{\mathrm{X}}\, \upsilon^{\kleg};\quad \gamma_c \gets \pi_{k_1} s_{\leg};\quad \kappa \gets \gamma_{\ckey}\, L_\rho(\tau) \label{line:full-scalars} \\
  z_{-} \gets z[-\upsilon/\omega];\quad z_{\circ} \gets z[-\upsilon];\quad z_{+} \gets z[-\omega\upsilon] \label{line:full-conj} \\
  \vartheta_{\leg} \gets \usebox{\wefullmaskbox} \label{line:full-mask} \\
  F^{k_1}_i \gets \pi_{k_1} \bigl( y_i,\ A_i(\tau) \bigr);\quad
  H^{\leg}_j \gets \bigl( s_{\leg} (r_j + A_j(\tau)) + \vartheta_{\leg,j},\ \ s_{\leg} (B_j(\tau) - y_j) \bigr)^{\top} \label{line:full-factors} \\
  ct_1,\, ct_2 \gets \text{the elements of } \GG_1 \text{ and of } \GG_2 \text{ below};\quad ct_2^\alpha \gets \alpha \cdot ct_2 \label{line:full-publish} \\
  \sK^\alpha \gets [\alpha\kappa]_T;\quad \hdr \gets m \oplus \Hkey\bigl( [\kappa]_T \bigr) \label{line:full-header} \\
  \pcreturn ct \defeq \bigl( \seed,\, ct_1,\, ct_2,\, ct_2^\alpha,\, \sK^\alpha,\, \hdr \bigr)
}

\medskip
\renewcommand{\arraystretch}{1.15}
\begin{tabular*}{\linewidth}{@{\extracolsep{\fill}}l@{\hspace{0.6em}}l@{\hspace{0.6em}}l@{\hspace{0.6em}}l@{}}
 & element & indices & role \\
\midrule
$ct_1$ & $\sF^{k_1}_i \defeq [F^{k_1}_i\, \Omega]_1 \in \GG_1^2$ & $k_1$, $i$ & left factors \\
 & $\Pquo_k \defeq [\zeta \tau^k]_1$ & $k \in [0, \Delta - 1]$ & quotient powers \\
 & $\Prem_k \defeq [\zeta' \tau^k]_1$ & $k \in [0, \Delta' - 1]$, $\nu \nmid k$ & remainder powers \\
 & $\sY_i \defeq [\zeta'' y_i]_1$ & $i$ & cancellation column \\
\midrule
$ct_2$ & $\sH^{\leg}_j \defeq [\Omega^{-1} H^{\leg}_j]_2 \in \GG_2^2$ & $\leg$, $j$ & right factors \\
 & $\sS^{\leg} \defeq [s_{\leg}]_2$ & $\leg$ & search bases \\
 & $\Bquo_{c,j} \defeq [\gamma_c\, Z_c(\tau)\, \tau^{j\Delta}/\zeta]_2$ & $c$, $j \in [0, J - 1]$ & quotient bases \\
 & $\Brem_{c,j} \defeq [\gamma_c\, Z_0(\tau)\, \tau^{j\Delta'}/\zeta']_2$ & $c$, $j \in [0, J' - 1]$ & remainder bases \\
 & $\sB^{0}_{a,e} \defeq [\pi\, \eta_a\, P(\upsilon)\, \upsilon^e/\zeta'']_2$ & $a \in [3]$, $e \in [0, n-2]$ & \tiebases{} \\
 & $\sB^{\iota}_{a,e} \defeq [\pi\, \eta_a\, \varepsilon^{\iota} \upsilon^e/\zeta'']_2$ & $(a, \iota) \in \cI$, $\iota \ne 0$, $e \in [0, 2n-2]$ & \crossbases{} \\
\end{tabular*}
\end{minipage}}
\caption{\scheme in full: setup and encryption. Notation is as in \cref{fig:enc} and \cref{sec:optimizations}, with $\mathrm{X}$ the kind of $\leg$ and the chunk index $j$ ranging as listed.}\label{fig:full-enc}
\end{figure}

\begin{figure}[!tb]
\centering\small
\fbox{\begin{minipage}{\dimexpr\linewidth-2\fboxsep-2\fboxrule\relax}
\procedure[linenumbering, width=\linewidth]{$\Dec(\pp, ct, w)$, for a witness $w$ of $S$ and $ct = (\seed, ct_1, ct_2, ct_2^\alpha, \sK^\alpha, \hdr)$}{
  M \gets \bigl( \Hseed(\seed, i, j) \bigr)_{i,j};\quad (\varphi_e)_{e<n} \gets \Forms(M);\quad \text{evaluate } \xi_c(w) \text{ for } c \in C \label{line:full-weights} \\
  \pcif \xikey(w) = 0 \pcthen \pcreturn \bot \\
  \hkey,\, \hoth,\, R_w \gets \text{the quotients and the remainder of } w \text{ (\cref{eq:qap-w})} \\
  x^\iota_a \gets \text{the cancellation polynomials of } w \text{ (\cref{eq:cancel})},\ \text{for } (a, \iota) \in \cI \\
  \sF^{k_1}_w \gets \textstyle\sum_i w_i\, \sF^{k_1}_i;\quad \sF^{\leg}_w \gets \textstyle\sum_{k_1} \xi_{(k_1, \leg)}(w)\, \sF^{k_1}_w;\quad \sY_w \gets \textstyle\sum_i w_i\, \sY_i \label{line:full-sums} \\
  \sP^{\mathrm{key},j}_w \gets \textstyle\sum_k h^{(j)}_{\mathrm{key},k}\, \Pquo_k;\quad
  \sP^{\mathrm{oth},j}_w \gets \textstyle\sum_k h^{(j)}_{\mathrm{oth},k}\, \Pquo_k;\quad
  \sP^{\mathrm{rem},j}_w \gets \textstyle\sum_k R^{(j)}_{w,k}\, \Prem_k \label{line:full-msm} \\
  (\Lambda, \Psi) \gets \Comb(ct_2);\quad (\Lambda^\alpha, \Psi^\alpha) \gets \Comb(ct_2^\alpha) \label{line:full-comb} \\
  \Lambda' \gets \Lambda^\alpha - \xikey(w) \cdot \sK^\alpha - c_w \cdot \Psi^\alpha \label{line:full-target} \\
  t \gets \text{the integer of least absolute value with } |t| \le B \text{ and } t \cdot \Psi^\alpha = \Lambda' \label{line:full-search} \\
  \pcif \text{there is no such integer} \pcthen \pcreturn \bot \\
  K \gets \xikey(w)^{-1} \cdot \bigl( \Lambda - (c_w + t) \cdot \Psi \bigr) \label{line:full-key} \\
  \pcreturn \hdr \oplus \Hkey(K)
}

\medskip
\procedure[linenumbering, width=\linewidth]{$\Comb(ct_2)$; on $ct_2^\alpha$, every element of $ct_2$ is replaced by its $\alpha$-copy}{
  \sH^{\leg}_w \gets \textstyle\sum_i w_i\, \sH^{\leg}_i;\quad
  \sB_w \gets \textstyle\sum_{(a,\iota) \in \cI} \sum_e x^\iota_{a,e}\, \sB^\iota_{a,e} \\
  \Lambda \gets \textstyle\sum_{\leg} e\bigl( \sF^{\leg}_w, \sH^{\leg}_w \bigr)
    - \sum_{j<J} e\bigl( \sP^{\mathrm{key},j}_w,\ \xikey(w)\, \Bquo_{\ckey,j} \bigr) \pcskipln \\
  \t\t {} - \textstyle\sum_{j<J} e\bigl( \sP^{\mathrm{oth},j}_w,\ \textstyle\sum_{c \ne \ckey} \xi_c(w)\, \Bquo_{c,j} \bigr) \pcskipln \\
  \t\t {} - \textstyle\sum_{j<J'} e\bigl( \sP^{\mathrm{rem},j}_w,\ \textstyle\sum_{c} \xi_c(w)\, \Brem_{c,j} \bigr)
    - e\bigl( \sY_w, \sB_w \bigr) \\
  \Psi \gets \textstyle\sum_{\leg} e\bigl( \sF^{\leg}_{w,1}, \sS^{\leg} \bigr) \\
  \pcreturn (\Lambda, \Psi)
}
\end{minipage}}
\caption{\scheme in full: decryption. Notation is as in \cref{fig:dec} and \cref{sec:optimizations}, with $x^\iota_{a,e}$ the coefficient of $T^e$ in $x^\iota_a$. In \cref{line:full-msm}, $h^{(j)}_{\mathrm{key}}, h^{(j)}_{\mathrm{oth}} \in \FF_q^{<\Delta}[X]$ for $j < J$ and $R^{(j)}_w \in \FF_q^{<\Delta'}[X]$ for $j < J'$ are the chunks of $\hkey$, $\hoth$, and $R_w$, as in $\hkey(X) = \sum_{j<J} X^{j\Delta}\, h^{(j)}_{\mathrm{key}}(X)$, with coefficients $h^{(j)}_{\mathrm{key},k}$, $h^{(j)}_{\mathrm{oth},k}$, and $R^{(j)}_{w,k}$ of $X^k$; $c_w$ is the center for $\mathcal{D}_g$ (\cref{sec:optimizations}).}\label{fig:full-dec}
\end{figure} 
\weSkipFloatBarriers{2}  \section{Deferred Proofs}\label{app:proofs}

\makeatletter
\makeatother

\subsection{The QAP Criterion, Correctness, and Efficiency}\label{app:calc}

\begin{proof}[of \cref{lem:qap}]
By the table of \cref{def:qap}, (a) holds exactly when $P_W(d_j) = 0$ for every $j$ and the values of $R_W$ sum to zero over each coset. Let $R(X)$ be any polynomial in $\FF_q^{<\nu\nc}[X]$, as every allowed polynomial is. The identity of (b) holds with this $R$ and some $h$ exactly when $Z(X)(X - \rho)$, the vanishing polynomial of $\mathcal{P}$, divides $P_W - W_{00} L_\rho - Z_0 R$, whose values are zero at $\rho$ and $d_0$, $P_W(d_j)$ at $d_j$, and $P_W(p) - Z_0(p)\, R(p)$ at a point $p$ of a coset. So it holds exactly when $P_W(d_j) = 0$ for every $j$ and $R = R_W$, the interpolant of the $\nu\,\nc$ values on the cosets; likewise that of (c), where $Z(X)$ divides $P_W - Z_0 R$. So (b) and (c) each hold exactly when $P_W(d_j) = 0$ for every $j$ and $R_W$ is allowed, and then only with $R = R_W$.

Finally, by the paragraph before \cref{lem:qap}, $R_W$ is allowed exactly when its values sum to zero over each coset. The degree bounds hold because $P_W$ has degree at most $2D - 2$, $L_\rho$ and $Z_0 R_W$ at most $D - 1$, and $Z$ exactly $D - 1$.
\end{proof}

\parhead{Ties}
Lemmas~\ref{lem:honest} and~\ref{lem:view} rest on the tie polynomials, the cancellation polynomials of $\Dec$, and the mask identity below. For a stack $\mathbf{W}$, the sums $W_U$ and $W_V$ and the tie polynomials $t_1$, $t_2$, and $t_3$ are those of \cref{sec:copy-set} (\cref{eq:tie-poly}). As $\sigma^j$ maps $a(T)$ to $a(\omega^j T)$ and $\omega^{n/2} = -1$, $P(T)$ divides the tie polynomials exactly if the stack satisfies the ties as \cref{sec:ext-weights} states them.

\parhead{Cancellation polynomials}
For a witness $w$, let $U_w$ and $V_w$ be the sums $W_U$ and $W_V$ with $\xi_c(w)$ for $W_c$, $U'_w$ and $V'_w$ these sums with the term of cell $c = (k_1, \leg)$ times $\omega^{2k_1}$, and $U''_w$ the sum $U_w$ with that term times $(-\omega)^{\kleg}$, and let $z^-_w$, $z^\circ_w$, $z^+_w$ be $w^\top z[-T/\omega]$, $w^\top z[-T]$, $w^\top z[-\omega T]$. Its \emph{cancellation polynomials} are
\begin{equation}\label{eq:cancel}
\begin{alignedat}{3}
  x^0_1 &\defeq ( U_w\, z^\circ_w - V_w\, z^-_w ) / P, &\quad
  x^+_1 &\defeq U'_w\, z^\circ_w - V'_w\, z^-_w, \\
  x^0_2 &\defeq ( U_w\, z^+_w - U_w(-\omega T)\, z^-_w ) / P, &\quad
  x^+_2 &\defeq U'_w\, z^+_w, &\quad
  x^-_2 &\defeq -U''_w\, z^-_w, \\
  x^0_3 &\defeq ( V_w\, z^+_w - U_w(-\omega T)\, z^\circ_w ) / P, &\quad
  x^+_3 &\defeq V'_w\, z^+_w, &\quad
  x^-_3 &\defeq -U''_w\, z^\circ_w .
\end{alignedat}
\end{equation}

\parhead{Masks}
By \cref{line:full-scalars,line:full-conj,line:full-mask} of $\Enc$ (\cref{fig:full-enc}), as $k_c = 2k_1 + \kleg$, the part without $\varepsilon$ of $\pi_{k_1} \vartheta_{\leg}$ is $\pi \bigl( \upsilon^{k_c} (\eta_1 z_\circ + \eta_2 z_+) - (-\omega\upsilon)^{k_c} (\eta_2 z_- + \eta_3 z_\circ) \bigr)$ for the kinds E and O and $\pi\, \upsilon^{k_c} (\eta_3 z_+ - \eta_1 z_-)$ for V. Summed with the $W_c$ of a stack $\mathbf{W}$, it gives $\pi \sum_a \eta_a\, t_a(\upsilon)$; summed with the $\xi_c(w)\, w^\top$ for a witness $w$, all parts give the second line, by \cref{eq:cancel} and with $x^-_1 \defeq 0$. This is the \emph{mask identity}, identically in the coins, with $\Theta^{\pm}$ free of $\varepsilon$:
\begin{equation}\label{eq:mask}
\begin{aligned}
  \textstyle \sum_{c = (k_1, \leg) \in C} \pi_{k_1} W_c\, \vartheta_{\leg} &= \textstyle \pi \sum_{a \in [3]} \eta_a\, t_a(\upsilon) + \varepsilon\, \Theta^{+} + \varepsilon^{-1} \Theta^{-} , \\
  \textstyle \sum_{c = (k_1, \leg) \in C} \pi_{k_1} \xi_c(w)\, w^\top \vartheta_{\leg} &= \textstyle \pi \sum_{a \in [3]} \eta_a \bigl( P\, x^0_a + \varepsilon\, x^+_a + \varepsilon^{-1} x^-_a \bigr)(\upsilon) .
\end{aligned}
\end{equation}

\begin{proof}[of \cref{lem:honest}]
As in \cref{sec:scheme}, the summed factors pair to the sum of $\xi_c(w)\, w_i w_j$ times $F^{k_1}_i H^{\leg}_j$, whose terms $\pm \gamma_c\, y(w)\, A_w(\tau)$ cancel: the search term $r(w)\, \psi$, the QAP term $\sum_c \xi_c(w)\, \gamma_c\, A_w(\tau)\, B_w(\tau)$, and the masks' term $y(w) \sum_c \xi_c(w)\, \pi_{k_1} w^\top \vartheta_{\leg}$. The first identity of \cref{eq:qap-w} serves the \keycopy{}, where $\gamma_{\ckey} L_\rho(\tau) = \kappa$, and the second the others; the powers suffice, as $\hkey, \hoth \in \FF_q^{<D}[X]$ with $D \le J\Delta$ and $R_w$ is allowed (\cref{lem:qap}), and each sum over the chunks encodes what one pairing with the whole polynomial would (\cref{sec:optimizations}). The tie polynomials of the \honeststack{}, which satisfies the ties (\cref{sec:weights-ties}), are $w$ times the numerators in \cref{eq:cancel}; as $w_0 = 1$, the divisions are exact, and the bases of $\sB_w$ suffice, as the factors lie in $\FF_q^{<n}[T]$. By the mask identity, the pairing of $\sY_w$ with $\sB_w$ encodes the masks' term. Finally, $\Psi$ encodes $\sum_c \xi_c(w)\, \pi_{k_1} y(w)\, s_{\leg} = \psi$, and $(\Lambda^\alpha, \Psi^\alpha) = \alpha \cdot (\Lambda, \Psi)$, as $\Comb$ is linear in $ct_2$ and $ct_2^\alpha = \alpha \cdot ct_2$.
\end{proof}

\begin{proof}[of \cref{lem:range}]
In the first case, $r(w)$ is the sum of the $1 + \wt(w)$ entries $r_i$ with $w_i = 1$, which are independent, each within $(g-1)/2$ of its mean $(g+3)/2$ and of variance $(g^2 - 1)/12$. The mean $(1 + \wt(w))(g+3)/2$ of $r(w)$ is a multiple of $1/2$, and $c_w$ is its floor, so $|r(w) - c_w| \le (1 + \wt(w))(g-1)/2 + 1/2 \le B$, and the second claim follows by Minkowski's inequality in $L^2$.

In the second case, let $a \defeq \lceil N(g+1)/2 \rceil$, $\Sigma \defeq \sum_{i \in [N]} r_i$, and $X \defeq r_\star + \sum_{i \in [N]} (w_i - 1/2)\, r_i$, a sum of independent terms with mean $\mu \defeq \frac{g+3}{2} (1 + \wt(w) - N/2)$. Then $r(w) = a + X + \Sigma/2 - \lfloor \Sigma/2 \rfloor$ and $c_w = \lfloor a + \mu \rfloor$, so $r(w) - c_w$ is $X - \mu$ plus two rounding errors, $\Sigma/2 - \lfloor \Sigma/2 \rfloor \le 1/2$ and $a + \mu - \lfloor a + \mu \rfloor < 1$. The term $r_\star$ of $X$ is within $(g-1)/2$ of its mean and each other term within $(g-1)/4$, so $|r(w) - c_w| \le (N+2)(g-1)/4 + 3/2 \le B$. The variance of $X$ is $(1 + N/4)(g^2 - 1)/12$, so the second claim follows by Minkowski's inequality in $L^2$.
\end{proof}

\parhead{The law of the search}\phantomsection\label{app:search-law}\Cref{sec:costs}'s expected search and its point at $2^{-40}$ rest on certified numerical bounds, valid for every witness, on the law of $t_0 \defeq r(w) - c_w$ under the \centeredsampler{}. By the proof of \cref{lem:range}, $t_0$ is $(r_\star - \frac{g+3}{2}) + \sum_{i \in [N]} (w_i - \frac12)(r_i - \frac{g+3}{2})$ plus two roundings with a total in $[0, 3/2)$, and $r_\star - \frac{g+3}{2}$ and each $r_i - \frac{g+3}{2}$ are uniform on the $g$ equally spaced values from $-\frac{g-1}{2}$ to $\frac{g-1}{2}$. Couple each $r_i - \frac{g+3}{2}$ to $g \tilde r_i$, and $r_\star - \frac{g+3}{2}$ to $g \tilde r_\star$, with $\tilde r_i$ and $\tilde r_\star$ uniform on $[-\frac12, \frac12]$ and each variable within $\frac12$ of its partner. Then $|t_0 - g(\tilde r_\star + \sum_{i \in [N]} (w_i - \frac12)\tilde r_i)| \le N/4 + 2$, and as each $\tilde r_i$ is symmetric, this sum has, for every witness, the law of $\varsigma Y$, with $\varsigma \defeq g\sqrt{(N+4)/48}$ and $Y \defeq (\sum_{i \in [N]} \tilde r_i + 2\tilde r_\star)/\sqrt{(N+4)/12}$. So for $x \ge 0$, $\Pr[|t_0| > x]$ lies between $2\Pr[Y > (x + N/4 + 2)/\varsigma]$ and $2\Pr[Y > (x - N/4 - 2)/\varsigma]$. With $\operatorname{sinc} x \defeq \sin(x)/x$, the characteristic function of $Y$ is $\operatorname{sinc}\bigl(s/\sqrt{(N+4)/3}\bigr)^N \operatorname{sinc}\bigl(2s/\sqrt{(N+4)/3}\bigr)$, and by the series $\log \operatorname{sinc} x = -\sum_{k \ge 1} \zeta(2k)\, x^{2k}/(k \pi^{2k})$, with $\zeta$ Riemann's zeta function, its logarithm is $-s^2/2 - a_4 s^4 - b_6 s^6 - \varrho(s)$, with $a_4 \defeq (N+16)/(20(N+4)^2)$, $b_6 \defeq (N+64)/(105(N+4)^3)$, and $0 \le \varrho(s) \le c_8\, s^8$ for $|s| \le 16$, where $c_8 \defeq 2(N+256)/\bigl(81 (N+4)^4 (1 - 1024/(3(N+4)))\bigr)$. The function $e^{-s^2/2}(1 - a_4 s^4 - b_6 s^6 + a_4^2 s^8/2)$ is the characteristic function of the standard normal density times $1 - a_4 \mathit{He}_4 + b_6 \mathit{He}_6 + \frac{a_4^2}{2} \mathit{He}_8$, with $\mathit{He}_k$ the probabilists' Hermite polynomials, and differs from that of $Y$ by at most $e^{-s^2/2}\bigl(c_8 s^8 + a_4 b_6 s^{10} + b_6^2 s^{12}/2 + (a_4 s^4 + b_6 s^6)^3/6\bigr)$ for $|s| \le 16$. Beyond $16$, the characteristic function of $Y$ is at most $e^{-N s^2/(2(N+4))}$ in absolute value up to $s = \pi\sqrt{(N+4)/3}$ and at most $(\sqrt{(N+4)/3}/s)^N$ after that, and that of the approximation is at most $e^{-s^2/4}(1 + 64 a_4 + 1728 b_6 + 32768 a_4^2)$. By the inversion formula of Gil-Pelaez~\cite{Gil51}, the two survival functions differ by at most $1/\pi$ times the integral over $s > 0$ of the absolute difference of the two characteristic functions divided by $s$, and so by less than $8.05 \cdot 10^{-18}$ at our parameters. We evaluate these bounds, each side widened by a further $1/g$, which only loosens them, in ball arithmetic with $192$ bits at every endpoint of the search's rounds up to the implementation's cap. They give the point at $2^{-40}$; the first endpoint's count of candidates plus each later increment times the bound at the endpoint before it bounds the capped search's expectation; and adding the bound beyond the cap's reach times the length of the longest search gives a bound on the uncapped expectation, which bounds decryption's.

\begin{proof}[of \cref{prop:efficiency}]
Parts (a) to (c) count \cref{fig:full-enc,fig:full-dec}: a sum over an index, or a single multiple in $\GG_2$, is an MSM unless it is a sum of pairings or has the bits of $w$ as coefficients, and \cref{line:full-target,line:full-key} of $\Dec$ take additions and four scalar multiplications in $\GG_T$. In $\FF_q$, $\Dec$ computes the forms at $w$ as $k(\ell + 1)$ products in $R_q$, by number-theoretic transforms, and $\Enc$ computes $z$ at its three points by prefix sums along each $M_{i,j}$. $\Enc$ computes the scalar of each published element with $O(1)$ more operations, and $\Dec$ the weights and the cancellation polynomials with $O(1)$ products of polynomials of degree below $n$, each from four products of their halves by fast Fourier transforms over the subgroup of $\mathcal{H}$ of order $2^{\lceil \log_2 n \rceil}$, which exists as $n \le D$. In the order of \cref{sec:qap}, the vanishing polynomial of a range of $\mathcal{H}$ is a product of $O(\log D)$ binomials $X^{2^b} - c$, and the interpolant of values on a range $\mathcal{P}'$ is $f / Z'$, where $Z'$ is that of $\mathcal{H} \setminus \mathcal{P}'$ and $f$ interpolates on $\mathcal{H}$ the values times $Z'$, and zero elsewhere. $\Setup$ keeps each $A_i$ and $B_i$ as its nonzero values on $\mathcal{P}$, so $\Enc$ evaluates them at $\tau$ by the Lagrange basis, and $\Dec$, which interpolates only $A_w$ and $B_w$, computes its polynomials by fast Fourier transforms over $\mathcal{H}$, multiplying by halves in the same way and dividing by one binomial at a time~\cite[Section~1]{Pol71}. $\Setup$ computes the powers of $\omega$ and of a generator of $\mathcal{H}$, the vanishing polynomials it needs at each point from their binomials, the interpolation weights from them by simultaneous inversion, and the stored values: $O(|S| + D \log D)$ operations, as $n \le D$, and $O(\log q)$ for the roots of unity.

For (d), search the intervals $|t| \le 4^j$ for $j = 0, 1, \dots$, the last cut at $B$, each with $O(2^j)$ operations by baby steps and giant steps~\cite[pp.~419--420]{Sha71}, and return the solution of least absolute value in the first interval that has one: a result $t$ is found after $O(\sqrt{|t|} + 1)$, and if there is none the search ends after the interval cut at $B$, with $O(\sqrt{B})$ operations. On a ciphertext of $\Enc$, $t$ is $r(w) - c_w$ or $0$ (\cref{thm:correct}), and $\sqrt{|r(w) - c_w|}$ has expectation at most $\bigl( g \sqrt{(N+4)/48} + 3/2 \bigr)^{1/2}$ by the second case of \cref{lem:range} and Jensen's inequality; with the independent entries of \cref{fig:enc} and the center of \cref{fig:dec} (\cref{sec:scheme}), the first case gives at most $\bigl( g \sqrt{(N+1)/12} + 1/2 \bigr)^{1/2}$, as $\wt(w) \le N$, of the same order.
\end{proof} 
\subsection{The Plan of the Security Proof and the Bound \texorpdfstring{$E$}{E}}\label{app:bound}
\begin{figure}[!tb]
\centering
\begin{tikzpicture}[font=\footnotesize, line width=0.5pt, >=stealth,
    box/.style={draw=black!55, rounded corners=2pt, align=center, inner sep=3.5pt,
      execute at begin node={\hyphenpenalty=10000 \exhyphenpenalty=10000 }},
    branch/.style={box, text width=3.35cm, minimum height=0.95cm}]
  \definecolor{figblue}{HTML}{3871C1}
  \definecolor{figrust}{HTML}{7A3D00}
  \node[box, minimum width=11.6cm] (view) at (0,0)
    {\textbf{view lemma:} a comparison that succeeds without being an identity gives\\ a nonzero passing stack in one sector that satisfies the ties, or a coincidence in $\FF_q$};
  \node[box] (cls) at (0,-1.2)
    {\textbf{classification} by rank:\\ the dimension of the row space of the stack};
  \foreach \x/\n/\col/\a/\b in {
      -3.95/l/black!55/{rank four or more}/{light: \mbox{$\operatorname{gp} \le g^{-4} < \theta$}},
      0/k/figrust/{rank at most three, a kernel vector in the row space}/{if heavy: a \shortmultiple{} of the vector, an MSIS solution},
      3.95/w/figblue/{rank at most three, no kernel vector}/{a combination of \honeststack{}s of witnesses: the extractor outputs one, or its recovery fails}} {
    \node[branch, draw=\col] (\n1) at (\x,-2.65) {\a};
    \node[branch, draw=\col] (\n2) at (\x,-4.3) {\b};
    \draw[->] (cls.south) -- (\n1.north);
    \draw[->] (\n1) -- (\n2);
  }
  \node[box, minimum width=11.6cm] (cnt) at (0,-6.3)
    {\textbf{counting:} light stacks, coincidences, and allowances: $E(2Q+2) \le (Q+1)/2^{\lambda}$;\\
     \textbf{extraction,} by the \stoppedcoupling{}:\\ $\Pr[\,b' = b \wedge \text{no witness extracted}\,] \le \tfrac12 + \tfrac12(E(2Q+2) + s_0 + s_1)$;\\
     on a system without witnesses, \textbf{hiding:} advantage at most $E(2Q+2) + s_0 + s_1$};
  \draw[->] (view) -- (cls);
  \foreach \n in {l2,k2,w2} { \draw[->] (\n.south) -- (cnt.north -| \n.south); }
\end{tikzpicture} 
\caption{The proof: view, classification by rank, counting. Light stacks are charged at $\theta$ in every branch. The bound holds within the query caps; $E(2Q+2)$ includes the extractor's failures to recover a row space.}\label{fig:roadmap}
\end{figure}
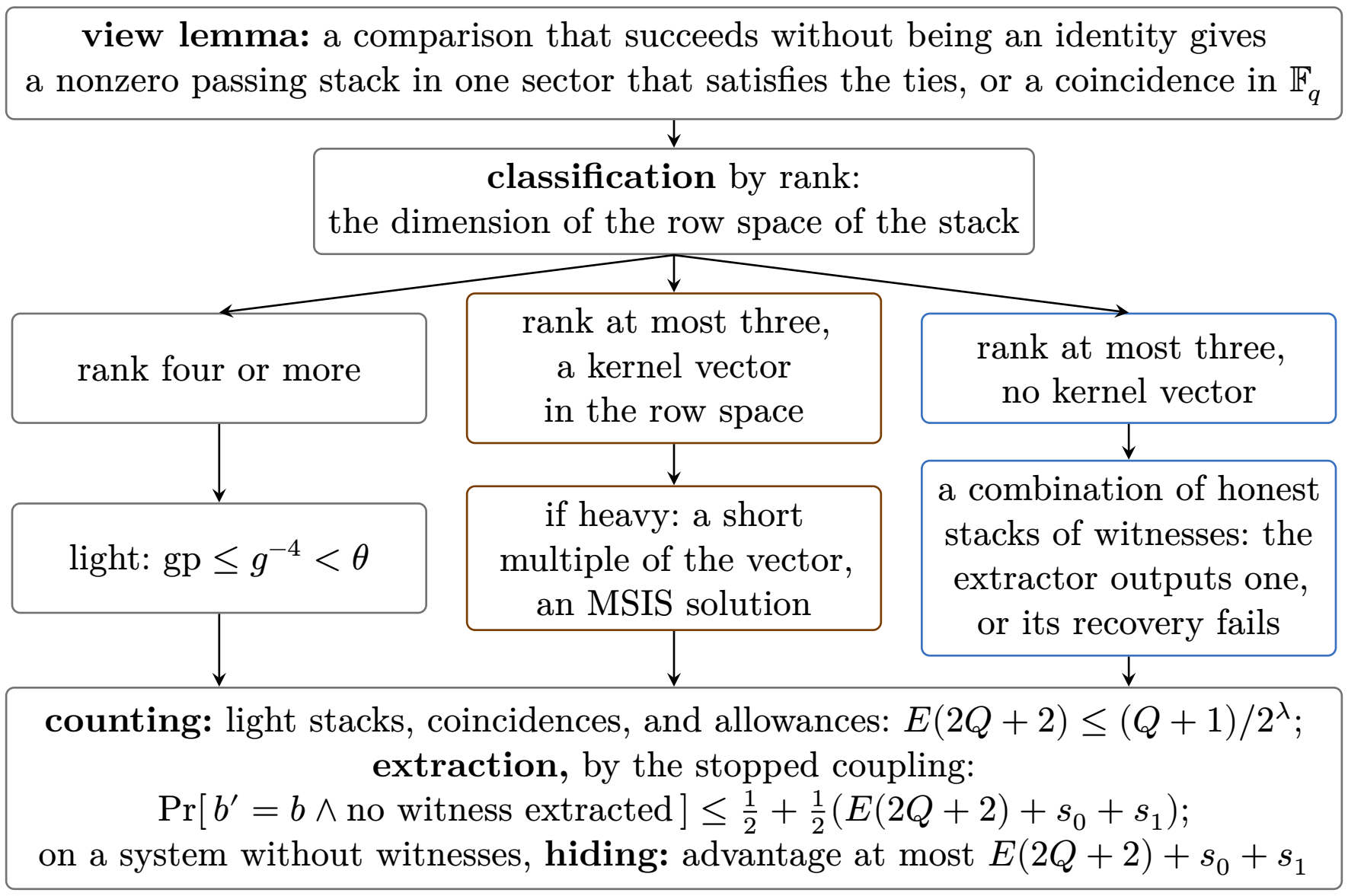

Let $\mathsf{Tgt}$ be the event that $\mathcal{A}$ queries $\Hkey$ at $[\kappa]_T$, $\mathsf{Fail}$ the event that $\mathcal{E}$ outputs no witness of $S$, and $f_i \defeq \Pr[\mathsf{Tgt} \wedge \mathsf{Fail}]$ in the experiment of \cref{def:we-ext} with $b = i$. The runs with $m_0$ and $m_1$ have the same law until $\mathsf{Tgt}$, and whether $\mathsf{Tgt}$ occurs is settled before its query is answered, so it suffices to bound $f_0$ and $f_1$ (\cref{lem:half}; \cref{fig:roadmap}). In the generic model, every element of $\GG_T$ that $\mathcal{A}$ forms without guessed labels is a known combination of $[\alpha\kappa]_T$ and of pairings of published elements, generators included, a rational function of the hidden coins, and a comparison asks whether such a combination equals $\chi\kappa$, with $\chi = 0$ when two elements are compared and $\chi = 1$ at the query of $\mathsf{Tgt}$. The terms without $\alpha$ form the \emph{\plainsector{}}, and those with $\alpha$ the \emph{\alphasector{}}. If a comparison is an identity in all coins but the \searchvector{} $r$, then in each sector it is one identity for each cell, and the view lemma reads it: the cell's coefficient matrix $W_c$ is symmetric, by the \symmetryterm{}; it passes the tests, by the \qapcriterion{}; $W_c r = u_c$ for a vector $u_c$ that $\mathcal{A}$ supplies; in the \plainsector{}, the corner of the \keycopy{}'s matrix is $\chi$; and the remaining mask terms make the stack satisfy the ties. Only $W_c r = u_c$ involves $r$, so such a comparison succeeds only on a correct guess of every $W_c r$, with probability at most the guessing probability of either sector's stack. A monitor marks the first comparison that succeeds without being an identity in the coins or, if none comes first, the query of $\mathsf{Tgt}$ (\cref{app:monitor}). This \emph{first mark} is \emph{good} if its comparison is an identity in all coins but $r$, with a nonzero heavy \markstack{}: the plain stack if nonzero, else the alpha stack. That stack is \emph{usable}, that is, passes and satisfies the ties (\cref{lem:view}), and has rank at most three (\cref{lem:heavy}); so its row space has a kernel vector, which a reduction turns into an MSIS solution (\cref{lem:reduction}), or contains a witness, which $\mathcal{E}$ finds except with probability at most $2/q$ (\cref{lem:classification,lem:extractor}). The bound $E$, at $u = 2Q + 2$, pays for every other first mark and for unlucky coins of the reduction and of $\mathcal{E}$ (\cref{tab:ledger}). Of the adversary's $L$ labels, $L_1$, $L_2$, and $L_T = 2$ are in $\GG_1$, $\GG_2$, and $\GG_T$, $\varepsilon_f \defeq (3D + 2n + 2K_1 + 67)/(q - D)$ bounds the probability of a coincidence in $\FF_q$ (\cref{lem:view}\ref{it:view-sz}), and $\mathcal{A}$ makes at most $Q_H \le 2^\lambda$ queries to $\Hkey$ (\cref{thm:extraction}).

\begin{definition}[Bound $E$]\label{def:ledger}
Let $\Ncmp(u) \defeq uL + u(u+1)/2$ and $N_{\mathsf{free}} \defeq 2^{\ell_{\mathsf{lab}}} - L - 3 \cdot 2^{\lambda+1} - 1$, which is positive by \ref{it:p-labels}. For an integer $u \ge 0$,
\begin{align*}
  E(u) \defeq{}& \Bigl( \tbinom{L_1}{2} + \tbinom{L_2}{2} + \tbinom{L_T}{2} + L \Bigr) \varepsilon_f
    + \Ncmp(u) (\theta + \varepsilon_f)
    + \Bigl( \tbinom{L + 3u}{2} + L + 3u \Bigr) \varepsilon_f \displaybreak[0] \\
  &+ (8u + 8)\, \varepsilon_f
    + (2L + 4u + 2^{\lambda+2}) / N_{\mathsf{free}}
    + 4u \cdot 2^{-126} + 2^{-128} .
\end{align*}
\end{definition}

\begin{lemma}[Generic bound]\label{lem:ledger}
$E$ is a polynomial of degree two in $u$ with nonnegative coefficients. If $E(2) \le 2^{-\lambda}$ and $E(2^{\lambda+1}) - E(2) \le 1 - 2^{-\lambda}$, then $E(2Q + 2) \le (Q+1)/2^\lambda$ for every integer $0 \le Q < 2^\lambda$. For $\lambda \ge 1$, the second premise requires $\theta < 2^{-(2\lambda + 1)}$.
\end{lemma}

\begin{proof}
Every term of $E(u)$ is a polynomial in $u$ with nonnegative coefficients, as $L \ge 1$, of degree two for $\Ncmp(u)$ and at most two otherwise. So $Q \mapsto E(2Q+2) - (Q+1)/2^\lambda$ is convex; by the premises it is at most $0$ at $Q = 0$ and at $Q = 2^\lambda - 1$, where it is $E(2^{\lambda+1}) - 1$, hence in between. For $\lambda \ge 1$, $E(2^{\lambda+1}) - E(2) \ge \bigl(\Ncmp(2^{\lambda+1}) - \Ncmp(2)\bigr)\theta > \frac12 (2^{2\lambda+2} - 6)\,\theta$, which for $\theta \ge 2^{-(2\lambda+1)}$ is at least $1 - \frac32\, 2^{-2\lambda} > 1 - 2^{-\lambda}$.
\end{proof}

\begin{table}[!htbp]
\caption{Events that cover $\mathsf{Tgt} \wedge \mathsf{Fail}$ in the run with $m_b$, by first mark; bad events and $\mathsf{D}$ are those of \cref{app:counting}.}\label{tab:ledger}
\centering\small\setlength{\tabcolsep}{4pt}
\begin{tabular}{@{}>{\raggedright\arraybackslash}p{0.41\linewidth}l>{\raggedright\arraybackslash}p{0.15\linewidth}l@{}}
\toprule
First mark of the run & Probability at most & Paid by, in $E(u)$ & Lemmas \\
\midrule
$\mathsf{init}$: two received elements coincide & $\bigl( \binom{L_1}{2} + \binom{L_2}{2} + \binom{L_T}{2} \bigr)\, \epsf$ & first term & \ref{lem:init} \\
charged: a bad event, no guess variable in $\mathsf{D}$ & $\Ncmp(u)\, (\theta + \epsf)$ & second term & \ref{lem:charged} \\
charged: a bad event, a guess variable in $\mathsf{D}$ & $\bigl( \binom{L + 3u}{2} + L + 3u \bigr)\, \epsf$ & third term & \ref{lem:charged} \\
$\mathsf{guess}$: the target label is guessed & $Q_H / \Nfree$ & fifth term & \ref{lem:names} \\
good, a kernel vector in the row space & $s_b + 2\epsf$ & $s_b$, fourth term & \ref{lem:heavy}, \ref{lem:reduction} \\
good, no kernel vector, no witness from $\mathcal{E}$ & $2/q$ & fourth term & \ref{lem:classification}, \ref{lem:extractor} \\
\bottomrule
\end{tabular}
\end{table}

\begin{proof}[of \cref{thm:extraction}\ref{it:ext-bound}, given the lemmas of \cref{tab:ledger}]
From here on, $u \defeq 2Q + 2$. A run in $\mathsf{Tgt}$ has a first mark, which is $\mathsf{init}$, $\mathsf{guess}$, charged, or good. So the rows cover $\mathsf{Tgt} \wedge \mathsf{Fail}$, and $f_i$ is at most the sum of their bounds for $b = i$. Each bound is at most what pays for it, with terms of $E(u)$ to spare, as $E$ and $\epsf$ are those of the Lean statement (\cref{sec:formal-verification}): the row $\mathsf{guess}$ is at most $2^{\lambda}/\Nfree$ (\cref{lem:names}), within the fifth term, and a good mark's two rows sum to $s_i + 2\epsf + 2/q \le s_i + (8u + 8)\, \epsf$. So $f_i \le E(u) + s_i$, and by \ref{it:p-generic} and \cref{lem:ledger}, $E(u) \le (Q+1)/2^\lambda$, which gives \ref{it:ext-bound}.
\end{proof} 
\subsection{Expressions, View, Classification, and Lifting}\label{app:setting}\label{app:view}\label{app:classification}\label{app:inverse}

From here on, the parameters, $S$, its evaluation points, and \scheme are those of \cref{thm:extraction}. The matrix $M$ is fixed: the bounds of \cref{app:setting,app:monitor} hold for each $M$, and \cref{app:reduction}, like the proof of \cref{thm:extraction}\ref{it:ext-bound}, takes $M$ uniform. The fourth term of $E(2)$ is $24\,\epsf$ and no term is negative; so $24\,\epsf \le E(2) \le 2^{-\lambda}$ by \ref{it:p-generic}, and $\epsf < 1$. A \emph{linear operation}, a call of $\gadd$ or $\gsmul$, returns $\mathsf{lab}_s\bigl( c_1\, \mathsf{lab}_s^{-1}(\mathsf{u}_1) + c_2\, \mathsf{lab}_s^{-1}(\mathsf{u}_2) \bigr)$, where $c_1 = c_2 = 1$ for $\gadd(s, \mathsf{u}_1, \mathsf{u}_2)$, and $(c_1, c_2, \mathsf{u}_2) \defeq (c, 0, \mathsf{u}_1)$ for $\gsmul(s, c, \mathsf{u}_1)$.

\parhead{Stacks and kernel vectors}
The stack of a vector $v \in \FF_q^{N+1}$ is $\mathbf{H}(v) \defeq (\xi_c(v)\, v v^\top)_{c \in C}$, the \emph{\honeststack{}} of a witness; $v$ is a \emph{kernel vector} if it is nonzero and $z(v) = 0$; a \emph{\shortmultiple{}} of $v$ is an $x \in \ZZ^{N+1}$ with $\|x\|_\infty \le \beta$ whose reduction modulo $q$ is a nonzero multiple of $v$; and the \emph{\searchvector{}} is the $r$ of $\Enc$ (\cref{sec:overview}).

\parhead{Coins and exponents}
Let $\mathbf{c}$ be the tuple of the values that $\Enc$ samples after the seed and $r$, from $y$ to $\varepsilon$; it is \emph{valid} if each entry lies in the set from which $\Enc$ samples it, which has at least $q - D$ elements, as $K_1 < n \le D$. The \emph{coins} of $\Enc$ are a uniform valid $\mathbf{c}$ and an independent $r \gets \mathcal{D}_g$. For valid $\mathbf{c}$ and every $r \in \FF_q^{N+1}$, $\Enc$ defines the exponent $\mathsf{X}(\mathbf{c},r)$ of each published element $\mathsf{X}$ (\cref{fig:full-enc}). The generators have exponent $1$, the coordinates $\sF^{k_1}_{i,1}$ and $\sF^{k_1}_{i,2}$ of $\sF^{k_1}_i$ have $\pi_{k_1} y_i$ and $\pi_{k_1} ( \Omega_{12}\, y_i + \Omega_{22}\, A_i(\tau) )$, and $\sH^{\leg}_{j,1}$ and $\sH^{\leg}_{j,2}$ of $\sH^{\leg}_j$ have $h^1_{\leg,j} - (\Omega_{12} / \Omega_{22})\, h^2_{\leg,j}$ and $h^2_{\leg,j} / \Omega_{22}$, for the entries $h^1_{\leg,j}$ and $h^2_{\leg,j}$ of $H^{\leg}_j$ (\cref{line:full-factors}); so $r$ enters only through the term $s_{\leg}\, r_j$ of the $\sH^{\leg}_{j,1}$ and their $\alpha$-copies. Each exponent is homogeneous in each of $\pi$, $\zeta$, $\zeta'$, $\zeta''$ and, jointly, in $(\Omega_{12}, \Omega_{22})$, written $\Omega$; its \emph{class} is the monomial of these degrees. For a \emph{sector} $\sect \in \{\mathrm{plain}, \alpha\}$ and $\mathsf{X}$ in $ct_2$, $\mathsf{X}^{\sect}$ is $\mathsf{X}$ or its $\alpha$-copy, of exponent $\alpha_{\sect}\, \mathsf{X}(\mathbf{c}, r)$, for $\alpha_{\sect} \defeq 1$ or $\alpha$.

\parhead{Expressions}
Let $\mathcal{X}_1$ consist of $G_1$ and the elements of $ct_1$, and $\mathcal{X}_2$ of $G_2$ and those of $ct_2$ and $ct_2^\alpha$. An \emph{expression} $\mathsf{E}$ of $\GG_1$ or $\GG_2$ is a vector in $\FF_q^{\mathcal{X}_1}$ or $\FF_q^{\mathcal{X}_2}$, read as $\sum_a \mathsf{E}_a\, a$; its \emph{value} is $\val(\mathsf{E}; \mathbf{c}, r) \defeq \sum_a \mathsf{E}_a\, a(\mathbf{c},r)$. One of $\GG_T$ is a pair $\mathsf{D} = \bigl( (\mathsf{D}_{ab})_{a \in \mathcal{X}_1, b \in \mathcal{X}_2}, \mathsf{D}_K \bigr)$, read as $\sum_{a,b} \mathsf{D}_{ab}\, e(a,b) + \mathsf{D}_K\, \sK^\alpha$, of value $\val(\mathsf{D}; \mathbf{c}, r) \defeq \sum_{a,b} \mathsf{D}_{ab}\; a(\mathbf{c},r)\, b(\mathbf{c},r) + \mathsf{D}_K\, \alpha\, \kappa$; $G_T$ is the pair $(G_1, G_2)$, and the pairing of $\mathsf{E}_1$ and $\mathsf{E}_2$ is $(\mathsf{E}_{1,a} \mathsf{E}_{2,b})_{a,b}$.

\begin{lemma}[Degrees]\label{lem:degree}
Let $\varpi \defeq \Omega_{22}\, \zeta\, \zeta'\, \zeta''\, \varepsilon$ and $d_f \defeq 3D + 2n + 2K_1 + 6$. For every expression $\mathsf{D}$ of $\GG_T$ and every $\chi \in \FF_q$, the function $\varpi \cdot \bigl( \val(\mathsf{D}; \mathbf{c}, r) - \chi\kappa \bigr)$ is a polynomial in the entries of $\mathbf{c}$ and of $r$ of total degree at most $d_f$. For expressions $\mathsf{E}, \mathsf{E}'$ of $\GG_1$ or of $\GG_2$, the same bound holds for $\varpi \cdot \bigl( \val(\mathsf{E}) - \val(\mathsf{E}') \bigr)$.
\end{lemma}

\begin{proof}
Here $\gamma_c = \pi_{k_1} s_{\leg}$ has degree $k_c + 3 \le n + 2$, the entries of $z_-$, $z_\circ$, and $z_+$ have degree below $n$ in $\upsilon$, $A_i$, $B_j$, and $L_\rho$ have degree below $D$, $Z_c$ has degree $D - 1$, or $D$ for $c = \ckey$, where $k_c = 0$, $Z_0$ has degree $N + 2$, and the five factors of $\varpi$ clear every denominator. As $n \le D$, every exponent of $\mathcal{X}_1$ has degree at most $d_1 \defeq \max\{D + 2K_1, \Delta\}$, and $\varpi$ times every exponent of $\mathcal{X}_2$, an $\alpha$-copy included, at most $d_2 \defeq D + n + 7 + (J-1)\Delta$: the $\Bquo_{c,j}$ reach $D + n + 6 + (J-1)\Delta$, and the $\Brem_{c,j}$ at most $N + n + 9 + (J'-1)\Delta' \le D + n + 7$, as $(J'-1)\Delta' \le \nu\nc = D - N - 2$, also when $\nc = 0$. So the first function has degree at most $d_1 + d_2$, also for its terms $\varpi\alpha\kappa$ and $\varpi\kappa$, of degrees $D + 8$ and $D + 7$, and the second at most $d_1 + 5$ or $d_2$; and $d_1 + d_2 = D + n + 7 + J\Delta + \max\{D + 2K_1 - \Delta,\, 0\} \le d_f$ by \ref{it:p-degree}, as $\Delta \ge 1$.
\end{proof}

\parhead{Stacks and identities}
Let $\mathsf{D}$ be an expression of $\GG_T$ and $\sect$ a sector. For a cell $c = (k_1, \leg)$, let $W^{\sect}_c(\mathsf{D})$ be the matrix with entries $\mathsf{D}_{ab}$ for $a = \sF^{k_1}_{i,1}$ and $b = (\sH^{\leg}_{j,1})^{\sect}$, in row $i$ and column $j$, and let $u^{\sect}_c(\mathsf{D})$ be the vector with entries $-\mathsf{D}_{ab}$ for $a = \sF^{k_1}_{i,1}$ and $b = (\sS^{\leg})^{\sect}$. The \emph{stack of $\mathsf{D}$ in sector $\sect$} is $\mathbf{W}^{\sect}(\mathsf{D}) \defeq (W^{\sect}_c(\mathsf{D}))_{c \in C}$. For $\chi \in \FF_q$ and $r \in \FF_q^{N+1}$, $\mathsf{D}$ \emph{is $\chi\kappa$ identically at $r$} if $\val(\mathsf{D}; \mathbf{c}, r) = \chi\, \kappa(\mathbf{c})$ for every valid $\mathbf{c}$, and \emph{formally} if this holds at every $r$. Two expressions of one group are \emph{formally equal} if their values agree at every valid $\mathbf{c}$ and every $r$. 

\begin{lemma}[View]\label{lem:view}
Let $\mathsf{D}$ be an expression of $\GG_T$, $\chi \in \FF_q$, and $r, r' \in \FF_q^{N+1}$.
\begin{enumerate}[label=(\alph*),leftmargin=*,nosep]
  \item\label{it:view-forward} \emph{Passing stacks.} If $\mathsf{D}$ is $\chi\kappa$ identically at $r$, then in each sector $\sect$ the stack $\mathbf{W}^{\sect}(\mathsf{D})$ passes and satisfies the ties, $W^{\sect}_c(\mathsf{D})\, r = u^{\sect}_c(\mathsf{D})$ for every cell $c$, and the corner $\bigl( W^{\mathrm{plain}}_{\ckey}(\mathsf{D}) \bigr)_{00}$ is $\chi$.
  \item\label{it:view-locus} \emph{Locus.} If $\mathsf{D}$ is $\chi\kappa$ identically at $r$ and $W^{\sect}_c(\mathsf{D})\, r' = u^{\sect}_c(\mathsf{D})$ for every sector and cell, then $\mathsf{D}$ is $\chi\kappa$ identically at $r'$. In particular, if $\mathsf{D}$ is $\chi\kappa$ identically at $r$ and has zero stacks, then $\mathsf{D}$ is $\chi\kappa$ formally, and $\chi = 0$.
  \item\label{it:view-sz} \emph{Coincidence.} If $\mathsf{D}$ is not $\chi\kappa$ identically at $r$, $\Pr[\val(\mathsf{D}; \mathbf{c}, r) = \chi\kappa] \le d_f/(q - D) \le \epsf$ for uniform valid $\mathbf{c}$.
  \item\label{it:view-nokey} \emph{No formal \keyscalar{}.} If $\chi \ne 0$, then $\mathsf{D}$ is not $\chi\kappa$ formally.
  \item\label{it:view-source} \emph{Source groups.} If $\mathsf{E}$ and $\mathsf{E}'$ are expressions of $\GG_1$, or of $\GG_2$, that are not formally equal, then for every $r$ their values at $(\mathbf{c}, r)$ are equal with probability at most $\epsf$ over $\mathbf{c}$.
\end{enumerate}
\end{lemma}

\begin{proof}
The valid tuples, at which $\varpi \ne 0$, form a product of sets of at least $q - D$ elements, and $q - D > d_f$, as $\epsf = (d_f + 61)/(q - D) < 1$. So by \cref{lem:degree} and the Schwartz--Zippel lemma~\cite{Sch80,Zip79}, $\mathsf{D}$ is $\chi\kappa$ identically at $r$ exactly if $\varpi \cdot ( \val(\mathsf{D}; \mathbf{c}, r) - \chi\kappa )$ is the zero polynomial in the entries of $\mathbf{c}$; otherwise it vanishes with probability at most $d_f/(q - D)$, which is \ref{it:view-sz}. Such an identity, multiplied by $\varpi$, may thus be compared coefficient by coefficient and evaluated at any point, also at an invalid $\tau \in \mathcal{P}$, as $\varpi$ is free of $\tau$. For \ref{it:view-forward} and \ref{it:view-locus}, let $\mathsf{D}$ be $\chi\kappa$ identically at $r$, and compare coefficients in three steps: the hidden basis leaves one coefficient matrix for each cell, the hidden scalars separate the sectors and the cells, and each cell's identity is read at the evaluation points, with the masks' terms giving the ties.

\emph{Step 1: the hidden basis.} The value $\chi\kappa - \mathsf{D}_K\, \alpha\kappa$ of the pairs of $\mathsf{D}$ is free of $\Omega$, so its terms with $\Omega_{12}^p / \Omega_{22}$, $p \ge 1$, sum to zero; this sum sets each coefficient on first coordinates against the matching one on second coordinates, and its derivative in $y_j$ at $y = B(\tau)$, where $h^2_{\leg,j}$ vanishes, separates them. The terms come only from pairs with the $(\sH^{\leg}_{j,1})^{\sect}$ or $(\sH^{\leg}_{j,2})^{\sect}$ and sum to $-\Omega_{12}/\Omega_{22}$ times $\sum_{\sect, a, \leg, j} \alpha_{\sect}\, \delta^{\sect}_{a,\leg,j}\, \tilde a\, s_{\leg} (B_j(\tau) - y_j)$, where $a \in \mathcal{X}_1$, $\tilde a$ is the exponent of $a$ at $\Omega_{22} = 0$, and $\delta^{\sect}_{a,\leg,j}$ is the coefficient of $\mathsf{D}$ on $a$ and $(\sH^{\leg}_{j,1})^{\sect}$, less, for $a = \sF^{k_1}_{i,1}$, that on $\sF^{k_1}_{i,2}$ and $(\sH^{\leg}_{j,2})^{\sect}$. So the derivative gives $\sum_{\sect, a, \leg} \delta^{\sect}_{a,\leg,j}\, \alpha_{\sect}\, s_{\leg}\, \tilde a|_{y = B(\tau)} = 0$. These polynomials, $\alpha_{\sect}\, s_{\leg}$ times $1$, $\zeta \tau^k$, $\zeta' \tau^k$, $\zeta'' B_i(\tau)$, $\pi_{k_1} B_i(\tau)$, or $\Omega_{12}\, \pi_{k_1} B_i(\tau)$, are linearly independent: the $B_i$ are, as $B(d_j) = e_j$, and the factors before $1$, $\tau^k$, and $B_i(\tau)$ have distinct monomials, as $s_{\leg}\, \pi_{k_1} = \gamma_c$ contains $\pi\, s_{\mathrm{X}}\, \upsilon^{k_c}$, which determines the cell $c$. So every $\delta^{\sect}_{a,\leg,j}$ is zero: in each sector, (i) $\mathsf{D}$ has the same coefficient on $\sF^{k_1}_{i,2}$ and $(\sH^{\leg}_{j,2})^{\sect}$ as on $\sF^{k_1}_{i,1}$ and $(\sH^{\leg}_{j,1})^{\sect}$, and (ii) $\mathsf{D}_{ab} = 0$ for $b = (\sH^{\leg}_{j,1})^{\sect}$ unless $a$ is some $\sF^{k_1}_{i,1}$.

\emph{Step 2: the class of $\kappa$.} Only the exponents of $ct_2^\alpha$ and $\sK^\alpha$ contain $\alpha$, as a factor, and the value of a pair is homogeneous of the degrees of the product of the classes of its elements. So in each sector $\sect$ the pairs of class $\pi$, the class of $\kappa$, have total value $\alpha_{\sect} \chi_{\sect} \kappa$, where $\chi_{\mathrm{plain}} \defeq \chi$ and $\chi_{\alpha} \defeq -\mathsf{D}_K$. They are those of the $\sF^{k_1}_{i,1}$, of class $\pi$, with the elements of class $1$, the $(\sH^{\leg}_{j,1})^{\sect}$, the $(\sS^{\leg})^{\sect}$, and, in the \plainsector{}, $G_2$; of the $\sF^{k_1}_{i,2}$, of class $\pi\,\Omega$, with the $(\sH^{\leg}_{j,2})^{\sect}$, of class $\Omega^{-1}$; and of the quotient and remainder powers and the $\sY_i$, of classes $\zeta$, $\zeta'$, and $\zeta''$, with their bases, of classes $\pi/\zeta$, $\pi/\zeta'$, and $\pi/\zeta''$. Fix $\sect$, and let $W_c \defeq W^{\sect}_c(\mathsf{D})$ and $u_c \defeq u^{\sect}_c(\mathsf{D})$. By (i), the hidden basis cancels; by the product $F^{k_1}_i H^{\leg}_j$ of \cref{sec:copies} their total value is $\alpha_{\sect} \bigl( \sum_c \gamma_c\, b_c + \Phi \bigr)$, where
\[
  b_c \defeq y^\top ( W_c\, r - u_c ) + y^\top ( W_c - W_c^\top )\, A(\tau) + \bigl( P_{W_c} + h_c Z_c + Z_0 R_c \bigr)(\tau) .
\]
Here $h_c$ is a polynomial and $R_c$ an allowed polynomial, from the pairs of the powers with the bases of cell $c$ (\cref{sec:optimizations}). The rest $\Phi$, the masks' terms $\sum_c \pi_{k_1}\, y^\top W_c\, \vartheta_{\leg}$ of the products and the pairs with the \cancellationbases{} and with $G_2$, is free of $s_{\mathrm{E}}$, $s_{\mathrm{O}}$, and $s_{\mathrm{V}}$. As $\kappa = \gamma_{\ckey} L_\rho(\tau)$, the coefficient of the monomial $\pi\, s_{\mathrm{X}}\, \upsilon^{k_c}$ of each $\gamma_c$ gives $b_{\ckey} = \chi_{\sect} L_\rho(\tau)$ and $b_c = 0$ for $c \ne \ckey$, as polynomials in $y$ and $\tau$; and then $\Phi = 0$.

\emph{Step 3: cells and ties.} The part of degree one in $y$ of these identities, at $\tau = d_0$, $\rho$, and $d_j$, where $A$ is $0$, $e_0$, and $e_j - e_0$, shows that $W_c\, r = u_c$ and that $W_c$ is symmetric. The part of degree zero is condition (c) of \cref{lem:qap} if $c \ne \ckey$, where $Z_c = Z$; if $c = \ckey$, where $Z_c(X) = Z(X)(X - \rho)$, its value at $\tau = \rho$ is $(W_c)_{00} = \chi_{\sect}$, and it is then condition (b). So $W_c$ passes. By the mask identity (\cref{eq:mask}), the terms of $\Phi$ with $\eta_a\, \varepsilon^0$ are $\pi\, \eta_a\, y^\top t_a(\upsilon)$, for the tie polynomials $t_a$ of the stack $(W_c)_c$, and multiples of $\pi\, \eta_a P(\upsilon)\, \upsilon^e y_i$ from the pairs of the $\sY_i$ with the \tiebases{} $\sB^0_{a,e}$. As $\Phi = 0$, $P(T)$ divides each $t_a(T)$: the ties hold.

\emph{Parts \ref{it:view-locus}, \ref{it:view-nokey}, and \ref{it:view-source}.} By (ii), $r$ enters $\val(\mathsf{D}; \mathbf{c}, r)$ only through the pairs behind the $W^{\sect}_c(\mathsf{D})$, so $\val(\mathsf{D}; \mathbf{c}, r') - \val(\mathsf{D}; \mathbf{c}, r) = \sum_{\sect} \alpha_{\sect} \sum_c \gamma_c\, y^\top W^{\sect}_c(\mathsf{D})\, (r' - r)$. This is zero when $W^{\sect}_c(\mathsf{D})\, r' = u^{\sect}_c(\mathsf{D})$, which is $W^{\sect}_c(\mathsf{D})\, r$ by \ref{it:view-forward}. If both stacks are zero, then so are the $u^{\sect}_c(\mathsf{D})$, every $r'$ qualifies, and $\chi$ is a corner of a zero matrix. For \ref{it:view-nokey}, if $\mathsf{D}$ is $\chi\kappa$ formally, then by \ref{it:view-forward} the product $W r$ does not depend on $r$, for $W \defeq W^{\mathrm{plain}}_{\ckey}(\mathsf{D})$, so $W = 0$ and $\chi = W_{00} = 0$. For \ref{it:view-source}, the pairing $\mathsf{D}$ of $\mathsf{E} - \mathsf{E}'$ with $G_2$, or of $G_1$ with $\mathsf{E} - \mathsf{E}'$, has zero stacks. So by \ref{it:view-locus} it is $0$ identically at some $r$ only if $\mathsf{E}$ and $\mathsf{E}'$ are formally equal; otherwise \ref{it:view-sz} applies at every $r$.
\end{proof} 

\parhead{Classification}
We follow the outline of \cref{sec:ext-weights}, in its notation; let $z_s \defeq \sigma^s(z) = z[\omega^s T]$.

\begin{lemma}[Consecutive conjugates]\label{lem:moore}
Let $h \in \FF_{q^n}^d$ have entries that are linearly independent over $\FF_q$, with $d \le n$. Then for every integer $s$ the vectors $\sigma^s(h), \sigma^{s+1}(h), \dots, \sigma^{s+d-1}(h)$ form a basis of $\FF_{q^n}^d$.
\end{lemma}

\begin{proof}
The matrix with these columns has the rows $(\beta, \beta^q, \dots, \beta^{q^{d-1}})$ for the entries $\beta$ of $\sigma^s(h)$, which are linearly independent over $\FF_q$, as those of $h$ are; so its determinant is nonzero~\cite[Lemma~3.51]{LN97}.
\end{proof}

\begin{lemma}[Classification]\label{lem:classification}
Let $\mathbf{W}$ be a stack of symmetric matrices that pass every Booleanity test and that satisfies the ties. Suppose that its row space $\mathcal{V}$ has dimension at most three and contains no kernel vector.
\begin{enumerate}[label=(\alph*),leftmargin=*,nosep]
  \item\label{it:cls-bool} There are a set $\mathcal{W} \subseteq \mathcal{V} \cap \bits^{N+1}$ of assignments and $\chi_w \in \FF_q$ with $\mathbf{W} = \sum_{w \in \mathcal{W}} \chi_w\, \mathbf{H}(w)$.
  \item\label{it:cls-gates} If $\mathbf{W}$ passes every constraint's test, every such representation has $\chi_w \ne 0$ only for witnesses $w$ of $S$.
\end{enumerate}
In particular, a nonzero \usablestack{} whose row space has dimension at most three and contains no kernel vector is a combination, with a nonzero coefficient, of \honeststack{}s of witnesses in its row space.
\end{lemma}

\begin{proof}
\emph{Step 1: one symmetric tensor.} The symmetric matrices $W_U$ and $W_V$ over $\FF_{q^n}$ determine the stack, and with $W_s \defeq W_U$, $W_V$, and $\sigma^{n/2+1}(W_U)$ for $s \in I \defeq \{n/2 - 1, n/2, n/2 + 1\}$, the ties are $W_s\, z_{s'} = W_{s'}\, z_s$ for $s, s' \in I$. Let $d \defeq \dim \mathcal{V}$, let the columns of $\mathcal{U} \in \FF_q^{(N+1) \times d}$ be a basis of $\mathcal{V}$, with rows $u_i \in \FF_q^d$, which span $\FF_q^d$, and let $h_s \defeq \mathcal{U}^\top z_s = \sigma^s(h_0)$. The entries of $h_0$, the values of $z$ on the basis, are linearly independent over $\FF_q$, as $\mathcal{V}$ contains no kernel vector; so $d \le n$, and by \cref{lem:moore} the $h_s$, $s \in I$, span $\FF_{q^n}^d$, as $d \le 3$. Each $W_s$ is symmetric with columns in the $\FF_{q^n}$-span of $\mathcal{V}$, so $W_s = \mathcal{U}\, \Xi_s\, \mathcal{U}^\top$ for a unique symmetric $\Xi_s$, and the ties read $\Xi_s\, h_{s'} = \Xi_{s'}\, h_s$. So if $\sum_{s \in I} \varrho_s h_s = 0$, then $(\sum_s \varrho_s \Xi_s)\, h_{s'} = \Xi_{s'} \sum_s \varrho_s h_s = 0$ for all $s'$, so $\sum_s \varrho_s \Xi_s = 0$, as the $h_{s'}$ span. Hence there is a linear map $\Xi$ from $\FF_{q^n}^d$ to symmetric matrices with $\Xi(h_s) = \Xi_s$ for $s \in I$, and the trilinear form $\Xi(h; u, u') \defeq u^\top\, \Xi(h)\, u'$ is symmetric in $u$ and $u'$ and, by the ties, in $h$ and $u'$, hence in all three. Booleanity at $j \in [N]$ is linear over $\FF_q$, so the $W_s$ satisfy it: $\Xi(h; u_j, u_j - u_0) = 0$ for $h = h_s$, hence for all $h$. Also $\Xi_s = \sigma^s(\Xi_{n-s})$ for $s \in \{n/2-1, n/2\}$, as $\sigma^n = \mathrm{id}$ and $\sigma^{n/2}$ fixes $W_V$, a polynomial in $T^2$.

\emph{Step 2: cubes of Boolean points, for $u_0 \ne 0$.} Extend $l_0 \defeq u_0$ by vectors $l_k \defeq u_{j_k}$, $k \in [d-1]$, to a basis of $\FF_q^d$. For $\delta \in \bits^{d-1}$ let $p_\delta \in \FF_q^d$ be the point with $\langle l_0, p_\delta \rangle = 1$ and $\langle l_k, p_\delta \rangle = \delta_k$, and let $w_\delta \defeq \mathcal{U} p_\delta$: the $w_\delta$ are the assignments in $\mathcal{V}$ that are Boolean at the $j_k$. By Booleanity at the $j_k$ and symmetry, $\Xi(l_a; l_b, l_c)$ depends only on the set $A \subseteq [d-1]$ of the nonzero indices among $a, b, c$, and the cube $\langle h, p_\delta \rangle \langle u, p_\delta \rangle \langle u', p_\delta \rangle$ of $p_\delta$ has the value $\delta^A \defeq \prod_{k \in A} \delta_k$ there. The matrix $(\delta^A)_{A, \delta}$ is invertible, as the monomials $\delta^A$ are a basis of the functions on the Boolean cube. So there are $c_\delta \in \FF_{q^n}$ with $\Xi(h) = \sum_{\delta} c_\delta\, \langle h, p_\delta \rangle\, p_\delta\, p_\delta^\top$; and the squares $p_\delta\, p_\delta^\top$ are linearly independent, as they have the values $\delta^A$ on the pairs $(l_a, l_b)$, whose sets $A$ are all the subsets of $[d-1]$, as $d - 1 \le 2$. Let $\bar z_\delta \defeq z(w_\delta) = \langle h_0, p_\delta \rangle$, nonzero as $w_\delta$ is no kernel vector; then $\langle h_s, p_\delta \rangle = \sigma^s(\bar z_\delta)$, so $\Xi_s = \sum_\delta c_\delta\, \sigma^s(\bar z_\delta)\, p_\delta p_\delta^\top$. Comparing the coefficients of the squares, which are over $\FF_q$, in $\Xi_s = \sigma^s(\Xi_{n-s})$ gives $c_\delta\, \sigma^s(\bar z_\delta) = \sigma^s(c_\delta)\, \bar z_\delta$ for $s \in \{n/2-1, n/2\}$: so $\sigma^{n/2-1}$ and $\sigma^{n/2}$, hence $\sigma$, fix $c_\delta / \bar z_\delta$, and $c_\delta = \chi_\delta\, \bar z_\delta$ with $\chi_\delta \in \FF_q$. So $W_s = \sum_\delta \chi_\delta\, \bar z_\delta\, \sigma^s(\bar z_\delta)\, w_\delta w_\delta^\top$: $\mathbf{W} = \sum_\delta \chi_\delta\, \mathbf{H}(w_\delta)$, as $U = z\, \sigma^{n/2-1}(z)$ and $V = z\, \sigma^{n/2}(z)$.

\emph{Step 3: a test sees each point by itself.} Suppose $\mathbf{W} = \sum_\delta \chi_\delta\, \mathbf{H}(w_\delta)$ with $\chi_\delta \in \FF_q$, and let $\varrho$ be a linear function of a matrix, with coefficients in $\FF_q$, vanishing on every $W_c$. So $\varrho$ vanishes on $W_s = \sum_\delta \chi_\delta\, \bar z_\delta\, \sigma^s(\bar z_\delta)\, w_\delta w_\delta^\top$ for $s \in I$; that is, $h_0^\top \mathcal{Q}\, h_s = 0$ for $\mathcal{Q} \defeq \sum_\delta \chi_\delta\, \varrho(w_\delta w_\delta^\top)\, p_\delta p_\delta^\top$ over $\FF_q$. As the $h_s$ span, $h_0^\top \mathcal{Q} = 0$; so $\mathcal{Q} = 0$, as the entries of $h_0$ are linearly independent over $\FF_q$, and $\chi_\delta\, \varrho(w_\delta w_\delta^\top) = 0$, as the squares are linearly independent. For an assignment $w$, $w w^\top$ passes the Booleanity test of $j$ exactly if $w_j \in \bits$, and every test exactly if $w$ is a witness (\cref{sec:linear-tests}). So the $\chi_\delta$ of Step~2 are nonzero only if $w_\delta \in \bits^{N+1}$, which is \ref{it:cls-bool}; and in a representation as in \ref{it:cls-bool}, where every $w \in \mathcal{W}$ is one of the $w_\delta$, $\chi_w \ne 0$ only for witnesses, which is \ref{it:cls-gates}.

\emph{Step 4: the case $u_0 = 0$.} Then $\mathcal{V}$ contains no assignment; we show $\Xi = 0$: the stack is zero, and both parts hold with $\mathcal{W} = \emptyset$. Booleanity reads $\Xi(h; u_j, u_j) = 0$, and some $u_j$ form a basis $l_1, \dots, l_d$ of $\FF_q^d$; so by symmetry $\Xi(l_a; l_b, l_c) = 0$ unless $a$, $b$, $c$ are distinct. Hence $\Xi = 0$ if $d \le 2$. If $d = 3$, let $t \defeq \Xi(l_1; l_2, l_3)$; then $\Xi(h; l_b, l_c)$ is $t$ times the coordinate of $h$ at $l_a$ for distinct $a$, $b$, $c$, so $\Xi_s = \sigma^s(\Xi_{n-s})$ gives $t\, h_s = \sigma^s(t)\, h_0$ for $s \in \{n/2-1, n/2\}$, as $\sigma^s(h_{n-s}) = h_0$. So $t = 0$, as $h_{n/2-1}, h_{n/2}$ are not both multiples of $h_0$ (\cref{lem:moore}).
\end{proof} 

The next lemma is an inverse Littlewood--Offord theorem as in Tao--Vu and Nguyen--Vu~\cites[Theorem~2.5]{TV09}[Theorem~2.1]{NV11}, but for all coordinates, modulo $q$, and with explicit constants.

\makeatletter
\providecommand{\invcond}[2]{\textup{(I#1)}\def\@currentlabel{(I#1)}\label{#2}}\makeatother
\begin{lemma}[Inverse theorem]\label{lem:inverse}
Let $q$ be prime. Every nonzero $v \in \FF_q^{N+1}$ with $v_0 = 0$ and $\gp(v^\top) > \theta$ has a \shortmultiple{} $x$, and $x_0 = 0$, if for $N$, $g$, $\theta$, and the integer $\beta$ there are an integer $b$ and positive reals $\varepsilon_{\mathrm{tr}}$, $R_{\mathrm{tr}}$, $\Lambda_1, \Lambda_2, \Lambda_3$, $\Gamma_1, \Gamma_2, \Gamma_3$, $A_{\mathrm{inv}}$, $P_1$, and $P_2$ that satisfy the following conditions, in which $\mu_{\mathrm{tr}} \defeq \theta - 100\,\varepsilon_{\mathrm{tr}}$, $\xi_{\mathrm{tr}} \defeq (1 + 2/g)^{254}$, $H_3 \defeq \xi_{\mathrm{tr}} / (g^3\mu_{\mathrm{tr}} - \xi_{\mathrm{tr}}/g)$, $H_1 \defeq \Gamma_1/(g\mu_{\mathrm{tr}})$, $H_2 \defeq \Gamma_2/(g^2\mu_{\mathrm{tr}})$, and $T_{\mathrm{tr}} \defeq R_{\mathrm{tr}}\, g \sqrt{N/3}$.
\invcond{1}{it:inv-range}~$b \ge 7$ and $2^b \le g < 2^{254}$.\hskip 0.5em plus 0.2em minus 0.2em\relax
\invcond{2}{it:inv-trim}~$R_{\mathrm{tr}} \ge 20$, $1000\, e^{-99 R_{\mathrm{tr}}^2/200} \le \varepsilon_{\mathrm{tr}} < \mu_{\mathrm{tr}}$, and $g^3 \mu_{\mathrm{tr}} > \xi_{\mathrm{tr}}/g$.\hskip 0.5em plus 0.2em minus 0.2em\relax
\invcond{3}{it:inv-geom}~$\Lambda_1 \ge 2\sqrt{3}\, R_{\mathrm{tr}} \sqrt{N}$, $\Lambda_2 \ge 27\sqrt{2}\, \pi R_{\mathrm{tr}}^2 N$, and $\Lambda_3 \ge 144\sqrt{2}\, \pi R_{\mathrm{tr}}^3 N^{3/2}$.\hskip 0.5em plus 0.2em minus 0.2em\relax
\invcond{4}{it:inv-reg}~$\Gamma_3 \ge \xi_{\mathrm{tr}} (1 + H_3/g)$; $A_{\mathrm{inv}} \ge \mu_{\mathrm{tr}}/(\mu_{\mathrm{tr}} - \varepsilon_{\mathrm{tr}})$; $P_2^{b-1} \ge (g^{-3} + \varepsilon_{\mathrm{tr}})/\mu_{\mathrm{tr}}$ and $P_1^{b-1} \ge (g^{-2} + 2\varepsilon_{\mathrm{tr}})/\mu_{\mathrm{tr}}$; $\Gamma_2^{b-1} \ge g^{-2}/\mu_{\mathrm{tr}}$ and $\Gamma_1^{b-1} \ge g^{-1}/\mu_{\mathrm{tr}}$; $\Gamma_2 \ge A_{\mathrm{inv}} \Lambda_1 \Gamma_3 P_2$ and $\Gamma_1 \ge A_{\mathrm{inv}} \max\{\Lambda_1 \Gamma_2, \Lambda_2 \Gamma_3\}\, P_1$.\hskip 0.5em plus 0.2em minus 0.2em\relax
\invcond{5}{it:inv-wrap}~$\Lambda_d \Gamma_d < \mu_{\mathrm{tr}} q$ for $d \in \{1,2,3\}$; $g^{-4} < \mu_{\mathrm{tr}}$; and $H_1 T_{\mathrm{tr}} < q$, $4 H_2 T_{\mathrm{tr}}^2 < q$, and $54 H_3 T_{\mathrm{tr}}^3 < q$.\hskip 0.5em plus 0.2em minus 0.2em\relax
\invcond{6}{it:inv-bound}~$2\beta < q$, $H_1 \le \beta$, $2 H_2\, q \le \beta^2$, and $6 \lfloor H_3 \rfloor q^2 \le \beta^3$.
\end{lemma}

\Cref{app:inverse-proof} has the proof.

\begin{lemma}[Heavy stacks]\label{lem:heavy}
Let the parameters satisfy \ref{it:p-inverse}, and let $\mathbf{W}$ be a heavy stack.
\begin{enumerate}[label=(\alph*),leftmargin=*,nosep]
  \item\label{it:heavy-rank} $\rank \mathbf{W} \le 3$.
  \item\label{it:heavy-short} Every kernel vector $v$ in the row space of $\mathbf{W}$ has a \shortmultiple{}.
  \item\label{it:heavy-pack} If $x$ is a \shortmultiple{} of a kernel vector, then $\hat x \in R^\ell$, with the $\hat x_j$ of \cref{def:forms} over $\ZZ$, solves MSIS for the matrix $A$ of the columns $1, \dots, \ell$ of $M$ (\cref{def:msis}): $0 < \|\hat x\|_\infty \le \beta$ and $A \hat x = 0 \bmod q$.
\end{enumerate}
\end{lemma}

\begin{proof}
\ref{it:heavy-rank} By \cref{lem:gp-rank}, a stack of rank at least four has $\gp \le g^{-4} < \theta$.
\ref{it:heavy-short} A kernel vector has $v_0 = 0$, as $\varphi_0(v) = v_0$, and $\gp(v^\top) \ge \gp(\mathbf{W}) > \theta$ by \cref{lem:gp-rank}, so \cref{lem:inverse} applies.
\ref{it:heavy-pack} Modulo $q$, $x$ is a nonzero multiple of a kernel vector, hence a kernel vector; so $x_0 = 0$, as $|x_0| \le \beta < q$. Then $A \hat x = 0 \bmod q$ by the remark after \cref{def:forms}, and $\|\hat x\|_\infty = \|x\|_\infty$, as the coefficients of $\hat x$ are $x_1, \dots, x_N$ and zeros.
\end{proof}

The reductions run $\mathsf{LIFT}$, the algorithm of \cref{lem:lift}, whose hypotheses hold by \ref{it:p-inverse} and \ref{it:p-lift}.

\begin{lemma}[Lifting step]\label{lem:lift}
Let $2\beta < q$ and $24\beta^4 < q^3$. There is a deterministic algorithm that returns, on input a nonzero $v \in \FF_q^{N+1}$ with $v_0 = 0$, a \shortmultiple{} of $v$, or $\bot$ if there is none. It makes $N - 1$ extensions, each with at most $32$ integer programs and $c_{\mathrm{HNF}} = O(1)$ Hermite normal forms, both in dimension at most four, and $O(N)$ further operations, on inputs of $O(N \log q)$ bits; its running time $(N-1)\bigl(32\, T_{\mathsf{IP}} + c_{\mathrm{HNF}}\, T_{\mathsf{HNF}} + O(N)\bigr)$, for the times $T_{\mathsf{IP}}$ and $T_{\mathsf{HNF}}$ of these programs and forms, is polynomial in $N$ and $\log q$.
\end{lemma}

\begin{proof}
Drop the coordinate $0$, reorder the others, and scale $v$ so that $v_1 = 1$. For $i \in [N]$ let $\mathcal{L}_i \subseteq \ZZ^i$ be the lattice of the $x$ congruent modulo $q$ to $x_1 (v_1, \dots, v_i)$, and call a vector $x$ \emph{bounded} if $\|x\|_\infty \le \beta$. As $2\beta < q$, the \shortmultiple{}s of $v$ are the nonzero bounded vectors of $\mathcal{L}_N$. The bounded vectors of $\mathcal{L}_i$ span a space of dimension at most three: four of them are linearly dependent, as each of their $4 \times 4$ minors is divisible by $q^3$, because modulo $q$ they lie on a line, and is at most $4!\, \beta^4 < q^3$ in absolute value.

Starting from $\mathcal{M}_1 = \ZZ$, the algorithm computes for $i = 2, \dots, N$ a basis of a lattice $\mathcal{M}_i \subseteq \mathcal{L}_i$ of rank at most three containing the bounded vectors of $\mathcal{L}_i$, and returns a nonzero one of $\mathcal{L}_N$, or $\bot$ if there is none. The first $i$ entries of a bounded vector $y$ of $\mathcal{L}_{i+1}$ form a bounded vector of $\mathcal{L}_i$, in $\mathcal{M}_i$; so $y$ lies in the lattice $\mathcal{N}$ generated by $q\, e_{i+1}$ and the basis vectors $x$ of $\mathcal{M}_i$, each extended by an entry in $[0, q)$ congruent to $x_1 v_{i+1}$.

Written in the at most four coefficients of these generators, which lie in $\mathcal{L}_{i+1}$, these $y$ are the integer solutions of $2(i+1)$ linear inequalities, and one lies outside the span of those found exactly if one of the at most four functionals in a basis of the integer functionals vanishing on this span is at least $1$ or at most $-1$ on it. So a round of at most eight integer programs~\cite[Section~1]{Len83} finds a further solution or shows that those found span all, in dimension $d \le 3$: at most four rounds, $32$ programs. Let $\mathcal{M}_{i+1}$ consist of the vectors of $\mathcal{N}$ in this span, which include all these $y$; the algorithm keeps the basis of it whose projection to $d$ coordinates, on which the bounded vectors found have a nonzero minor $h$, is in Hermite normal form~\cite[Section~2, Theorem~2]{KB79}. Its entries are at most $d \cdot d!\, \beta^d \le 18\beta^3$ in absolute value, as a $d \times d$ minor of bounded vectors is at most $d!\, \beta^d$: at these coordinates they are at most the projected lattice's determinant, which divides $h$, and by Cramer's rule the others are combinations of these $d$ whose coefficients are such minors divided by $h$. By Cramer's rule, a bounded vector has coefficients at most $24\beta \max\{18\beta^3, q\}^3$; so all numbers have $O(\log q)$ bits. Hermite normal forms also give the functionals and $\mathcal{M}_{i+1}$, and the cited algorithms are polynomial.
\end{proof} 
\subsection{Runs, Marks, and Counting}\label{app:monitor}\label{app:counting}\label{app:coupling}

A monitor replays a run with expressions in place of elements and marks the first label of two expressions not formally equal, or the first query of $\Hkey$ at $[\kappa]_T$. The analyses of the reductions and the extractor use this mark in real runs, where the calls of $\mathcal{A}$ depend on the coins of $\Enc$; so \cref{lem:charged} uses a symbolic game, which answers by formal equality and agrees with a run until values collide. This is Shoup's simulation argument~\cite{EC:Shoup97}, with Schwartz's lemma for the coins, but the guessing probability for the \searchvector{} $r$.

\parhead{Runs}
The \emph{run} $\mathsf{R}_m$ is the experiment of \cref{def:we-ext} until $\mathcal{A}$ outputs, with $\Enc$ encrypting $m$ and no bit $b$; the $L$ labels given to $\mathcal{A}$ are \emph{received}. On every run with $m_0$ or $m_1$, $\mathcal{A}$ makes at most $Q < 2^\lambda$ calls to $\Ogg$, rejected ones included, and at most $Q_H \le 2^\lambda$ queries to $\Hkey$; we may assume it on every sequence of answers, as stopping $\mathcal{A}$ before a call or query beyond these changes no such run. The label of $[\kappa]_T$ is the \emph{target label}.

\parhead{The monitor}
A label that $\mathcal{A}$ guesses gets a new \emph{guess variable} \ref{it:mon-operand}; so a \emph{generic expression} is an expression over $\mathcal{X}_1$ and $\mathcal{X}_2$ enlarged by guess variables of $\GG_1$ and $\GG_2$, plus, in $\GG_T$, a linear combination of guess variables of $\GG_T$. Let $\mathsf{X}^\circ$ be $\mathsf{X}$ with every guess variable set to zero. Generic expressions are \emph{formally equal} if the coefficients of each product of guess variables in them are, the empty product included, and $\mathsf{X}$ \emph{has a guess variable} if it is not formally equal to $\mathsf{X}^\circ$. The \emph{monitor}, a function of a finished run for the analysis only, reads the calls of $\mathcal{A}$ to $\Ogg$ and its queries to $\Hkey$, in order, and keeps for each group a table of \emph{entries}, labels with generic expressions. It stops when it sets a \emph{mark}, the run's \emph{first mark}; a run in $\mathsf{Tgt}$ has one \ref{it:mon-target}.
\begin{enumerate}[label=(M\arabic*),leftmargin=*,nosep,topsep=0pt plus 3pt]
  \item\label{it:mon-start} \emph{Start.} The received labels are entered with the expressions of their elements. The mark is $\mathsf{init}$ if two received elements of a group have one label and expressions that are not formally equal.
  \item\label{it:mon-operand} \emph{Operand.} An operand string of a linear operation or pairing, answered or rejected, that is a label of its group and has no entry, a \emph{guessed label}, is entered with a new guess variable.
  \item\label{it:mon-result} \emph{Result.} A \emph{result} is that of an answered linear operation or pairing, computed from the entries of its operands, or, at the first answered name request on a label $\mathsf{u}$, the negative of the entry of $\mathsf{u}$, at the other label of its name. A result $\mathsf{F}$ at a label without an entry is entered; at a label with an entry $\mathsf{X}$ not formally equal to $\mathsf{F}$, the mark is $\mathsf{coll}$, or $\mathsf{diff}(\mathsf{F}^\circ - \mathsf{X}^\circ)$ if the group is $\GG_T$ and $\mathsf{F} - \mathsf{X}$ has no guess variable.
  \item\label{it:mon-target} \emph{Target.} The first query of $\Hkey$ at the target label sets $\mathsf{key}(\mathsf{X}^\circ)$ if the label has an entry $\mathsf{X}$, but $\mathsf{coll}$ if $\mathsf{X}$ has a guess variable, and $\mathsf{guess}$ if it has no entry.
\end{enumerate}

\parhead{Good and charged marks}
A mark $\mathsf{diff}(\mathsf{D})$ or $\mathsf{key}(\mathsf{D})$ \emph{has the expression} $\mathsf{D}$ and the multiple $\chi = 0$ or $1$. Its \emph{\markstack{}} is $\mathbf{W}^{\mathrm{plain}}(\mathsf{D})$ if this stack is nonzero and $\mathbf{W}^{\alpha}(\mathsf{D})$ otherwise, and its \emph{\marksector{}} is this stack's sector. The mark is \emph{good} if $\mathsf{D}$ is $\chi\kappa$ identically at the run's \searchvector{} $r$ and its \markstack{} is nonzero and heavy; this stack is then usable (\cref{lem:view}\ref{it:view-forward}), of rank at most three (\cref{lem:heavy}\ref{it:heavy-rank}). A first mark is \emph{charged} if it is $\mathsf{coll}$, or has an expression and is not good; so a first mark is $\mathsf{init}$, $\mathsf{guess}$, charged, or good. 

\parhead{The coupling}
Let $\mathsf{R}_\$$ be $\mathsf{R}_m$ with a uniform string $h$ in place of the \maskedmessage{} $m \oplus \Hkey([\kappa]_T)$, so $\Enc$ does not query $\Hkey$. In $\mathsf{R}_m$ the value of $\Hkey$ at the target label is uniform and enters only the \maskedmessage{} and answers to queries there. So $\mathsf{R}_m$ has the law of $\mathsf{R}_\$$ with $\Hkey$ answering $h \oplus m$ at the target label, and on the same coins they agree up to and including the first query of $\mathcal{A}$ there, without its answer. This part, with the label injections and the coins of $\Enc$, determines $\mathsf{Tgt}$ and the first mark; so their probabilities may be computed in $\mathsf{R}_\$$, where $h$ and $\Hkey$ are independent of the label injections and the coins of $\Enc$.

\begin{lemma}[Factor one half]\label{lem:half}
In the experiment of \cref{def:we-ext}, $\Pr[\, b' = b \wedge \mathsf{Fail} \,] \le \frac12 + \frac14 (f_0 + f_1)$.
\end{lemma}

\begin{proof}
By the coupling, the coins of $\mathsf{R}_\$$, which are independent of $b$, determine whether $\mathsf{Tgt}$ occurs and, if not, the output $b'$~\cite[full version, proof of Lemma~2]{EC:BelRog06}. So $\Pr[b' = b \wedge \neg\mathsf{Tgt}] \le \frac12 (1 - p)$ for $p \defeq \Pr[\mathsf{Tgt}]$, and $f \defeq \Pr[\mathsf{Tgt} \wedge \mathsf{Fail}] = \frac12 (f_0 + f_1) \le p$. Hence $\Pr[b' = b \wedge \mathsf{Fail}] \le \tfrac12 (1 - p) + f \le \tfrac12 + \tfrac12 f$.
\end{proof}

\begin{lemma}[Initial coincidences]\label{lem:init}
A run's first mark is $\mathsf{init}$ with probability at most $\sum_{s \in \{1,2,T\}} \binom{L_s}{2}\, \epsf$.
\end{lemma}

\begin{proof}
The mark $\mathsf{init}$ requires two received elements of one group with equal exponents and expressions that are not formally equal. For a fixed pair and every $r$ this has probability at most $\epsf$ over $\mathbf{c}$: in $\GG_1$ or $\GG_2$ by \cref{lem:view}\ref{it:view-source}, and for the pair $G_T$, $\sK^\alpha$, with values $1$ and $\alpha\kappa$, by \cref{lem:view}\ref{it:view-sz}; and $\GG_s$ has $\binom{L_s}{2}$ pairs.
\end{proof}

\parhead{The symbolic game}
In the \emph{symbolic game}, $\mathcal{A}$ runs as in $\mathsf{R}_\$$, but $\Ogg$ knows only the images of the label injections, the sets of labels: it keeps the monitor's table by \ref{it:mon-start} to \ref{it:mon-result}, giving each received element and each result the label of the entry formally equal to it or else a uniformly random label of its group without an entry, which exists by \ref{it:p-labels}. It rejects calls as in \cref{def:ggm}, so a name request unless it is on a received label of $\GG_T$ or the answer to a linear operation in $\GG_T$ or a pairing, and answers one with the name of its label and its result's. So the game uses no coin of $\Enc$, which allows a union bound; the \emph{evaluation}, made afterwards, samples $(\mathbf{c}, r)$ as $\Enc$ does and an independent uniform value in $\FF_q$ for every guess variable, giving every entry a value. A \emph{collision} is a pair of entries $\mathsf{X}$, $\mathsf{X}'$ of one group, not formally equal, with equal values; it \emph{occurs} when the later of the two is made, and has $\mathsf{D} \defeq \mathsf{X} - \mathsf{X}'$ and $\chi \defeq 0$. A \emph{key value} is an entry $\mathsf{X}$ of $\GG_T$ of value $\kappa$; it has $\mathsf{D} \defeq \mathsf{X}$ and $\chi \defeq 1$. Either event is \emph{good} if $\mathsf{D}$ lies in $\GG_T$, has no guess variable, and meets the conditions of a good mark at the evaluation's $r$, and \emph{bad} otherwise.

\begin{lemma}[Charged marks]\label{lem:charged}
The first mark of a run is charged with probability at most $\Ncmp(u)\, (\theta + \epsf) + \bigl( \binom{L + 3u}{2} + L + 3u \bigr)\, \epsf$.
\end{lemma}

\begin{proof}
\emph{Step 1: the two games agree until a collision.} Sample the label injections of $\mathsf{R}_\$$, independent of $h$ and $\Hkey$, lazily, given their images, deciding a label's exponent when it gets an entry~\cites[proof of Theorem~1]{EC:Shoup97}{EC:BonBoy04a}. With the evaluation first, this is the symbolic game with value equality in place of formal equality, except that $\mathsf{R}_\$$ redraws a new guess variable's value if an entry already has it, which is a collision. So on the same randomness, until a collision, both games make the same calls and queries with equal answers, the monitor's table in $\mathsf{R}_\$$ is that of the symbolic game, and in each group its entries have distinct values, their labels' exponents, and no other element has a label yet. The first collision's call is the same; its answer may differ.

\vspace{0pt plus 4pt}\emph{Step 2: a charged first mark gives a bad event.} Let the first mark of $\mathsf{R}_\$$ be charged. No two received elements collide, as this sets $\mathsf{init}$, and by Step~1 the monitor sets no mark until a collision occurs at a call or $\mathcal{A}$ queries $\Hkey$ at the target label. If a collision comes first, the first one is bad. Suppose that it is good, with later entry $\mathsf{X}$ and earlier entry $\mathsf{X}'$. As $\mathsf{X} - \mathsf{X}'$ lies in $\GG_T$ and has no guess variable, $\mathsf{X}$ is not a new guess variable, which occurs in its difference with every earlier entry; so $\mathsf{X}$ is a result. In $\mathsf{R}_\$$ it has the value of $\mathsf{X}'$, hence the label of $\mathsf{X}'$, and the monitor sets $\mathsf{diff}(\mathsf{X}^\circ - \mathsf{X}'^\circ)$, a good first mark: a contradiction. If the query comes first, it sets the first mark, which is not $\mathsf{guess}$; so \ref{it:mon-target} reads the entry of the target label, a key value by Step~1, and a bad one: the mark is $\mathsf{coll}$ if it has a guess variable, and else a $\mathsf{key}(\mathsf{D})$ that is not good.

\vspace{0pt plus 4pt}\emph{Step 3: a bad event is unlikely.} Fix a symbolic game and in it two entries, or one entry; we bound the probability over the evaluation that they form a bad event.
\emph{(i) $\mathsf{D}$ has a guess variable: at most $\epsf$.} The coefficient in $\mathsf{D}$ of some product of guess variables is not formally zero: a constant, or an expression of $\GG_1$ or $\GG_2$ whose value at every $r$ is then zero with probability at most $\epsf < 1$ over $\mathbf{c}$ (\cref{lem:view}\ref{it:view-source}). So for every $r$, $\varpi \cdot (\val(\mathsf{D}) - \chi\kappa)$ is a nonzero polynomial in $\mathbf{c}$ and the guess variables, of total degree at most $d_f + 1$ by \cref{lem:degree}, as $\varpi$ has degree five, and by Schwartz's lemma it vanishes with probability at most $(d_f + 1)/(q - D) \le \epsf$.
\emph{(ii) $\mathsf{D}$ has none: at most $\theta + \epsf$.} Over $\mathbf{c}$, the bound is $\epsf$ in $\GG_1$ or $\GG_2$ (\cref{lem:view}\ref{it:view-source}) and in $\GG_T$ at an $r$ at which $\mathsf{D}$ is not $\chi\kappa$ identically (\cref{lem:view}\ref{it:view-sz}). Every other $r$ satisfies $W_c\, r = u_c$ for the \markstack{} $\mathbf{W}$ of $\mathsf{D}$ and its vectors $u_c$ (\cref{lem:view}\ref{it:view-forward}); if there is one, $\mathbf{W}$ is nonzero, as $\mathsf{D}$ is not $\chi\kappa$ formally (\cref{lem:view}\ref{it:view-locus}, \ref{it:view-nokey}), and light, as the event is bad, so these $r$ have probability at most $\gp(\mathbf{W}) \le \theta$.

\vspace{0pt plus 4pt}\emph{Step 4: counts.} A call adds at most three entries, two guessed operands and a result; of the at most $L + 3Q$ entries, at most $3Q$ have a guess variable, and case~(i) covers at most $\binom{L + 3Q}{2}$ pairs and $3Q \le L + 3u$ key values, as $u = 2Q + 2$. In case~(ii) a collision at call $j$ pairs its result, as $\mathsf{D}$ has no guess variable (Step~2), with one of at most $L + 3j - 1$ other entries, and a key value is a result or one of the two received elements of $\GG_T$: at most $\sum_{j=1}^{Q} (L + 3j - 1) + Q + 2 \le \Ncmp(u)$ events. A union bound gives the claim.
\end{proof}

\begin{lemma}[A guessed target label]\label{lem:names}
If $\mathcal{A}$ makes at most $Q_H \le 2^\lambda$ queries to $\Hkey$, the first mark of a run is $\mathsf{guess}$ with probability at most $Q_H / \Nfree \le 2^{\lambda}/\Nfree$.
\end{lemma}

\begin{proof}
In $\mathsf{R}_\$$, sample $\mathsf{lab}_T$ lazily, deciding whether a string is a label, and of which element, only for the two received labels of $\GG_T$ and, at each call, at most two operands and a result's label: at most $2 + 3Q$ strings, of which those that are labels have entries until the first mark. So $\mathsf{guess}$ is set at a query to $\Hkey$ before which $[\kappa]_T$ has no label. At such a query, given the other randomness of $\mathsf{R}_\$$, which is independent of $\mathsf{lab}_T$, and the sampler's earlier decisions, which together determine the query, its label is uniform among the $U \ge 2^{\ell_{\mathsf{lab}}} - 2 - 3Q \ge \Nfree$ undecided strings, so it is the query's string with probability at most $1/U$. A union bound over the at most $Q_H$ queries gives the first bound; the second holds as $Q_H \le 2^\lambda$.
\end{proof}

\parhead{Records}
The reductions and the extractor follow the answered calls and keep for a label of each group a computable \emph{record} of an expression: a reduction its value at a point and its formula in the calls, the extractor its sketch data. They start from the received labels, give an operand without a record that of $0$, which it keeps, compute the record of a result from those of its operands, and keep it if its label has none. \emph{Until the monitor sets a mark, a label with a record has an entry $\mathsf{X}$, and the record is that of an expression formally equal to $\mathsf{X}^\circ$; an entry whose label has no record is a guess variable that no answered call had as an operand, so it occurs in no result.} By induction: $\mathsf{X} \mapsto \mathsf{X}^\circ$ commutes with linear combinations and pairings, and a result that sets no mark meets no entry or one formally equal to it. So at a first mark $\mathsf{diff}(\mathsf{D})$, set by a result $\mathsf{F}$ at a label with the entry $\mathsf{X}$, the label has a record, at the latest from this call if it is one of its operands, as $\mathsf{F} - \mathsf{X}$ has no guess variable, and the result's record minus the label's is that of $\mathsf{D}$, up to formal equality. Formally equal expressions of $\GG_T$ have equal stacks, by \cref{lem:view}\ref{it:view-forward} for their difference at every $r$, so equal sketches. 
\subsection{The Reductions}\label{app:reduction}

Each reduction $\mathcal{B}_b$ of \cref{thm:extraction} tries up to two expressions from a \privatemonitor{} and, if one is a good first mark's, recovers the row space of its \markstack{} from three random combinations of the rows, gradients in $r$, which cost $O(1)$ operations for each call of $\mathcal{A}$~\cite{BS83}, while an expression has $|\mathcal{X}_1| |\mathcal{X}_2| + 1$ coefficients:
\begin{enumerate}[label=(R\arabic*),leftmargin=*,nosep]
  \item\label{it:red-run} \emph{Run.} On a challenge $A \in R_q^{k \times \ell}$, sample a uniform seed $\seed$ and set $\Hseed(\seed, i, j)$ to $A_{i,j}$ for $i \in [k]$ and $j \in [\ell]$; encrypt $m_b$ with this seed, and run $\mathcal{A}$ on the ciphertext, ignoring its output and sampling the label injections lazily; $\mathcal{A}$ queries the ROs directly; sampling them lazily instead gives the same law and an algorithm without oracle access.
  \item\label{it:red-coins} \emph{Coins.} Sample independently, also of the run, a \emph{\privatetestpoint{}} $\mathsf{a}_{\mathrm{p}}$ and, for $l \in [3]$ and each sector $\sect$, a \emph{\recoverypoint{}} $\mathsf{a}^{(l,\sect)}$: uniformly random valid tuples, each with a uniform vector of $\FF_q^{N+1}$ in place of $r$.
  \item\label{it:red-monitor} \emph{Private monitor.} Keep records with the values at $\mathsf{a}_{\mathrm{p}}$ (\cref{app:monitor}) to the end of the run. Then run \ref{it:red-grad} and \ref{it:red-kernel} in turn with $\mathsf{D} \defeq \mathsf{F} - \mathsf{X}$ for the first result $\mathsf{F}$ in $\GG_T$ whose label has a record $\mathsf{X}$ of another value, and with $\mathsf{D}$ the record of the target label, each if it exists; then output $\bot$.
  \item\label{it:red-grad} \emph{Gradients.} For each $l$ and $\sect$, compute $\nabla^{(l,\sect)}(\mathsf{D}) \in \FF_q^{N+1}$, the gradient in $r$ at $\mathsf{a}^{(l,\sect)}$ of the value of the part of $\mathsf{D}$ in sector $\sect$: with the exponents taken there, its entry $j$ is $\alpha_{\sect} \sum_{\leg} s_{\leg} \sum_{a \in \mathcal{X}_1} a\, \mathsf{D}_{a b}$ for $b = (\sH^{\leg}_{j,1})^{\sect}$.
  \item\label{it:red-kernel} \emph{Output.} Find a kernel vector $v$ in the span of the $\nabla^{(l,\sect)}(\mathsf{D})$ of the \plainsector{} or, if there is none, of the \alphasector{}. If $v$ exists and $\mathsf{LIFT}(v)$ (\cref{app:inverse}) returns $x$, output $\hat x$ (\cref{lem:heavy}\ref{it:heavy-pack}) and stop.
\end{enumerate}

\begin{lemma}[Reductions]\label{lem:reduction}
Let $\mathcal{A}$ be an adversary and $b \in \bits$.
\begin{enumerate}[label=(\alph*),leftmargin=*,nosep]
  \item\label{it:red-law} \emph{Law.} For a uniform challenge, the run of $\mathcal{A}$ in \ref{it:red-run} has the law of the run with $m_b$ of the experiment, as the values that \ref{it:red-run} sets are uniform; the coins of \ref{it:red-coins} are independent of it.
  \item\label{it:red-sound} \emph{Soundness.} Every output of $\mathcal{B}_b$ other than $\bot$ is a solution of MSIS for its challenge, the columns $1, \dots, \ell$ of $M$.
  \item\label{it:red-detect} \emph{Detection.} For every run whose first mark has an expression $\mathsf{D}$, an expression of \ref{it:red-monitor} is formally equal to $\mathsf{D}$, except with probability at most $\epsf$ over $\mathsf{a}_{\mathrm{p}}$.
  \item\label{it:red-recover} \emph{Recovery.} For every run whose first mark is good, with expression $\mathsf{D}$ and \marksector{} $\sect$, the vectors $\nabla^{(l,\sect)}(\mathsf{D})$ lie in the row space of the \markstack{}, and span it except with probability at most $\epsf$ over the \recoverypoint{}s; and if $\sect$ is the \alphasector{}, the vectors $\nabla^{(l,\mathrm{plain})}(\mathsf{D})$ are zero.
\end{enumerate}
Consequently, in the run with $m_b$ and a uniform $M$, the probability that the first mark is good and the row space of its \markstack{} contains a kernel vector is at most $s_b + 2\epsf$.
\end{lemma}

\begin{proof}
\ref{it:red-law} holds by construction, and \ref{it:red-sound} by \cref{lem:heavy}\ref{it:heavy-pack}, as $\mathsf{LIFT}(v)$ returns only \shortmultiple{}s of $v$.

\ref{it:red-detect} By the invariant of \cref{app:monitor}, at a first mark $\mathsf{key}(\mathsf{D})$ the record of the target label is formally equal to $\mathsf{D}$, as an entry without a guess variable has a record and records are never replaced. At a first mark $\mathsf{diff}(\mathsf{D})$, so is the result's record minus the label's, the first expression of \ref{it:red-monitor} unless $\val(\mathsf{D}; \mathsf{a}_{\mathrm{p}}) = 0$, as no earlier result in $\GG_T$ meets a record of another value. As $\mathsf{D}$ is then not formally zero \ref{it:mon-result}, $\varpi \cdot \val(\mathsf{D})$ is a nonzero polynomial of total degree at most $d_f$ in $\mathbf{c}$ and $r$ (\cref{lem:degree}), so by Schwartz's lemma it vanishes at $\mathsf{a}_{\mathrm{p}}$ with probability at most $d_f/(q - D) \le \epsf$.

\ref{it:red-recover} As $\mathsf{D}$ is $\chi\kappa$ identically at the \searchvector{}, the hidden-basis step in the proof of \cref{lem:view} gives $\mathsf{D}_{ab} = 0$ for $b = (\sH^{\leg}_{j,1})^{\sect'}$ unless $a$ is some $\sF^{k_1}_{i,1}$, whose exponent is $\pi_{k_1} y_i$. So $\nabla^{(l,\sect')}(\mathsf{D}) = \alpha_{\sect'} \sum_{c,i} \gamma_c\, y_i\, W_{c,i}$, with the coins of its \recoverypoint{} and the rows $W_{c,i}$ of the $W^{\sect'}_c(\mathsf{D})$: a vector of this stack's row space, zero if the stack is, as the plain one is if the \marksector{} is the \alphasector{}. Expressions formally equal to $\mathsf{D}$ have its values and stacks (\cref{app:monitor}), so the same vectors. Let $\phi$ be a linear form nonzero on the row space $\mathcal{V}$ of the \markstack{}. Then $\phi(\nabla^{(l,\sect)}(\mathsf{D})) / (\alpha_{\sect} \pi) = \sum_{c,i} \phi(W_{c,i})\, (\gamma_c / \pi)\, y_i$ is a nonzero polynomial of degree at most $n + 2$ in the coins: its monomial $s_{\mathrm{X}}\, \upsilon^{k_c} y_i$, for the kind $\mathrm{X}$ of $c$, which determines $c$ and $i$ (as in that step), has coefficient $\phi(W_{c,i})$, nonzero for some $(c, i)$ as these rows span $\mathcal{V}$. As $\alpha_{\sect} \pi \ne 0$ at valid coins, $\phi(\nabla^{(l,\sect)}(\mathsf{D})) = 0$ with probability at most $(n+2)/(q - D)$. As $\dim \mathcal{V} \le 3$, the three vectors span $\mathcal{V}$ unless one lies in the span of the earlier ones while that span is not $\mathcal{V}$, so that such a $\phi$ vanishes on it; as the \recoverypoint{}s are independent, the bound holds given the earlier ones: probability at most $3(n+2)/(q - D) \le \epsf$, as $n \le D$.

For the consequence, fix such a run. Except with probability $2\epsf$, by \ref{it:red-detect} an expression of \ref{it:red-monitor} is formally equal to $\mathsf{D}$, and by \ref{it:red-recover} and its proof its vectors of \ref{it:red-grad} are the $\nabla^{(l,\sect)}(\mathsf{D})$ and span $\mathcal{V}$ in the \marksector{} and, if this is the \alphasector{}, vanish in the plain one. So for it \ref{it:red-kernel} finds a kernel vector $v \in \mathcal{V}$ and $\mathsf{LIFT}$ a \shortmultiple{} of it (\cref{lem:heavy}\ref{it:heavy-short}), and $\mathcal{B}_b$ outputs a solution, for it or earlier \ref{it:red-sound}: probability $s_b$ by \ref{it:red-law}.
\end{proof}

Running times count \emph{operations}: arithmetic operations and equality tests in $\FF_q$; additions, subtractions, halvings, and comparisons of integers of absolute value below $2^{\ell_{\mathsf{lab}}}$; XORs and writes of strings of at most $\ell_{\mathsf{lab}}$ bits or elements of $\FF_q$; uniform samples from an explicitly given finite set, or from the elements of $\FF_q$ or strings of length $\ell_{\mathsf{lab}}$ that are not keys of a table; oracle calls; and insertions into and lookups in tables keyed by labels, names, oracle inputs, or elements of $\FF_q$, where an insertion at a key that is present keeps and returns the stored value. They also count the length of the algorithm's description, advice included, one for each element of $\FF_q$ or string of length $\ell_{\mathsf{lab}}$, with code packed into such strings. The oracle calls of $\mathcal{A}$ count in its running time $t$, the \liftingstep{} counts bit operations on its integers, and the constants in $O()$ are absolute.

\begin{proposition}[Running time of the reductions]\label{prop:reduction-time}
Under the hypotheses of \cref{thm:extraction}, if $\mathcal{A}$ runs in time $t$ and makes at most $Q_H$ queries to each of $\Hseed$ and $\Hkey$, then $\mathcal{B}_0$ and $\mathcal{B}_1$ run in time at most
\begin{equation}\label{eq:cr}
  t + 148\, Q + (d_R + 5)\, Q_H + C_R + 2 T_{\mathsf{LIFT}} , \quad C_R \defeq 8 I + 3 L + 2 C_{\mathrm{post}} + C_{\mathrm{set}} + C_{\mathrm{desc}} ,
\end{equation}
where $I$, $C_{\mathrm{post}}$, $C_{\mathrm{set}}$, and $C_{\mathrm{desc}}$, sums given in the proof, and hence $C_R$, do not depend on $\mathcal{A}$, and $T_{\mathsf{LIFT}}$, the time of the \liftingstep{}, is polynomial in $N$ and $\log q$ (\cref{lem:lift}, by \ref{it:p-inverse} and \ref{it:p-lift}). We call this bound $t_R$; it is $t + O(Q + d_R\, Q_H + \mu) + \poly(D, \log q)$, as $C_R = O((N + n) D + \mu)$, $N < D$, and $n \le D$.
\end{proposition}

\begin{proof}
Let $\mathcal{B}_b$ sample the coins of \ref{it:red-coins} first and follow the calls as $\mathcal{A}$ makes them, with tables of labels, exponents, records, rejected strings, names, and answers of $\Hkey$ and $\Hseed$; by \cref{lem:reduction}\ref{it:red-law} this changes no output. Its costs beyond the running times of $\mathcal{A}$ and $\mathsf{LIFT}$ are the following, by the term that pays for them.\par
{\small
\begin{itemize}[label={},leftmargin=1.5em,itemindent=-1.5em,nosep,topsep=2pt]
  \item $148\, Q$. A call to $\Ogg$ has one result (of a linear operation or pairing, or the negative of a name request) and takes up to $148$ operations: $14$ for its lookups, exponent, value at $\mathsf{a}_{\mathrm{p}}$, node, label, and test of \ref{it:red-monitor}; $6 \cdot 19$ for the six passes of \ref{it:red-grad}, each with three operations forwards and, for each expression, a gradient computed backwards through the calls~\cite[Theorem~2]{BS83} with four operations and four for the derivatives of the received operands; $2 \cdot 8$ for two operand strings not in the tables, to decide whether each is a label and to sample its exponent; $2$ for two records of $0$; $2$ for the name.
  \item $(d_R + 5)\, Q_H$: $3$ for a query to $\Hkey$ (a lookup, a sample, and an insertion); and $d_R + 2$ for a query to $\Hseed$ ($d_R$ samples, a lookup, and an insertion).
  \item $8 I$, for the exponents of all published elements at eight points, the coins of $\Enc$ and of \ref{it:red-coins}, $I$ operations each. With $N' \defeq N + 1$ and $n_2$ the number of elements of $ct_2$, $I$ is the sum of $3N' + 19$ for the samples and $r_0$; $\frac72 n + \Delta + \Delta' + J + J' + 3$ for the powers of $\upsilon$, $-\omega\upsilon$, and $\tau$; $9\,k\,d_R\,N' + 3n$ for $z$ at its three points, from $M$; $5K_1 + 3K_2 + \frac32 n$ for the scalars of the legs and cells; $13K_2 N' + 17K_2$ for the masks; $(4N' + 6)(D - 1) + 2$ for $A_i$, $B_i$, $Z$, $Z_0$, and $L_\rho$ at $\tau$ by Horner's rule, from their coefficients, which depend only on $S$ and are built into $\mathcal{B}_b$; and $(2K_1 + 24K_2 + 4) N' + \Delta + (1 - 1/\nu) \Delta' + 3n (J + J') + 13n + 3K_2 + n_2 + 22$ for the published elements.
  \item $3 L$, for the label of each received element, with tables for each group: a sample $\mathsf{u}$ that is not a key, an insertion of $\mathsf{u}$ at its exponent $x$, which returns the label $\mathsf{u}'$ kept there, and one of $x$ and its record at $\mathsf{u}'$; so equal exponents get one label.
  \item $2\, C_{\mathrm{post}}$, after the run, for each expression: $3N'$ for the gradients of the \alphasector{}; $2(18N' + 3)$ for two row echelon forms; $2\bigl((9\,k\,d_R + 5) N' + 21\,k\,d_R + 33\bigr)$ for the intersections with the kernel of $z$ in both sectors; and $N'$ for the packing.
  \item $C_{\mathrm{set}}$: the sum of $1$ for the seed; $2k\ell + k(d_R + 2)$ for $M$, that is, $k\ell$ insertions of the challenge, then the $k(\ell + 1)$ queries of $\Enc$, a lookup each in columns $1$ to $\ell$ and $d_R + 2$ operations each in column $0$; $6 + \lceil \mu/\ell_{\mathsf{lab}} \rceil$ for the \maskedmessage{}, that is, $3$ for the label of $[\kappa]_T$ and its exponent, with no record, $3$ for the answer of $\Hkey$ there, and an XOR for each block of at most $\ell_{\mathsf{lab}}$ bits of $m_b$; and $256$ for $G_T$ and $\sK^\alpha$, counted as results.
  \item $C_{\mathrm{desc}}$, for the description of $\mathcal{B}_b$ beyond that of $\mathcal{A}$: the sum of $(2N' + 4) D + 1$ for the coefficients of $A_i$, $B_i$, $Z$, $Z_0$, and $L_\rho$ and for the evaluation points; $\lceil \mu/\ell_{\mathsf{lab}} \rceil$ for $m_b$; and the length of its code, a constant.
\end{itemize}}
\noindent Hence $C_R = O((N + n) D + \mu)$, as $k\, d_R < n \le D$ and $J + J' \le 3D$, with $J' = 1$ if $\nc = 0$ (\cref{sec:optimizations}).
\end{proof} 
\subsection{The Extractor}\label{app:extractor}

\parhead{Sketches}
$\mathcal{E}$ samples, for every cell $c$, $l \in [3]$, and sector $\sect$, an independent uniform \emph{\probevector{}} $v^{(l,\sect)}_c \in \FF_q^{N+1}$. The \emph{sketches} of an expression $\mathsf{D}$ of $\GG_T$ in sector $\sect$ are the vectors $\sum_c W^{\sect}_c(\mathsf{D})\, v^{(l,\sect)}_c$, linear in $\mathsf{D}$. The record of a label, its \emph{sketch data}, are in $\GG_T$ the sketches of its expression, not its $|\mathcal{X}_1| |\mathcal{X}_2| + 1$ coefficients; in $\GG_1$ the $K_1$ vectors $a_{k_1}$ of its coefficients on the $\sF^{k_1}_{i,1}$; and in $\GG_2$, with $b_{\leg,\sect}$ the vector of its coefficients on the $(\sH^{\leg}_{j,1})^{\sect}$, the $9n$ numbers $\langle b_{\leg,\sect}, v^{(l,\sect)}_c \rangle$ for $c = (k_1, \leg)$. A pairing has $W^{\sect}_c = a_{k_1} b_{\leg,\sect}^\top$ and the sketches $\sum_c \langle b_{\leg,\sect}, v^{(l,\sect)}_c \rangle\, a_{k_1}$.

\parhead{Tests}
A \emph{test} of three vectors computes the reduced row echelon form of their span $\mathcal{V}'$, with coordinates in the order $0, \dots, N$. Its \emph{candidates} are the vectors of $\mathcal{V}'$ with entry $1$ at coordinate $0$ and entries in $\bits$ at the other pivots: at most four, as the pivot entries determine a vector of $\mathcal{V}'$, and none unless $0$ is a pivot. It finds the first candidate that is a witness of $S$, or nothing. For every result in $\GG_T$, after its operands have records and before its own is kept, $\mathcal{E}$ tests the sketches of the result's record and, if its label already has one, those of their difference, in the \plainsector{} and then in the \alphasector{}. It outputs the first witness found, or $\bot$.

\begin{lemma}[Extractor]\label{lem:extractor}
\begin{enumerate}[label=(\alph*),leftmargin=*,nosep]
  \item\label{it:extl-sound} Every output of $\mathcal{E}$ is $\bot$ or a witness of $S$.
  \item\label{it:extl-book} If the first mark of a run has an expression $\mathsf{D}$ and $\mathcal{E}$ outputs $\bot$, then the tests of the sketches of $\mathsf{D}$ in both sectors find nothing.
  \item\label{it:extl-span} Let $\mathbf{W}$ be a nonzero \usablestack{} whose row space $\mathcal{V}$ has dimension at most three and contains no kernel vector, and let the $v^{(l)}_c$ be independent and uniform. Then a test of the three vectors $\sum_c W_c\, v^{(l)}_c$, $l \in [3]$, outputs a witness, except with probability at most $2/q$.
\end{enumerate}
Consequently, the probability that the first mark of a run is good, the row space of its \markstack{} contains no kernel vector, and $\mathcal{E}$ outputs no witness is at most $2/q$.
\end{lemma}

\begin{proof}
\ref{it:extl-sound} A test finds only witnesses.
\ref{it:extl-book} As $\mathcal{E}$ outputs $\bot$, its tests found nothing. At a mark $\mathsf{diff}(\mathsf{D})$, the result's label has a record and the difference that $\mathcal{E}$ tested has the sketches of $\mathsf{D}$ (\cref{app:monitor}). At a mark $\mathsf{key}(\mathsf{D})$, $\mathsf{D}$ is $\mathsf{X}^\circ$ for an entry $\mathsf{X}$ without a guess variable: a received element, whose stacks are zero, or a result, whose sketches $\mathcal{E}$ tested. A test of zero vectors finds nothing.

\ref{it:extl-span} As the $W_c$ are symmetric, the linear map $(v_c)_c \mapsto \sum_c W_c\, v_c$ has image $\mathcal{V}$, so these vectors are independent and uniform on $\mathcal{V}$, of dimension $d \le 3$, and span it with probability $\prod_{0 \le i < d} (1 - q^{i-3}) > 1 - 2/q$~\cite[(1) and Lemma~2.3]{FG15}. Then the test finds a witness: $\mathcal{V}$ contains one (\cref{lem:classification}), which is a candidate.

For the consequence, fix a run as stated, with expression $\mathsf{D}$. Its \markstack{} meets the conditions of \ref{it:extl-span} (\cref{app:monitor}), and the \probevector{}s are independent of the run; so except with probability $2/q$ the test of the sketches of $\mathsf{D}$ in the \marksector{} finds a witness, and then $\mathcal{E}$ outputs a witness, by \ref{it:extl-book} and \ref{it:extl-sound}.
\end{proof}
\begin{proof}[of \cref{thm:extraction}\ref{it:ext-sound} and \ref{it:ext-time}]
Part \ref{it:ext-sound} is \cref{lem:extractor}\ref{it:extl-sound}. For \ref{it:ext-time}, a call makes at most three records and one difference, of $O(K_1 N + n)$ entries each, by $O(1)$ table and $O(K_1 N + n)$ field operations: a linear combination or, for a pairing, by the formula above, $K_1$ sums over the legs and a combination of the vectors $a_{k_1}$ for each $l$ and sector. A test costs $O(N)$ for the row echelon form and $O(N + |S|)$ for each of at most four candidates, and each of at most $Q$ results has at most four tests. The start samples $9n(N + 1)$ numbers and enters $L$ labels, whose records are written out when a call first uses them, within its $O(K_1 N + n)$ operations. So $\mathcal{E}$ runs in time $O\bigl( (Q+1)(K_1 N + n + |S|) + nN + L \bigr)$, which is the bound, as $K_1 \le n$ and $N \ge 1$.
\end{proof} 
\subsection{Proof of the Inverse Theorem}\label{app:inverse-proof}

This subsection proves \cref{lem:inverse}; symbols other than the lemma's are local to it. Reading the nonzero coordinates of $v$ one at a time, we keep a saturated integer lattice of rank at most three whose reduction modulo $q$ contains the prefix read, with a fiber at most $3\varepsilon_{\mathrm{tr}}$ less dense than the densest set on which that prefix is constant modulo $q$. To add a coordinate, we trim that set for the longer prefix to an ellipsoid: a count of lattice points shows that its congruence cannot wrap around modulo $q$, so small minors make it hold over the integers, which gives the next lattice; a regularization bounds the product of fiber density and height that the count needs. Each step starts from the densest set of its own prefix and inherits only a lattice, so losses do not accumulate. Minkowski's theorem in the last lattice gives the \shortmultiple{}.

\parhead{Boxes and prefixes}
For a finite set $I$, the \emph{box} on $I$ is $[2, g+1]^I$, and $r_I$ is uniform on it, as $r_1, \dots, r_N$ are independent and uniform on $[2, g+1]$ for both samplers (\cref{sec:guessing-probability}). The \emph{density} of a set $F$ of box points is $\Pr[F] \defeq |F|/g^{|I|}$; a condition on the entries at $I' \subseteq I$ has the same density in the boxes on $I'$ and $I$. For $u \in \FF_q^I$, $\gp(u^\top)$ is the largest density of a \emph{modular fiber} $\{ r : u^\top r = t \bmod q \}$, and $\gp(v^\top)$ is that of \cref{def:gp}, as $v_0 = 0$. Renumbering $1, \dots, N$, let $v_1, \dots, v_{N_0}$ be the nonzero ones among $v_1, \dots, v_N$, and $v_{[i]} \defeq (v_1, \dots, v_i)$; a \shortmultiple{} of $v_{[N_0]}$ (defined as for $v$), extended by zeros, is one of $v$ with $x_0 = 0$. Conditioning on $r_{i+1}$ shows that $\gp(v_{[i]}^\top)$ does not increase with $i$, and $g + 1 < q$ makes $x \mapsto v_1 x \bmod q$ injective on $[2, g+1]$; so
\begin{equation}\label{eq:inv-chain}
  \theta < \gp(v^\top) = \gp(v_{[N_0]}^\top) \le \gp(v_{[i]}^\top) \le 1/g \qquad \text{for } i \in [N_0] .
\end{equation}

\parhead{Lattices}
A lattice $\Theta \subseteq \ZZ^I$ of rank $d$ has basis matrices $B \in \ZZ^{I \times d}$, unique up to $B \mapsto BU$ for a unimodular $U$; $B_J$ denotes the rows $J \subseteq I$ of $B$. It is \emph{saturated} if $\Theta = \operatorname{span}_{\RR}(\Theta) \cap \ZZ^I$, as are the integer points of any subspace; two saturated lattices of equal rank, one inside the other, are equal. For $d \ge 1$, its \emph{height} $h(\Theta)$ is the largest $|\det B_J|$ over the sets $J$ of $d$ indices, and it has a \emph{largest minor at} $J$ if $|\det B_J| = h(\Theta)$. Its \emph{fibers} are the sets $\{ r : B^\top r = t \}$ of box points, for $t \in \ZZ^d$; its \emph{largest fiber density} $m(\Theta)$ is the largest density of one, at most $g^{-d}$, as fixing the entries outside a set $J$ with $\det B_J \ne 0$ leaves at most one value of $r_J$, as in Odlyzko's lemma~\cite[Lemma~3.1]{NW22}; and it is \emph{regular} if $d \in [3]$ and $m(\Theta)\, h(\Theta) \le \Gamma_d/g^d$. As $(BU)^\top = U^\top B^\top$, none of these depends on $B$. The reduction of $\Theta$ modulo $q$ is $\bar\Theta \defeq \{ x \bmod q : x \in \Theta \} = B\, \FF_q^d$. A nonempty $F \subseteq \ZZ^I$ has \emph{affine dimension} $\dim F$, that of the span of its differences, and \emph{normal lattice} $F^\perp$ of the integer vectors whose inner product with the points of $F$ is constant, saturated of rank $|I| - \dim F$, as the orthogonal complement of a span of integer vectors has a basis of integer vectors. A set $F$ of box points lies in a fiber of $F^\perp$, so $\Pr[F] \le m(F^\perp)$, and if it lies in a fiber of $\Theta$, then $\Theta \subseteq F^\perp$.

\begin{lemma}[Integer lattices]\label{lem:lattice-facts}
Let $\Theta \subseteq \ZZ^I$ be a lattice of rank $d$ with a basis matrix $B$.
\begin{enumerate}[label=(\alph*),leftmargin=*,nosep]
  \item\label{it:facts-minor} If $\Theta$ has a largest minor at $J$, then $|y_i| \le \|y_J\|_1$ for all $y \in \operatorname{span}_{\RR}(\Theta)$ and $i \in I$.
  \item\label{it:facts-split} If $\Theta$ is saturated, then $B^\top \ZZ^I = \ZZ^d$.
  \item\label{it:facts-small} If an integer matrix $\Delta$ with rows indexed by $I$ has rank $k$ and $k \times k$ minors below $q$ in absolute value, then every $c \in \ZZ^I$ with $c^\top \Delta = 0 \bmod q$ is congruent modulo $q$ to a $u \in \ZZ^I$ with $u^\top \Delta = 0$.
\end{enumerate}
\end{lemma}

\begin{proof}
\ref{it:facts-minor} As $B_J$ is invertible, Cramer's rule writes each row of $B$ as a combination of the rows of $B_J$ with coefficients, ratios of $d \times d$ minors of $B$ (or zero) to $\det B_J$, of absolute value at most $1$; and $y_i$ is the same combination of the entries of $y_J$.
\ref{it:facts-split} As $\Theta$ is saturated, the columns of $B$ extend to a basis of $\ZZ^I$~\cite[Chapter~I, Theorem~I, Corollary~3]{Cas97}, the columns of a unimodular matrix $(B, B')$, and the first $d$ rows of its integer inverse form a left inverse of $B$, so $B^\top \ZZ^I = \ZZ^d$.
\ref{it:facts-small} As $\Delta$ has rank $k$ and a nonzero $k \times k$ minor below $q$ in absolute value, it has rank $k$ modulo $q$, so the reductions of these $c$ form a subspace of $\FF_q^I$ of dimension $|I| - k$. It contains the reduction of the saturated lattice of the $u \in \ZZ^I$ with $u^\top \Delta = 0$, of rank $|I| - k$, whose basis matrices have integer left inverses by \ref{it:facts-split}; so that reduction has dimension $|I| - k$ too, and the two are equal.
\end{proof}

\begin{lemma}[Prefix lattices]\label{lem:prefix}
Let $\Theta \subseteq \ZZ^I$ be a saturated lattice of rank $d \ge 1$ with a basis matrix $B$ and a largest minor at $J$. Order $I$ as $i_1, \dots, i_{|I|}$ with $J$ first, and for $d \le k \le |I|$ let $I_k \defeq \{ i_1, \dots, i_k \}$, and let $\Theta_k \subseteq \ZZ^{I_k}$ be the integer points of the column span of the submatrix $B_{I_k}$, with height $h_k$ and largest fiber density $m_k$.
\begin{enumerate}[label=(\alph*),leftmargin=*,nosep]
  \item\label{it:prefix-chain} $\Theta_d = \ZZ^{I_d}$, so $h_d = 1$ and $m_d = g^{-d}$; $\Theta_{|I|} = \Theta$; and each $\Theta_k$ has rank $d$ and a largest minor at $J$.
  \item\label{it:prefix-step} For $d \le k < |I|$ there is an integer $s_k \ge 1$ with $h_{k+1} = s_k h_k$ and $m_{k+1} \le m_k \lceil g/s_k \rceil/g$. If $s_k \ge g$, the entry at $i_{k+1}$ is constant on each fiber of $\Theta_{k+1}$; if $s_k < g$, then $m_{k+1}^b h_{k+1}^{b-1} \le m_k^b h_k^{b-1}$.
  \item\label{it:prefix-three} If $d = 3$ and $m(\Theta) \ge \mu_{\mathrm{tr}}$, then $h(\Theta) \le H_3$ and $\Theta$ is regular.
\end{enumerate}
\end{lemma}

\begin{proof}
For $d \le k \le |I|$, $B_{I_k}$ contains the invertible $B_J$, so $\Theta_k$ has rank $d$, and $B_{I_k} = YC$ for a basis matrix $Y$ of $\Theta_k$ and an integer matrix $C$ with $\det C \ne 0$. So the fibers of $\Theta_k$ are the level sets of $r \mapsto B_{I_k}^\top r$; the group $G_k \defeq B_{I_k}^\top \ZZ^{I_k} = C^\top \ZZ^d$ has index $|\det C|$ in $\ZZ^d$, by \cref{lem:lattice-facts}\ref{it:facts-split}; and the minors of $B_{I_k}$ are those of $Y$ times $\det C$, the largest being $|\det B_J| = h(\Theta)$. Hence $h_k\, [\ZZ^d : G_k] = h(\Theta)$, and $\Theta_k$ has a largest minor at $J$. As $B_J$ is invertible, $\Theta_d = \ZZ^{I_d}$, whose fibers are points; and $\Theta_{|I|} = \Theta$, as $\Theta$ is saturated. This proves \ref{it:prefix-chain}.

\ref{it:prefix-step} Let $s_k$ be the order of $B_{i_{k+1}}^\top$ modulo $G_k$, a subgroup of finite index. Then $G_{k+1} = G_k + \ZZ B_{i_{k+1}}^\top$, so $[\ZZ^d : G_k] = s_k\, [\ZZ^d : G_{k+1}]$ and $h_{k+1} = s_k h_k$. For box points $(r', x)$ on $I_{k+1}$, with $x$ the entry at $i_{k+1}$, $B_{I_{k+1}}^\top (r', x) = t$ requires $t - x B_{i_{k+1}}^\top \in G_k$, as $B_{I_k}^\top r' \in G_k$; this holds for $x$ in at most one class modulo $s_k$, which has at most $\lceil g/s_k \rceil$ elements in $[2, g+1]$, and at most one if $s_k \ge g$. Averaging over the uniform $x$ gives $m_{k+1} \le m_k \lceil g/s_k \rceil/g \le m_k$, which proves the last claim for $s_k = 1$. For $2 \le s \defeq s_k < g$, as $h_{k+1}^{b-1} = s^{b-1} h_k^{b-1}$, it follows from $\lceil g/s \rceil/g < 1/s + 1/g \le s^{-1+1/b}$: for $s \ge 2^b$, $1/s + 1/g < 2/s \le s^{-1+1/b}$; for $s < 2^b$, $1/s + 1/g \le 1/s + 2^{-b}$ by \ref{it:inv-range}, and $1 + s\, 2^{-b} \le s^{1/b}$, as the concave function $s^{1/b} - 1 - s\, 2^{-b}$ vanishes at $s = 2^b$ and is positive at $s = 2$, where $2^{1/b} - 1 > (\ln 2)/b > 2^{1-b}$, as $b \ge 7$.

\ref{it:prefix-three} Let $h \defeq h(\Theta)$. By \ref{it:prefix-chain} and \ref{it:prefix-step}, and as $\lceil g/s \rceil/g < 1/s + 1/g$, $m(\Theta) \le g^{-3} \prod_k (1/s_k + 1/g)$ over the $n'$ steps with $s_k \ge 2$, whose $s_k$ multiply to $h$. Expanded, the product is $1/h$ plus terms with a factor $1/g$ and other factors at most $1/2$, which sum to at most $(\frac12 + \frac1g)^{n'} - 2^{-n'} \le \frac{n'}{g} (\frac12 + \frac1g)^{n'-1} \le (1 + \frac2g)/g \le \xi_{\mathrm{tr}}/g$, by the mean value theorem and as $n' (\frac12 + \frac1g)^{n'-1}$ is largest at $n' = 2$, since $(1 + \frac1{n'})(\frac12 + \frac1g) < 1$ for $n' \ge 2$, by \ref{it:inv-range}. So $\mu_{\mathrm{tr}} \le m(\Theta) \le g^{-3} \xi_{\mathrm{tr}} (1/h + 1/g)$, whence $h \le \xi_{\mathrm{tr}}/(g^3 \mu_{\mathrm{tr}} - \xi_{\mathrm{tr}}/g) = H_3$ by \ref{it:inv-trim}, and $m(\Theta)\, h \le g^{-3} \xi_{\mathrm{tr}} (1 + H_3/g) \le \Gamma_3/g^3$ by \ref{it:inv-reg}.
\end{proof}

\parhead{Trimming}
For a lattice $\Theta \subseteq \ZZ^I$ of rank $d \in [3]$, with $|I| \le N$ and a basis matrix $B$, let $\mathcal{T}_\Theta$ be the trimming ellipsoid of the $y \in \RR^d$ with $(y - y_\Theta)^\top \Sigma_\Theta^{-1} (y - y_\Theta) \le R_{\mathrm{tr}}^2$, where $y_\Theta \defeq \frac{g+3}{2} B^\top \mathbf{1}$, for the all-ones vector $\mathbf{1}$, is the mean of $B^\top r_I$ and $\Sigma_\Theta \defeq \frac{g^2}{12} B^\top B$. A set $F$ of points of a box on indices containing $I$ is \emph{trimmed by} $\Theta$ if $B^\top r_I \in \mathcal{T}_\Theta$ for all $r \in F$, and \emph{$F$ trimmed by $\Theta$} is the set of these $r \in F$. Neither depends on $B$: replacing $B$ by $BU$ replaces $B^\top r_I - y_\Theta$ by $U^\top (B^\top r_I - y_\Theta)$ and $\Sigma_\Theta$ by $U^\top \Sigma_\Theta U$.

\begin{lemma}[Trimming]\label{lem:trim}
Trimming a set by such a lattice $\Theta$ lowers its density by at most $\varepsilon_{\mathrm{tr}}$.
\end{lemma}

\begin{proof}
Let $B$ be a basis matrix of $\Theta$, $X \defeq B^\top r_I - y_\Theta$, and $\Sigma \defeq \Sigma_\Theta$. Each $r_i - \frac{g+3}{2}$ is uniform on $g$ points spaced by $1$ around $0$; adding an independent uniform variable on the real interval $[-\frac12, \frac12]$, whose moment generating function is at least $1$, makes it uniform on an interval of length $g$, so $\mathbb{E}\, e^{t (r_i - (g+3)/2)} \le \sinh(y)/y \le e^{y^2/6}$ for $y = tg/2$, as $(2k+1)! \ge 6^k k!$. By independence, $\mathbb{E}\, e^{u^\top X} \le e^{(g^2/24) \|B u\|^2} = e^{u^\top \Sigma u/2}$ for $u \in \RR^d$. For a standard Gaussian vector $Z$ of $\RR^d$, independent of $r$, Fubini's theorem then gives $\mathbb{E}\, e^{99 X^\top \Sigma^{-1} X/200} = \mathbb{E}\, e^{(99/100)^{1/2} Z^\top \Sigma^{-1/2} X} \le \mathbb{E}\, e^{99 \|Z\|^2/200} = 10^d$; so $X^\top \Sigma^{-1} X > R_{\mathrm{tr}}^2$ with probability at most $10^d e^{-99 R_{\mathrm{tr}}^2/200} \le \varepsilon_{\mathrm{tr}}$, by Markov's inequality and \ref{it:inv-trim}, and trimming removes only such points.
\end{proof}

\begin{lemma}[Trimmed sets are sparse]\label{lem:sparse}
Let $\Theta \subseteq \ZZ^I$, with $|I| \le N$, be a lattice of rank $d \in [3]$ with a basis matrix $B$, $F$ a nonempty set of box points on $I \cup \{i'\}$, $i' \notin I$, trimmed by $\Theta$, and $\Phi(r) \defeq (B^\top r_I, r_{i'})$.
\begin{enumerate}[label=(\alph*),leftmargin=*,nosep]
  \item\label{it:sparse-wrap} If $\dim \Phi(F) = d + 1$ and $\Phi(F)$ is in a coset of a subgroup of index $q$ of $\ZZ^{d+1}$, then $\Pr[F] \le \Lambda_d\, g^d\, h(\Theta)\, m(\Theta)/q$.
  \item\label{it:sparse-transition} If $F^\perp$ has a rank $d'$ with $1 \le d' < d$ and a largest minor at a set $J \subseteq I$, and the restriction to $I$ of each vector of $F^\perp$ lies in $\operatorname{span}_{\RR}(\Theta)$, then $\Pr[F]\, h(F^\perp) \le \Lambda_{d-d'}\, g^{d-d'}\, h(\Theta)\, m(\Theta)$.
\end{enumerate}
\end{lemma}

\begin{proof}
The box points $r$ with $\Phi(r) = (y, x)$ have density $\Pr[B^\top r_I = y]/g$, so $\Pr[F] \le |\Phi(F)|\, m(\Theta)/g$; we count $\Phi(F)$, which lies in $\mathcal{C} \defeq \mathcal{T}_\Theta \times \{ x \in \RR : 2 \le x \le g+1 \}$, after a projection to $k$ coordinates. Write $B = (B_1, B_2)$, with $B_2$ the last $k - 1$ columns of $B$, and $y = (y_1, y_2)$ accordingly.

In \ref{it:sparse-wrap}, $k \defeq d + 1$, so $B_1$ is empty; let $J \defeq \emptyset$, $\det (B_1)_J \defeq 1$, $\delta \defeq q$, and $p$ the identity. Part \ref{it:sparse-transition} does not depend on $B$, which we choose as follows, with $k \defeq d + 1 - d'$. As $Y_J$, for a basis matrix $Y$ of $F^\perp$, is invertible and its columns lie in $B_J \RR^d$ by hypothesis, $B_J$ has rank $d'$. Extending a basis of the saturated lattice of the $z \in \ZZ^d$ with $B_J z = 0$ to one of $\ZZ^d$ in which it comes last~\cite[Chapter~I, Theorem~I, Corollary~3]{Cas97}, and changing $B$ accordingly, we may take $(B_2)_J = 0$, so that $(B_1)_J$ is invertible; $F$ stays trimmed.

\emph{Image normals.} The injective map $\Phi^*(a) \defeq (B a_{[d]}, a_{d+1})$ has $\Phi^*(a)^\top r = a^\top \Phi(r)$, and each vector of $F^\perp$ is a $\Phi^*(a)$, as its restriction to $I$ lies in $B\, \RR^d$; so $\Phi^*$ maps the real vectors orthogonal to the differences of $\Phi(F)$ onto those of $F$, the span of $F^\perp$, and $\Phi(F)^\perp$ has rank $d'$ and $\dim \Phi(F) = k$. A basis matrix $V$ of $\Phi(F)^\perp$, with row blocks $V_1$, $V_2$, and $V_3$ of $d'$, $k-1$, and $1$ rows, maps under $\Phi^*$ to $Y C$ for an integer matrix $C$ with $\det C \ne 0$; at the rows $J$, this reads $(B_1)_J V_1 = Y_J C$, so $|\det (B_1)_J| \cdot |\det V_1| \ge h(F^\perp)$.

\emph{Projected lattice.} Let $\delta \defeq |\det V_1|$ and $p(y, x) \defeq (y_2, x)$, which is injective on $\ker V^\top$, as $V_1$ is invertible; $\ker V^\top$ contains the differences of $\Phi(F)$. As $V^\top \ZZ^{d+1} = \ZZ^{d'}$ by \cref{lem:lattice-facts}\ref{it:facts-split}, $(y_2, x) \mapsto V_2^\top y_2 + x V_3^\top$ maps $\ZZ^k$ onto $\ZZ^{d'}/V_1^\top \ZZ^{d'}$, of order $\delta$, with kernel $p(\ker V^\top \cap \ZZ^{d+1})$.

\emph{Counting and volume.} In both parts, then, $p(\Phi(F))$ has $|\Phi(F)|$ points and affine dimension $k$, and lies in a coset of a subgroup of index $\delta$ of $\ZZ^k$ and in $p(\mathcal{C})$. So $|\Phi(F)| \le k!\, \operatorname{vol}(p(\mathcal{C}))/\delta + k \le (k+1)!\, \operatorname{vol}(p(\mathcal{C}))/\delta$, by Blichfeldt's inequality~\cite[(1.1)]{HW08}, for the image of the convex hull of $p(\Phi(F))$ under an affine map that takes the coset to $\ZZ^k$ and divides volumes by $\delta$, and as that hull contains a simplex with vertices in the coset, of volume at least $\delta/k!$. Here $p(\mathcal{C})$ is the product of an interval of length $g - 1$ and the projection of $\mathcal{T}_\Theta = y_\Theta + R_{\mathrm{tr}} \Sigma^{1/2} \mathbb{B}^d$ to the coordinates of $y_2$, the ellipsoid $(y_\Theta)_2 + R_{\mathrm{tr}} \Sigma_{22}^{1/2} \mathbb{B}^{k-1}$ of volume $\operatorname{vol}(\mathbb{B}^{k-1}) R_{\mathrm{tr}}^{k-1} (\det \Sigma_{22})^{1/2}$, where $\Sigma \defeq \Sigma_\Theta$, its block at $y_2$ is $\Sigma_{22} = \frac{g^2}{12} B_2^\top B_2$, and $\mathbb{B}^e$ is the unit ball of $\RR^e$. By the Cauchy--Binet formula, $\det (B_2^\top B_2)$ is the sum of the $\det((B_2)_{J''})^2$ over the sets $J'' \subseteq I \setminus J$ of $k - 1$ indices, as $(B_2)_J = 0$; and $|\det (B_1)_J \det (B_2)_{J''}| = |\det B_{J \cup J''}| \le h(\Theta)$. Altogether, $\Pr[F]\, |\det (B_1)_J|\, \delta$ is at most $g^{k-1} h(\Theta)\, m(\Theta)$ times $(k+1)!\, \operatorname{vol}(\mathbb{B}^{k-1})\, R_{\mathrm{tr}}^{k-1} \bigl( \binom{N}{k-1} / 12^{k-1} \bigr)^{1/2}$, and as $\binom{N}{k-1} \le N^{k-1}/(k-1)!$, this factor is at most $2\sqrt{3}\, R_{\mathrm{tr}} \sqrt{N}$, $\sqrt{2}\, \pi R_{\mathrm{tr}}^2 N$, and $\frac{20\pi}{9\sqrt{2}} R_{\mathrm{tr}}^3 N^{3/2}$ for $k - 1 = 1, 2, 3$, hence at most $\Lambda_{k-1}$ by \ref{it:inv-geom}. This is \ref{it:sparse-wrap}, and it gives \ref{it:sparse-transition}, as $|\det (B_1)_J|\, \delta \ge h(F^\perp)$.
\end{proof}

\begin{lemma}[Regularization]\label{lem:regularize}
Let $\Theta \subseteq \ZZ^I$, $|I| \le N$, be a saturated lattice of a rank $d \in [3]$ with $m(\Theta) \ge \mu_{\mathrm{tr}} + (3 - d)\, \varepsilon_{\mathrm{tr}}$. Then $\Theta$ lies in a regular saturated $\Theta' \subseteq \ZZ^I$ of rank in $[d, 3]$ with $m(\Theta') \ge m(\Theta) - (3 - d)\, \varepsilon_{\mathrm{tr}}$.
\end{lemma}

Along the prefix lattices of \cref{lem:prefix}, $m^b h^{b-1}$ does not increase at steps of order below $g$, and a step of order at least $g$ would put a unit vector into the normal lattice of a dense set. So $m(\Theta)\, h(\Theta)$ stays small unless the dense set of some prefix satisfies more integer equations: at the last prefix where it does, \cref{lem:sparse}\ref{it:sparse-transition} bounds the product, and if it does at the end, they give a lattice of higher rank, regularized by induction.

\begin{proof}
We use downward induction on $d$. For $d = 3$, the lattice $\Theta' \defeq \Theta$ is regular, by \cref{lem:prefix}\ref{it:prefix-three}. Let $d \le 2$, and take $B$, $J$, and the $I_k$, $\Theta_k$, $h_k$, $m_k$, and $s_k$ of \cref{lem:prefix}; by its part \ref{it:prefix-step}, $m_k \ge m(\Theta) \ge \mu_{\mathrm{tr}} + (3-d)\, \varepsilon_{\mathrm{tr}}$ for all $k$.

For $k = d, \dots, |I|$, let $F_k$ be a densest fiber of $\Theta_k$, trimmed by $\tilde\Theta_{k-1}$ if $k > d$ and $F_{k-1}^\perp \ne \Theta_{k-1}$, so that $\Pr[F_k] \ge m_k - \varepsilon_{\mathrm{tr}}$ (\cref{lem:trim}). Here, if $F_k^\perp \ne \Theta_k$, then $F_k^\perp \supseteq \Theta_k$ has a rank $d_k \in [d+1, 3]$, as rank four or more would give $\Pr[F_k] \le g^{-4} < \mu_{\mathrm{tr}}$, by \ref{it:inv-wrap}; and $m(F_k^\perp) \ge \Pr[F_k] \ge \mu_{\mathrm{tr}} + (2-d)\, \varepsilon_{\mathrm{tr}} \ge \mu_{\mathrm{tr}} + (3-d_k)\, \varepsilon_{\mathrm{tr}}$, so the induction hypothesis gives a regular saturated $\tilde\Theta_k \supseteq F_k^\perp$ with $m(\tilde\Theta_k) \ge \Pr[F_k] - (2-d)\, \varepsilon_{\mathrm{tr}} \ge m_k - (3-d)\, \varepsilon_{\mathrm{tr}}$.

\emph{Promotion.} If $F_{|I|}^\perp \ne \Theta$, take $\Theta' \defeq \tilde\Theta_{|I|} \supseteq \Theta$.

\emph{The last prefix with more equations.} Otherwise, let $j$ be the last $k < |I|$ with $F_k^\perp \ne \Theta_k$, or $j \defeq d - 1$ if there is none. For $j < k < |I|$, $s_k < g$: else the entry at $i_{k+1}$ is constant on $F_{k+1}$ (\cref{lem:prefix}\ref{it:prefix-step}), so $e_{i_{k+1}} \in F_{k+1}^\perp = \Theta_{k+1}$, a nonzero vector that vanishes at $J$, against \cref{lem:lattice-facts}\ref{it:facts-minor} and \cref{lem:prefix}\ref{it:prefix-chain}. As $m(\Theta) \ge \mu_{\mathrm{tr}}$, \cref{lem:prefix}\ref{it:prefix-step} then gives $\mu_{\mathrm{tr}} (m(\Theta)\, h(\Theta))^{b-1} \le m(\Theta)^b h(\Theta)^{b-1} \le m_{j+1}^b h_{j+1}^{b-1}$, so $\Theta' \defeq \Theta$ is regular once $m_{j+1}^b h_{j+1}^{b-1} \le \mu_{\mathrm{tr}} (\Gamma_d/g^d)^{b-1}$, as $b \ge 2$.

\emph{The bound at $j + 1$.} If $j = d - 1$, this condition reads $g^{-d} \le \mu_{\mathrm{tr}} \Gamma_d^{b-1}$ (\cref{lem:prefix}\ref{it:prefix-chain}), which is \ref{it:inv-reg}. If $j \ge d$, let $\tilde d$ be the rank of $\tilde\Theta_j$. \Cref{lem:sparse}\ref{it:sparse-transition} applies to $\tilde\Theta_j$, $i_{j+1}$, and $F_{j+1}$: $F_{j+1}^\perp = \Theta_{j+1}$ has rank $d < \tilde d$ and a largest minor at $J \subseteq I_j$, and its vectors restricted to $I_j$ lie in $\operatorname{span}_{\RR}(\Theta_j) \subseteq \operatorname{span}_{\RR}(F_j^\perp) \subseteq \operatorname{span}_{\RR}(\tilde\Theta_j)$. With the regularity of $\tilde\Theta_j$ and as $\Pr[F_{j+1}] \ge m_{j+1} - \varepsilon_{\mathrm{tr}} \ge m_{j+1}/A_{\mathrm{inv}}$ by $m_{j+1} \ge \mu_{\mathrm{tr}}$ and \ref{it:inv-reg}, it gives $m_{j+1} h_{j+1} \le A_{\mathrm{inv}} \Pr[F_{j+1}]\, h_{j+1} \le A_{\mathrm{inv}} \Lambda_{\tilde d - d}\, g^{\tilde d - d}\, h(\tilde\Theta_j)\, m(\tilde\Theta_j) \le A_{\mathrm{inv}} \Lambda_{\tilde d - d} \Gamma_{\tilde d}/g^d$. And $m_{j+1} \le m_j \le \Pr[F_j] + \varepsilon_{\mathrm{tr}} \le m(F_j^\perp) + \varepsilon_{\mathrm{tr}} \le g^{-d-1} + \varepsilon_{\mathrm{tr}} \le \mu_{\mathrm{tr}} P_d^{b-1}$, as $F_j^\perp$ has rank above $d$, by \ref{it:inv-reg}. So, by the last two conditions of \ref{it:inv-reg}, as $(d, \tilde d)$ is $(2, 3)$, $(1, 2)$, or $(1, 3)$, $m_{j+1}^b h_{j+1}^{b-1} \le \mu_{\mathrm{tr}} (A_{\mathrm{inv}} \Lambda_{\tilde d - d} \Gamma_{\tilde d} P_d/g^d)^{b-1} \le \mu_{\mathrm{tr}} (\Gamma_d/g^d)^{b-1}$.
\end{proof}

\begin{proof}[of \cref{lem:inverse}]
For $i = 1, \dots, N_0$ in turn, we find a regular saturated lattice $\Theta \subseteq \ZZ^{[i]}$ of a rank $d \in [3]$ with $v_{[i]} \in \bar\Theta$ and $m(\Theta) \ge \gp(v_{[i]}^\top) - 3\varepsilon_{\mathrm{tr}}$; so $m(\Theta) > \mu_{\mathrm{tr}}$ by \cref{eq:inv-chain}, and $h(\Theta) \le H_d$, for $d \le 2$ by regularity, as $H_d = \Gamma_d/(g^d \mu_{\mathrm{tr}})$, and for $d = 3$ by \cref{lem:prefix}\ref{it:prefix-three}. For $i = 1$, $\ZZ^{[1]}$ serves: its reduction $\FF_q$ contains $v_{[1]}$, its fibers are points, so $m = 1/g \ge \gp(v_{[1]}^\top)$, and $h = 1 \le \Gamma_1$, by \ref{it:inv-reg} and $\mu_{\mathrm{tr}} < 1/g$.

\emph{Adding a coordinate.} Let $i < N_0$, and let $\Theta$ have a largest minor at $J$. Restriction to $J$ maps $\Theta$ injectively (\cref{lem:lattice-facts}\ref{it:facts-minor}) onto a lattice of determinant $h(\Theta)$ in $\RR^J$. As the unit ball of $\|\cdot\|_1$ has volume $2^d/d!$, the successive minima $\lambda_1 \le \dots \le \lambda_d$ of that lattice for $\|\cdot\|_1$ have $\lambda_1 \cdots \lambda_d \le d!\, h(\Theta)$, by Minkowski's second theorem~\cite[Chapter~VIII, Theorem~V]{Cas97}; and by Mahler's lemma~\cite[Chapter~V, Lemma~8]{Cas97}, for linearly independent vectors that attain them, the lattice has a basis of $\ell_1$ norms at most $\lambda_1$, $\lambda_2$, and $\max\{\lambda_3, (\lambda_1 + \lambda_2 + \lambda_3)/2\} \le \frac32 \lambda_3$. So, by the same part, $\Theta$ has a basis matrix $B$ with $\prod_j \|B e_j\|_\infty \le h(\Theta)$, $2h(\Theta)$, or $9h(\Theta)$ for $d = 1, 2, 3$.

As $v_{[i]} \in \bar\Theta$, $v_{[i]} = B c' \bmod q$ for some $c' \in \FF_q^d$; with $c \defeq (c', v_{i+1})$ and $\Phi(r) \defeq (B^\top r_{[i]}, r_{i+1})$, $v_{[i+1]}^\top r = c^\top \Phi(r) \bmod q$. Let $F$ be a densest modular fiber $\{ r : v_{[i+1]}^\top r = t \bmod q \}$ of $v_{[i+1]}$, trimmed by $\Theta$, so that $\Pr[F] \ge \gp(v_{[i+1]}^\top) - \varepsilon_{\mathrm{tr}} > \theta - \varepsilon_{\mathrm{tr}}$ by \cref{lem:trim} and \cref{eq:inv-chain}. As $v_{i+1} \ne 0$, $\Phi(F)$ lies in a coset of the subgroup $\{ y \in \ZZ^{d+1} : c^\top y = 0 \bmod q \}$ of index $q$, so $k \defeq \dim \Phi(F) \le d$: otherwise \cref{lem:sparse}\ref{it:sparse-wrap}, regularity, and \ref{it:inv-wrap} would give $\Pr[F] \le \Lambda_d \Gamma_d/q < \mu_{\mathrm{tr}}$.

Let $\Delta$ be the matrix of the differences of the points of $\Phi(F)$, of rank $k$. By the Cauchy--Schwarz inequality $(e_j^\top y)^2 \le (e_j^\top \Sigma_\Theta e_j)(y^\top \Sigma_\Theta^{-1} y)$, with $e_j^\top \Sigma_\Theta e_j \le \frac{g^2 N}{12} \|B e_j\|_\infty^2$, entries $j$ of two points of $\mathcal{T}_\Theta$ differ by at most $2 R_{\mathrm{tr}}\, g \sqrt{N/12}\, \|B e_j\|_\infty = T_{\mathrm{tr}} \|B e_j\|_\infty$, which bounds the entries of the row $j \in [d]$ of $\Delta$; those of its last row are at most $g - 1 < T_{\mathrm{tr}}$, as $R_{\mathrm{tr}} \ge 20$. As $T_{\mathrm{tr}}$ and the $\|B e_j\|_\infty$ are at least $1$, the $k \times k$ minors of $\Delta$, at most $k!$ times a product of $k$ of these bounds, are at most $d!\, T_{\mathrm{tr}}^d \prod_j \|B e_j\|_\infty$, which is at most $H_1 T_{\mathrm{tr}}$, $4 H_2 T_{\mathrm{tr}}^2$, or $54 H_3 T_{\mathrm{tr}}^3$, below $q$ by \ref{it:inv-wrap}. As $c^\top \Delta = 0 \bmod q$, \cref{lem:lattice-facts}\ref{it:facts-small}, for an integer vector that reduces to $c$, gives $c = u \bmod q$ for a $u \in \ZZ^{d+1}$ with $u^\top \Delta = 0$. Then $(B u_{[d]}, u_{d+1})^\top r = u^\top \Phi(r)$ is constant on $F$, so $(B u_{[d]}, u_{d+1}) \in F^\perp$, and it reduces to $(B c', v_{i+1}) = v_{[i+1]}$; so $v_{[i+1]} \in \overline{F^\perp}$ and $F^\perp \ne \{0\}$. Then $m(F^\perp) \ge \Pr[F] > \theta - \varepsilon_{\mathrm{tr}} \ge \mu_{\mathrm{tr}} + 2\varepsilon_{\mathrm{tr}}$, so $F^\perp$ has rank in $[3]$, by \ref{it:inv-wrap}, and \cref{lem:regularize} gives a regular saturated $\Theta^+ \supseteq F^\perp$ with $m(\Theta^+) \ge m(F^\perp) - 2\varepsilon_{\mathrm{tr}} \ge \gp(v_{[i+1]}^\top) - 3\varepsilon_{\mathrm{tr}}$ and $v_{[i+1]} \in \overline{\Theta^+}$: the lattice for $i + 1$.

\emph{The \shortmultiple{}.} Let $\Theta$ be the lattice for $N_0$, $B$ a basis matrix of it with a largest minor at $J$, and $z_0 \in \ZZ^d$ with $B z_0 = v_{[N_0]} \bmod q$, so $z_0 \ne 0 \bmod q$. The $z \in \ZZ^d$ congruent modulo $q$ to multiples of $z_0$ form a subgroup of index $q^{d-1}$, so the $B_J z$ form a lattice of determinant $h(\Theta)\, q^{d-1}$ in $\RR^J$, and $d!\, h(\Theta)\, q^{d-1} \le \beta^d$ by \ref{it:inv-bound}: $h(\Theta) \le H_1 \le \beta$, $2 h(\Theta)\, q \le 2 H_2\, q \le \beta^2$, or $6 h(\Theta)\, q^2 \le 6 \lfloor H_3 \rfloor q^2 \le \beta^3$, as heights are integers. The closed $\ell_1$ ball of radius $\beta$ in $\RR^J$ is convex and symmetric, and its volume $(2\beta)^d/d!$ is at least $2^d h(\Theta)\, q^{d-1}$. Thus Minkowski's theorem~\cite[Section~30]{Min10}, applied after a linear map taking the lattice to $\ZZ^d$, gives a nonzero lattice point $B_J z$ in this ball. Then $x \defeq B z$ is congruent modulo $q$ to a multiple of $v_{[N_0]}$, a nonzero one as $0 < \|x\|_\infty \le \|x_J\|_1 \le \beta < q$ (\cref{lem:lattice-facts}\ref{it:facts-minor}, \ref{it:inv-bound}): $x$ is a \shortmultiple{} of $v_{[N_0]}$.
\end{proof}  
\section{Comparison with Hair and Sahai}\label{app:hair-sahai}

We compare with Hair and Sahai's constructions set up beyond what their theorems allow, and extend their proof to show that these settings still meet the security level. Each choice is in their favor, and the paragraph that makes it gives its reason:
\begin{itemize}[leftmargin=*, topsep=2pt, itemsep=1pt]
\item the least parameters that their bound certifies for $\lambda$-bit security, where their theorems fix larger ones: the threshold $R$ of their Theorem~5.1, and the degree and weight degree of their Theorem~6.1, which their proof leaves free;
\item a group whose order has $\Theta(\lambda)$ bits, where their theorems take $p \ge 2^m$, as their proof needs only three conditions on the order;
\item our decryption search in place of their list of multiples;
\item our fixed-base scalar multiplication, $11$ affine additions at our parameters, where their paper takes $O(\log p)$ group operations~\cite[Section~3.1]{HS26};
\item counts that are lower bounds: the published elements, the scalar multiplications that make them, and the group operations of decryption, but not $Z$ or their arithmetic in $\FF_p$, and with the dimension $k$ of $S$ bounded from below.
\end{itemize}
 Their section, equation, and statement numbers are those of the third version of their paper~\cite{HS26}, and logarithms have base two.

\parhead{Their scheme}
Their constructions take $N'$ quadratic equations in $N$ Boolean unknowns and give a space $S$ of $m \times m$ matrices over $\FF_p$ with a basis of $k$ matrices. A witness $w = (1, w_1, \dots, w_N)$ gives a matrix $u v^\top \in S$ with $v$ Boolean, and on a false statement every nonzero matrix of $S$ has rank at least $d$: a gap of one versus $d$. The left factor $u$ stacks $K$ copies $\xi(w)\, v$ of $v$, one for each \emph{weight} $\xi$, a polynomial in $w$, so $m = cK$ for the length $c$ of $v$. A bit's ciphertext has $m + k + 1$ group elements, made by $m + k$ scalar multiplications~\cite[Section~3.1]{HS26}. Decryption combines $m$ of them along $u$ into $H$ and $k$ along the coordinates of $u v^\top$ into $Z$, and decides whether $Z = (v^\top r)\, H$, where $v^\top r$ is the sum of the $n$ entries of a hidden vector $r$ at the ones of $v$, each uniform on $\{1, \dots, m\}$. Hair and Sahai list $H, 2H, \dots, m^2 H$; in their favor, we give them our search (\cref{sec:rank-template}), by baby steps and giant steps outward from the mean of $v^\top r$. It takes $O(\sqrt{m}\, n^{1/4})$ group operations in expectation on an encryption of zero, and $O(\sqrt{m}\, (n \lambda)^{1/4})$ to decide an encryption of one, stopped at the distance $O(m \sqrt{n \lambda})$ from the mean beyond which $v^\top r$ lies with probability below $2^{-\lambda}$. A string is encrypted one bit at a time, each bit under a fresh group~\cite[Definition~2.1 and Section~7]{HS26}, so a $\lambda$-bit key takes $\lambda$ ciphertexts. Our concrete counts of their costs are lower bounds: they count the published elements, the scalar multiplications that make them, and the group operations of $H$ and the search, but not $Z$ or their arithmetic in $\FF_p$, which is polynomial in $m$, $N$, $N'$, and $\log p$~\cite[Sections~5.4 and~6.5]{HS26}.

\parhead{The level and the group}
For one bit and $Q$ group operations on a false statement, their bound on the advantage is $\binom{m+k+Q+2}{2} (m^{-d} + 2/p)$, plus at most $4Qp/(2^\ell - m - k - 3Q - 2)$ for strings guessed as labels of $\ell \ge 2 \log p$ bits~\cite[proof of Theorem~3.5]{HS26}. A hybrid over its $\lambda$ ciphertexts bounds a key's advantage by $\lambda$ times this, and the key has $\lambda$-bit security if that is at most $(Q+1)/2^\lambda$ for every $Q < 2^\lambda$; as the bound is convex in $Q$ and the level linear, $Q = 0$ and $Q = 2^\lambda - 1$ suffice. Whatever $k$ is, their bound meets the level only if
\begin{equation}\label{eq:hs-necessary}
  m^d > \lambda\, 2^{2\lambda - 1} ,
\end{equation}
and if $k \le m^2$ and $p \ge \lambda\, 2^{2\lambda + 6} m^4$, it holds when $m^{d-4} \ge \lambda\, 2^{2\lambda + 3}$. One table~\cite[Section~4]{HS26}, a space of dimension between $1 + N + \binom{N}{2} - N'$ and $1 + N + \binom{N}{2}$, on the order of $N^2$ for systems of linear size, has $d = 2$ and meets no level, as $\binom{m+2}{2} m^{-2} > 1/2$. Their theorems take $p \in [2^m, 2^{m+1})$, but their proof needs only $p > m$~\cite[Lemma~3.4]{HS26}, $p$ above every sample, and $2$ of multiplicative order above $N$ modulo $p$: their Lemma~5.2 uses $p > 2^N$ only to make $2^0, \dots, 2^N$ distinct, and their Lemma~6.2 holds over every field. A prime that fails the last condition divides $\prod_{j=1}^{N} (2^j - 1) < 2^{N^2}$, so fewer than $N^2/(b-1)$ primes of $b$ bits fail it. For $N \le \lambda^{O(1)}$, some prime of $\Theta(\lambda)$ bits thus meets all three conditions and the bound on $p$ above; in their favor we take one, which extends their theorems to a far smaller group.

\parhead{Their basic construction}
Their Theorem~5.1, the construction of \cref{tab:sizes}, is that of their first arXiv version, with their third version's range for $r$; Campanelli proved that construction extractable for one ciphertext, in Maurer's GGM~\cite{Cam26}. At the parameters below, and for a key of $\lambda$ ciphertexts, neither of which his proof covers, we bound only its hiding, by their bound. It has $v = w$, so $c = N + 1$ and $n = 1 + \wt(w)$, where $\wt(w)$ counts the ones among $w_1, \dots, w_N$~\cite[Section~5]{HS26}. For a threshold $R$, each sample $t \in \{0, \dots, 2NR\}$ has $R$ forms $\ell_{j,t}(w) = \sum_{i=0}^{N} (2^{j-1} t)^i w_i$ and, as weights, the $\binom{2R}{R}$ monomials of degree at most $R$ in them, so $m = (N + 1)(2NR + 1) \binom{2R}{R}$ and $d = R + 1$~\cite[(13) and Proposition~5.3]{HS26}; their proof holds for every $R$, and their Theorem~5.1 takes $R = \lfloor \log N \rfloor$. As a polynomial in $t$, each $\ell_{j,t}(w)$ is nonzero, with constant term one, and of degree at most $N$, so at least $NR + 1$ samples are roots of none of their forms. The weights of these samples are nonzero, so $H$, a sum of independent published elements, has at least $(NR + 1) \binom{2R}{R}\, n$ nonzero terms, and computing it takes at least that many group operations, less one.

Let $\lambda \le N \le \lambda^{O(1)}$ and $N' \le \lambda^{O(1)}$. As $4^R/(2\sqrt{R}) \le \binom{2R}{R} \le 4^R$, $\log m = 2R + 2 \log N + O(\log \lambda)$, and, with $k \le m(N+1)$~\cite[Section~5.4]{HS26}, \cref{eq:hs-necessary} and the sufficient condition put their least $R$ at $\sqrt{\lambda + (\log N)^2/4} - (\log N)/2 + O(\log \lambda) = \sqrt{\lambda} + O(\log \lambda)$, where $\log m = 2\sqrt{\lambda} + \log N + O(\log \lambda)$. A bit's ciphertext then has between $m$ and $m(N+3)$ elements, encryption takes at least $m$ scalar multiplications, and decryption at least $m/(2N+2) - 1$ group operations. With elements of $\Theta(\lambda)$ bits and group operations of $O(1)$ field operations, the ciphertext's bits and the field operations of encryption and decryption for a $\lambda$-bit key are thus each $2^{2\sqrt{\lambda} + O(\log N)}$, that is, $2^{(2+o(1))\sqrt{\lambda}}$, and $O(N^3\, 2^{\Theta(\sqrt{\lambda})})$ also with their arithmetic in $\FF_p$. The factor $N^3$ in \cref{tab:sizes} is their growth at a fixed $R$: then $m = \Theta(N^2)$ and, for $N' = O(N)$, $k = \Theta(N^3)$ by \cref{prop:hs-dimension} and $k \le m(N+1)$, so a bit takes $\Theta(N^3)$ published elements and fixed-base scalar multiplications, and decryption $O(N^3)$ group operations, their arithmetic in $\FF_p$ aside.

\begin{proposition}\label{prop:hs-dimension}
For $p$ as above and $2 \le R \le (N-2)/2$, $k \ge R \binom{N+1-2R}{3} - N' K$.
\end{proposition}
\begin{proof}[Sketch]
The tests cut at most $N' K$ dimensions from the stacks' span. For $t \in \{0, \dots, 2NR\}$, the weights include $\ell_{1,t}\, \ell_{1+\delta,t}$, ${0 \le \delta < R}$. Interpolation in $t$ and \mbox{Vandermonde-type} inversion over $\delta$ give, for $1 \le i \le R < a < b < s - R \le N + 1 - R$, a combination of stack entries whose only monomial on $\{0,1\}^N$ with an index at most $R$ is $w_i w_a w_b w_{s-i}$; these are independent.
\end{proof}

\parhead{The construction of their Theorem~6.1}
This construction, behind their main theorem~\cite[Sections~6 and~7]{HS26}, has a degree $e \ge 1$ and a weight degree $h \ge 1$, their $d$ and $r$. Its $v$ lists the $c = B_e(N) \defeq \sum_{j \le e} \binom{N}{j}$ products of at most $e$ unknowns, so $n = B_e(\wt(w))$. Its gap is one versus $d = \binom{h + 2e}{e}$, and each sample $t \in \{0, \dots, N(d^2 + h)\}$ has $d^2$ forms $\ell_{j,t}$ and, as weights, the $\binom{d^2 + h}{h}$ monomials of degree at most $h$ in them, so
\[
  m = B_e(N) \bigl( N(d^2 + h) + 1 \bigr) \binom{d^2 + h}{h}
\]
rows~\cite[(26), (27), and (33)]{HS26}. Their proof of the gap~\cite[Section~6.4]{HS26} holds for every $h \ge 1$: $h$ enters it only as the degree of the weights, which sets the number of samples, and through their Lemma~6.2, which gives $d$. Their (26) sets $h = \max\{1, \lfloor \sqrt{\log(N+2)} \rfloor\}$ so that the gap grows with $N$ while $m$ stays polynomial in $N$ at a fixed degree~\cite[Section~6.5]{HS26}; in their favor, we let $h$ vary.

In the same range of $N$ and $N'$, take $h = 1$ and $e = \lceil \frac12 \log \lambda \rceil + 1$. Then $d \ge \binom{2e}{e} \ge 4^e/(2\sqrt{e}) \ge 2\lambda/\sqrt{e}$ and $\log m \ge \log \lambda$, so, as $k \le mc \le m^2$, the sufficient condition holds for large $\lambda$; and $\log m \le (e+1) \log(N+1) + 4 \log d + 2$, with $\log d \le 2e + 1$, gives $m \le N^{\frac12 \log \lambda + O(1)}$. Conversely, let $(e, h)$ meet \cref{eq:hs-necessary} with $m \le N^{\log \lambda}$. Then $\log d \ge \log \lambda - O(\log \log \lambda)$ and $d \le 2^{h + 2e}$, while $\log m \ge \log B_e(N) \ge e \log(N/e)$ and $\log m \ge \log \binom{d^2 + h}{h} \ge h \log d$. So $e \ge \frac14 \log \lambda$, or $h \ge \frac13 \log \lambda$ for large $\lambda$, and either way $m \ge N^{\Omega(\log \lambda)}$. Their least $m$ is thus $N^{\Theta(\log \lambda)}$, at a degree $\Theta(\log \lambda)$: if $m \le N^{O(\log \lambda)}$, the same two bounds on $\log m$ give $e, h = O(\log \lambda)$, and $d \le (3(2 + h/e))^e$ would be $\lambda^{o(1)}$ for $e = o(\log \lambda)$. No fixed degree is best: at a fixed $e$, $d \le (h + 2e)^e$ and $h \le \log m$, so \cref{eq:hs-necessary} gives $(\log m + 2e)^e \log m > 2\lambda - 1$, and $m = 2^{\Omega(\lambda^{1/(e+1)})}$, above $N^{O(\log \lambda)}$ for large $\lambda$. A bit's ciphertext has between $m$ and $2m^2$ elements, encryption takes at least $m$ scalar multiplications, and our search $\Omega(\sqrt{m})$ group operations in expectation, so at their least $m$ all three costs are $N^{\Theta(\log \lambda)}$, with or without their arithmetic in $\FF_p$.

\parhead{At $\lambda = 100$}
Our statement has $N = \num{391693}$ and $N' \le 2N + 1$. We take $p = q$, BN254's group order and the field of our statement: $2$ has order $(q-1)/78 > N$ modulo $q$, $q$ exceeds $m$ and every sample below, and elements are $32$ bytes, as in BN254's $\GG_1$, where their theorems' $p \ge 2^m$ would make an element over $m$ bits, over $9.7$\,TiB at the $R = 4$ below. With $p = q$, their list errs on a bit with probability about $m^2/q$, below $2^{-161}$ at that $R$, in place of the smaller error that their theorems' larger prime gives; our search, stopped as above, errs with probability below $2^{-\lambda}$. By \cref{eq:hs-necessary}, the level needs $m^d > 100 \cdot 2^{199} > 2^{205.6}$, so the basic construction fails for $R \le 3$, where $m^{R+1} < 2^{176.3}$. At $R = 4$, a gap of one versus five, $K = \num{219348150}$ and $m = \num{85917354266100} \approx 2^{46.3}$. To certify the level at $q$, we bound $k$ by interpolation in $t$. As functions of $w$, the stacks' entries are weights times products of at most $2e$ unknowns ($e = 1$ for the basic construction); for $f$ forms at each sample, the weights of one monomial of degree $a$ in them are, over all samples, polynomials of degree at most $aN$ in $t$ with functions of $w$ as coefficients, and so span at most $aN + 1$ functions. For weights of degree at most $h'$, $k \le W B_{2e}(N)$ with $W = \sum_{a \le h'} \binom{a + f - 1}{f - 1} (aN + 1)$. At $R = 4$, $W = 224N + 70$, and their bound meets the level at $q$ at $Q = 0$ and $Q = 2^{100} - 1$, so throughout. By \cref{prop:hs-dimension}, $k > 2^{55.14}$ at $R = 4$, and $m + k > 2^{55.1}$ at every $R \ge 4$, with $m > 2^{56.8}$ from $R = 9$ on.

\looseness=-1 For a $100$-bit key the basic construction thus takes over $1.1 \times 10^{11}$\,GiB, against our $7.88$\,GiB. We convert their group operations at our rate in $\GG_1$ without the shared inversion (\cref{sec:costs}): an affine addition takes $6$ field multiplications, a squaring among them, and $6$ field additions~\cite{EFD}, and a scalar multiplication $11$ affine additions, as each element they publish is a known multiple of their generator $[1]$. Encryption then takes them over $2^{67.8}$ multiplications ($2.6 \times 10^{20}$ in \cref{tab:sizes}). For $H$ and our search alone, decryption takes over $2^{54.5}$ for a witness with $\lfloor N/2 \rfloor$ ones at which no form vanishes, so that all $K$ weights are nonzero, and over $2^{53.5}$ for every witness with $\lfloor N/2 \rfloor$ ones ($1.2 \times 10^{16}$ in \cref{tab:sizes}). Their lists of multiples, $H, 2H, \dots, m^2 H$, computed in full, would take over $2^{101.8}$ field multiplications for a $100$-bit key, over eighteen orders of magnitude more group operations than our search, even on encryptions of one. Each count comes with as many additions. Ours are about $2^{34.2}$ and, for a witness with $\lfloor N/2 \rfloor$ ones, $2^{33.0}$ in expectation: over ten, ten, and six orders of magnitude fewer.

For the construction of their Theorem~6.1, $(e, h) = (1, 1)$ and $(1, 2)$, gaps of one versus three and four, fail whatever $k$ is, as $m^d < 2^{194.4}$. At $(1, 3)$, a gap of one versus five, each of the $28N + 1$ samples has $25$ forms and $\num{3276}$ weights, $m = \num{14073259420813320} \approx 2^{53.6}$, $W = \num{3701502126}$, so $k < 2^{68}$, and their bound meets the level at $q$. Every other $(e, h)$ that meets it has more rows, as $m$ grows with $e$ and $h$; at degree two, $(2, 1)$, a gap of one versus ten, has $m \approx 2^{68.1}$ and, by the same bound on $k$, meets the level in a group of order above $2^{297}$; at $q$, that bound is too weak to certify it. The argument of \cref{prop:hs-dimension} applies to this construction at $e = 1$ and $h \ge 2$, with the $d^2$ forms of a sample in place of the $R$ forms: then $v = w$, its tests are $N'$ for each copy~\cite[(25)]{HS26}, its weights include the products $\ell_{1,t}\, \ell_{1+\delta,t}$ for $0 \le \delta < d^2$, and its samples include $0, \dots, 2N$. At $(1, 3)$, with $K = (28N + 1) \cdot \num{3276}$ copies, it gives $k \ge 25 \binom{N - 49}{3} - N' K > 2^{57.6}$, so a $100$-bit key takes over $7.0 \times 10^{11}$\,GiB and over $2^{70.4}$ ($1.5 \times 10^{21}$) multiplications to encrypt, over ten orders of magnitude more than ours. Counted as for the basic construction, decryption takes over $2^{61.8}$ for a witness with $\lfloor N/2 \rfloor$ ones at which no form vanishes, and over $2^{58.6}$ ($4.5 \times 10^{17}$) for every witness with $\lfloor N/2 \rfloor$ ones, as at least $3N + 1$ samples are roots of none of its forms. \Cref{tab:sizes} leaves this construction out because no extraction proof is known for it, and our lower bounds on its size, encryption, and decryption for every witness with $\lfloor N/2 \rfloor$ ones exceed the basic construction's.

Additions are $1.0$ to $1.8$ times the multiplications in each total of either scheme. Neither scheme's counts include point decompression, as both sizes are of compressed elements; counting it for us alone keeps decryption over five orders of magnitude apart. Hashing and memory traffic, left out of both counts, are small for us (\cref{sec:costs}) and for them would only add to costs we bound from below.

\parhead{One bit and a key}
\looseness=-1 Their scheme as stated carries one bit per ciphertext: the decryptor's two elements, $H$ and $Z$, depend on the witness, so neither can serve as a key. Our key in the corner (\cref{sec:rank-template}) would likely remove this limit. In each of their constructions, the copy of weight $1$ has the entry $w_0^2 = 1$ for every witness, so the encryptor could add $\kappa$ times that entry of each basis matrix; a witness's combination would then encode $\kappa$ plus the search term, which a published check of $[\kappa]$ or a second, key-free copy lets the decryptor remove. An adversary's element would encode $\kappa$ only along a nonzero matrix, of rank at least $d$ on a false statement, as in their Lemma~3.4, but neither their paper nor ours proves this variant. A key would then cost one or two of their one-bit ciphertexts in place of $100$, and their costs would still be over eight orders of magnitude above ours in size and encryption and over four in decryption.

\parhead{Jin's construction}
Jin's theorems~\cite{Jin26} take a group of prime order $p$ with $2^{\lambda} < p < 2^{\lambda + 1}$ and, for a verifier circuit of $v$ gates, a matrix of $M \ge 3v^2$ rows with $\log_2 p > \frac32 M \log_2 M$; a ciphertext holds at least $M$ elements of this group for each encrypted bit~\cite[Theorem~5.1, Lemma~5.3, and Appendix~A.6]{Jin26}. Taking $v = N$, far fewer gates than known Boolean circuits for the Groth16 verifier have, gives $M > 4.6 \times 10^{11}$, so $\lambda > 2^{44}$, and a $100$-bit key takes over $10^{17}$\,GiB.

\end{document}